\documentclass[preprint,12pt]{elsarticle}

\usepackage{amssymb}

\usepackage{lineno,hyperref}

\usepackage{graphicx}
\usepackage{amsmath}
\usepackage{amssymb}
\usepackage{psfrag}
\usepackage{setspace}
\usepackage{subfig}
\usepackage{wrapfig}
\usepackage{sidecap}

\usepackage{hyperref}

\newcommand{\bs}[1]{\boldsymbol{#1}}
\def\RR{ \mathbb R}
\def\bet{\bs{\eta}}

\def\bt{\bs{\Theta}}

\usepackage{epsf}
\newcommand{\ee}{\end{equation}}
\newcommand{\be}{\begin{equation}}
\newcommand{\ec}{\end{center}}
\newcommand{\bc}{\begin{center}}
\newcommand{\eea}{\end{eqnarray}}
\newcommand{\bea}{\begin{eqnarray}}
\newcommand{\bd}{\begin{description}}
\newcommand{\ed}{\end{description}}
\newcommand{\bi}{\begin{itemize}}
\newcommand{\ei}{\end{itemize}}
\newcommand{\pa}{\partial}

\newcommand{\bpsi}{\bs{\Psi}}
\newcommand{\bmu}{\bs{\mu}}
\newcommand{\btheta}{\bs{\Theta}}

\newcommand{\bW}{\bs{W}}
\newcommand{\bphi}{\bs{\Phi}}
\newcommand{\dpsi}{d_{\Psi}}
\newcommand{\dth}{d_{\Theta}}

\usepackage[dvipsnames]{xcolor}

\newcommand{\tcb}{\textcolor{black}}

\newcommand{\tcg}{\textcolor{black}}
\newcommand{\tcbr}{\textcolor{black}}
\newcommand{\tcbl}{\textcolor{black}}

\newcommand{\tcr}{\textcolor{black}}

\newcommand{\refeq}[1]{Equation (\ref{#1})}
\newcommand{\reffig}[1]{Figure \ref{#1}}
\newcommand{\refsec}[1]{Section \ref{#1}}

\journal{Journal of }

\usepackage{amsmath}
\usepackage{algorithm} 
\usepackage{algpseudocode} 

\usepackage{rotating} 

\usepackage{tikz} 
\usetikzlibrary{shapes,arrows}  
\usetikzlibrary{fit,backgrounds} 

\begin{document}

\begin{frontmatter}

\title{Multimodal, high-dimensional, model-based, Bayesian inverse problems with applications in biomechanics}



\author[rvt]{Isabell M. Franck}
\ead{franck@tum.de}
\author[rvt]{P.S. Koutsourelakis\corref{cor1}}
\ead{p.s.koutsourelakis@tum.de}
\cortext[cor1]{Corresponding Author. Tel: +49-89-289-16690}
\address[rvt]{Professur  f\"ur Kontinuumsmechanik, Technische Universit\"at M\"unchen, Boltzmannstrasse 15, 85747 Garching (b. M\"unchen), Germany}
\ead[url]{http://www.contmech.mw.tum.de}

\begin{abstract}
This paper is concerned with the numerical solution of model-based, Bayesian inverse problems. 
We are particularly interested in cases where the cost of each likelihood evaluation (forward-model call) is expensive and the number of unknown (latent) variables is high. This is the setting in many problems in computational physics where forward models with nonlinear PDEs are used and the parameters to be calibrated involve spatio-temporarily varying coefficients, which upon discretization give rise to a high-dimensional vector of unknowns.

One of the consequences of the well-documented ill-posedness of inverse problems is the possibility of multiple solutions. While such information is contained in the posterior density in Bayesian formulations, the discovery of a single mode, let alone multiple, is a formidable task.
The goal of the present paper is two-fold. On one hand, we propose approximate,  adaptive inference strategies using mixture densities to capture multi-modal posteriors, and on the other, to extend our work in \cite{franck_sparse_2015} with regards to effective dimensionality reduction techniques that reveal low-dimensional subspaces where the posterior variance is mostly concentrated.
We validate the model proposed  by employing  Importance Sampling   which confirms that the bias introduced is small and  can be efficiently corrected if the analyst wishes to do so. 
 We demonstrate the performance of the proposed strategy   in nonlinear elastography where the identification of the mechanical properties of biological materials can inform non-invasive, medical diagnosis. The discovery of multiple modes (solutions) in such problems is critical in achieving the diagnostic objectives.
\end{abstract}


\end{frontmatter}


\section{Introduction}\label{sec:Introduction}

Model-based (or model-constrained), inverse problems appear in many scientific fields and their solution represents a fundamental challenge in the context of model calibration and system identification \cite{biegler_large-scale_2010}.
Bayesian formulations offer a rigorous setting for their solution as they account for various sources of uncertainty that is unavoidably present in these problem. Furthermore, they possess a great advantage over deterministic alternatives as apart from point-estimates, they provide quantitative metrics of the uncertainty in the unknowns  encapsulated in the {\em posterior distribution} \cite{gelman_bayesian_2003}.
%

An application of particular, but not exclusive, interest for this paper  involves  the identification of the mechanical properties of  biological materials, in the context non-invasive medical diagnosis ({elastography}). While in certain cases mechanical properties can also be measured directly by excising multiple tissue samples, non-invasive procedures offer obvious advantages in terms of ease, cost and reducing risk of complications to the patient.  Rather than x-ray techniques which capture variations in density, the identification of stiffness, or mechanical properties in general, can potentially  lead to earlier and more accurate diagnosis \cite{ISI:000263259100006,liver2006},  provide valuable insights that  differentiate between modalities of the same pathology \cite{cur12gen}, monitor the progress of treatments and  ultimately lead to patient-specific treatment strategies.

All elastographic  techniques consist of the following three basic steps \cite{doyley_model-based_2012} : \textbf{1)} excite the tissue using a (quasi-)static, harmonic or transient source, \textbf{2)} (indirectly) measure tissue deformation (e.g. displacements, velocities)  using an imaging technique such as ultrasound  \cite{Ophir:1991}, magnetic resonance \cite{mut95mag} or optical tomography \cite{khalil_tissue_2005}, and
 \textbf{3)} infer the mechanical properties from this data using a suitable continuum mechanical model of the tissue's deformation. 
Perhaps the most practical such imaging technique due to its lower relative cost and increased portability  is ultrasound elasticity imaging \cite{sarvazyan_elasticity_2011,doyley_elastography:_2014}.
The pioneering work of   Ophir and coworkers \cite{Ophir:1991}
 followed by several clinical studies \cite{Garra:1997,Bamber:2002,Thomas:2006,par11im}
 have demonstrated that the resulting strain images typically improve the diagnostic accuracy over ultrasound alone. Furthermore, technological advances have led to ultra-portable ultrasound transducers, attachable to smartphones/tablets \cite{schleder_diagnostic_2013,driver_doctors_2016}.  As  the rate of data acquisition increases and the cost decreases, it becomes increasingly important to develop tools that leverage the capabilities of physics-based models in order to produce quickly and accurately diagnostic estimates as well as  quantify the confidence in them.
  Apart from breast cancer, there is a wealth of evidence indicating the potential of elastography-based techniques in detecting a variety of other pathologies such as prostate \cite{kro98ela,hoy08tis} and liver cancer \cite{asbach_assessment_2008},  characterizing blood clots \cite{schmitt_noninvasive_2007}, brain imaging \cite{hamhaber_vivo_2010}, atherosclerosis \cite{ohayon_biomechanics_2013}, osteopenia \cite{shore_transversely_2011}.

In this paper we advocate a probabilistic,  {\em indirect or iterative} procedure (in contrast to {\em direct elastography}  \cite{ISI:000275699600006}) which admits an inverse problem formulation and involves the discrepancy 
 between observed and model-predicted displacements \cite{ISI:000223500200013,doyley_enhancing_2006,ISI:000275756200016,ISI:000280774700004,doyley_model-based_2012}. . 
 Several other problems which involve complex forward models (i.e. expensive likelihood)  and the identification of {\em spatially varying} model parameters share similar characteristics such as permeability estimation for soil transport processes that can assist in the detection of contaminants, oil exploration and carbon sequestration  \cite{wang_bayesian_2004,wang_hierarchical_2005,dostert_coarse-gradient_2006}
 
 The solution of such  model calibration problems  in the Bayesian framework is hampered by two main difficulties. The first affects the computational efficiency of such methods and stems from the poor scaling of traditional Bayesian inference tools,  with respect to the dimensionality of the unknown parameter vector - another instance of the {\em curse-of-dimensionality}. In problems such as the one described above, the model parameters of interest (i.e. material properties) exhibit spatial variability which requires fine discretizations  in order to be captured. This variability can also span different scales \cite{mallat_wavelet_2008,koutsourelakis_multi-resolution_2009}. Standard Markov Chain Monte Carlo (MCMC, \cite{green_bayesian_2015}) techniques require an exorbitant number of likelihood evaluations (i.e. solutions of the forward model) in order to  converge  \cite{roberts_exponential_1996,roberts_optimal_1998,mattingly_diffusion_2012,pillai_optimal_2012}. As each of these calls implies the solution of very large systems of (non)linear, and potentially transient,  equations, it is generally of interest to minimize their number  particularly in time-sensitive applications. 
Advanced sampling schemes, involving adaptive MCMC \cite{lee_markov_2002,holloman_multi-resolution_2006,chopin_free_2012} and Sequential Monte Carlo (SMC, \cite{moral_sequential_2006,koutsourelakis_multi-resolution_2009,del_moral_adaptive_2010}) exploit the physical  insight and the use of multi-fidelity solvers in order to expedite the inference process. Nevertheless, the number of forward calls can still be in the order of tens of thousands if not much more.
Several attempts have also been directed towards using emulators or surrogates or reduced-order models of various kinds \cite{marzouk_stochastic_2007,bui-thanh_model_2008,rosic_sampling-free_2012,bilionis_solution_2014,
chen_sparse-grid_2015,lan_emulation_2016} but such a task is severely hindered by the high- dimensionality. 
The use of first and second-order derivatives has also been advocated either in a standard MCMC format or by developing advanced sampling strategies. These are generally available by solving appropriate {\em adjoint problems}  which are well-understood in the context of deterministic formulations \cite{flath_fast_2011,girolami_riemann_2011,martin_stochastic_2012,bui-thanh_solving_2014,petra_computational_2014}. 
More recent treatments, attempt to exploit lower-dimensional structure of the target posterior by identifying subspaces where either most of the probability mass is contained \cite{franck_sparse_2015} or where maximal sensitivity is observed \cite{cui_likelihood-informed_2014,spantini_optimal_2015,cui_scalable_2015,cui_dimension-independent_2016}. This enables inference tasks  that are carried out on spaces of significantly reduced dimension and  are not hampered by the aforementioned difficulties. Generally all such schemes construct such approximations around the MAP point by employing local information (e.g.  gradients) and therefore are not suitable for multi-modal or highly non-Gaussian  posteriors.

The latter represents the second challenge that we attempt to address in this paper. That is, 
 the identification of multiple posterior modes. 
In the context of elastography, multi-modality can originate from  anisotropic material  \cite{chatelin_anisotropic_2014}, wrong/missing information from images/measurements \cite{fang_compositional-prior-guided_2010} or the imagaing modality employed \cite{fromageau_estimation_2007}. In all cases, each mode in the posterior can lead to  different diagnostic conclusions and it is therefore very important to identify them and correctly assess their posterior probabilities.
 The majority of Bayesian strategies for the solution of computationally intensive inverse problems operates under the assumption of a unimodal posterior or focuses on the approximation of a single mode of the posterior. Some numerical inference tools based on SMC or other tempering mechanisms  \cite{gill_dynamic_2004, li_adaptive_2015, feroz_multimodal_2008} have been developed but  require a very large  number of forward model calls particularly when the dimension of unknowns increases.  We note finally that the treatment of multi-modal densities in high-dimensions has attracted significant interest in atomistic  simulation  in the context of free energy computations \cite{lelievre_free_2010,bilionis_free_2012} but in such problems (apart from other distinguishing features) the cost per density evaluation (i.e. one MD time-step) is smaller than in our setting. 

In this paper we propose a Variational Bayesian (VB) strategy that extends our previous work \cite{franck_sparse_2015}. Therein we have shown how accurate approximations of the true posterior can be attained by identifying a  low-dimensional subspace where posterior uncertainty is concentrated. This has led to computational schemes that require only a few tens of forward model runs in the problems investigated. Nevertheless, our previous work  was based on the assumption of a unimodal posterior  which we propose overcoming in this paper by employing a mixture of multivariate Gaussians.  Such mixtures have been employed in various statics and machine learning applications (e.g. speaker identification \cite{reynolds_speaker_2000}, data clustering \cite{jain_data_1999}) and in combination with Variational Bayesian inference techniques as well \cite{choudrey_variational_2003,beal_variational_2003,kuusela_gradient-based_2009}. Nevertheless, all these problems were characterized by inexpensive likelihoods, relatively low-dimensions and multiple data/measurements. 
In contrast to that, the inverse problems considered here are based on a  single experiment and a single observation vector. Furthermore, we propose an adaptive algorithm based on information-theoretic criteria  for the identification of the number of the required mixture components (section \ref{sec:method}).
We present the parametrization of the proposed model in section  \ref{sec:method} where we also discuss a Variational-Bayesian Expectation-Maximization \cite{beal_variational_2003} scheme for performing inference and learning.  In section \ref{sec:NumericalExamples} we present numerical illustrations involving a simple toy-example and  an example in the context of elastography.  
\section{Methodology}  \label{sec:method}

This section discusses the methodological framework advocated. We begin (\refsec{sec:forward}) with a generic presentation of the forward model and in particular with models arising from the discretization of nonlinear PDEs such as in our motivating application of nonlinear elastography. In \refsec{sec:bayes} we present a Bayesian mixture model  that can identify lower-dimensional subspaces where most of the posterior mass is concentrated as well as accounts for multiple modes. The prior assumptions for the model  parameters are summarized in \refsec{sec:Prior}. In \refsec{sec:Variational} we discuss a Variational Bayesian Expectation-Maximization scheme for computing efficiently approximations to the posterior for a fixed number of mixture components, and in \refsec{sec:NrMoG} we discuss a scheme for determining the appropriate number of such components. Finally, in \refsec{sec:IS} we discuss how to assess the accuracy of the  approximation computed  as well as a way to 
correct for any bias if this is deemed to be necessary.

\subsection{Forward model - Governing equations}
\label{sec:forward}

Canonical formulations of model-based, inverse problems  postulate the existence of a forward model that typically arises from the discretization of governing equations, such as PDEs/ODEs, and which can be expressed in the residual form as follows:
\be
    \bs{r} (\bs{u}; \bpsi)=\bs{0}.
    \label{eq:res}
\ee
The residual vector  $\bs{r}: \RR^n \times \RR^{\dpsi} \to \RR^n$ depends on the state vector  $\bs{u} \in \RR^n$ (forward-problem unknowns) and $\bpsi \in \RR^{\dpsi}$, the vector of unobserved (latent) model parameters (inverse-problem unknowns). The aforementioned equation is complemented by a set of (noisy) observations/measurements $\hat{\bs{y}} \in \RR^{d_y}$ which pertain to the model output $\bs{Y}(\bs{u}(\bpsi))=\bs{y}(\bpsi)$:
\be
\hat{\bs{y}}= \bs{y}(\bpsi)+noise.
\ee
The unknowns $\bpsi$ in the problems considered arise from the discretization of {\em spatially varying} parameters and we refer to them as {\em material parameters} in view of the biomechanics applications discussed in \refsec{sec:ex2}.  
Throughout this work we assume that the noise term pertains only to measurement/observation errors which we model with a zero-mean, uncorrelated  Gaussian vector $\bs{z} \in \RR^{d_y}$:
\be
  \bs{\hat{y}} = \bs{y(\bpsi)} + \bs{z}, \quad \bs{z} \sim \mathcal{N}(\bs{0},\tau^{-1} \bs{I}_{d_y}). \label{eq:objective}
\ee
The  precision $\tau$ of the observation noise will be assumed unknown and will be inferred from the data along with $\bpsi$. Other noise distributions (e.g. to account for faulty sensors) can be readily considered   \cite{jin_variational_2012}.
We note that the difference  between observations and model predictions would in general contain model errors arising from the discretization of the governing equations and/or the inadequacy of the model itself to capture the underlying physical process. While the former source can be reduced by considering very fine discretizations (at the cost of increasing the dimensionality of the state vector $\bs{u}$ and potentially $\bpsi$), the latter requires a much more thorough treatment which exceeds the scope of this work \cite{kennedy_bayesian_2001,higdon_computer_2008,bayarri_framework_2007,berliner_modeling_2008,koutsourelakis_novel_2012,strong_when_2014,sargsyan_statistical_2015}.


In the traditional route, the  likelihood $p(\bs{\hat{y}} | \bpsi, \tau)$,  implied by \refeq{eq:objective}:  
\be
p(\bs{\hat{y}} | \bpsi, \tau) \propto \tau^{d_y/2} e^{-\frac{\tau}{2} || \bs{\hat{y}} -\bs{y}(\bpsi) ||^2 }.
\label{eq:likelihood}
\ee
is complemented by priors $p_{\Psi}(\bpsi)$ and $p_{\tau}(\tau)$ which, with the  application of the Bayes' rule, lead to the definition of the posterior density $p(\bpsi, \tau  | \bs{\hat{y}})$ which is proportional to:
\be
p(\bpsi, \tau  | \bs{\hat{y}}) \propto p(\bs{\hat{y}} | \bpsi, \tau)~p_{\Psi}(\bpsi)~ p_{\tau}(\tau).
\ee
The intractability of the map $\bs{y}(\bpsi)$ precludes the availability of closed-form solutions for the posterior and necessitates the use of various sampling schemes such as those discussed in the introduction. This task is seriously impeded by a) the need for repeated solutions of the discretized forward problem (\refeq{eq:res}) each of which can be quite taxing computationally,
 b) the high dimensionality of the vector of unknowns $\bpsi$ which hinders the efficient search (e.g. by sampling) of the latent parameter space and further increases the computational burden.
 The goal of the proposed Variational Bayesian scheme is to alleviate  these difficulties by proposing adequate approximations and dimensionality-reduction techniques that are seamlessly integrated in the inference framework. Furthermore, we attempt to overcome  well-known limitations that have to do with the multimodality of the posterior and which further exacerbate these problems. Multimodality is inherently related to the ill-posedness of inverse problems and its potential can increase when the dimension of the vector of unknowns increases and/or the noise is amplified.

\subsection{Bayesian Mixture Model}
\label{sec:bayes}

In this section and in view of the aforementioned desiderata  we introduce the augmented formulation of the Bayesian inverse problem. 
\bi
\item In order to capture multiple modes of the posterior (if those are present) we  introduce the {\em discrete, latent variable} $s$ which takes integer values between $1$ and $S$. The latter represents the number of modes identified, each of which will be  modeled with a multivariate  Gaussian. The cardinality of the model i.e. $S$  is learned in a manner that is described in the sequel.
\item In order to identify a lower-dimensional representation of the unknowns $\bpsi$, we define the latent variables $\bt \in \RR^{\dth}$ such that $\dth \ll \dpsi$. The premise here is that while $\bpsi$ is high-dimensional, its posterior can be adequately represented on a subspace of dimension $\dth$ that captures most of the variance. As we have argued in \cite{franck_sparse_2015} these latent variables can give rise to a PCA-like representation of the form:
\be
\bpsi=\bs{\mu}+\bs{W} \bt+\bet
\ee
where $\bs{\mu} \in \RR^{\dpsi}$ is the mean vector and the columns of the orthogonal matrix $\bs{W} \in \RR^{\dpsi \times \dth}$ ($\bs{W}^T \bs{W}=\bs{I}_{d_y}$) span the aforementioned subspace with reduced coordinates $\bt$. The vector  $\bet \in \RR^{\dpsi}$ captures the residual variance (noise) that complements the main effects.  

\begin{figure}[!t]
  \centering
  \def\svgwidth{\columnwidth}
  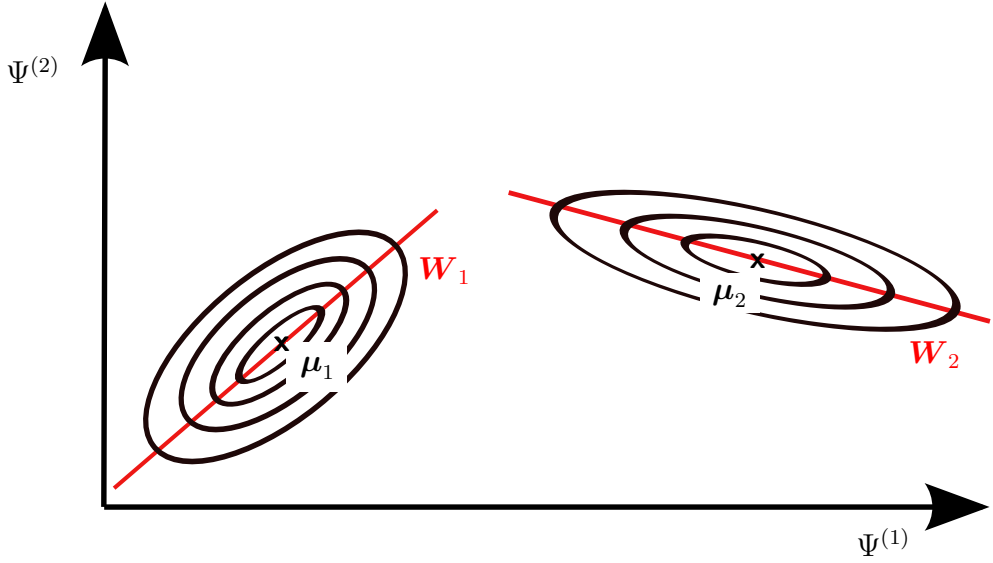
   \caption{Illustration of the multimodal representation for $S=2$ in 2D i.e. when $\bs{\Psi} = \{ \Psi^{(1)}, \Psi^{(2)}\}$.} 
  \label{fig:MoGReducedBases}  
\end{figure}

In view of the multimodal approximation and since each mode implies a different mean and a different lower-dimensional subspace (Figure \ref{fig:MoGReducedBases}), we advocate in this work $S$ expansions of the form:
\be
\bpsi_s=\bs{\mu}_s+\bs{W}_s \bt+\bet, \quad s=1,2,\ldots,S
\label{eq:psis}
\ee
where the notation $\bpsi_s$ implies the representation of $\bpsi$ within mode $s$. As with multiple modes there can also be multiple subspaces where the variance is concentrated, so it is necessary/important to distinct the $\bs{W}_s$. In principle, the dimension $\dth$ of the reduced subspaces can also vary with $s$ but we do not consider this here for simplicity of notation.

\ei

\begin{figure}
 \centering
   \def\svgwidth{0.8\columnwidth}
   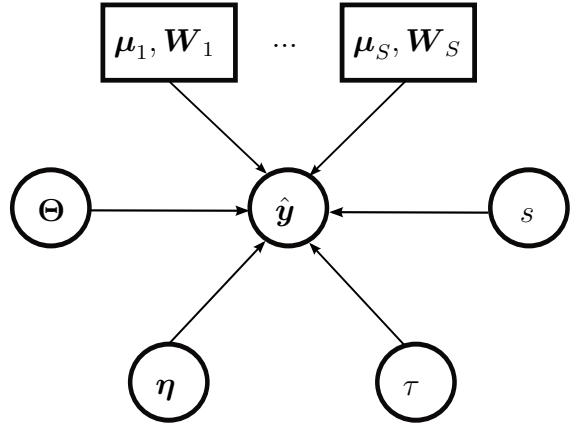    
 \caption{ Graphical representation of the proposed generative probabilistic model. Circles denote random variables, solid rectangles, model parameters and arrows denote dependencies \cite{murphy_machine_2012}.}
\label{fig:plate}
\end{figure}

For reasons that will become apparent in the following, we distinguish between {\em latent variables}:  $s$, $\bt$, $\bet$ and $\tau$, and model parameters:  $\bs{\mu}=\{ \bs{\mu}_j\}_{j=1}^S$,  $\bs{W}=\{ \bs{W}_j\}_{j=1}^S$. We seek point estimates for the latter and (approximations) of the actual (conditional) posterior for the former. 
 The discussion thus far suggests that that the likelihood of \refeq{eq:likelihood}
 takes the form:
\be
    p(\bs{\hat{y}} | s, \bt, \bet, \tau, \bs{\mu}_s, \bs{W}_s) \propto \tau^{d_y/2} e^{-\frac{\tau}{2} || \bs{\hat{y}} -\bs{y}(\bs{\mu}_s+\bs{W}_s \bt+\bet) ||^2 }.
    \label{eq:likelihoodExtended}
\ee
A graphical illustration of the proposed probabilistic generative model proposed is in \reffig{fig:plate}.

Following the standard Bayesian formalism, one would complement the aforementioned likelihood with priors on the model parameters $p(\bs{\mu}, \bs{W})$ and the latent variables $p(\bt,\bet,s, \tau)$, in order to obtain the joint posterior (given $S$):
\be
    \begin{array}{ll} 
    p( s, \bt, \bet, \tau,  \bs{\mu},\bs{W} | \bs{\hat{y}})  \propto & p(\bs{\hat{y}} | s, \bt, \bet, \tau, \bs{\mu}_s, \bs{W}_s)  ~p(\bt,\bet,s, \tau)~p(\bs{\mu},\bs{W}).
    \end{array}
    \label{eq:jointpost}
\ee
We discuss the specific form of the priors (and associated hyperparameters) in Subsection \ref{sec:Prior} as well as the inference/learning strategy we propose in \ref{sec:Variational}.
We note at this stage however,  that given this posterior, one would  obtain a {\em mixture} representation of the unknown material parameters $\bpsi$, as implied by \refeq{eq:psis}. In particular, given values  for $ (\bs{\mu},\bs{W})=\{ \bs{\mu}_j, \bs{W}_j \}_{j=1}^S$, it immediately follows that the {\em posterior} $p(\bpsi | \bs{\mu},\bs{W}, \bs{\hat{y}} )$ (given $S$) of $\bpsi$ is:
\be
\begin{array}{ll}
   p(\bpsi | \bs{\mu},\bs{W}, \bs{\hat{y}}, \bs{T} ) & = \sum_{s=1}^S \int  p(\bpsi,  s, \bt, \bet, \tau, | \bs{\mu},\bs{W}, \bs{\hat{y}}  ) ~d\bt~d\bet~d\tau \\ 
   & = \sum_{s=1}^S \int  p(\bpsi |  s, \bt, \bet, \tau, \bs{\mu}_s,\bs{W}_s ) ~p( s, \bt, \bet, \tau, | \bs{\mu},\bs{W}, \bs{\hat{y}} )~d\bt~d\bet~d\tau \\
   & = \sum_{s=1}^S \int  \delta(\bpsi - (\bs{\mu}_s+\bs{W}_s \bt + \bet))~p( s, \bt, \bet, \tau, | \bs{\mu},\bs{W}, \bs{\hat{y}} )~d\bt~d\bet~d\tau \\
\end{array}
%
    \label{eq:postmix}
\ee
where the conditional posterior $p( s, \bt, \bet, \tau, | \bs{\mu},\bs{W}, \bs{\hat{y}} )$ is found from \refeq{eq:jointpost}.
We discuss in \refsec{sec:Variational} how the posterior on the latent variables is approximated as well as the values (point estimates) for the model parameters $ \bs{\mu},\bs{W}$ are computed.


\subsection{Priors}
\label{sec:Prior}

We assume that, a priori, the  precision $\tau$ of the observation noise is independent of the remaining latent variables $\bt,\bet,s$  i.e.:
\be
p(\bt,\bet,s, \tau) = p(\bt,\bet,s) ~p_{\tau}(\tau).
\ee
In particular, we employ:

\bi
  \item a Gamma prior on $\tau$: We employ a (conditionally) conjugate Gamma prior:
      \be
	p_{\tau}(\tau) \equiv  Gamma(a_0,b_0).  \label{eq:tauPrior}
      \ee 
      We use $a_0 = b_0 = 0$ which results in a non-informative Jeffreys' prior that is scale-invariant.

  \item  We assume that $\bt$ and $\bet$ are a priori, {\em conditionally independent} i.e. that  $p(\bt,\bet,s)=p(\bt,\bet|s) p_s(s) =p_{\Theta}(\bt|s) p_{\eta}(\bet|s)  p_s(s)$. We discuss each of these terms  below: 
  \bi
	\item  We assume that each component $s$ is, a priori, equally likely, which implies:
	    \be
	      p_s(s) = \frac{1}{S}, \quad s \in [1: S].
	    \ee
	    Hierarchical priors can be readily be adopted (e.g. \cite{beal_variational_2003}) but we consider  here the simplest possible scenario. An interesting extension would involve infinite models with Dirichlet Process priors \cite{antoniak_mixtures_1974,
	    maceachern_estimating_1998} which would enable the number of components $S$ to be automatically determined. In this work, a less elegant, but quite effective adaptive scheme for determining $S$ is proposed  in \refsec{sec:NrMoG}.
	
	\item A Gaussian prior on $\bs{\Theta}$:
	    \\The role of the  latent variables $\bs{\Theta}$  is to capture the most significant  variations of $\bs{\Psi}_s$ around its mean $\bs{\mu}_s$. By significant we mean in this case the directions along which, the largest posterior uncertainty is observed. Naturally these are closely related to the matrices $\bs{W}_j$ and represent the reduced coordinates along the subspace spanned by their column vectors. We assume therefore that, a priori, these are independent, have zero mean and follow a multivariate Gaussian:
	    \be
		p_{\Theta}(\bt|s) = \mathcal{N}(\bs{0}, \bs{\Lambda}_{0,s}^{-1}) 
		\label{eq:priortheta}
	    \ee
	    where $\bs{\Lambda}_{0,s} = diag(\lambda_{0,s,i})$, $i = 1,...,\dth$ express prior variances along each of the latent principal directions. 
	  	
	\item A Gaussian prior on $\bet$:
	
	    As the role of the latent variables is to capture any residual variance (that is not accounted for by $\bt$), we assume that, {\em a priori}, $\bet$ can be modeled by a multivariate Gaussian that has zero mean and an isotropic covariance:
	    \be
		p_{\eta}(\bet|s)= \mathcal{N}(\bs{0}, \lambda_{0,\eta,s}^{-1} \bs{I}_{\dpsi}).
	    \ee
  \ei
\ei	

For the model parameters $\bs{\mu},\bs{W}$, we assume that, a priori, the parameters associated with each component $j=1,\ldots, S$ are independent. In particular:
 \bi
 \item Prior on each $\bs{\mu}_j$ for $j\in 1:S$:
      \\
      In general such priors must encapsulate not only the information/beliefs available a priori to the analyst but also reflect the physical meaning of the parameters  $\bpsi$.
      We are motivated by applications in elastography  where the goal is to identify inclusions that correspond to tumors and generally have very different properties from the surrounding tissue (\cite{wellman_breast_1999, krouskop_elastic_1998}). 
      The vector  $\bpsi$ represents the spatial discretization of the material parameters  i.e. each of its entries corresponds to the value of the material parameter at a certain point in the physical domain. This structure  is inherited by  $\bs{\mu}_j$  and for this reason we employ a  hierarchical prior that penalizes jumps between neighboring locations (on the spatial domain) \cite{calvetti_hypermodels_2008} in a manner controlled by appropriately selected hyperparameters. The model was discussed in detail in \cite{franck_sparse_2015} and is included for completeness in \ref{app:muPrior}.
    
  \item Prior specification on each $\bs{W}_j$ for $j\in 1:S$:
	\\ We require that each $\bs{W}_j$ is orthonormal i.e. $\bs{W}_j^T \bs{W}_j=\bs{I}_{\dth}$, where $\bs{I}_{\dth}$ is the $\dth-$dimensional identity matrix. This is equivalent to employing a uniform prior  on the Stiefel manifold $V_{\dth}(\RR^{\dpsi})$. 


 \ei

\subsection{Variational Approximation} \label{sec:Variational}

\begin{figure}[H]
  \centering
  \def\svgwidth{0.6\columnwidth}
  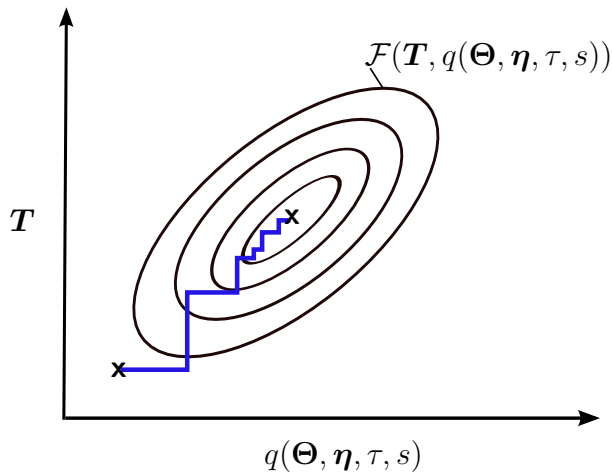        
  \caption{Schematic illustration of the advocated Variational Bayesian Expectation-Maximization (VB-EM, \cite{beal_variational_2003}).}
   \label{fig:maxF} 
\end{figure}

We note that inference (exact or approximate) for all the model parameters described previously would pose a formidable task particularly with regard to $\bs{\mu}$ and $\bs{W}$ which scale with $\dpsi \gg 1$. For that purpose, we advocate a hybrid approach whereby Maximum-A-Posteriori (MAP) point estimates of the high-dimensional parameters  $\bs{T}=( \bs{\mu}, \bs{W})=\{\bmu_j, \bW_j \}^S_{j=1}$ are obtained  and the posterior of the remaining (latent) variables $s,\bt, \bet, \tau$ is approximated. To that end we make use of the Variational Bayesian Expectation-Maximization scheme (VB-EM, \cite{beal_variational_2003, franck_sparse_2015}) which provides a lower bound $\mathcal{F}$ on the log of the marginal posterior of $\bs{T}=( \bs{\mu}, \bs{W})$. This can be iteratively maximized by a generalized coordinate ascent (\reffig{fig:maxF})  which alternates between finding optimal approximations $q(s,\bt,\bet, \tau)$ of the exact (conditional) posterior $p(s,\bt, \bet, \tau | \bs{\hat{y}}, \bs{T})$ and 
optimizing with respect 
to $\bs{T}$.



On the basis of the discussion above and the separation between latent variables ($s, \bt, \bet, \tau$) and model parameters $\bs{T}$,  we can rewrite \refeq{eq:jointpost} (for a given $S$) as follows:
\be
    p( s, \bt, \bet, \tau, \bs{T} | \bs{\hat{y}}) =\frac{  p(\bs{\hat{y}} | s, \bt, \bet, \tau, \bs{T})  ~p_{s}(s)~p_{\Theta}(\bt|s)~p_{\eta}(\bet|s)~ p_{\tau}(\tau)~p_T(\bs{T}) }{p(\bs{\hat{y}})}.
    \label{eq:jointpost1}
\ee
We note that both sides of the equation above depend implicitly on $S$ i.e. the total number of components in the model. This is especially important  for the model evidence term $p(\bs{\hat{y}})$ which we discuss in \refsec{sec:NrMoG}. We nevertheless omit $S$ from the expressions in order to simplify the notation.

Furthermore the {\em conditional} posterior of $(s, \bt, \bet, \tau)$ given $\bs{T}$ is:
\be
p(s, \bt, \bet, \tau | \bs{T}, \bs{\hat{y}}) = \frac{p( s, \bt, \bet, \tau, \bs{T} | \bs{\hat{y}})}{p(\bs{T} |\bs{\hat{y}})}
\label{eq:condpost}
\ee
where $p(\bs{T} |\bs{\hat{y}})$ is the (marginal) posterior of the model parameters $\bs{T}$.

For an arbitrary density $q(\bt, \bet, \tau,s )$ and by employing Jensen's inequality, it can be shown that \cite{franck_sparse_2015}: 
\be
 \begin{array}{ll} 
 \log p( \bs{T} | \bs{\hat{y}} ) & = \log \sum_{s=1}^S \int  p( \bs{T}, \bs{\Theta}, \bet, \tau,s | \hat{\bs{y}} )~d\bt ~d\bet~d\tau\\
 & = \log \sum_{s=1}^S \int q(\bt,\bet, \tau,s ) \frac{ p( \bs{T}, \bs{\Theta}, \bet, \tau, s | \hat{\bs{y}} )}{q(\bt,\bet,\tau, s)} ~d\bt ~d\bet~d\tau\\
 &  \ge \sum_{s=1}^S \int q(\bt,\bet,\tau,s) \log \frac{ p( \bs{T}, \bs{\Theta}, \bet, \tau, s | \hat{\bs{y}} )}{q(\bt,\bet,\tau, s)} ~d\bt~d\bet~ d\tau\\
 & = \mathcal{F}(q(\bt, \bet, \tau,s), \bs{T}).
 \end{array}
   \label{eq:loglike1}
\ee

We note here that the variational lower bound $\mathcal{F}$ has a direct connection with the Kullback-Leibler (KL) divergence between $q(\bt, \bet, \tau,s)$ and the (conditional) posterior $p(\bt, \bet, \tau, s | \bs{T}, \bs{\hat{y}})$.
In particular, if we denote by $E_q[.]$ expectations with respect to $q$, then:
\be
\begin{array}{ll}
  KL\left( q(\bt, \bet, \tau, s)  || p(\bt, \bet, \tau, s | \bs{\hat{y}} , \bs{T}) \right) & = -E_q \left[ \log \frac{p(\bt, \bet, \tau,s | \bs{\hat{y}} , \bs{T})}{ q(\bt, \bet, \tau,s) } \right] \\
  & = -E_q \left[ \log  \frac{p(\bs{T}, \bt, \bet, \tau, s | \bs{\hat{y}}  ) }{p( \bs{T} | \bs{\hat{y}} )~q(\bt, \bet, \tau, s) } \right] \\
  & = \log p( \bs{T} | \bs{\hat{y}} )-\mathcal{F}(q(\bt, \bet, \tau, s), \bs{T}).
  \end{array}
  \label{eq:KLDIVF}
\ee
The Kullback-Leibler divergence is by definition non-negative and becomes zero when $q(\bt, \bet, \tau, s)   \equiv p(\bt, \bet, \tau,s | \bs{\hat{y}} , \bs{T}) $. Hence, for a given $\bs{T}$, constructing a good approximation to the conditional posterior (in the KL divergence sense) is equivalent to maximizing the lower bound $\mathcal{F}(q(\bt, \bet, \tau,s), \bs{T})$ with respect to $q(\bt, \bet, \tau, s)$. Analogously, maximizing $\mathcal{F}$ with respect to $\bs{T}$ (for a given $q(\bt, \bet, \tau,s)$ leads to  (sub-)optimal MAP estimates \cite{franck_sparse_2015}.
This suggests an iterative scheme that alternates between:
\bi
\item \textbf{VB-Expectation} step: Given the current estimate of $\bs{T}$, find the $q(\bt, \bet, \tau,s)$ that maximizes $\mathcal{F}$.
\item \textbf{VB-Maximization} step:  Given the current  $q(\bt, \bet, \tau,s)$, find $\bs{T}$ that maximizes $\mathcal{F}$.
\ei
As in standard EM schemes \cite{neal_view_1998}, relaxed versions of the aforementioned partial optimization problems can be considered that improve upon the current $\mathcal{F}$ rather than finding the optimum at each iteration. 

Using \refeq{eq:jointpost1}, the lower bound $\mathcal{F}$ can be expressed as:
\be
    \begin{split}
    \mathcal{F}(q(\bt, \bet, \tau, s), \bs{T})  = & E_q\left[ \log \frac{  p(\bs{\hat{y}} | s, \bt, \bet, \tau, \bs{T})  ~p_{s}(s)~p_{\Theta}(\bt|s)~p_{\eta}(\bet|s)~ p_{\tau}(\tau)~p_T(\bs{T}) }{p(\bs{\hat{y}}) ~q(\bt, \bet, \tau, s)}\right]   \\
	= &E_q\left[ \log \frac{  p(\bs{\hat{y}} | s, \bt, \bet, \tau, \bs{T})  ~p_{s}(s)~p_{\Theta}(\bt|s)~p_{\eta}(\bet|s)~ p_{\tau}(\tau) }{q(\bt, \bet, \tau, s)}\right]   \\
	  & + \log p_T(\bs{T})-\log p(\bs{\hat{y}})\\
	= & \mathcal{\hat{F}}(q(\bt, \bet, \tau, s), \bs{T}) + \log p_T(\bs{T}) -\log p(\bs{\hat{y}}).
    \end{split}
   \label{eq:Fvar1}
\ee
We will omit the term $ -\log p(\bs{\hat{y}})$ as it does not depend on $q$ nor $\bs{T}$. It is apparent that the challenging term in $\mathcal{\hat{F}}$ involves  the likelihood, i.e.:
\be
    \begin{array}{ll}
    \mathcal{\hat{F}}(q(\bt, \bet, \tau, s), \bs{T})  &  =E_q\left[ \log \frac{  p(\bs{\hat{y}} | s, \bt, \bet, \tau, \bs{T})  ~p_{s}(s)~p_{\Theta}(\bt|s)~p_{\eta}(\bet|s)~ p_{\tau}(\tau) }{q(\bt, \bet, \tau, s)}\right] \\
    & = E_q \left[ \frac{d_y}{2} \log \tau - \frac{\tau}{2}  || \bs{\hat{y}} -\bs{y}(\bs{\mu}_s+\bs{W}_s \bt+\bet) ||^2  \right] \\
    & +E_q \left[ \log \frac{p_{s}(s)~p_{\Theta}(\bt|s)~p_{\eta}(\bet|s)~ p_{\tau}(\tau) }{q(\bt, \bet, \tau, s)}\right].
    \end{array}
    \label{eq:fvarhat}
\ee
The intractability of the map $\bs{y}(.)$ precludes an analytic computation of the expectation with respect to $q$, let alone the optimization with respect to this. While stochastic approximation techniques in the context of VB inference have been suggested \cite{titsias_doubly_2014} to carry out this task, these would require repeated forward solves (i.e. evaluations of $\bs{y}(.)$) which would render them impractical.
For that purpose, as in our previous work \cite{franck_sparse_2015}, we invoke an {\em approximation} by using a first-order Taylor series expansion  of $\bs{y}$ (given $s$) at $\bs{\mu}_s$ i.e.:
\be
  \bs{y} (\bs{\mu}_s + \bs{W}_s \bs{\Theta}+\bet)  = \bs{y}(\bs{\mu}_s) + \bs{G}_s \left( \bs{W}_s\bs{\Theta} + \bet \right)+ \mathcal{O}(||\bs{W}_s\bt+\bet||^2)         \label{eq:FirstTaylorDisp}
\ee
where $ \bs{G}_s=\frac{\pa \bs{y} }{\pa \bs{\Psi} }|_{\bpsi=\bs{\mu}_s}$ is the gradient of the map at $\bs{\mu}_s$. We will discuss rigorous validation strategies of the approximation error thus introduced in \refsec{sec:IS}. Truncating \refeq{eq:FirstTaylorDisp} to first-order, the term $|| \hat{\bs{y}}- \bs{y}(\bs{\mu}_s + \bs{W}_s \bt +\bet )||^2$ in the exponent of the likelihood becomes:
\be
\begin{array}{ll}
    || \hat{\bs{y}}- \bs{y}(\bs{\mu}_s + \bs{W}_s \bt+\bet)||^2 &= || \hat{\bs{y}}- \bs{y}(\bs{\mu}_s) - \bs{G}_s \bs{W}_s\bt -\bs{G}_s \bet ||^2 \\
	  & = || \hat{\bs{y}}- \bs{y}(\bs{\mu}_s)||^2-2(\hat{\bs{y}}- \bs{y}(\bs{\mu}_s))^T \bs{G}_s \bs{W}_s\bt\\
	  & +\bs{W}_s^T \bs{G}_s^T \bs{G}_s \bs{W}_s: \bt \bt^T \\
	  & -2 \bet^T \bs{G}_s^T \left( \hat{\bs{y}}- \bs{y}(\bs{\mu}_s) -\bs{G}_s\bs{W}_s \bt \right) \\
	  & + \bet^T \bs{G}_s^T \bs{G}_s \bet.
    \label{eq:l1}
\end{array}
\ee
We introduce a second {\em approximation} in terms of the family of $q$'s over which we wish to optimize by using a {\em mean-field} decomposition  (\cite{peierls_minimum_1938,opper_advanced_2001}) of the form:
\be 
\begin{array}{ll}
  q(\bs{\Theta}, s, \tau ) & \approx q(\bs{\Theta}, \bet,s)q(\tau) \\
  & =q(\bs{\Theta}, \bet|s)q(s)~q(\tau)  \\
  & \approx q(\bs{\Theta}| s)q(\bet | s) q(s) q(\tau).
  \end{array}
\label{eq:MeanFieldApp}
\ee
In the first line, $\tau$ is assumed to be a posteriori independent of the remaining latent variables on the premise that the measurement noise precision is determined by the experimental conditions and is not directly dependent on the latent variables. In the the third line,  we assume that $\bt$ and $\bet$ are {\em conditionally} independent given $s$\footnote{This implies that $\bt$ and $\bet$ are actually dependent as one would expect.}. 
The latter assumption is justified by the role of $\bt$ and $\bet$ in the representation of $\bpsi$ (\refeq{eq:psis}) expressing the main effects around the mean and the residual ``noise'' respectively. As such, it is also  reasonable to  assume that the means of $\bt$ and $\bet$ are zero a posteriori i.e. $E_{q(\bt|s)}[\bt]=\bs{0}$ and $E_{q(\bet|s)}[\bet]=\bs{0}$. Furthermore we employ an {\em isotropic}  covariance for $\bet$  i.e. $E_{q(\bet|s)}[\bet \bet^T]=\lambda_{\eta,s}^{-1} \bs{I}_{\dpsi}$ where $\lambda_{\eta,s}^{-1}$ represents the (unknown) variance. 

If we denote the expectations with respect to $q(\tau)$, $q(\bt|s)$ and $q(\bet|s)$ with $<.>_{\tau}$, $<.>_{\bt|s}$ and $<.>_{\bet|s}$, then \refeq{eq:fvarhat} becomes \footnote{we omit constants that do not affect the optimization.}:
\be
\begin{array}{llr}
\mathcal{\hat{F}}(q(\bt, \bet, \tau, s), \bs{T}) 
& =  \frac{d_y}{2}< \log \tau>_{\tau} &  (\textrm{\em \footnotesize  $<\log p(\bs{\hat{y}} | s, \bt, \bet, \tau, \bs{T})>$})\\
	  & - \frac{<\tau >_{\tau}}{2} \sum_s q(s) || \hat{\bs{y}}- \bs{y}(\bs{\mu}_s)||^2  & \\
	  &  +<\tau >_{\tau} \sum_s q(s) (\hat{\bs{y}}- \bs{y}(\bs{\mu}_s))^T \bs{G}_s \bs{W}_s<\bt>_{\bt|s} & (\textrm{\em \footnotesize $=0$ since $ <\bt>_{\bt|s}=\bs{0}$ }) \\ 
	  & - \frac{<\tau >_{\tau}}{2} \sum_s q(s) \bs{W}_s^T \bs{G}_s^T \bs{G}_s \bs{W}_s: <\bt \bt^T >_{\bt|s} &  \\
	  &  +<\tau>_{\tau} \sum_s q(s) <\bet>^T_{\bet|s} \bs{G}_s^T ( \hat{\bs{y}}- \bs{y}(\bs{\mu}_s)) & (\textrm{\em \footnotesize $=0$ since $ <\bet>_{\bet|s}=\bs{0}$ })  \\
	  & -<\tau>_{\tau} \sum_s q(s) <\bet>^T_{\bet|s} \bs{G}_s^T \bs{G}_s\bs{W}_s <\bt>_{\bt|s}  &   (\textrm{\em \footnotesize $=0$  since $ <\bet>_{\bet|s}=\bs{0}$ })\\ 
	  & - \frac{<\tau >_{\tau}}{2} \sum_s q(s)\bs{G}_s^T \bs{G}_s : <\bet\bet^T>_{\bet|s} & \\
	  & + \sum_s q(s)\log \frac{1}{S}  					&  (\textrm{\em \footnotesize  $<\log p_s(s)>$}) \\
 	  & + (a_0-1)<\log \tau>_{\tau} -b_0 <\tau>_{\tau}  			& (\textrm{\em \footnotesize  $<\log p_{\tau}(\tau)>$}) \\
 	  & + \sum_s q(s) (\frac{1}{2} \log | \bs{\Lambda}_{0,s} |-\frac{1}{2}  \bs{\Lambda}_0: <\bt \bt^T>_{\bt|s}) & (\textrm{\em \footnotesize  $<\log p_{\Theta}(\bt|s)>$}) \\
 	  & + \sum_s q(s) (\frac{\dpsi}{2} \log \lambda_{0,\eta,s} -\frac{\lambda_{0,\eta,s}}{2} \bs{I}:<\bet \bet^T>_{\bet|s}) 	& (\textrm{\em \footnotesize  $<\log p_{\eta}(\bet|s)>$}) \\
 	  & - \sum_s q(s) \int q(\bt |s) \log q(\bt|s)~d\bt & (\textrm{\em \footnotesize  $-<\log q(\bt| s)>$}) \\
 	  & - \sum_s q(s) \int q(\bet |s) \log q(\bet|s)~d\bt & (\textrm{\em \footnotesize  $-<\log q(\bet| s)>$}) \\
 	  & - \sum_s q(s) \log q(s) & (\textrm{\em \footnotesize  $-<\log q(s)>$})  \\
 	  & - <\log q(\tau) >_{\tau}. & (\textrm{\em \footnotesize  $-<\log q(\tau)>$}) 	 
\end{array}
\label{eq:fvarhatElaborated}
\ee

Despite the long expression, the optimization of $\hat{\mathcal{F}}$ in the \textbf{VB-Expectation} step can be done {\em analytically} and we find that the optimal $q$ (given $\bs{T}$) is:
\be
    \begin{array}{l}
    q^{opt}(\bt|s) \equiv \mathcal{N}(\bs{0}, \bs{\Lambda}_s^{-1}) \\
    q^{opt}(\bet|s) \equiv \mathcal{N}(\bs{0}, \lambda_{\eta,s}^{-1} \bs{I}_{\dpsi})  \\
    q^{opt}(\tau)\equiv Gamma(a,b)
    \end{array}
    \label{eq:qopt1}
\ee
where:
\be
    \boldsymbol{\Lambda}_s = \boldsymbol{\Lambda}_{0,s}+<\tau>_{\tau} \boldsymbol{W}_s^T \boldsymbol{G}_s^T \boldsymbol{G}_s \boldsymbol{W}_s 
    \label{eq:thetaupd}
\ee
\be
    \lambda_{\eta,s}=\lambda_{0,\eta,s}+\frac{1}{\dpsi} <\tau>_{\tau} trace(\bs{G}_s^T \bs{G}_s)
    \label{eq:etaupd}
\ee
\be
  a = a_0 + d_y/2  
    \label{eq:taua}
\ee
\be 
  b = b_0 + \frac{1}{2}\sum_{s=1}^S q(s) \left( ||\bs{\hat{y}} - \bs{y}(\bs{\mu}_s)||^2 + \bs{W}_s^T \bs{G}_s^T \bs{G}_s \bs{W}_s:  \bs{\Lambda}_s^{-1} +\lambda_{\eta,s}^{-1}~trace(\bs{G}_s^T\bs{G}_s)\right).
  \label{eq:taub}
\ee
Furthermore, for the latent variable $s$ we find that:
\be
  q^{opt}(s) \propto e^{ c_s }
\ee
  where:
\be
  c_s=\frac{1}{2} \log \frac{|\bs{\Lambda}_{0,s}|}{|\bs{\Lambda}_{s}|} + \frac{\dpsi}{2}\log \frac{\lambda_{0,\eta,s}}{\lambda_{\eta,s}} - \frac{<\tau>_{\tau}}{2} ||\hat{\boldsymbol{y}} - \boldsymbol{y}(\boldsymbol{\mu}_{s})||^2 \label{eq:qs}
\ee
and  $<\tau>_{\tau}=E_{q(\tau)}[\tau]=\frac{a}{b}$. 
The  normalization constant for $q(s)$  can be readily found by imposing the condition $\sum_{s=1}^S q^{opt}(s) =1$ which yields:
\be
  q^{opt}(s)=\frac{ e^{c_s}}{\sum_{s'} e^{c_{s'}} }.
\label{eq:qs}
\ee
While the optimal $q'$s are inter-dependent, we note in the expression above that the posterior probability of each mixture component $s$, as one would expect, increases as the mean-square error $||\hat{\boldsymbol{y}} - \boldsymbol{y}(\boldsymbol{\mu}_{s})||^2$ gets smaller. More interestingly perhaps, $q^{opt}(s)$ increases as the determinant of the posterior precision matrix $\bs{\Lambda}_{s}$ decreases i.e. as the posterior variance associated with the reduced coordinates $\bt$  of component $s$ increases. The same effect is observed for the posterior residual variance $\lambda_{\eta,s}^{-1}$. This implies that, ceteris paribus, mixture components with larger posterior variance will have a bigger weight in the overall posterior.


For the optimal $q^{opt}$ (given $\bs{T}$) in the equations above, the variational lower bound $\hat{\mathcal{F}}$ takes the following form (terms independent of $q^{opt}$ or $\bs{T}$ are omitted - for details see \ref{app:Lowerbound}):
\be
  \begin{split}
     \mathcal{\hat{F}} (q^{opt},\bs{T}) =& \sum_{s=1}^S  q^{opt}(s)  \bigg(  - \frac{<\tau >_{\tau}}{2}|| \hat{\bs{y}}- \bs{y}(\bs{\mu}_s)||^2  
		  +   \frac{1}{2} \log\frac{| \bs{\Lambda}_{0,s}|}{| \bs{\Lambda}_s|}  +  \frac{\dpsi}{2} \log \frac{\lambda_{0,\eta,s}}{\lambda_{\eta,s}}  -  \log q^{opt}(s)  \bigg)  \\
	      & +a \log(<\tau>_{\tau})
     \label{eq:lowerboundhat}
  \end{split}
\ee
%
and
\be
     \mathcal{F}(q^{opt}, \bs{T}) = \mathcal{\hat{F}}(q^{opt},\bs{T}) + \log p_T(\bs{T})
     \label{eq:lowerbound}
\ee
where $Z(a,b)=\frac{\Gamma(a)}{b^{a}}$ is the normalization constant of a $Gamma$ distribution with parameters $a,b$. This can be computed at each full iteration of VB-EM in order to monitor convergence. 

While it is difficult again to gain insight in the expression above due to the inter-dependencies between the various terms, we note that the smaller the mean-square error of $||\hat{\boldsymbol{y}} - \boldsymbol{y}(\boldsymbol{\mu}_{s})||^2$ becomes (i.e. the better the mean $\bmu_s$ is able to reproduce the measurements), the more the lower bound increases. In addition we can see that the lower bound increases as the variance of the mixture components $\bs{\Lambda}_s^{-1}, \lambda_{\eta,s}^{-1}$ gets larger, meaning the more variance they can capture.

For the \textbf{VB-Maximization} step, it can be readily established from \refeq{eq:fvarhatElaborated} that the optimization of $\mathcal{F}$ with respect to $\bs{\mu}$ (given $q$) involves the following set of {\em uncoupled} optimization problems \cite{franck_sparse_2015}:
\be
    \underset{\boldsymbol{\mu}_j}{max} \,  \mathcal{F}_{\mu_j} =  -\frac{<\tau>}{2} ||\bs{\hat{y}} - \boldsymbol{y}(\boldsymbol{\mu}_j)||^2  + \log p(\bs{\mu}_j), \quad j=1,\ldots, S.
    \label{eq:muupd}
\ee
Since the objectives are identical for each $j$, we can deduce that $\bs{\mu}_j$ should correspond to (different or identical) local maxima of $\mathcal{F}$. This implies that in the posterior approximation constructed, each Gaussian in the mixture is  associated with a (regularized - due to the prior) local optimum in the least-square solution of the inverse problem. The search for multiple local optima, and more importantly their  number, is discussed in the next section.

The determination of the optimal $\bmu_j$ is performed using first-order derivatives of $\frac{\pa \mathcal{F}_{\mu_j}}{\pa  \bmu_j}$. 
Since  $\log p(\bs{\mu}_j)$ and its derivative $\frac{\pa \log p(\bs{\mu}_j)}{\pa \bmu_j}$ are analytically unavailable, we employ an additional layer (inner loop) of Expectation-Maximization to deal with the hyperparameters in the prior of $\bs{\mu}_j$. The details were discussed in \cite{franck_sparse_2015} and are included in \ref{app:muPrior} for completeness.


Considering the \textit{computational cost} of these operations,  we point out that the updates of $\bmu_j$ are the most demanding  as they require calls to the forward model to evaluate $\bs{y}(\bmu_j^{(n)})$ and the derivatives $\bs{G}^{(n)}_j = \frac{\pa \bs{y}}{\pa \bpsi_j}|_{\bpsi_j=\mu_j^{(n)}}$. For the computation of the derivatives $\bs{G}_j$ we employ the adjoint formulations which offer great savings when $\dpsi \gg 1$ and $\dpsi \gg d_y$ \cite{papadimitriou_direct_2008}. As discussed in detail in \cite{franck_sparse_2015}, the latter condition can be removed as long as a direct solver is used for the solution of the forward problem. In this case, the cost of the solution of the adjoint equations is even less that that of the forward solution.


The remaining aspect of the \textbf{VB-Maximization} step, involves the optimization with respect to the $\bW$ (given $q$). As with $\bmu$, it suffices to consider only the 
 the terms in \refeq{eq:fvarhatElaborated} that depend on $\bW$ (which we denote by $\mathcal{F}_{W_j}$) and which again lead to a set of $S$ uncoupled problems:
\be
  \underset{\boldsymbol{W}_j}{max} \,  \mathcal{F}_{W_j}  = -\frac{<\tau>}{2}\boldsymbol{W}_j^T \boldsymbol{G}_j^T\boldsymbol{G}_j \boldsymbol{W}_j : \boldsymbol{\Lambda}_j^{-1}+ \log p(\bs{W}_j), \quad j=1,\ldots, S.
  \label{eq:wupd}
\ee
The first term prefers directions corresponding to the smallest eigenvectors of $\bs{G}_j^T\bs{G}_j$, where $ \bs{G}_j=\frac{\pa \bs{y} }{\pa \bs{\Psi}_j }|_{\bpsi_j=\bs{\mu}_j}$ is the gradient of the map at $\bs{\mu}_j$.
As discussed previously in \refsec{sec:Prior}, the prior $p(\bs{W}_j)$ enforces the orthogonality of the basis vectors in $\bs{W}_j$. To solve this constrained optimization problem, we use the iterative algorithm of \cite{wen_feasible_2013}, which employs a Cayley transform to enforce the constraint. It makes use of first-order derivatives of $\mathcal{F}_{W_j}$ and as such does not required any additional forward model runs.

%


With regards to the number of columns $\dth$ in $\bs{W}_j$ (which is equal to the dimension of $\bt$), we assume that this is the same across all mixture components $S$. We had developed an information-theoretic criterion in \cite{franck_sparse_2015} which can also be employed here. This allows the adaptive determination of $\dth$ by measuring the information gain, here denoted by $I(\dth,j)$ for each mixture component $j$, that each new  dimension in $\bt$ furnishes. When these fall below a
threshold $I_{max}$ (in our examples we use $I_{max} = 1\%$) i.e.:
\be
 I(\dth)=   \max_j I(\dth,j) ~{\leq} ~I_{max}
 \label{eq:infgain}
\ee
we assume that the number of $\bt$ is sufficient. A detailed discussion on the estimation of $\dth$ using the information gain criteria $I(\dth,j)$ is given in \ref{app:IGBases} and \cite{franck_sparse_2015}.

%
Following the previous discussion in \refeq{eq:postmix}, we note that once the (approximate) posterior $q(\bt, \bet, \tau, s)$ and the optimal model parameters $\bs{T}$ have been computed, we obtain a {\em multimodal} posterior approximation for the material parameters $\bpsi$, which is given by:
\be
\begin{array}{ll}
 p(\bpsi | \bs{\mu},\bs{W}, \bs{\hat{y}}, \bs{T} ) & = \sum_{s=1}^S \int  \delta(\bpsi - (\bs{\mu}_s+\bs{W}_s \bt + \bet))~p( s, \bt, \bet, \tau, | \bs{\mu},\bs{W}, \bs{\hat{y}} )~d\bt~d\bet \\
 & \approx \sum_{s=1}^S \int  \delta(\bpsi - (\bs{\mu}_s+\bs{W}_s \bt + \bet))~q( s, \bt, \bet,   )~d\bt~d\bet \\
 & = \sum_{s=1}^S q(s) \int  \delta(\bpsi - (\bs{\mu}_s+\bs{W}_s \bt + \bet))~q( \bt, \bet|s)~d\bt~d\bet \\
 & =\sum_{s=1}^S q(s) ~q_s(\bpsi) =q(\bpsi)
 \end{array}
\label{eq:qpost}
\ee
where each component in the last mixture is given by:
\be
\begin{array}{ll}
q_s(\bpsi) & =\int  \delta(\bpsi - (\bs{\mu}_s+\bs{W}_s \bt + \bet))~q( \bt, \bet|s)~d\bt~d\bet \\
& \equiv \mathcal{N}(\bs{\mu}_s, \bs{D}_s),
\label{eq:postmix2}
\end{array}
\ee
i.e. a multivariate Gaussian with mean $\bs{\mu}_s$ and covariance $\bs{D}_s$ where:
\be
\bs{D}_s=\bs{W}_s\bs{\Lambda}_s^{-1} \bs{W}_s^T+\lambda_{\eta,s}^{-1} \bs{I}_{\dpsi}.
\label{eq:postcov}
\ee

Based on \refeq{eq:qpost}, one can evaluate the posterior mean and covariance of $\bpsi$ as follows:
\be
      \begin{array}{ll}
      E_q[\bpsi] & =  E\left[ E_q[\bpsi | s]\right] = E\left[ \bs{\mu}_s\right]= \sum_{s=1}^S q(s)  \bs{\mu}_s \\
      Cov_q[\bpsi] & =E_q[\bpsi \bpsi^T]- E_q[\bpsi] E_q^T[\bpsi] = E\left[ E_q[\bpsi \bpsi^T|s] \right] - E_q[\bpsi] E_q^T[\bpsi] \\
      & = \sum_{s=1}^S q(s) (\bs{D}_s+\bs{\mu}_s\bs{\mu}_s^T)-\left(\sum_{s=1}^S q(s)  \bs{\mu}_s \right)\left(\sum_{s=1}^S q(s)  \bs{\mu}_s \right)^T \\
      & =  \sum_{s=1}^S q(s)\bs{D}_s+ \sum_{s=1}^S q(s)  \bs{\mu}_s  \bs{\mu}_s^T -\left(\sum_{s=1}^S q(s)  \bs{\mu}_s \right)\left(\sum_{s=1}^S q(s)  \bs{\mu}_s^T \right).
      \end{array}
      \label{eq:qpost1}
\ee
Posterior moments of any order or posterior probabilities can be readily computed as well.

Note that if $\bs{\Lambda}_s^{-1}$ is diagonalized, e.g. $\bs{\Lambda}_s^{-1} =\bs{U}_s \bs{\hat{\Lambda}}_s^{-1} \bs{U}_s^T$ where $\bs{\hat{\Lambda}}_s^{-1}$ is diagonal and $\bs{U}_s$ contains the eigenvectors of $\bs{\Lambda}_s^{-1}$ then:
\be
  \begin{split}
      \bs{D}_s &=\bs{W}_s\bs{U}_s \bs{\hat{\Lambda}}_s^{-1} \bs{U}_s^T  \bs{W}_s^T+\lambda_{\eta,s}^{-1} \bs{I}_{\dpsi} \\
	       &= \hat{\bs{W}}_s  \bs{\hat{\Lambda}}_s^{-1} \hat{\bs{W}}_s^T+\lambda_{\eta,s}^{-1} \bs{I}_{\dpsi}.
  \end{split}
\ee
Each $\hat{\bs{W}}_s$ is also orthogonal and contains the $\dth$ principal directions of posterior covariance of $\bpsi_s$. We see therefore, that in the VB-E step, it suffices to consider an approximate posterior $q(\btheta |s)$ with a diagonal covariance, e.g. $\bs{\Lambda}_s = diag(\lambda_{s,i}), i=1,...,\dth$. As a consequence the update equation for $\bs{\Lambda}_s$ (\refeq{eq:thetaupd}) reduces to:
\be
    \lambda_{s,i} = \lambda_{0,s,i} + <\tau>_{\tau} \boldsymbol{w}_{s,i}^T \boldsymbol{G}_s^T \boldsymbol{G}_s \boldsymbol{w}_{s,i}
    \label{eq:thetaupd2}
\ee
where $\boldsymbol{w}_{s,i}$ is the $i^{th}$ column vector of $\bs{W}_s$.




We note that in all the aforementioned expressions we assumed that the number of components $S$ is given and fixed. Nevertheless, if for some $s$, $q^{opt}(s)$ is zero (or negligible), the corresponding component will have no (posterior) contribution. In Algorithm \ref{alg:FixedSAlgorithm} we  summarize the main steps of the algorithm for a fixed  $S$. Steps $5-7$ correspond to the aforementioned VB-Expectation and steps $2$ and $4$ to the VB-Maximization step. In the next section we discuss an adaptive strategy for determining $S$. 

\begin{algorithm}[H]
\caption{Algorithm for fixed $S$} \label{alg:FixedSAlgorithm}
    \begin{algorithmic}[1]
    \While{$\mathcal{F}_{\mu}$ in \refeq{eq:muupd}, has not converged}
      \State For $j=1:S$: Optimize $\bs{\mu}_j$ using \refeq{eq:muupd}
      \While{$\mathcal{F}$ in  \refeq{eq:lowerbound}, has not converged}
	  \State For $j=1:S$: Optimize $\bs{W}_j$ using \refeq{eq:wupd}
	  \State For $s=1:S$: Update $q(\bt|s)\equiv \mathcal{N}(\bs{0}, \bs{\Lambda}_s^{-1})$ using \refeq{eq:thetaupd2} 
	  \State For $s=1:S$: Update $q(\bet|s)\equiv \mathcal{N}(\bs{0}, \lambda_{\eta,s}^{-1}\bs{I}_{\dpsi})$ using \refeq{eq:etaupd}
	  \State Update  $q(\tau) \equiv Gamma(a,b)$  and $q(s)$ using Equations (\ref{eq:taua}, \ref{eq:taub}, \ref{eq:qs})
      \EndWhile
    \EndWhile
    \end{algorithmic}
\end{algorithm}
\tikzstyle{decision} = [diamond, aspect=2.5, draw, fill=blue!20, 
    text width=5.5em, text badly centered, node distance=2.5cm, inner sep=0pt] 
\tikzstyle{block} = [rectangle, draw, fill=blue!20,
    text width=15em, text centered, rounded corners, minimum height=3em]
\tikzstyle{blocksmall} = [rectangle, draw, fill=blue!20, 
    text width=8em, text centered, rounded corners, minimum height=2em]  
\tikzstyle{blockverysmall} = [rectangle, draw, fill=blue!20, 
    text width=5em, text centered, rounded corners, minimum height=0.6em]   
\tikzstyle{blockVB} = [rectangle, draw, fill=green!20, 
    text width=3.5em, text centered,  minimum height=5em]       
\tikzstyle{line} = [draw, very thick, color=black!50, -latex']
\tikzstyle{cloud} = [draw, ellipse,fill=red!20, node distance=2.5cm,
    minimum height=2em]
\tikzstyle{fancytitle} =[fill=green!40, text=black]

\tikzstyle{decision} = [diamond, aspect=2.5, draw, fill=blue!20, 
    text width=5.5em, text badly centered, node distance=2.5cm, inner sep=0pt] 
\tikzstyle{block} = [rectangle, draw, fill=blue!20,
    text width=15em, text centered, rounded corners, minimum height=3em]
\tikzstyle{blocksmall} = [rectangle, draw, fill=blue!20, 
    text width=8em, text centered, rounded corners, minimum height=2em]  
\tikzstyle{blockverysmall} = [rectangle, draw, fill=blue!20, 
    text width=5em, text centered, rounded corners, minimum height=0.6em]   
\tikzstyle{blockVB} = [rectangle, draw, fill=green!20, 
    text width=3.5em, text centered,  minimum height=5em]       
\tikzstyle{line} = [draw, very thick, color=black!50, -latex']
\tikzstyle{cloud} = [draw, ellipse,fill=red!20, node distance=2.5cm,
    minimum height=2em]
\tikzstyle{fancytitle} =[fill=green!40, text=black]

\subsection{Finding the required number of mixture components $S$}\label{sec:NrMoG}

A critical component of the framework proposed is the cardinality $S$ of the model  i.e. the number of modes in the approximation of the posterior. The mean $\bs{\mu}_j$ of each Gaussian component is optimal when it corresponds to a local maximum of the objective in \refeq{eq:muupd}, but suboptimal solutions can be found by using suboptimal $\bs{\mu}_j$. A consistent way of carrying out this model selection task, within the advocated Bayesian framework, is to compute or approximate the model evidence term $p(\bs{\hat{y}})$ in \refeq{eq:jointpost1} for various values of $S$. This can be followed by selecting the one that gives the largest $p(\bs{\hat{y}})$ or performing model averaging with probabilities proportional to these terms for each  values of $S$ \cite{bishop_pattern_2006, beal_variational_2003}. Nevertheless computing  $p(\bs{\hat{y}})$ is impractical  as it requires integrating over all parameters including the high-dimensional $\bs{T}$ i.e. a fully Bayesian treatment of the $\bs{\mu}$ and $\bs{W}$. In the formulation 
presented thus far however, we computed point estimates by maximizing the variational bound $\mathcal{F}$ to the log posterior $p(\bs{T} | \bs{\hat{y}})$ (\refeq{eq:loglike1}). One might be inclined to compute this  $\mathcal{F}$ (assuming it is a good approximation of $p(\bs{T} | \bs{\hat{y}})$) for different values of $S$ and use it to identify the optimal $S$. We note though that such terms are not comparable as they depend on the number of parameters in $\bs{T}$ which changes with $S$. As a result such comparisons would be meaningless. One could potentially employ one of the well-known approximate validation metrics, e.g. AIC or BIC,  which penalize the log posterior ($p(\bs{T} | \bs{\hat{y}})$ or $\mathcal{F}$) with the number of parameters, but these are known to be  only in limiting cases valid for large data  \cite{beal_variational_2003,robert_bayesian_2007}.

 Furthermore, we note that if two components ($S=2$) with the same $\bs{\mu}_1=\bs{\mu}_2$ (and as a result $\bs{G}_1=\bs{G}_2$, and $\bs{W}_1=\bs{W}_2$, $\bs{\Lambda}_1=\bs{\Lambda}_2$) are considered, then $q(s=1)=q(s=2)=\frac{1}{2}$. Even though a mixture of these two identical components gives rise to a single Gaussian (\refeq{eq:postmix2}), it is obvious  that the second component  provides no new information regarding the  posterior. This is because the posterior $p(s | \bs{\hat{y}})$ (and its approximation $q(s)$) accounts for the {\em relative} plausibility (as compared to the other components) that the component $s$ could have given rise to a $\bpsi$ (that in turn gave rise to $\bs{y}(\bpsi)$) that matches the observations $\bs{\hat{y}}$.  Hence $s$ is in a sense a hyper-parameter in the prior specification of $\bs{\Psi}$. 
 For this purpose, we propose an adaptive algorithm (Algorithm \ref{alg:adapt}) that proposes new components (component birth) and removes those (component death) that do not furnish new information. 

 \begin{algorithm}[H]
\caption{Adaptive algorithm for the determination of appropriate $S$} \label{alg:adapt}
\begin{algorithmic}[1]
\State Initialize with $S=S_0$ (e.g. $S_0=1$), $L=0$, $iter=0$ and call Algorithm \ref{alg:FixedSAlgorithm}.
\While{$L < L_{max}$} 
    \State $iter \leftarrow  iter+1$
    \State (Component Birth) Propose $\Delta S$ new mixture components and initialize $\mu_j$ for $j=S+1,\ldots, S+\Delta S$ according to \refeq{eq:muinit} 
    \State Call Algorithm \ref{alg:FixedSAlgorithm}
    \State (Component Death) Delete any of the new components that satisfy the component death criterion in \refeq{eq:compdeath1} 
    \State Compute $q(s)$ of surviving components (\refeq{eq:qs}), remove any components with $q(s) < q_{min}$  \footnotemark   and update $S$ 
    \If{None of the $\Delta S$ new components remain active}
	\State $L \leftarrow L+1$;
    \Else
	\State $L\leftarrow0$;
    \EndIf	
\EndWhile
\end{algorithmic}
\end{algorithm}
 \footnotetext{Throughout this work we use $q_{min}=1\times 10^{-3}$.}

 We discuss in detail the steps above that contain new features as compared to Algorithm \ref{alg:FixedSAlgorithm}:
 \bi
  \item Steps 2 and 8-12:\\
      The overall algorithm is terminated when $L_{max}$ successive attempts to add new mixture components have failed (in all examples discussed $L_{max}=3$). $L$ counts the number of successive failed attempts to add new components and $iter$ the total number of component birth attempts. During each of those, $\Delta S$ new mixture components are proposed (component birth) and optimized. Since the $\bs{\mu}$-updates of each mixture component imply a certain number of forward model solutions, the termination criterion could be alternatively expressed in terms of the maximum allowable number of such forward calls. \footnote{See \reffig{fig:FlowchartMoGToy} where for $iter=2$, none of the $\Delta S=3$ mixture components survive. Here $L$ increases from $0$ to $1$. 
      }

  \item Step 4 (Component Birth):\label{sec:Step4}\\
       Given $S$ mixture components, we propose the addition of $\Delta S$ new components. Their means $\bs{\mu}_{j_{new}}$, for $j_{new} = S+1, \ldots, S+\Delta S$ are initialized by perturbing the mean of one of the pre-existing $S$ components as follows: We pick the mixture component $j_{parent} \in {1,\ldots,S}$ that has the smallest contribution in the lower bound $\mathcal{\hat{F}}$ in \refeq{eq:lowerbound} and therefore provides the worst fit to the data \footnote{If a specific mixture component has already been used as a parent in a previous unsuccessful attempt, the next worst mixture component is used.}. 
       We initialize  $\bmu_{j_{new}}$ randomly as follows:
       \be
	    \bs{\mu}_{j_{new}}= \bs{\mu}_{j_{parent}}+  \bW_{j_{parent}} \bt+ \alpha  \bs{\eta}
	    \label{eq:muinit}
      \ee
      where $\bt$ is sampled from the posterior $q(\bt|s=j_{parent})$ and $\bs{\eta}$ is sampled from $q(\bs{\eta}| s=j_{parent})$. The role of $\alpha$ is to amplify the perturbations. The value of $\alpha=10$ was used throughout this work. Very large  $\alpha$ increase the possibility of finding a new mode but increase the number of $\bs{\mu}-$updates and therefore the number of forward model calls. 
      The remaining  model parameters for each new component are initialized  based on the parent and are updated with according to  the VB-EM scheme discussed in Step 5.

%
 
  \item Step 5: \\
      Whereas the VB-EM scheme discussed in the previous section has to be run every time new components are proposed (i.e. $S$ changes), we note here that the updates for the pre-existing  components require only very few new (if any) forward-model runs. This is because updates for pre-existing $\bs{\mu}_j$ (\refeq{eq:muupd}) are only required if $<\tau>$ changes. While $<\tau>$ is affected by all components $S$ (old and new, \refeq{eq:taub}), it generally does not change significantly after the first few components. 
 
  \item Step 6 (Component Death): \label{sec:Step7}
      \\We employ an information-theoretic criterion that measures the discrepancy ('similarity distance') $d_{j_{old},j_{new}}$  between a new component $j_{new} \in \{S+1,\ldots,S+\Delta S\} $  and an existing one $j_{old} \in \{1,\ldots,S\}$. If this is smaller than a prescribed threshold $d_{min}$, {\em for any of the existing components $j_{old}$}, then the component $j_{new}$ is removed as the two mixture components are too close to each other. In other words, the component death criterion may be stated as:
      \be
	  \textrm{if $\exists j_{old}$ such that $d_{j_{old},j_{new}} < d_{min}$}.
	  \label{eq:compdeath1}
      \ee
      Throughout this work, we use $d_{min}=0.01$\footnote{Nevertheless, as shown in the numerical examples, a much lower value of $d_{min}=10^{-8}$ would have yielded identical results.} and define $d_{j_{old},j_{new}}$ as follows:
      \be
	  d_{j_{old},j_{new}}=\frac{ KL(q_{j_{old}} || q_{j_{new}} ) }{\dpsi}
	  \label{eq:kldist}
      \ee
      where the $KL$ divergence between two multivariate Gaussians $q_{j_{old}}(\bpsi)$ and $q_{j_{new}}(\bpsi)$ (\refeq{eq:postmix2}) can be analytically computed as:
      \be
	  \begin{array}{ll}
	  KL(q_{j_{old}}(\bpsi) || q_{j_{new}}(\bpsi)) = & \frac{1}{2} \log |\bs{D}_{j_{new}}|+  \frac{1}{2} \bs{D}_{j_{new}}^{-1}:\bs{D}_{j_{old}} \\	 		
	  & +\frac{1}{2}(\bs{\mu}_{j_{old}}-\bs{\mu}_{j_{new}})^T \bs{D}_{j_{new}}^{-1} (\bs{\mu}_{j_{old}}-\bs{\mu}_{j_{new}})  \\
	  & -\frac{1}{2}  \log |\bs{D}_{j_{old}}|-\frac{\dpsi}{2}.
	  \end{array}
      \ee
       We note that such a discrepancy metric takes into account the whole distribution and not just the locations of the modes $\bs{\mu}_{j_{old}}$, $\bs{\mu}_{j_{new}}$.  
       The denominator $\dpsi$ of the KL divergence normalizes it with respect to the number of dimensions and therefore $d_{j_{old},j_{new}}$ is the average KL divergence over the $\dpsi$ dimensions. Consequently $d_{min}$ expresses the minimum acceptable averaged KL distance per dimension.
     
\begin{figure}[H]
	  \centering{
	  \captionsetup[subfigure]{labelformat=empty}
		  {\includegraphics[width=0.65\textwidth]{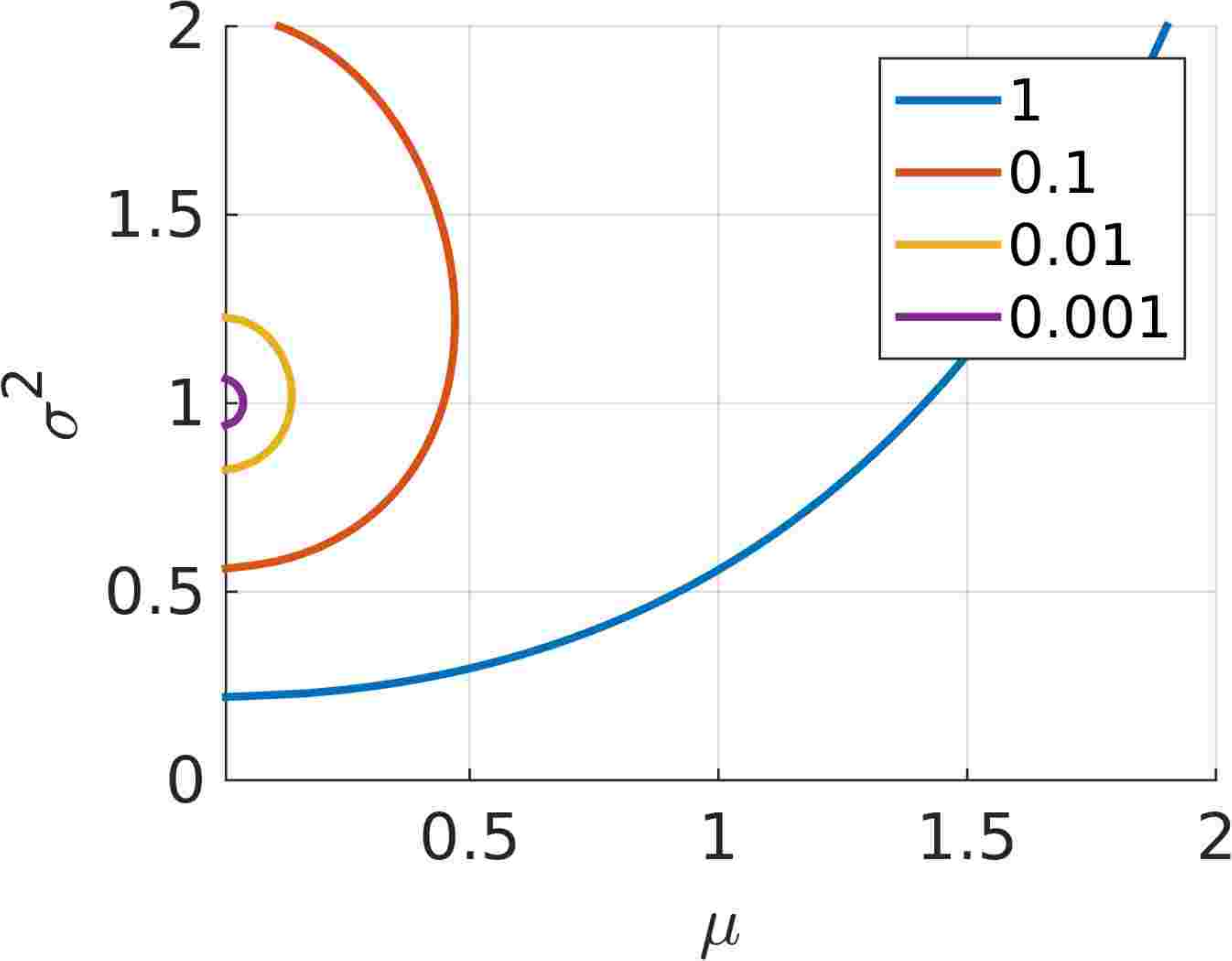}} 
		  \hspace{0.1cm}
	    }
	    \caption{Contour lines of the  KL-divergence between $ \mathcal{N}(0,1)$ and $\mathcal{N}(\mu,\sigma^2)$ with respect to  $\mu,\sigma^2$. Any Gaussian with $( \mu,\sigma^2)$ within the yellow line, would be be deleted according to the criterion defined. }
	  \label{fig:GaussianMixtureKLDiv}
     \end{figure}

       In \reffig{fig:GaussianMixtureKLDiv} we plot for illustration purposes the contour lines of the KL divergence of various one-dimensional Gaussians $\mathcal{N}(\mu, \sigma^2)$ (as a function of their mean $\mu$ and variance $\sigma^2$) with respect to  the   standard Gaussian $\mathcal{N}(0,1)$.
       We note that other distance metrics, e.g. by employing the Fisher information matrix, could have equally been used.
       
     We can take advantage of the low-rank decomposition $\bs{D}_s$ in \refeq{eq:postcov} in order to efficiently compute the inverse of  $\bs{D}_{j_{new}}^{-1}$ as:
      \be
	  \begin{array}{ll}
	  \bs{D}_{s}^{-1} & =\ (\bs{W}_s\bs{\Lambda}_s^{-1} \bs{W}_s^T+\lambda_{\eta,s}^{-1} \bs{I}_{\dpsi})^{-1} \\
	  & = \lambda_{\eta,s} \bs{I}_{\dpsi} - \lambda_{\eta,s}^2 \bs{W}_s  \underbrace{ (\bs{\Lambda}_s +\lambda_{\eta,s} \bs{I})^{-1}}_{diagonal} \bs{W}_s^T.
	  \end{array}
      \ee
      Similarly, the determinants can be readily  computed as:
      \be
      \begin{array}{ll}
      |\bs{D}_{s}| & =|\bs{W}_s\bs{\Lambda}_s^{-1} \bs{W}_s^T+\lambda_{\eta,s}^{-1} \bs{I}_{\dpsi}|\\
      & = | \bs{\Lambda}_s +\lambda_{\eta,s} \bs{I}| ~|\bs{\Lambda}_s^{-1}| ~\lambda_{\eta,s}^{-\dpsi}.
      \end{array}
      \ee
     

%
%

 \ei

\subsection{Validation - Combining VB approximations with Importance Sampling}
\label{sec:IS}

The  framework advocated is based on  two approximations: a)  linearization of the response (\refeq{eq:FirstTaylorDisp}) and, b)  the mean-field decomposition of the approximating distribution (\refeq{eq:MeanFieldApp}). This unavoidably introduces bias and the approximate  posterior  will deviate from the exact.  In order to assess the quality of the approximation but also  to correct for any bias in the posterior estimates, we propose using Importance Sampling (IS) \cite{liu_monte_2001}. In particular, we employ the approximate conditional posterior $q$ as the Importance Sampling density. 
Since the performance of IS can decay rapidly in high dimensions \cite{beal_variational_2003} and  due to the fact that $\bs{\eta}$ has a negligible effect in the inferred posterior, we propose using $p(\bt,s | \bs{\hat{y}}, \bs{T})$ as the target density. According to \refeq{eq:condpost}:
\be
\begin{array}{ll}
p(\bt,s | \bs{\hat{y}}, \bs{T}) &  = \int p(\bt,s, \tau | \bs{\hat{y}}, \bs{T}) ~d\tau \\
 & \propto \int  p( \bs{\hat{y}} |s, \bt, \tau, \bs{T}) p_{\tau}(\tau) ~p_{\Theta}(\bt|s) p_s(s) ~d\tau \\
&  \propto \int \tau^{d_y/2} e^{-\frac{\tau}{2} || \bs{\hat{y}}-\bs{y}(\bs{\mu}_s+\bs{W}_s \bt)||^2 }  p_{\tau}(\tau)  ~d\tau~p_{\Theta}(\bt|s) p_s(s) \\
& = \frac{\Gamma(a_0+ d_y/2)}{(b_0+\frac{ ||\hat{\bs{y}}- \bs{y}(\bs{\mu}_s + \bs{W}_s \btheta )||^2}{2} )^{a_0+d_y/2}} p_{\bt}(\bt|s) p_s(s)
\end{array}
\ee
where the Gamma prior $p_{\tau}(\tau)$ is from \refeq{eq:tauPrior} and MAP estimates of $\bs{\mu}, \bs{W}$ are used. In cases where non-conjugate priors for $\tau$ are employed, the IS procedure detailed here has to be performed in the joint space ($\bt, s, \tau)$.

Given $M$ samples $(\btheta^{(m)}, s^{(m)})$ drawn from the mixture of Gaussians $q(\bt,s)$ in \refeq{eq:qopt1} and \refeq{eq:qs}, IS reduces to computing the unnormalized weights $w^{(m)}$ as follows:
\be
    w^{(m)}= \frac{p(\btheta^{(m)}, s^{(m)} | \bs{\hat{y}}, \bs{T}) }{q(\btheta^{(m)}, s^{(m)})}.
\ee
With $w^{(m)}$ (asymptotically) unbiased estimates, the expectations of any integrable function $g(\bpsi)$ with respect to the exact posterior can be computed as:
\be
    \begin{array}{ll}
    E [ g(\bpsi)] & =\sum_{s=1}^S \int g(\bs{\mu}_s+\bs{W}_s \bt) ~ p(\bt,s | \bs{\hat{y}}, \bs{T}) ~d\bt \\
    & = \sum_{s=1}^S \int g(\bs{\mu}_s+\bs{W}_s \bt) ~ \frac{ p(\bt,s | \bs{\hat{y}}, \bs{T}) }{q(\bt,s)} q(\bt,s) ~d\bt \\
    & =  \sum_{m=1}^M \hat{w}^{(m)} g(\bs{\mu}_{s}+\bs{W}_{s} \btheta^{(m)})
    \end{array}
\ee
where the $\hat{w}^{(m)}$ are the normalized IS weights ($\sum_{m=1}^M \hat{w}^{(m)}=1$):
\be
    \hat{w}^{(m)}= \frac{ w^{(m)}}{ \sum_{m'=1}^M w^{(m')}}.
\ee

%
%
In the following examples we employ estimators such as these to compute the asymptotically (as $M \to \infty$) {\em exact} posterior mean (i.e. $g(\bpsi)=\bpsi$), posterior variances 
as well as posterior quantiles.

Furthermore in order to assess the overall accuracy of the approximation and to provide a  measure of comparison with other inference strategies (past and future),  we report  the (normalized) Effective Sample Size (ESS). This measures the degeneracy in the population of samples as quantified by their variance \cite{liu_blind_1995}:
\be
      ESS=\frac{1}{M} \frac{ (\sum_{m=1}^M w^{(m)} )^2 }{ \sum_{m=1}^M (w^{(m)} )^2 } = \frac{1}{M} \frac{1}{\sum_{m=1}^M  (\hat{w}^{(m)})^2}.
    \label{eq:ess}
\ee
The ESS takes values between $1/M$ and $1$ \cite{liu_monte_2001}. If $q(\bt, s)$ coincides with the exact posterior then all the importance weights $w^{(m)}$ are equal ($\hat{w}^{(m)}=1/M$) and the ESS attains its maximum value of $1$. On the other hand, if $q(\bt,s)$ provides a poor approximation then the expression for the $ESS$ is dominated by the largest weight $w^{(m)}$ and yields $ESS=1/M \to 0$ (as $M \to \infty$).
 The normalized ESS can be compared with that of  MCMC \cite{robert_monte_2013}:
\be
    ESS_{MCMC}= \frac{1}{1+2\sum_{k=1}^M (1-\frac{k}{M})\rho(k)} \to \frac{1}{1+2\sum_{k=1}^{\infty} \rho(k)}
    \label{eq:mcmcess}
\ee
where $\rho(k)$ is the autocovariance between MCMC states that are $k$ steps apart.
Finally, we note that if there are additional modes in the exact posterior that have  not been discovered by $q(\bt,s)$, the ESS could still be misleadingly large (for large but finite sample sizes $M$). This however is a general problem of Monte Carlo-based techniques i.e. they cannot reveal (unless $M\to \infty$) the presence of modes in the target density unless these modes are visited by samples. 
\section{Numerical Illustrations} \label{sec:NumericalExamples}

We consider two numerical illustrations. The primary goal of the first example is to provide insight into the adaptive search algorithm for determining $S$ and for that reason we analyse a one-dimensional, multimodal density. The second example pertains to the motivating application of elastography. We demonstrate how the proposed framework can reveal the presence of multiple modes and, when justified, can identify low-dimensional approximations for each of these modes  with a limited number of forward calls. 
An overview of the most important quantities/dimensions of the following two examples is contained in Table \ref{tab:NumberOverview}. 

\begin{table}[h]
\begin{center}
  \begin{tabular}{  c  c  c }
    \hline
      & Example 1 & Example 2 \\ \hline\hline
    Dimension of observables: 		$d_{y}$ & $1$ & $5100$ \\ \hline
    Dimension of latent variables: 	$\dpsi$ & $1$ & $2500$ \\ \hline
    Dimension of reduced latent variables: $\dth$ & $1$ & $11$ \\ \hline
    No. of forward calls 		& $<200$ & $<1200$ \\
    \hline
 
  \end{tabular}   
  \caption{Summary of the number of observables, forward calls and the dimensionality reduction in the following two examples.}
  \label{tab:NumberOverview}  
\end{center}  
\end{table}


\subsection{Toy Example}

\begin{figure}[H]
	\centering{
	\captionsetup[subfigure]{labelformat=empty}
		{\includegraphics[width=0.85\textwidth]{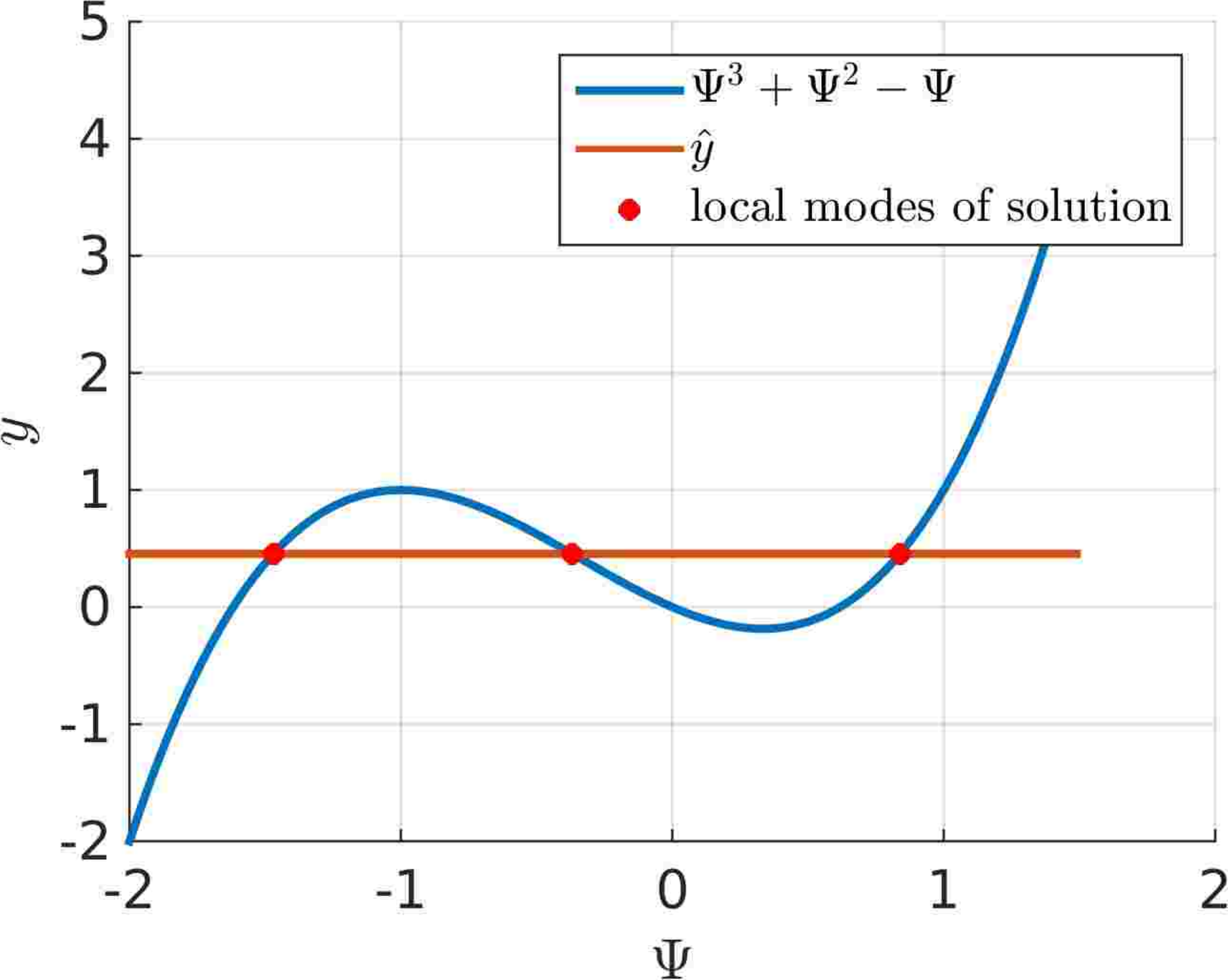}} 
		\hspace{0.1cm}
	  }
	  \caption{Polynomial $y = \Psi^3 +\Psi^2-\Psi$. It can be seen that that for the measurement at $\hat{y} = 0.45$ three possible solutions exist.}
	 \label{fig:Polynomial}
\end{figure}

Our goal in this first example is solely to illustrate  the features and capabilities of the adaptive search algorithm for determining the number of mixture components $S$. For that purpose we selected a one-dimensional example (in order to remove any effects from the dimensionality reduction) that can be semi-analytically investigated and exhibits a multimodal posterior.  We assume that the model equation is of the form:
\be
y(\Psi) = \Psi^3 +\Psi^2-\Psi, \quad \Psi \in \RR
\ee
and is depicted in \reffig{fig:Polynomial}. Let $\Psi_{exact}= 0.8$ be the reference solution for which $y(\Psi_{exact})=0.352$. With the addition of noise it is assumed that the actual measurement is  $\hat{y}=0.45$. This is shown with a horizontal line in \reffig{fig:Polynomial}, where for $\hat{y}=0.45$ three modes for $\Psi$ exist. The Gaussian prior on the $\Psi$ has zero mean and a variance of $\lambda_0 =1\times 10^{-10}$. 

\begin{figure}[!t]
\tikzstyle{decision} = [diamond, aspect=2.5, draw, fill=blue!20, 
    text width=5.5em, text badly centered, node distance=2.5cm, inner sep=0pt] 
\tikzstyle{block} = [rectangle, draw, fill=blue!20,
    text width=15em, text centered, rounded corners, minimum height=3em]
\tikzstyle{blocksmall} = [rectangle, draw, fill=blue!20, 
    text width=8em, text centered, rounded corners, minimum height=2em]  
\tikzstyle{blockVB} = [rectangle, draw, fill=green!20, 
    text width=3.5em, text centered,  minimum height=4em]  
\tikzstyle{blocknumbering} = [rectangle, 
    text width=5.5em, text centered,  minimum height=5em]      
\tikzstyle{line} = [draw, very thick, color=black!50, -latex']
\tikzstyle{lineblue} = [draw, very thick, color=blue!80, -latex']
\tikzstyle{cloud} = [draw, ellipse,fill=red!20, node distance=2.5cm,
    minimum height=2em]
\tikzstyle{fancytitle} =[fill=green!40, text=black]
    \vspace{-2.5cm}
    \hspace{-3.5cm}
    \makebox[\textwidth][c]{ 
\begin{minipage}{0.9\textwidth}    
      \begin{tikzpicture}[scale=2, node distance = 2.5cm, auto] 
	  \node [blocksmall](init) at (2.5,0) {\small initialize values};
	  
	  \node [blockVB] (VB1_1) at (0,-1)  {\small s=1 \\ \scriptsize$\mu=0.84$};
	  \node [blockVB, right of=VB1_1, node distance=3.5cm ] (VB1_2) {\small  s=2\\ \scriptsize$\mu=-0.37$};
	  \node [blockVB, right of=VB1_2, node distance=3.5cm , cross out, draw=red,very thick ] (VB1_3) {\small  s=3 \\ \scriptsize$\mu=0.84$}; 
	  \node [blockVB, right of=VB1_3, node distance=3.5cm , cross out, draw=red,very thick] (VB1_4)  {\small  s=4 \\ \scriptsize$\mu=-0.37$};
	  
	  \node [blockVB] (VB2_1) at (0,-2.5)  {\small  s=1 
	  \\ \scriptsize$\mu=0.84$};	  
	  \node [blockVB, right of=VB2_1] (VB2_2) {\small  s=2 
	  \\\scriptsize$\mu=-0.37$};	  
	  \node [blockVB, right of=VB2_2, node distance=1.8cm, cross out, draw=red,very thick] (VB2_3)  {\small  s=3 \\ \scriptsize$\mu=-0.37$};	  
	  \node [blockVB, right of=VB2_3, node distance=1.8cm] (VB2_4)  {\small  s=4 \\\scriptsize$\mu=-1.74$};
	  \node [blockVB, right of=VB2_4, node distance=1.8cm, cross out, draw=red,very thick] (VB2_5)  {\small  s=5 \\\scriptsize$\mu=-1.74$};	  
	  
	  \node [blockVB] (VB3_1) at (0,-4.0)  {\small  s=1 
	  \\ \scriptsize$\mu=0.84$};	  
	  \node [blockVB, right of=VB3_1] (VB3_2) {\small  s=2 
	  \\\scriptsize$\mu=-0.37$};	  
	  \node [blockVB, right of=VB3_2] (VB3_3)  {\small  s=3 
	  \\\scriptsize$\mu=-1.74$};	  
	  \node [blockVB, right of=VB3_3, node distance=1.8cm, cross out, draw=red,very thick] (VB3_4)  {\small  s=4 \\\scriptsize$\mu=-1.74$};
	  \node [blockVB, right of=VB3_4, node distance=1.8cm, cross out, draw=red,very thick] (VB3_5)  {\small  s=5 \\\scriptsize$\mu=-0.37$};
	  \node [blockVB, right of=VB3_5, node distance=1.8cm, cross out, draw=red,very thick] (VB3_6)  {\small  s=6 \\\scriptsize$\mu=-1.74$};
	  
	  \node [blockVB] (VB4_1) at (0,-5.5)  {\small  s=1 
	  \\ \scriptsize$\mu=0.84$};	 
	  \node [blockVB, right of=VB4_1] (VB4_2) {\small  s=2 
	  \\\scriptsize$\mu=-0.37$};	  
	  \node [blockVB, right of=VB4_2] (VB4_3)  {\small  s=3 
	  \\\scriptsize$\mu=-1.74$};	  
	  \node [blockVB, right of=VB4_3, node distance=1.8cm, cross out, draw=red,very thick] (VB4_4)  {\small  s=4 \\\scriptsize$\mu= 0.84$};
	  \node [blockVB, right of=VB4_4, node distance=1.8cm, cross out, draw=red,very thick] (VB4_5)  {\small  s=5 \\\scriptsize$\mu=-0.37$};
	  \node [blockVB, right of=VB4_5, node distance=1.8cm, cross out, draw=red,very thick] (VB4_6)  {\small  s=6 \\\scriptsize$\mu= 0.84$};
	  
	  \node [blockVB] (VB5_1) at (0,-7.0)  {\small  s=1 
	  \\ \scriptsize$\mu=0.84$};	 
	  \node [blockVB, right of=VB5_1] (VB5_2) {\small  s=2 
	  \\\scriptsize$\mu=-0.37$};	  
	  \node [blockVB, right of=VB5_2] (VB5_3)  {\small  s=3 
	  \\\scriptsize$\mu=-1.74$};	  
	  \node [blockVB, right of=VB5_3, node distance=1.8cm, cross out, draw=red,very thick] (VB5_4)  {\small  s=4 \\\scriptsize$\mu=-0.37$};
	  \node [blockVB, right of=VB5_4, node distance=1.8cm, cross out, draw=red,very thick] (VB5_5)  {\small  s=5 \\\scriptsize$\mu=-0.37$};
	  \node [blockVB, right of=VB5_5, node distance=1.8cm, cross out, draw=red,very thick] (VB5_6)  {\small  s=6 \\\scriptsize$\mu=-1.74$};

	  \node [blocknumbering](headlinenr) at (-1.1,0.5) {\small $iter$};
	  \node [blocknumbering, left of=VB1_1, node distance=2.2cm](VBMoG1)  {\small 0};	  
	  \node [blocknumbering, left of=VB2_1, node distance=2.2cm](VBMoG2)  {\small 1};
	  \node [blocknumbering, left of=VB3_1, node distance=2.2cm](VBMoG3)  {\small 2};
	  \node [blocknumbering, left of=VB4_1, node distance=2.2cm](VBMoG3)  {\small 3};
	  \node [blocknumbering, left of=VB5_1, node distance=2.2cm](VBMoG3)  {\small 4};
	  
	  \path [lineblue] (init) -- (VB1_1);
	  \path [lineblue] (init) -- (VB1_2);
	  \path [lineblue] (init) -- (VB1_3);
	  \path [lineblue] (init) -- (VB1_4);
	  
	  \path [line] (VB1_1) -- (VB2_1);
	  \path [line] (VB1_2) -- (VB2_2);
	  \path [lineblue] (VB1_2) -- (VB2_3);	  
	  \path [lineblue] (VB1_2) -- (VB2_4);
	  \path [lineblue] (VB1_2) -- (VB2_5);	  

	  \path [line] (VB2_1) -- (VB3_1);
	  \path [line] (VB2_2) -- (VB3_2);
	  \path [line] (VB2_4) -- (VB3_3);	  
	  \path [lineblue] (VB2_4) -- (VB3_4);	  
	  \path [lineblue] (VB2_4) -- (VB3_5);
	  \path [lineblue] (VB2_4) -- (VB3_6);
	  
	  \path [line] (VB3_1) -- (VB4_1);
	  \path [line] (VB3_2) -- (VB4_2);
	  \path [line] (VB3_3) -- (VB4_3);	  
	  \path [lineblue] (VB3_1) -- (VB4_4);	  
	  \path [lineblue] (VB3_1) -- (VB4_5);
	  \path [lineblue] (VB3_1) -- (VB4_6);
	  
	  \path [line] (VB4_1) -- (VB5_1);
	  \path [line] (VB4_2) -- (VB5_2);
	  \path [line] (VB4_3) -- (VB5_3);	  
	  \path [lineblue] (VB4_3) -- (VB5_4);	  
	  \path [lineblue] (VB4_3) -- (VB5_5);
	  \path [lineblue] (VB4_3) -- (VB5_6);
	  

	  \node [blocknumbering](headlinenr) at (6.3,0.5) {\small $L$};
	  \node [blocknumbering, right of=VB1_4, node distance=2.2cm](M1)  {\small 0};	  
	  \node [blocknumbering, below of=M1, node distance =3.0cm](M2)  {\small 0};
	  \node [blocknumbering, below of=M2, node distance=3.0cm](M3)  {\small 1};
	  \node [blocknumbering, below of=M3, node distance=3.0cm](M4)  {\small 2};
	  \node [blocknumbering, below of=M4, node distance=3.0cm](M5)  {\small 3};

      \end{tikzpicture}
\end{minipage}%
}
\caption{Evolution of Algorithm  \ref{alg:adapt} for Example 1 with $S_0=4$ and $\Delta S=3$.  Green boxes correspond to surviving mixture components, whereas the ones that are deleted are marked with a red cross.   The rows are numbered based on $iter$ and the value of $L$ is reported on the right. The  mean $\mu_j$ of each component is also reported in each box.} 
\label{fig:FlowchartMoGToy}
\end{figure}
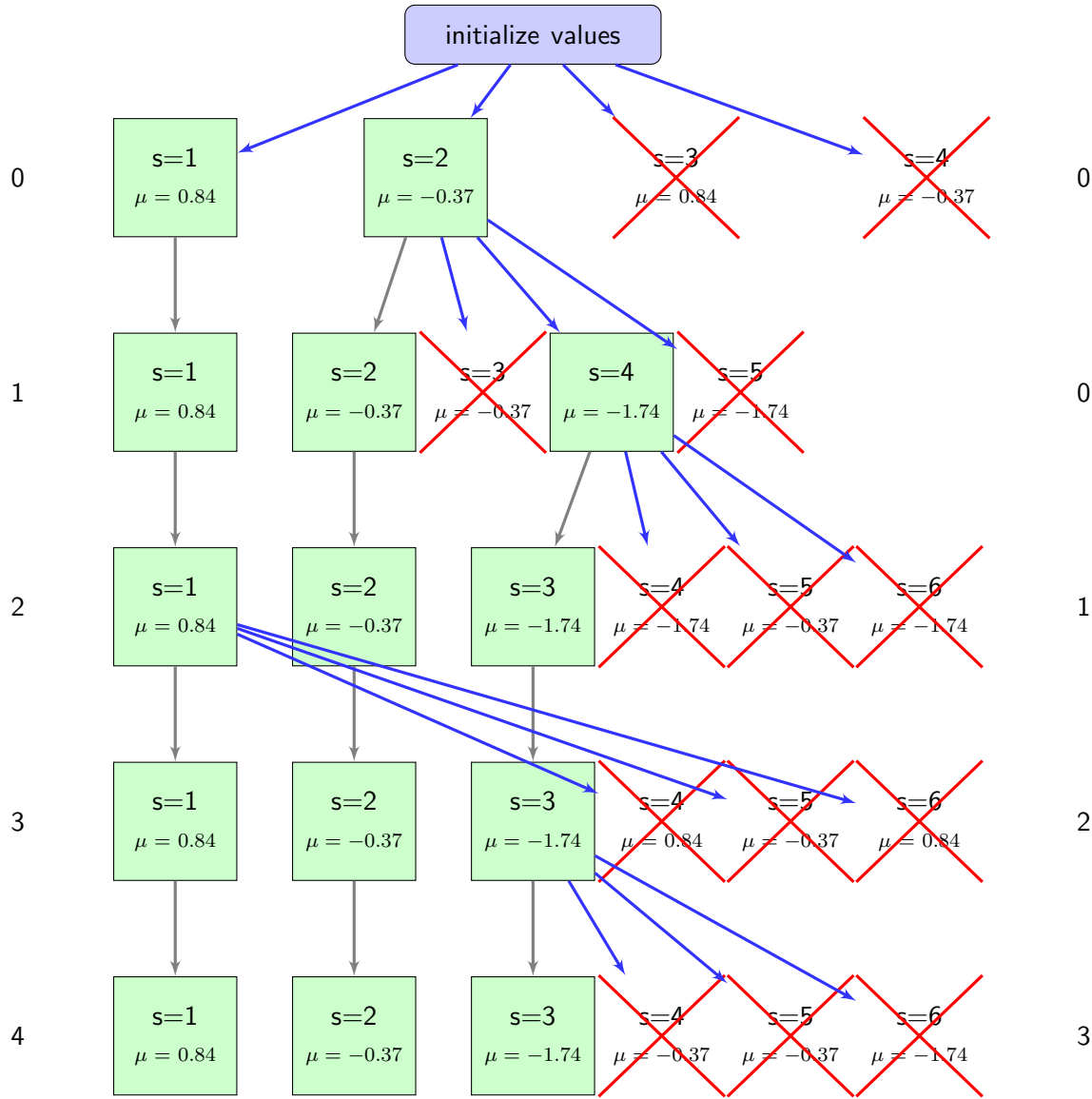

As this is a one-dimensional example, the dimensionality reduction aspects are invalid and $\bet$, in \refeq{eq:psis}, is also unnecessary. We initialize the adaptive Algorithm \ref{alg:adapt} with $S_0=4$ and propose/add $\Delta S=3$ components at each iteration $iter$. We   summarize the results produced by successive iterations in \reffig{fig:FlowchartMoGToy}. 
Two mixture components  are identified at initialization (out of the $S_0=4$ proposed). Proposed components at subsequent iterations that do not survive are marked with a red cross. 

%

Table \ref{tab:KLDivMoGToy} contains the values of the normalized $KL$-based discrepancy metric (\refeq{eq:kldist}) for all pairs of the $6$ mixture components at $iter=2$ (\reffig{fig:FlowchartMoGToy}). As it can be seen by the values, components $4$, $5$ and $6$ satisfy the component death criterion (\refeq{eq:compdeath1}) and are therefore removed.

\begin{figure}[H]
	\centering{
		{\includegraphics[width=0.850\textwidth]{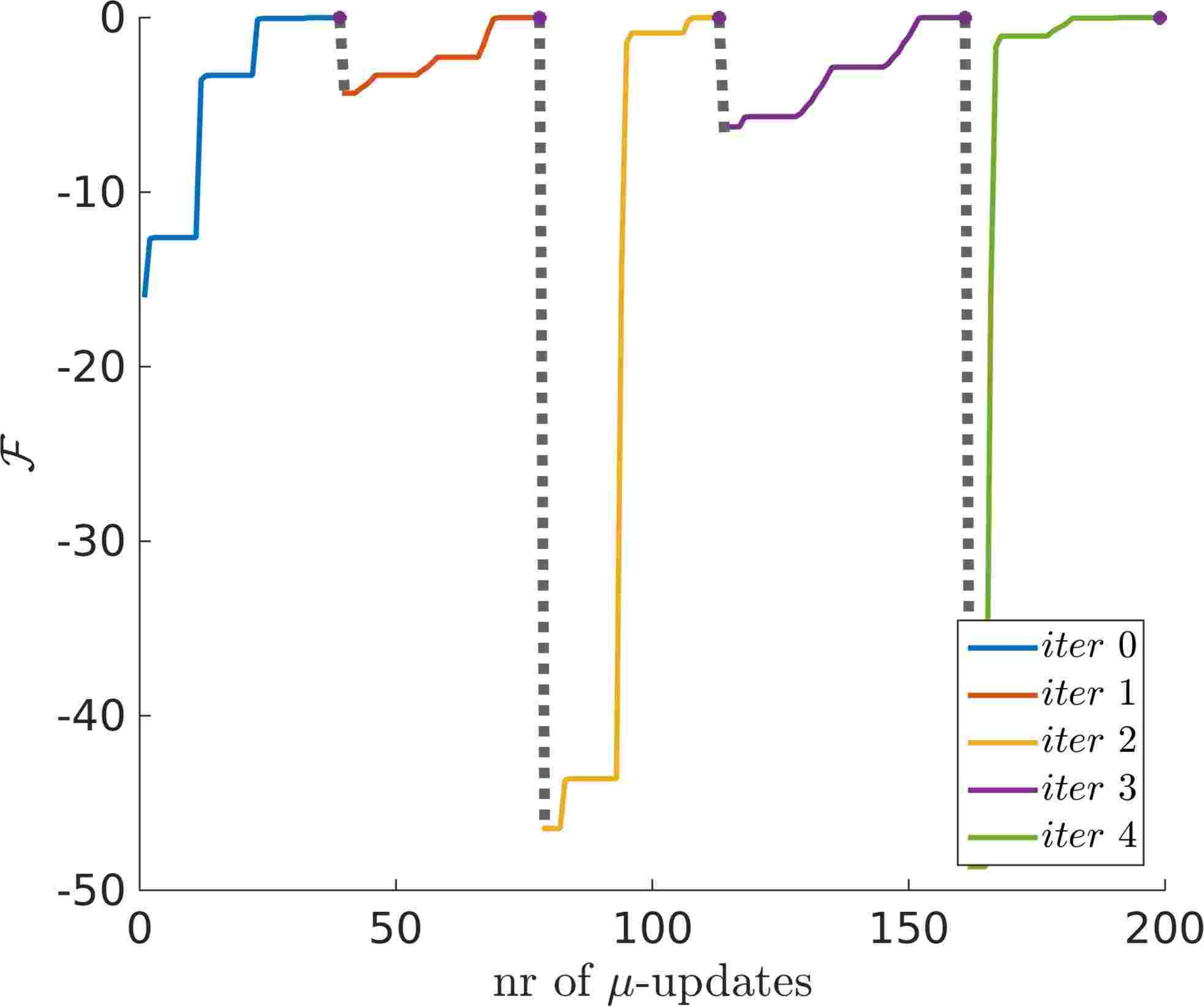}} 
		\hspace{0.1cm}
	  }
 	  \caption{Evolution of $\mathcal{F}$ (\refeq{eq:lowerbound}) over the number of $\bs{\mu}$ updates (which is equal to the number of forward calls) for Example 1.
Each color corresponds to a different value of $iter$. The number of $\bs{\mu}$ updates associated with mixtures that are subsequently deleted, is also included.
}
	 \label{fig:CostToy}
\end{figure}

\begin{table}[h]
\begin{center}
  \begin{tabular}{ | c | c | c | c| c| c| c|}
    \hline
	 & s=1 & s=2 & s=3 & s=4 & s=5 & s=6 \\ \hline\hline	 
    s=1 & 0	& 61.97& 824  &	824  & 61.97&	824 \\ \hline
    s=2 & 	& 0    &188.4 &	188.4&	$\bs{1.2\times 10^{-10}}$	&188.4 \\ \hline
    s=3 & 	&     &0     &$\bs{2.3\times 10^{-09}}$&	51.59	&$\bs{2.3\times 10^{-09}}$	\\ \hline
    s=4 & 	& 	&	&0	&51.59	&$\bs{2.3\times 10^{-09}}$	 \\ \hline
    s=5 & 	& 	&	&	&0	& 188.4	 \\
    \hline
  \end{tabular}   
  \caption{Normalized KL divergence (\refeq{eq:kldist}) 
      between each pair of  mixture components. Pairs  which are very similar (see also the means in \reffig{fig:FlowchartMoGToy}) have a very small KL divergence (shown in bold).}
  \label{tab:KLDivMoGToy}  
\end{center}  
\end{table}
The three components that persist have the following posterior  probabilities:
\be
q(s=1) = 0.24, \quad q(s=2)=0.50, \quad  q(s=3)=0.26.
\ee
The  Gaussians (\refeq{eq:qpost}) associated with each component are:
\be
\begin{split}
    q(\Psi|s=1) &=\mathcal{N}(0.84, 0.00135)  \\
    q(\Psi|s=2) &=\mathcal{N}(-0.37, 0.00590) \\
    q(\Psi|s=3) &=\mathcal{N}(-1.74, 0.00162).
\end{split}
\ee
The algorithm terminates after $L=L_{max}=3$ unsuccessful, successive proposals (at $iter=4$) and the overall cost in terms of forward calls (i.e.  evaluations of $y(\Psi)$ and its derivative) was $200$.   Since forward  model calls are required everytime any $\bs{\mu}_j$ is updated (\refeq{eq:muupd}), we plot the evolution of $\mathcal{F}$ (\refeq{eq:lowerbound}) with respect to the total number of $\bs{\mu}$-updates (including those for components that end up being deleted) in \reffig{fig:CostToy}. 

%

\begin{figure}[H]
	\centering{
	\includegraphics[width=0.65\textwidth]{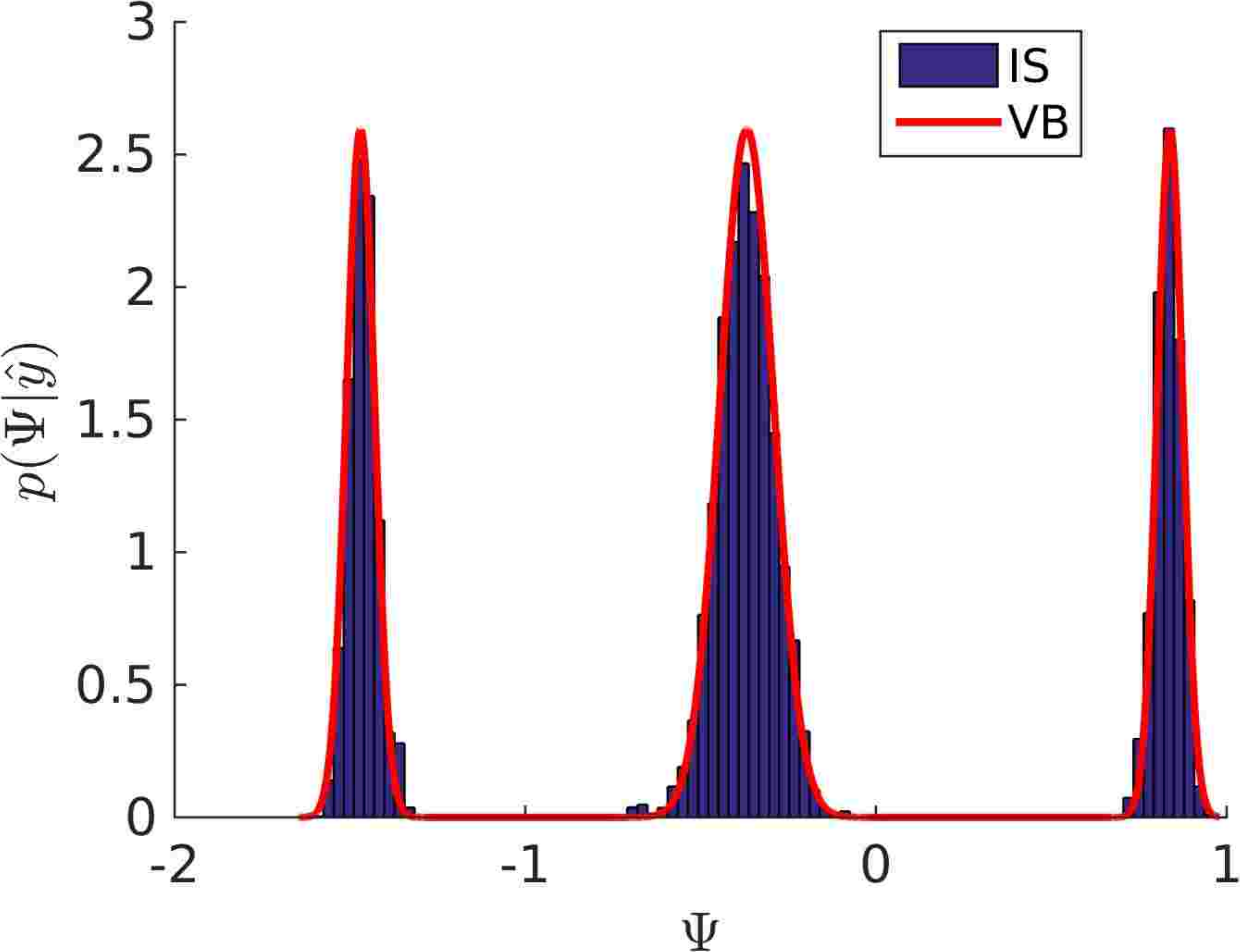} }
	\caption{Exact (IS) and approximated (VB) posterior probability distribution, which show excellent agreement.}
	\label{fig:ToyExample}
\end{figure}

To validate the results we carry out Importance Sampling as described in \refsec{sec:IS}. The Effective Sample Size (\refeq{eq:ess}) was $ESS=0.96$ which is very close to $1$. In \reffig{fig:ToyExample} the approximate posterior (\refeq{eq:postmix2}) is compared with the exact posterior (IS), and excellent agreement is observed. One can see that not only the locations (mean) and the variances of the mixture components but also their corresponding probability weights are captured correctly.

 
\begin{figure}[H]
	\centering{
	\subfloat[][{Normalized histogram.}]
	{\includegraphics[width=0.3\textwidth]{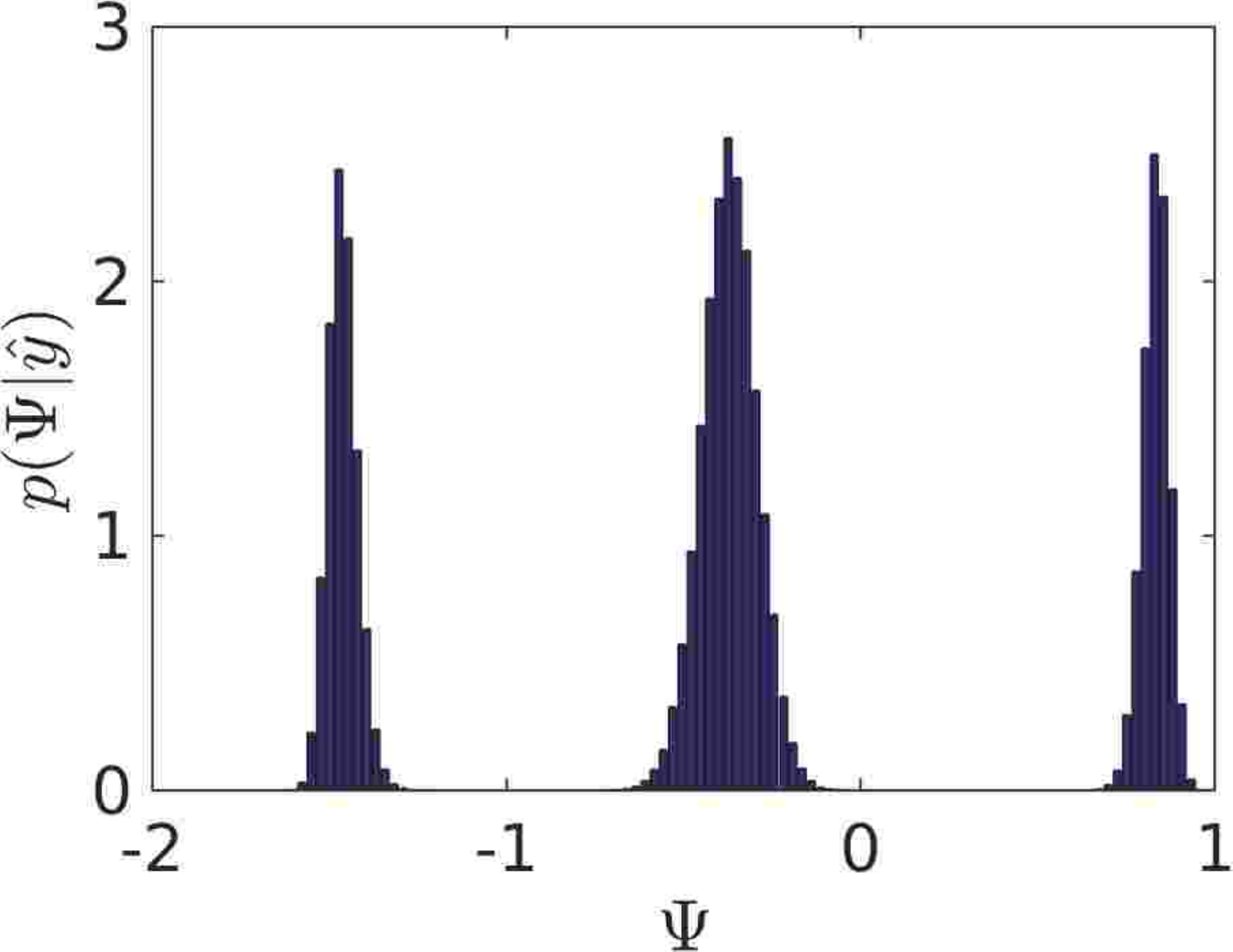} }}
	\subfloat[][{$\Psi$ over different samples (part of it).}] 
	{\includegraphics[width=0.3\textwidth]{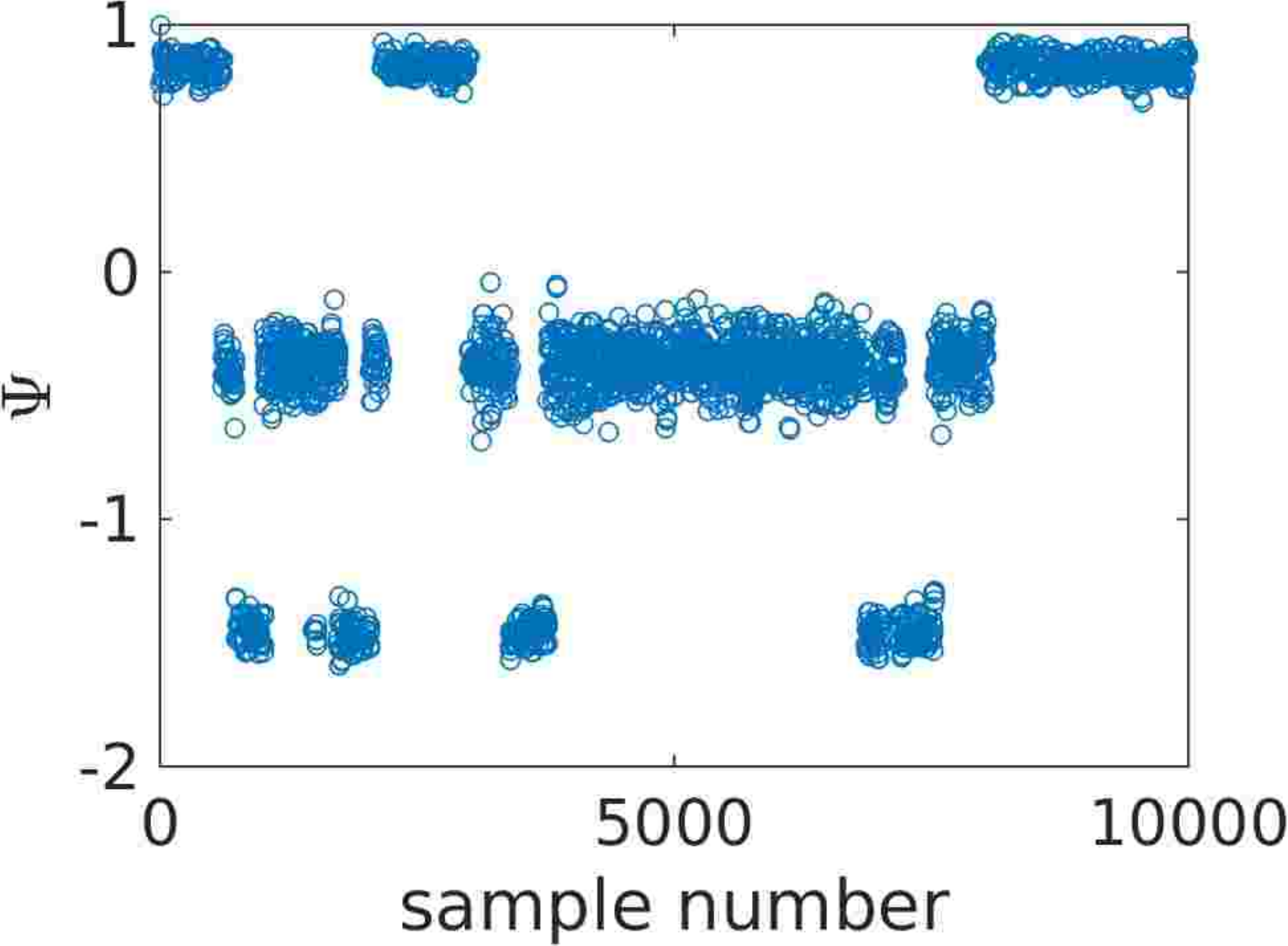}} 
	\subfloat[][{Autocovariance.}] 
	{\includegraphics[width=0.30\textwidth]{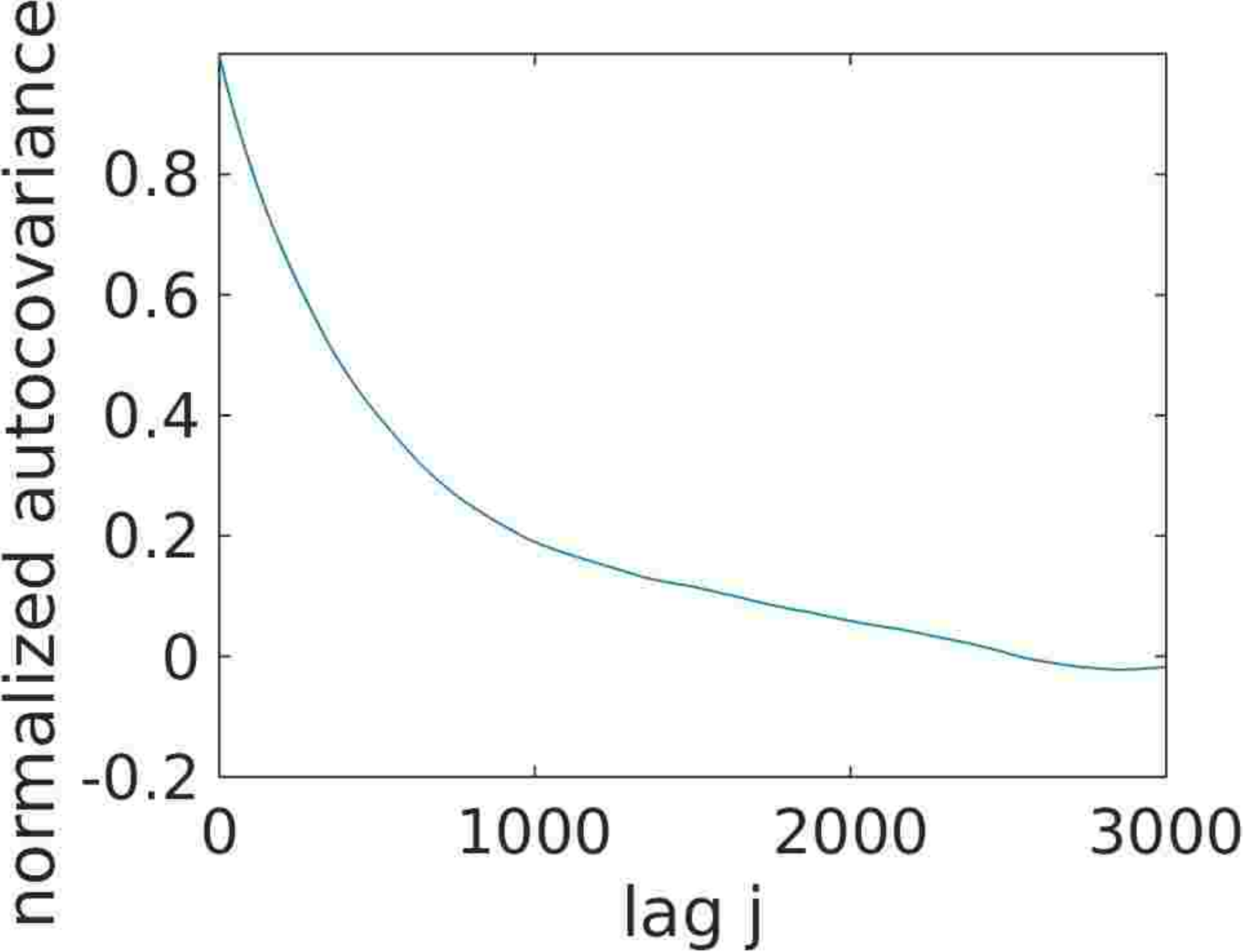}} 
	\caption{Left: Posterior distribution obtained with random walk MCMC with $10^{6}$ MCMC samples which coincides with \reffig{fig:ToyExample}. Middle: Evolution of the state $\Psi$ per MCMC step. Right: Normalized autocovariance which decays slowly and results in a small ESS.}
	\label{fig:ToyExampleMCMC}
\end{figure}

For comparison purposes, and as the cost per forward model  evaluation in this problem is negligible, we also performed random-walk MCMC with a Gaussian proposal density with standard deviation $0.35$ that yielded an average acceptance ratio of $20\%$. The results are depicted in \reffig{fig:ToyExampleMCMC}. The corresponding ESS was $ESS_{MCMC} = 1\times 10^{-3}$, i.e. roughly $1000$ times more expensive (in terms of forward model evaluations) than the proposed strategy. 

\subsection{Elastography}
\label{sec:ex2}

\begin{figure}[H]
	\centering
	\subfloat[][{Problem configuration.}] 
	{\def\svgwidth{0.45\columnwidth}
	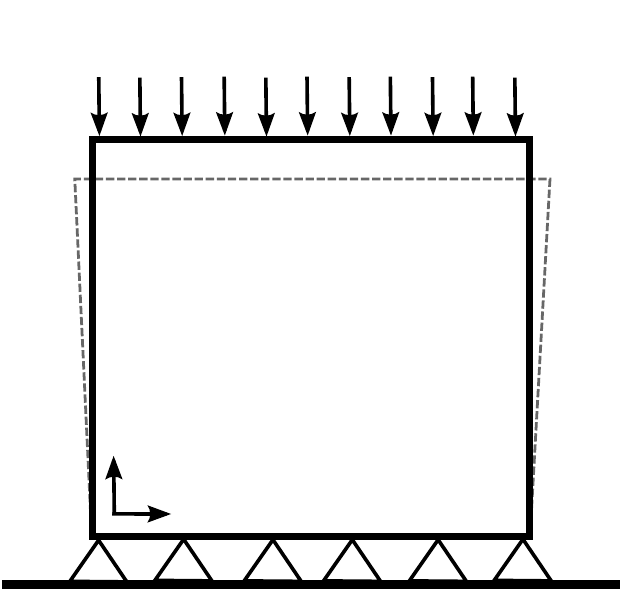  }
	\subfloat[][{Reference configuration of $\psi$ in the log scale.}] 
	{\includegraphics[width=0.40\textwidth]{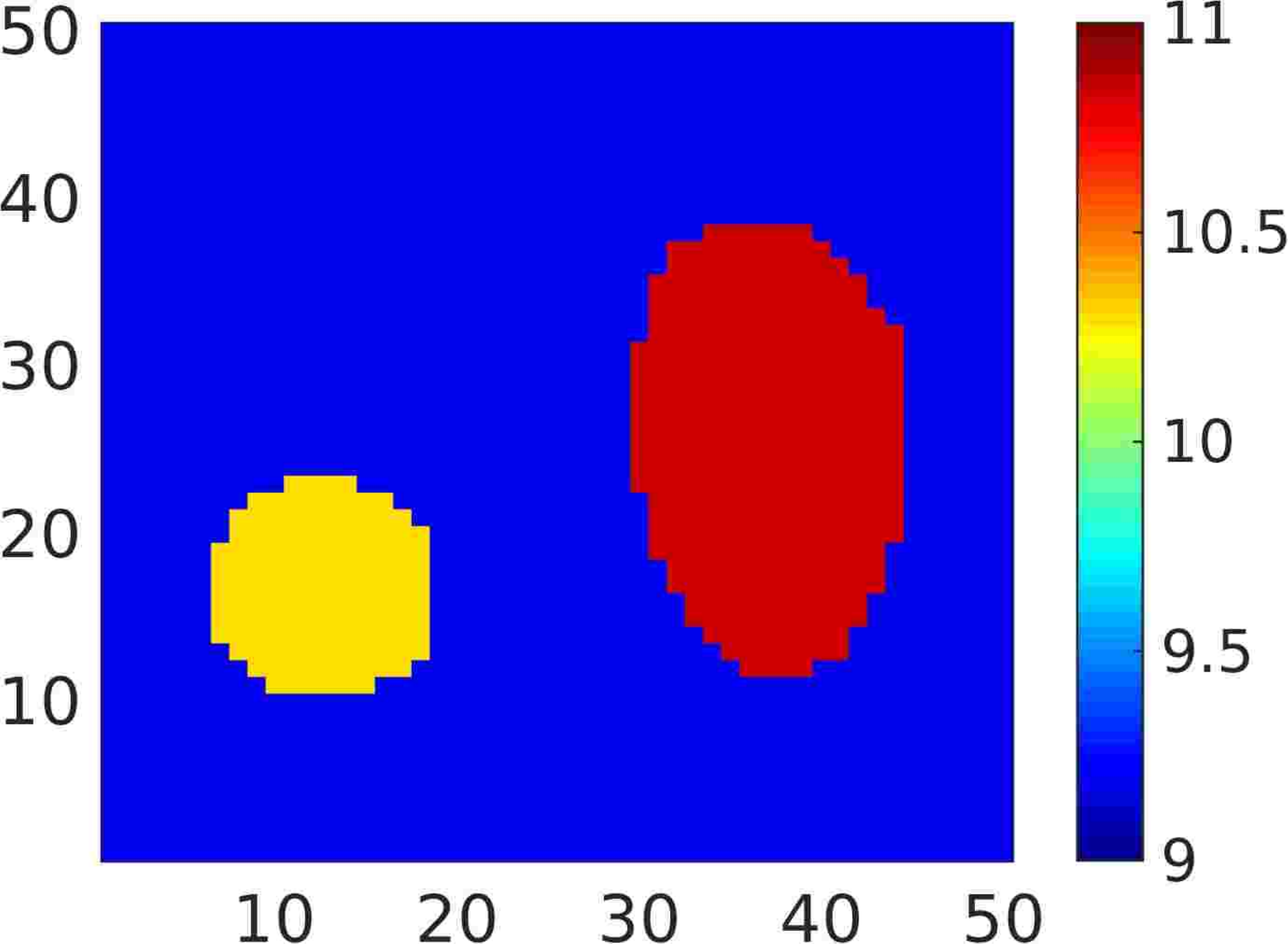} }
	
	\caption{Problem and reference configuration.}
	\label{fig:ForwardLarge}
\end{figure}

In the motivating problem  of  nonlinear elastography, we simulate a scenario of applying a  quasi-static pressure  (e.g. with the ultrasound wand) and using  the pre- and post-compression images to infer the material properties of the underlying tissue. 
We consider a two-dimensional domain $\Omega_0 =[0,50] \times [0,50]$  shown in \reffig{fig:ForwardLarge}. The governing equations consist of the conservation of linear momentum\footnote{Dependencies on the spatial variables $\bs{x} \in \Omega_0$ have been suppressed for simplicity.}:
\begin{equation}
       \bigtriangledown \cdot (\boldsymbol{FS})  = 0   \quad in \quad \Omega_0
\end{equation}
where $\boldsymbol{F}=\bs{I}+\nabla \bs{u}$ is the deformation map, $\bs{u}$ is the displacement field and  $\bs{S}$ is the second Piola-Kirchhoff stress.  
We assume Dirichlet boundary conditions along the bottom boundary  (\reffig{fig:ForwardLarge}) i.e.:
\be
  \bs{u} = \bs{0}   \quad on ~x_1=[0,50], ~x_2=0 
\ee
and the following Neumann conditions on the remaining boundaries:
\be
\begin{array}{l}
  \bs{FS}\cdot \bs{N} = \left[\begin{array}{c} 0 \\ -100 \end{array} \right], \quad on ~x_1=[0,50], ~x_2=50 \\
  \bs{FS}\cdot \bs{N} = \bs{0} \quad on ~x_1=0 \textrm{ and } x_1=50,  ~x_2 \in [0,50]. \\
  \end{array}
\ee
A nonlinear, elastic constitutive law (stress-strain relation) is adopted of the form:
\begin{equation}
      \boldsymbol{S} = \frac{\pa U}{\pa \textbf{E}} 
      \label{eq:const}
\end{equation}
where $\bs{E}=\frac{1}{2}(\bs{F}^T\bs{F}-\bs{I})$ is the Lagrangian strain tensor and  $U(\bs{E},\psi)$ is the strain energy density function which depends (apart from $\bs{E}$) on the  the material parameters. 
 In this example we employ the  St. Venant-Kirchhoff model \cite{holzapfel_nonlinear_2000,gladilin_nonlinear_2008, yanovsky_unbiased_2008} that corresponds to the  following strain energy density function $U$:
\be
 U =  \frac{\nu \psi}{2(1+\nu)(1-2\nu)}[tr(\mathbf{E})]^2+ \frac{\psi}{2(1+\nu)} ~ tr(\mathbf{E}^2). 
 \label{eq:mr}
\ee

The St. Venant-Kirchhoff model is an extension of the linear elastic material model to the nonlinear regime i.e.  large deformations. 
In this example $\nu = 0.3$ and the Young modulus $\psi$ is assumed to vary in the problem domain i.e. $\psi(\bs{x})$. In particular we assume the presence of two inclusions (tumors, \reffig{fig:ForwardLarge}). In the larger, elliptic inclusion the Young modulus is $\psi = 50000$ (red),  in the smaller, circular inclusion $\psi = 30000$ (yellow/orange) and in the remaining material $\psi =10000$ (blue). The contrast  $\frac{\psi_{inclusion}}{\psi_{matrix}}\approx 4-5$  coincides with experimental evidence on actual tissue \cite{wellman_breast_1999, krouskop_elastic_1998}. 
 We generate synthetic data $\bs{\hat{y}}$ by using a $100 \times 50$ mesh and collecting the displacements at the interior points. These are in turn contaminated by zero mean, isotropic, Gaussian noise resulting in a signal-to-noise-ratio (SNR) of $1000$.
 The forward solver used in the solution of the inverse problem consists of a regular grid with $50\times 50$ quadrilateral finite elements. We assume that within each finite element, $\psi$  is constant, resulting in a $2500$ dimensional vector of inverse-problem unknowns  $\bpsi$. We note that in the discretized form, the resulting algebraic equations are nonlinear (geometric and material nonlinearities) and the state vector (forward-problem unknowns) i.e. the displacements are of dimension $5100$.

As in \cite{franck_sparse_2015} for each mixture component we employ an adaptive learning scheme for the reduced coordinates $\Theta_i$ which are added one-by-one in such a way that they have a posteriori progressively smaller variances. For that reason we define the prior precisions $\lambda_{0,s,i}$ such that they are gradually larger. Given $\lambda_{0,s,1}$, which is assumed to be the same for all mixture components $s$, we define the {\em prior} precisions as follows \cite{franck_sparse_2015}: 
\be
  \lambda_{0,s,i} = max(\lambda_{0,s,1}, \lambda_{s,i-1} - \lambda_{0,s,i-1}), \quad i=2,3, \ldots, \dth
  \label{eq:priorthnumsec}
\ee
where $\lambda_{s,i-1}$ corresponds to the posterior precision for the previous reduced coordinate $\Theta_{i-1}$ of the same  component $s$. This implies that, a priori, the next reduced coordinate 
will have at least the precision of the previous one as long as it is larger than the threshold $\lambda_{0,s,1}$. For the prior of $\bet$ we use $\lambda_{0,\eta,s} = \max_i(\lambda_{0,s,i})$ as $\bet$ represents the residual variance which is a priori smaller than the smallest variance of the reduced coordinates $\bt$. The results presented in the following were obtained for $\lambda_{0,s,1}=1$ for all $s$ and the material parameters are plotted in log scale. 

\begin{figure}[H]
	\centering
	\subfloat[][{ $\bmu_1 = \bmu_{{parent}}$ }] 
	{\includegraphics[width=0.40\textwidth]{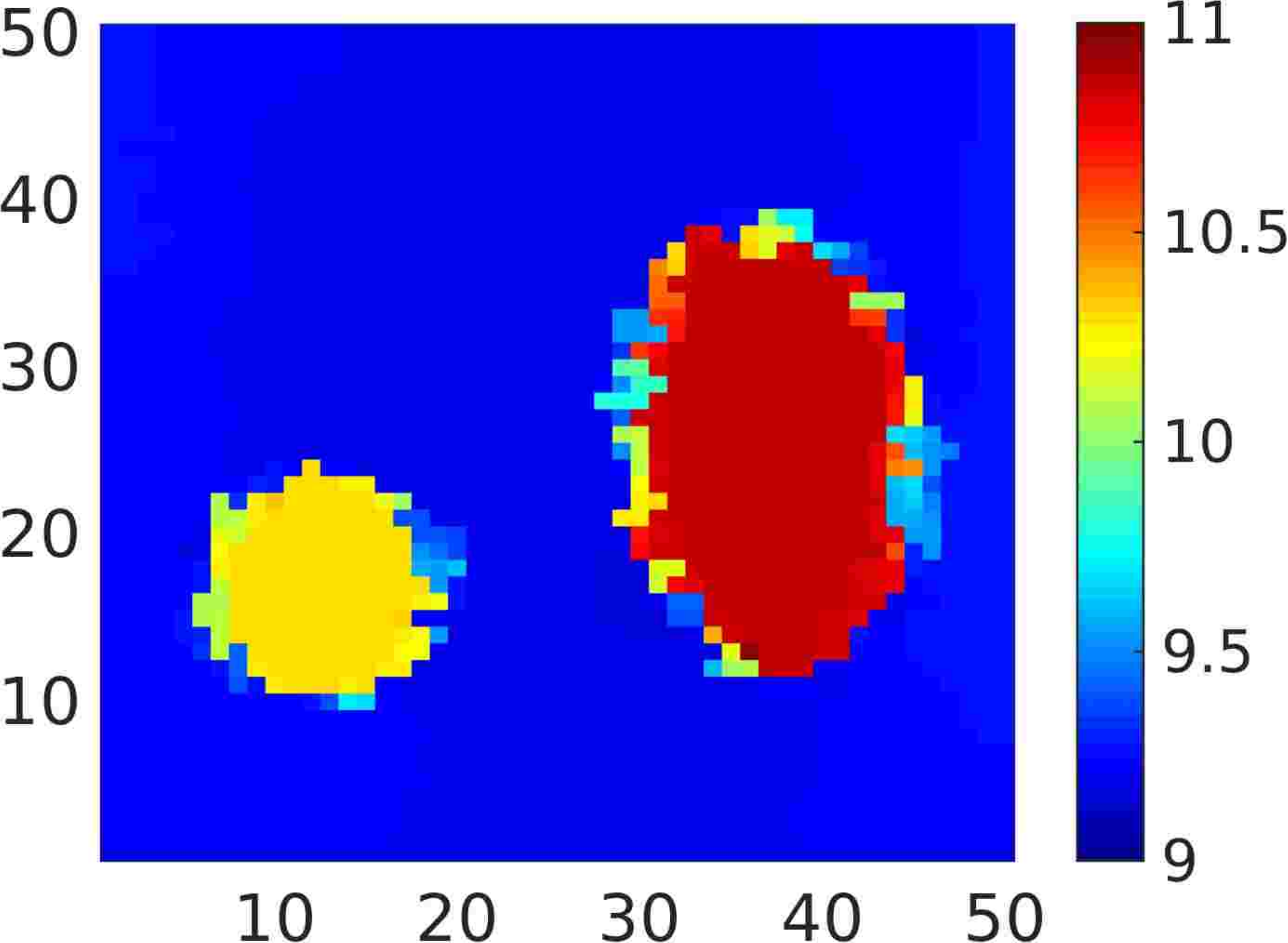}} 
	\hspace{0.1cm}
	\subfloat[][{ $\bmu_{2}$}] 
	{\includegraphics[width=0.40\textwidth]{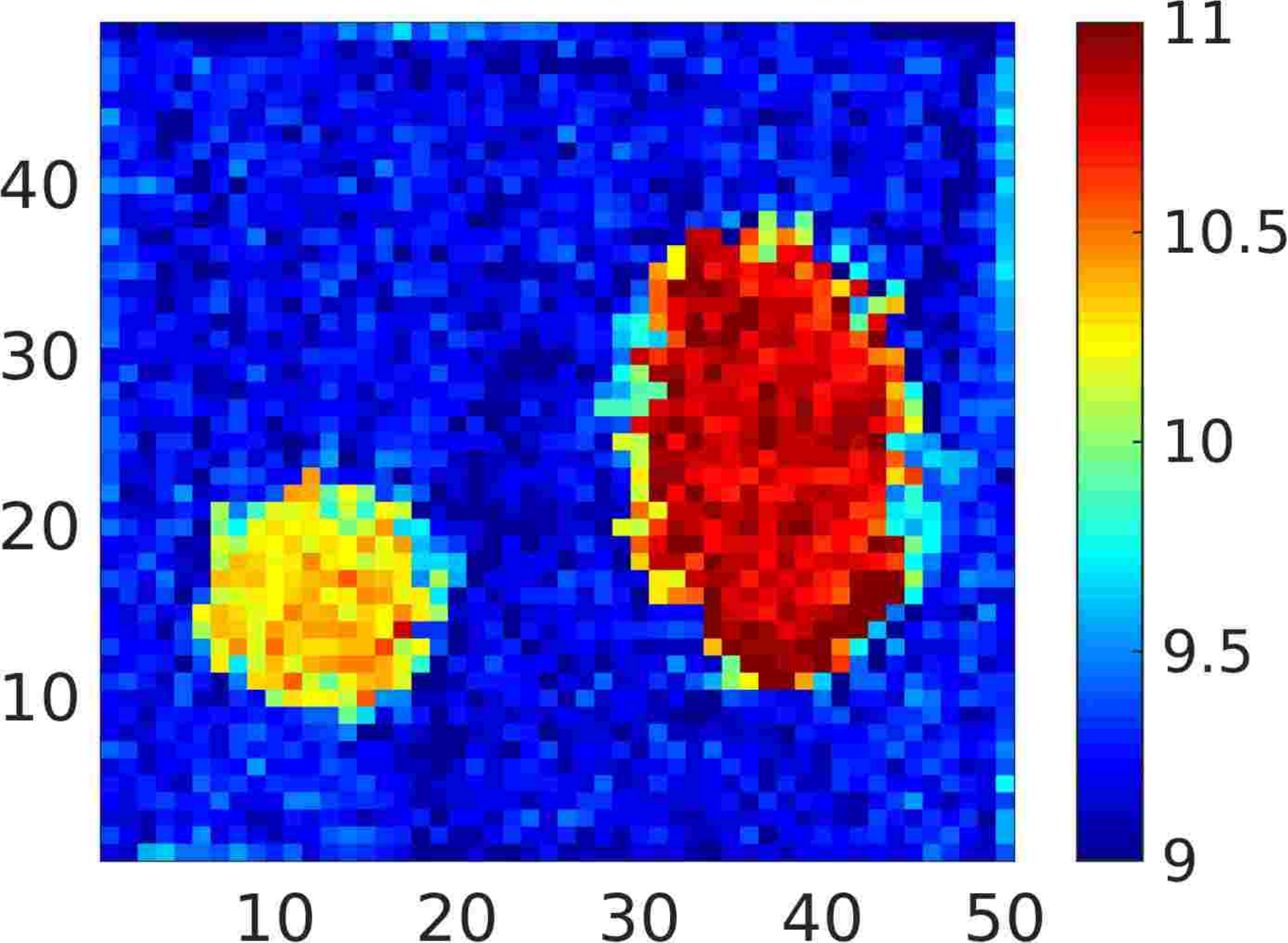}} \\
	\subfloat[][{ $\bmu_{3}$}] 
	{\includegraphics[width=0.40\textwidth]{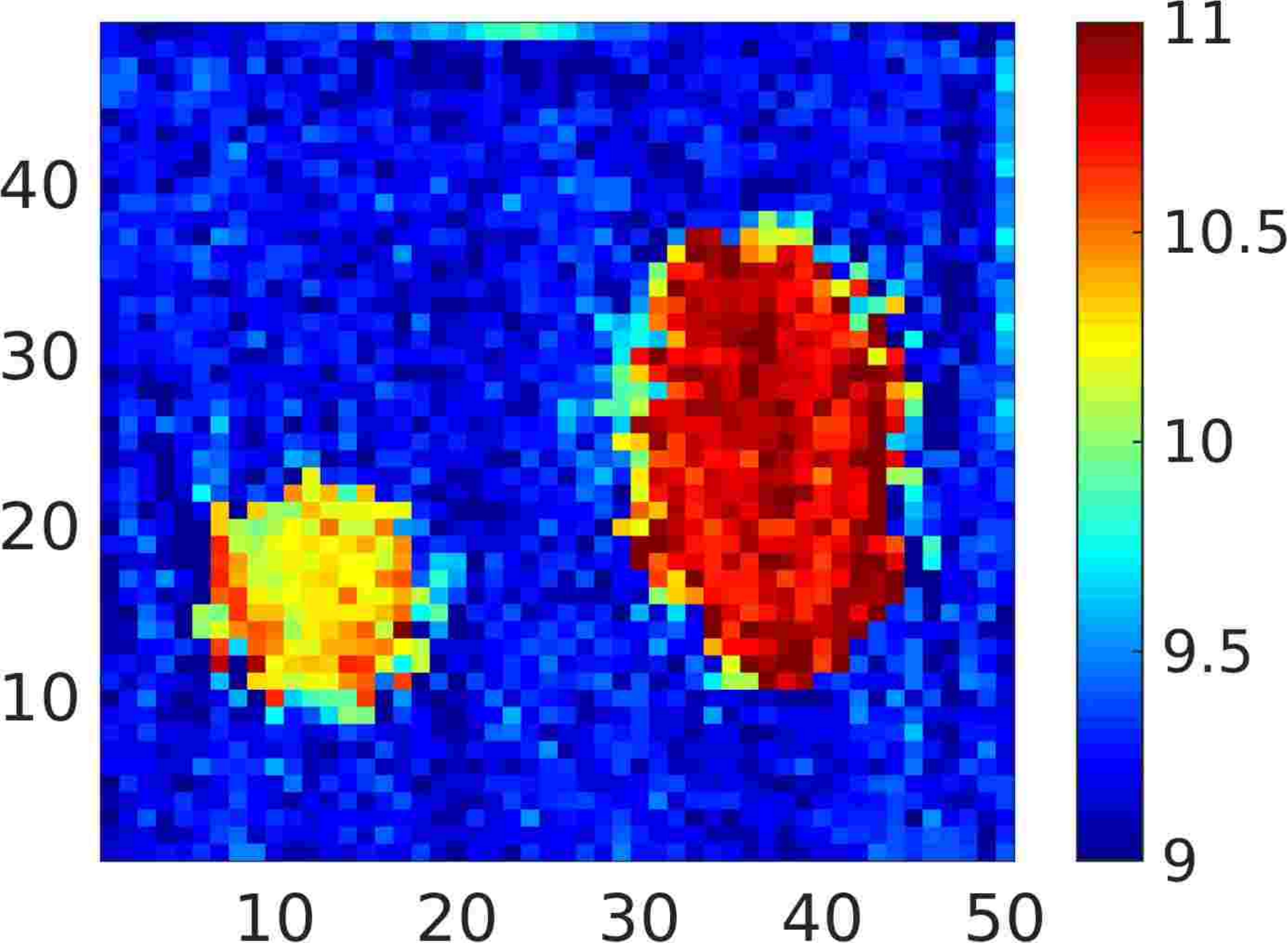}} 
	\hspace{0.1cm}
	\subfloat[][{ $\bmu_{4}$}] 
	{\includegraphics[width=0.40\textwidth]{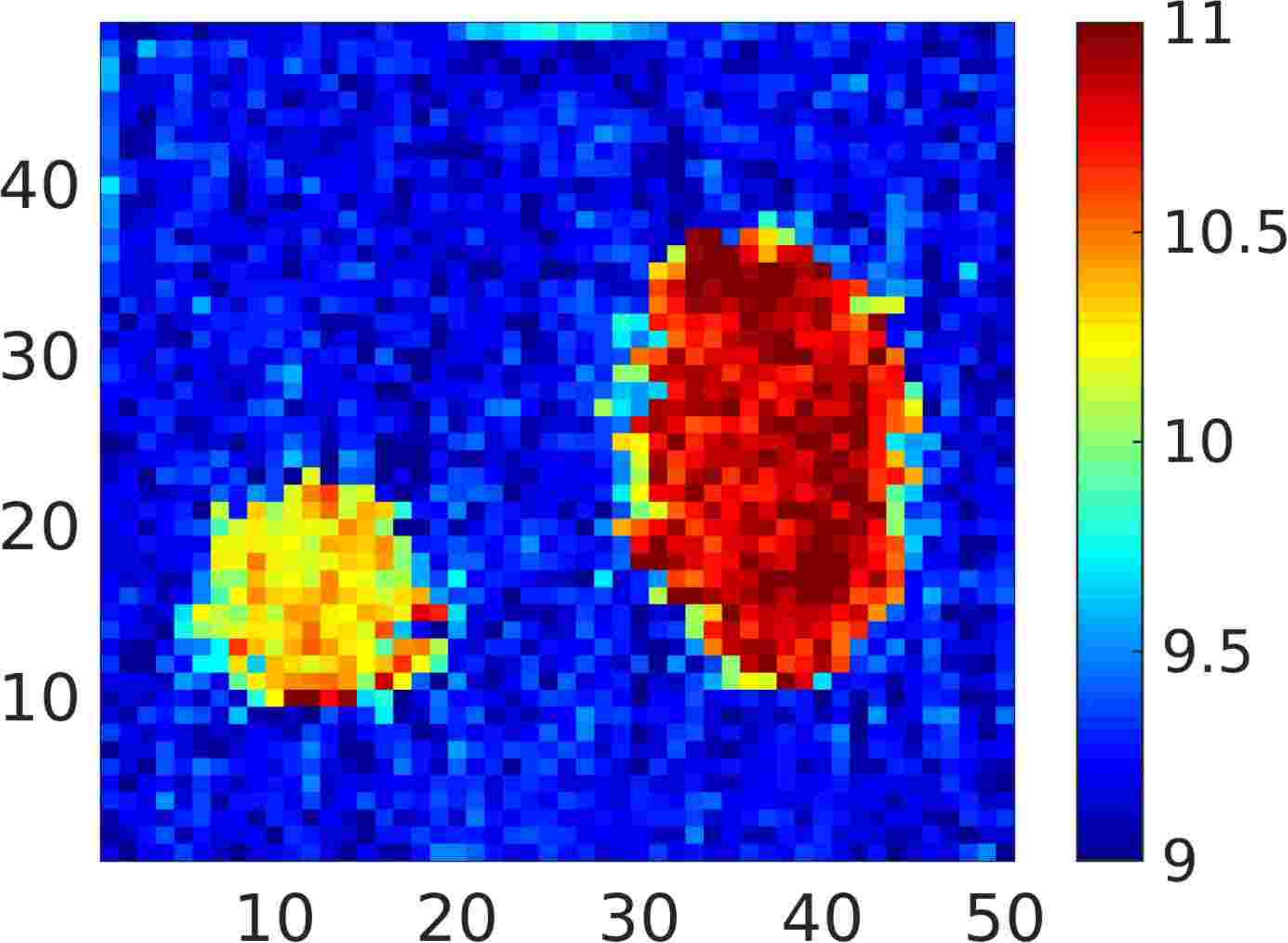}} 
		
	\caption{In a) the converged $\bs{\mu}_1$ is depicted and in b), c) and d) three perturbations (\refeq{eq:muinit}) used to initialize the means for the $\Delta S$ new proposed components (in log scale).}
	\label{fig:InitializedMu}
\end{figure}

The algorithm is initialized with four components i.e. $S_0=4$, and $\Delta S=3$ new components are proposed at each iteration $iter$ (Algorithm \ref{alg:adapt}).
\reffig{fig:InitializedMu} depicts the mean $\bs{\mu}_1$ identified upon convergence ($iter=0$) of an active component. Furthermore, it shows three perturbations, obtained according to \refeq{eq:muinit},  which were used  as initial values for the means of  the $\Delta S=3$ new components proposed at $iter=1$.   
%

\begin{figure}[H]{
	\hspace{-1.5cm}
		{\includegraphics[width=1.2\textwidth]{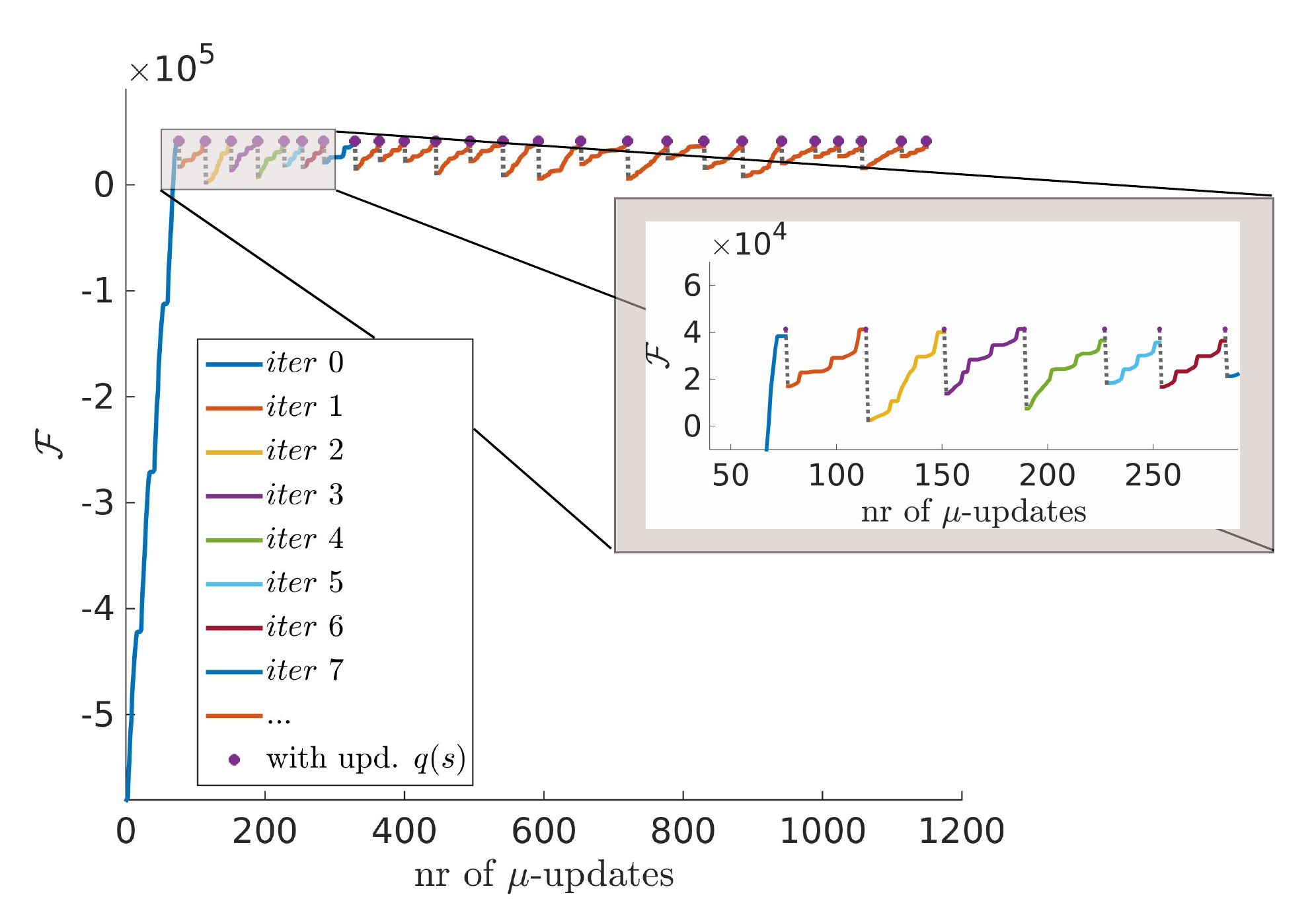}} 
	  }
	  \caption{
	  Evolution of $\mathcal{F}$ (\refeq{eq:lowerbound})  over the number of $\bs{\mu}$ updates (which is equal to the number of forward calls) for Example 2.
Each color corresponds to a different value of $iter$. The number of $\bs{\mu}$ updates associated with mixtures that are subsequently deleted, is also included.
%
	  }
	 \label{fig:Cost}
\end{figure}

\reffig{fig:Cost} depicts the evolution of the variational lower bound  $\mathcal{F}$, (\refeq{eq:lowerbound}) per $\bs{\mu}$-update i.e. per call to the forward model solver. In total the algorithm performed $iter=24$ iterations which entailed proposing $S_0 +24 \times \Delta S =76$ new mixture components (until $L=L_{max}=3$ was reached). For each of the $76$ mixture components, the number of required forward calls ranged from $7$ to $34$.
The total number of such calls was $1200$. 


\begin{figure}[H]{
	\centering
	\captionsetup[subfigure]{labelformat=empty}
		\subfloat[][{ $q(s=1) =0.318$ }] 
		{\includegraphics[width=0.30\textwidth]{FiguresIF/PlotYoungModulusMeanVBMoG0.pdf}} 
		\hspace{0.1cm}
		\subfloat[][{ $q(s=2) =0.156$}] 
		{\includegraphics[width=0.30\textwidth]{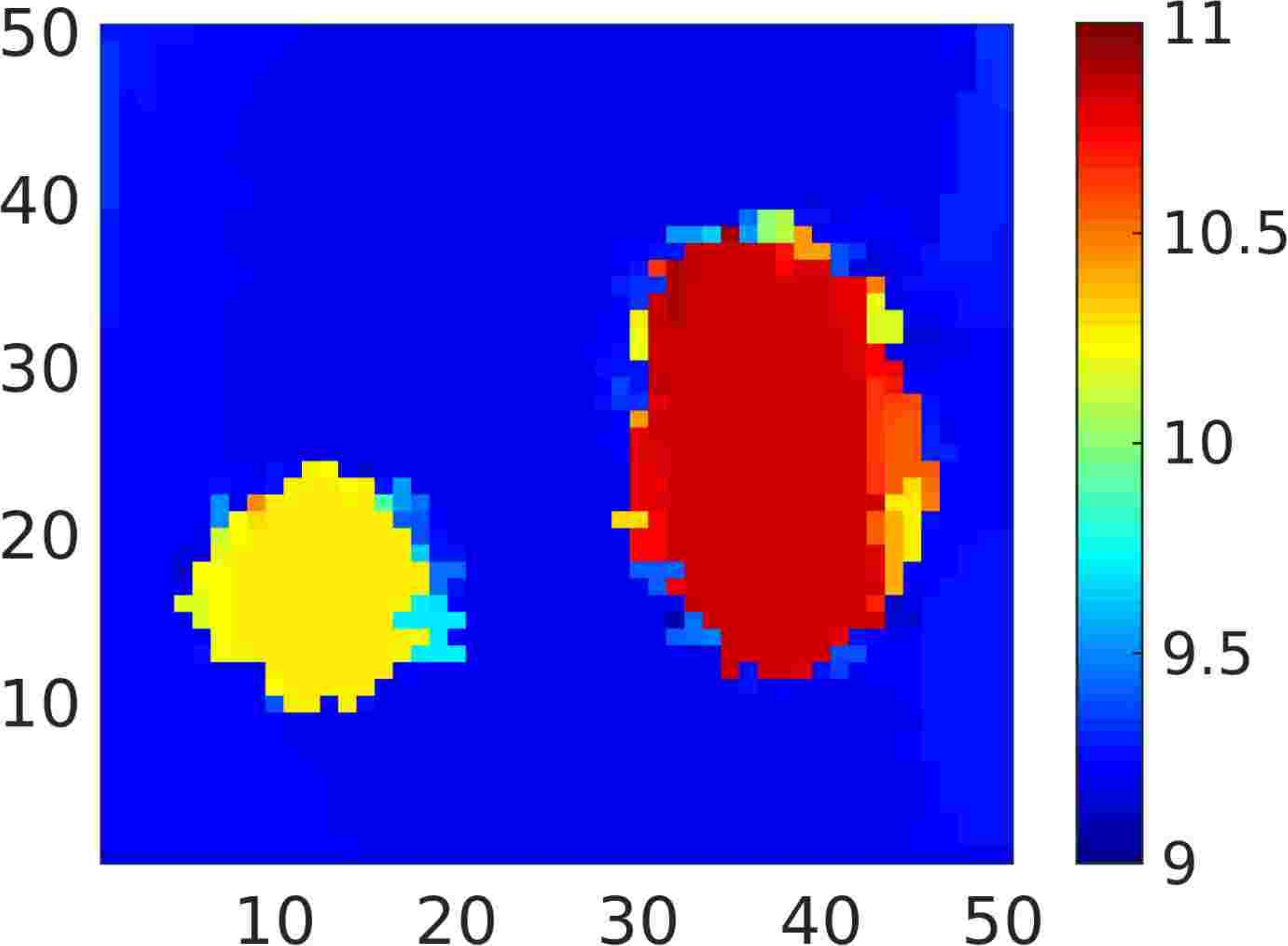}} 
		\hspace{0.1cm}
		\subfloat[][{ $q(s=3) =0.213$}] 
		{\includegraphics[width=0.30\textwidth]{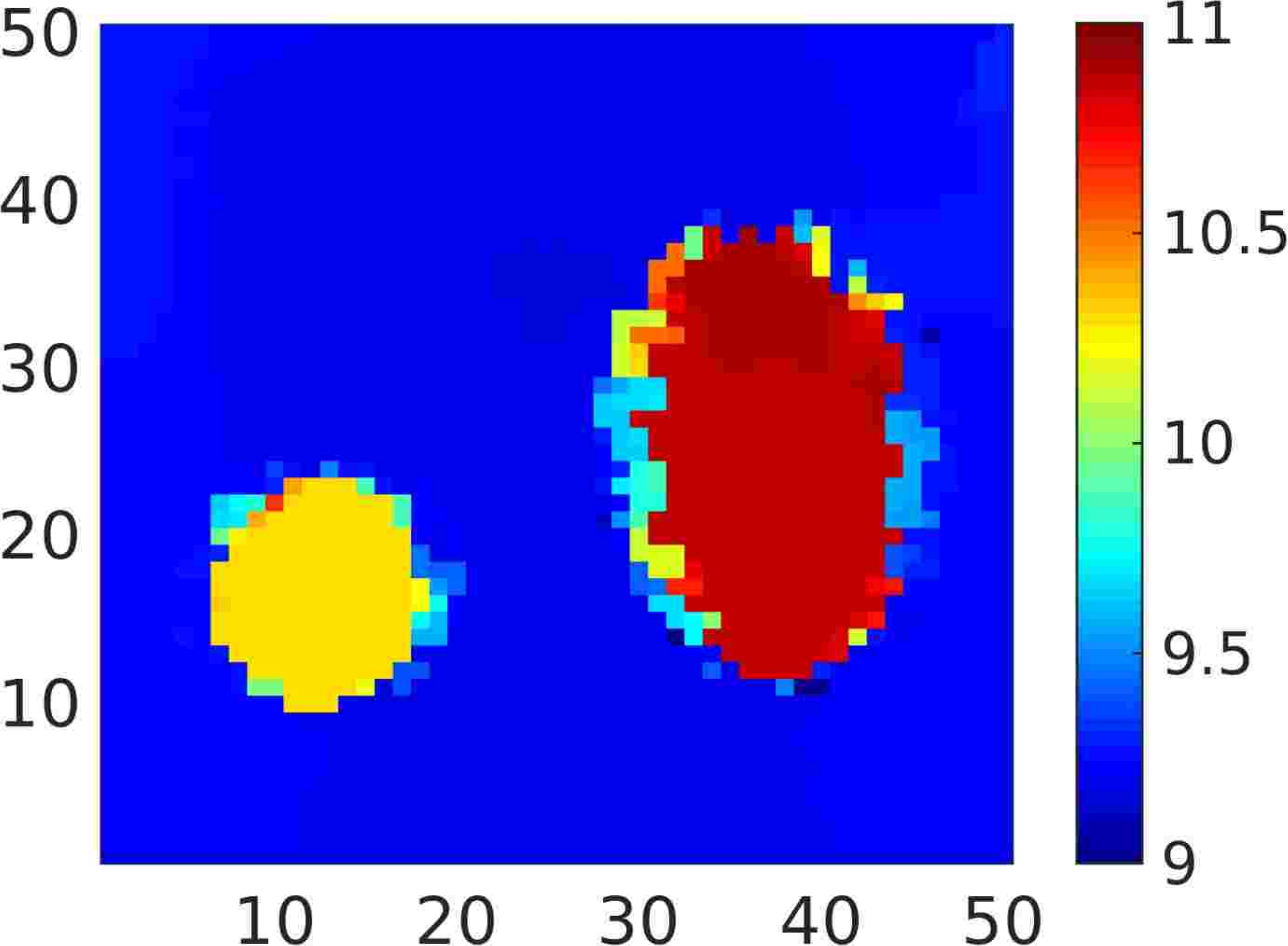}} 
		\hspace{0.3cm}
		\\
		\subfloat[][{ $q(s=4) =0.011$ }] 
		{\includegraphics[width=0.30\textwidth]{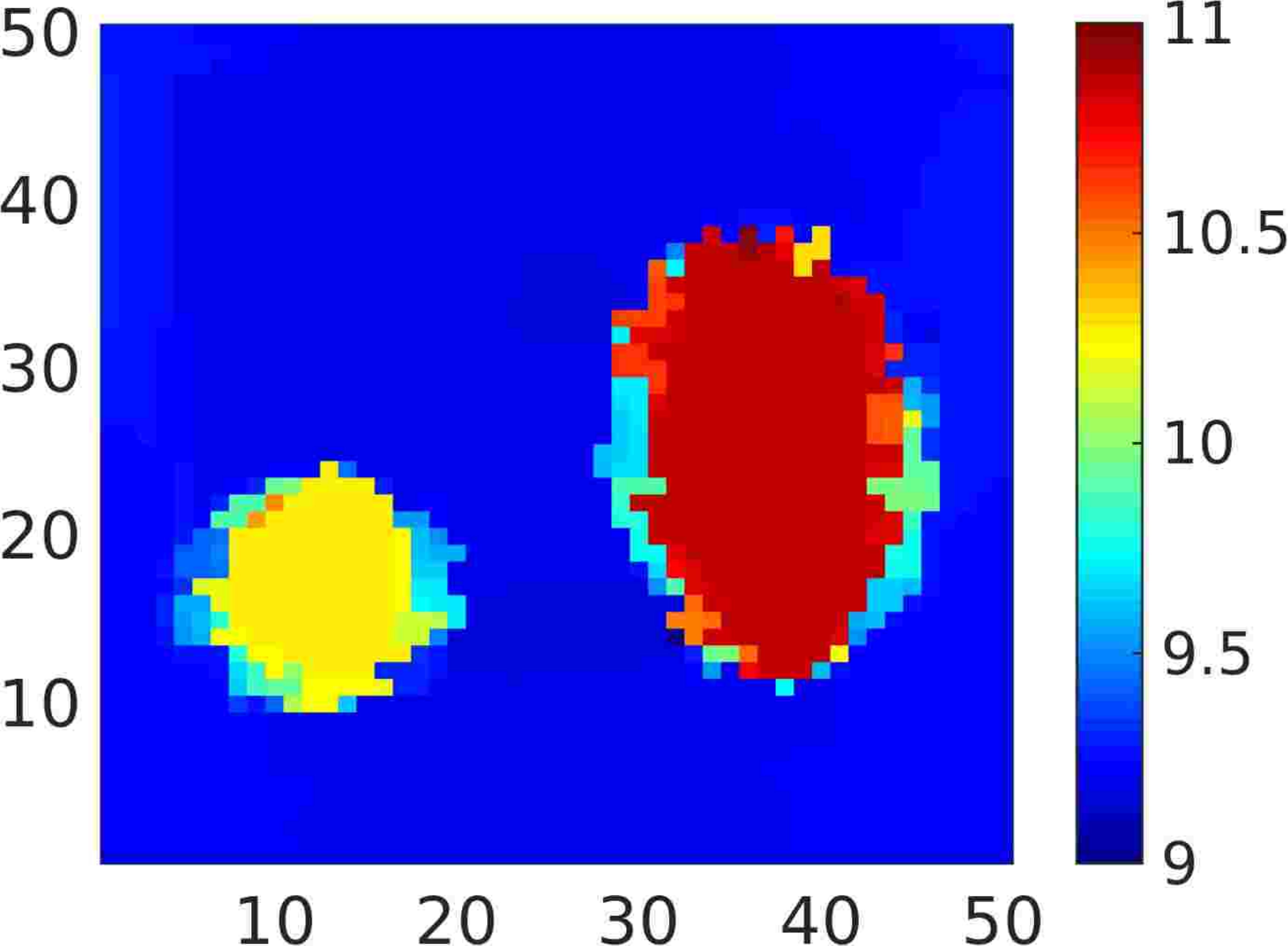}} 
		\hspace{0.1cm}
		\subfloat[][{ $q(s=5) =0.092$}] 
		{\includegraphics[width=0.30\textwidth]{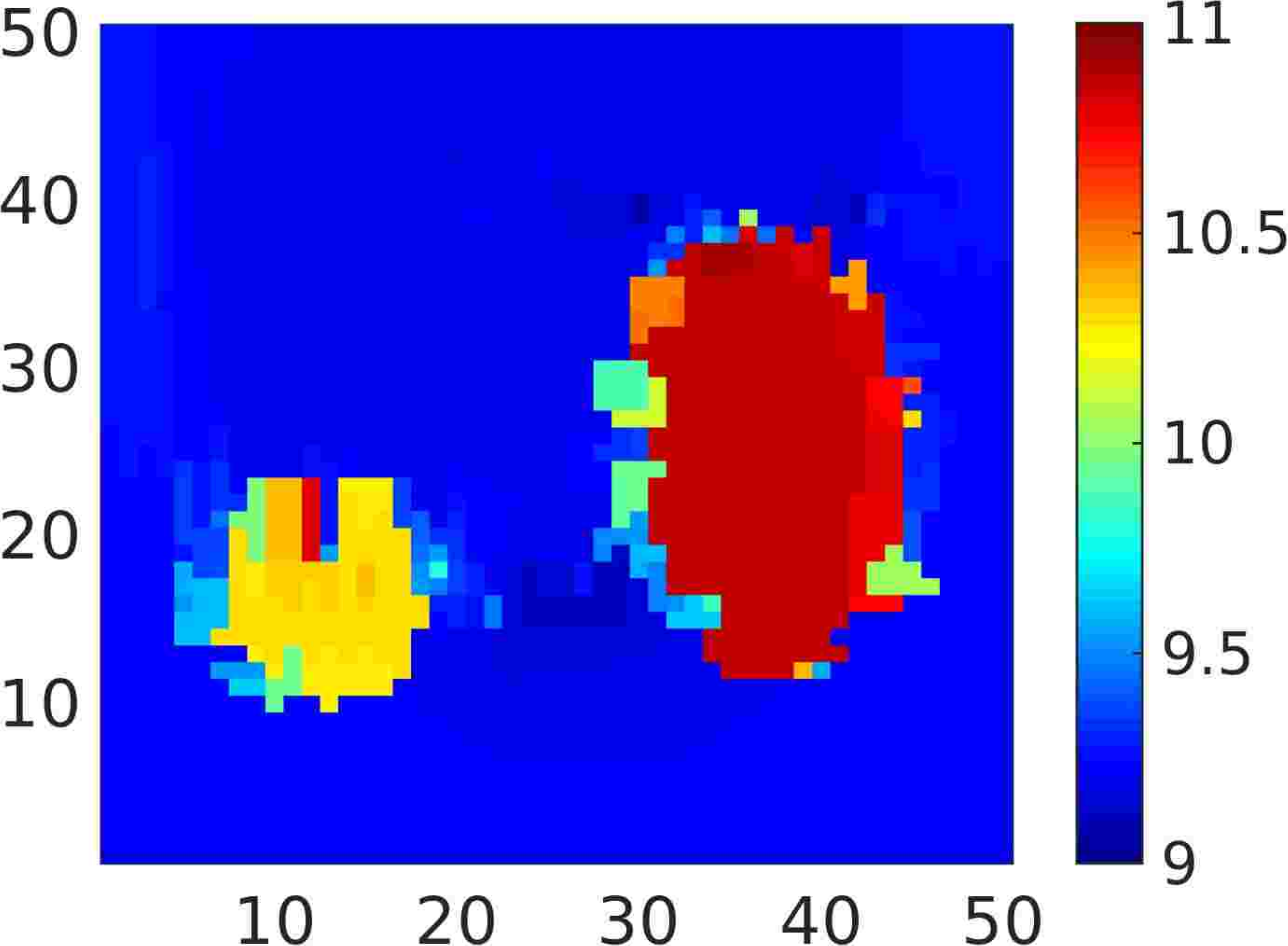}} 
		\hspace{0.1cm}
		\subfloat[][{ $q(s=6) =0.160$}] 
		{\includegraphics[width=0.30\textwidth]{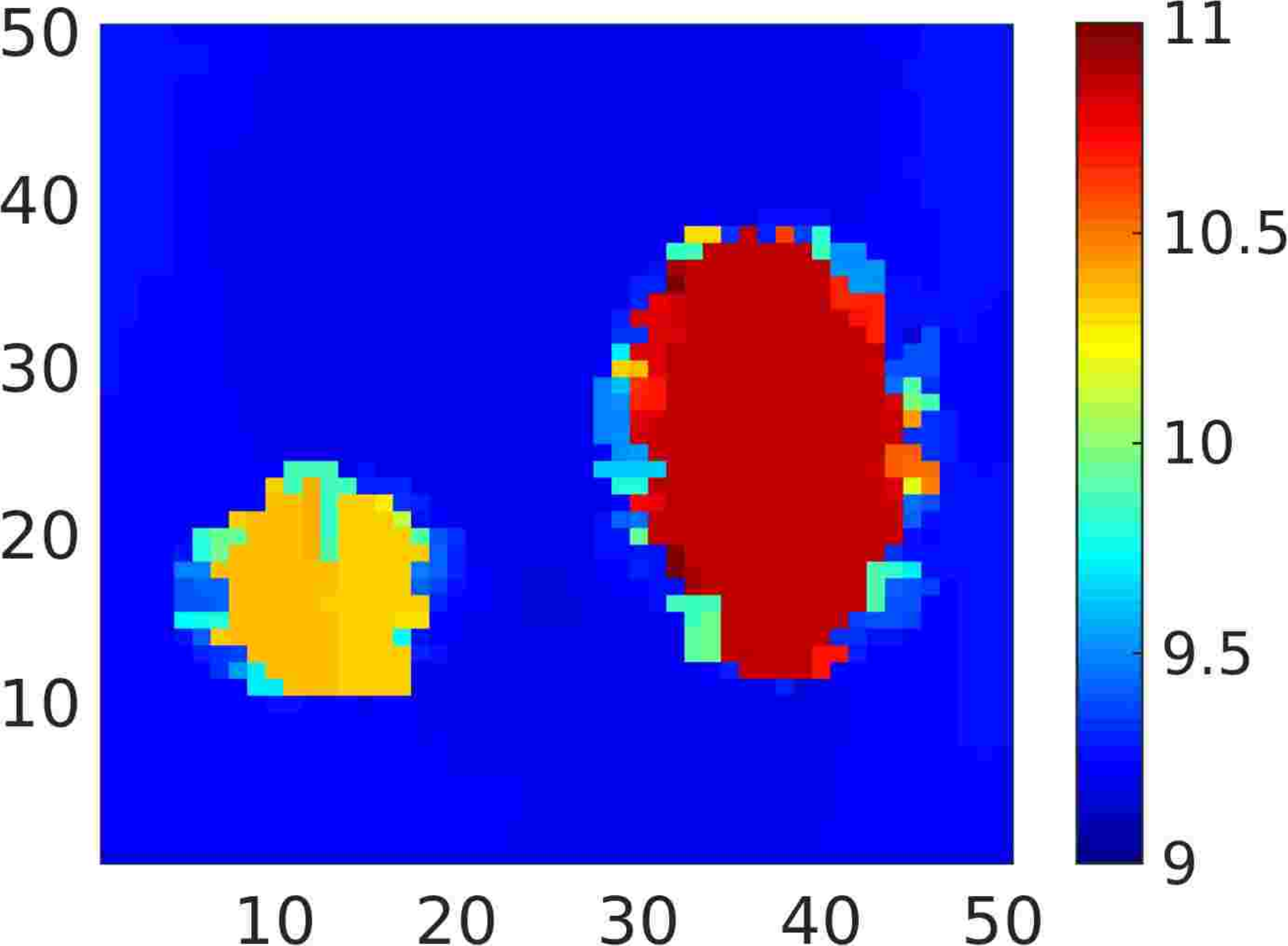}} 
		\hspace{0.3cm}
		\\
		\subfloat[][{ $q(s=7) =0.049$ }] 
		{\includegraphics[width=0.30\textwidth]{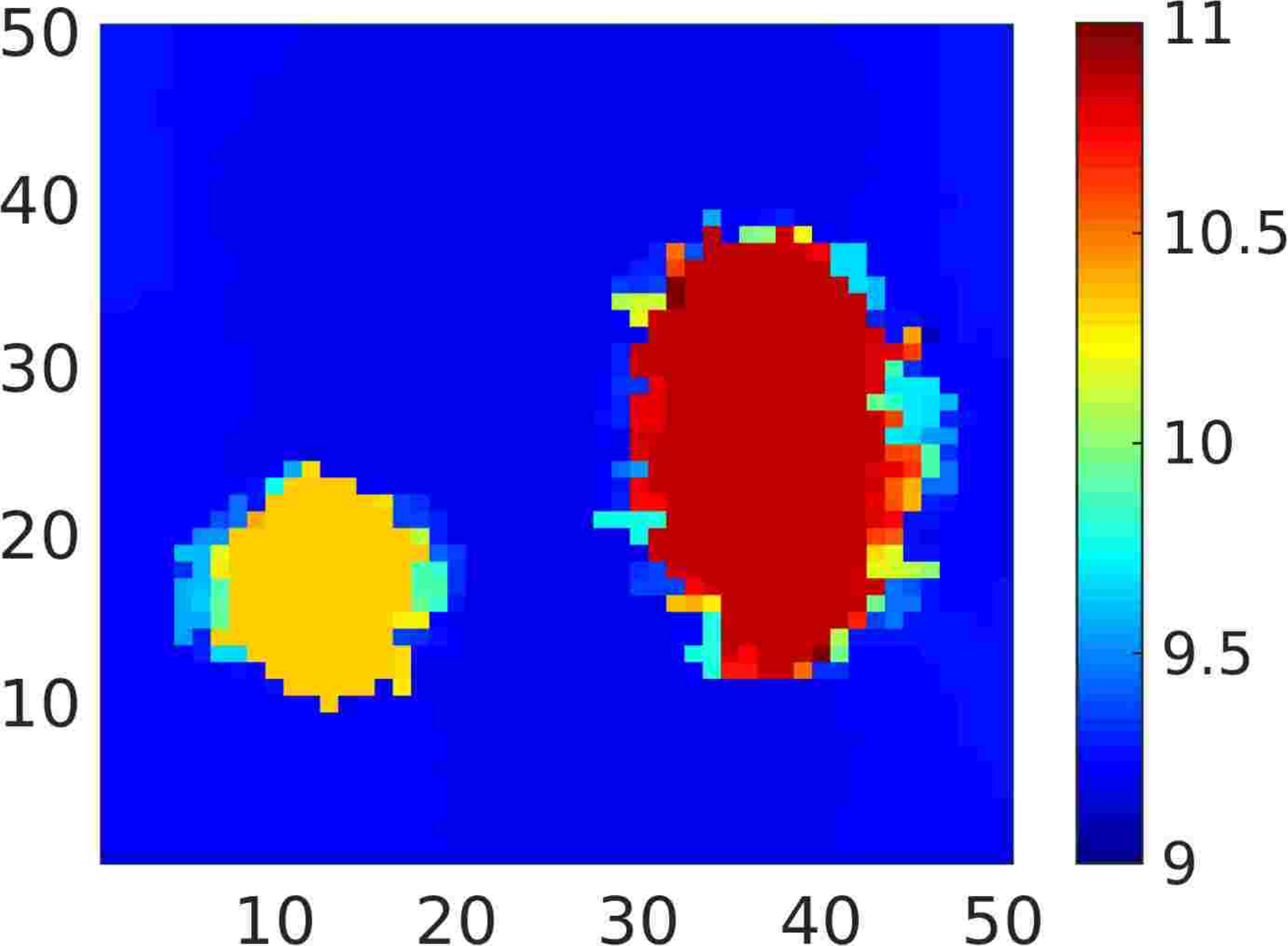}} 
		\hspace{0.1cm}
		\subfloat[][{$s=8$: deactivated}] 
		{\includegraphics[width=0.30\textwidth]{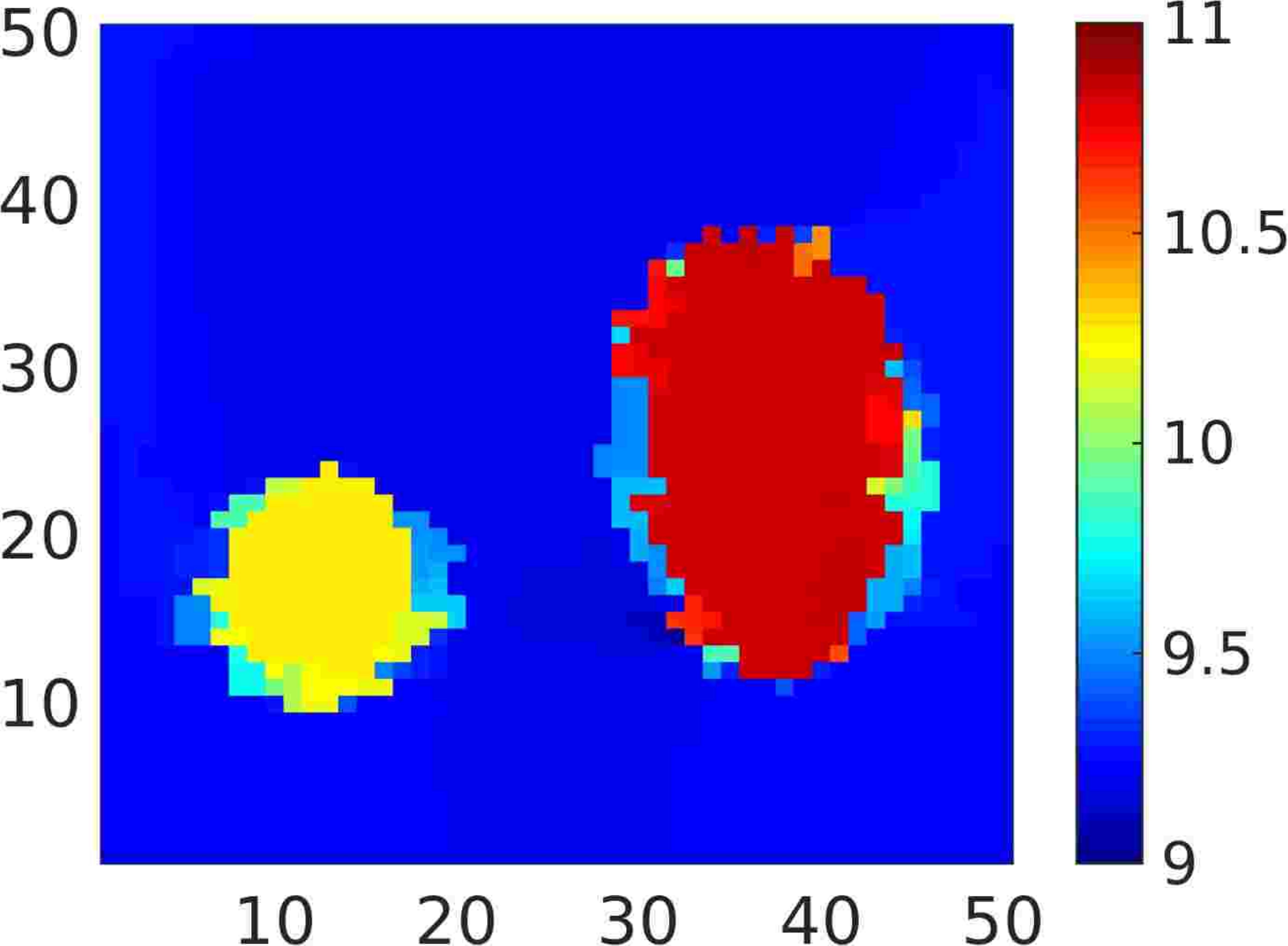}} 
		\hspace{0.1cm}
		\subfloat[][{$s=9$: deactivated}] 
		{\includegraphics[width=0.30\textwidth]{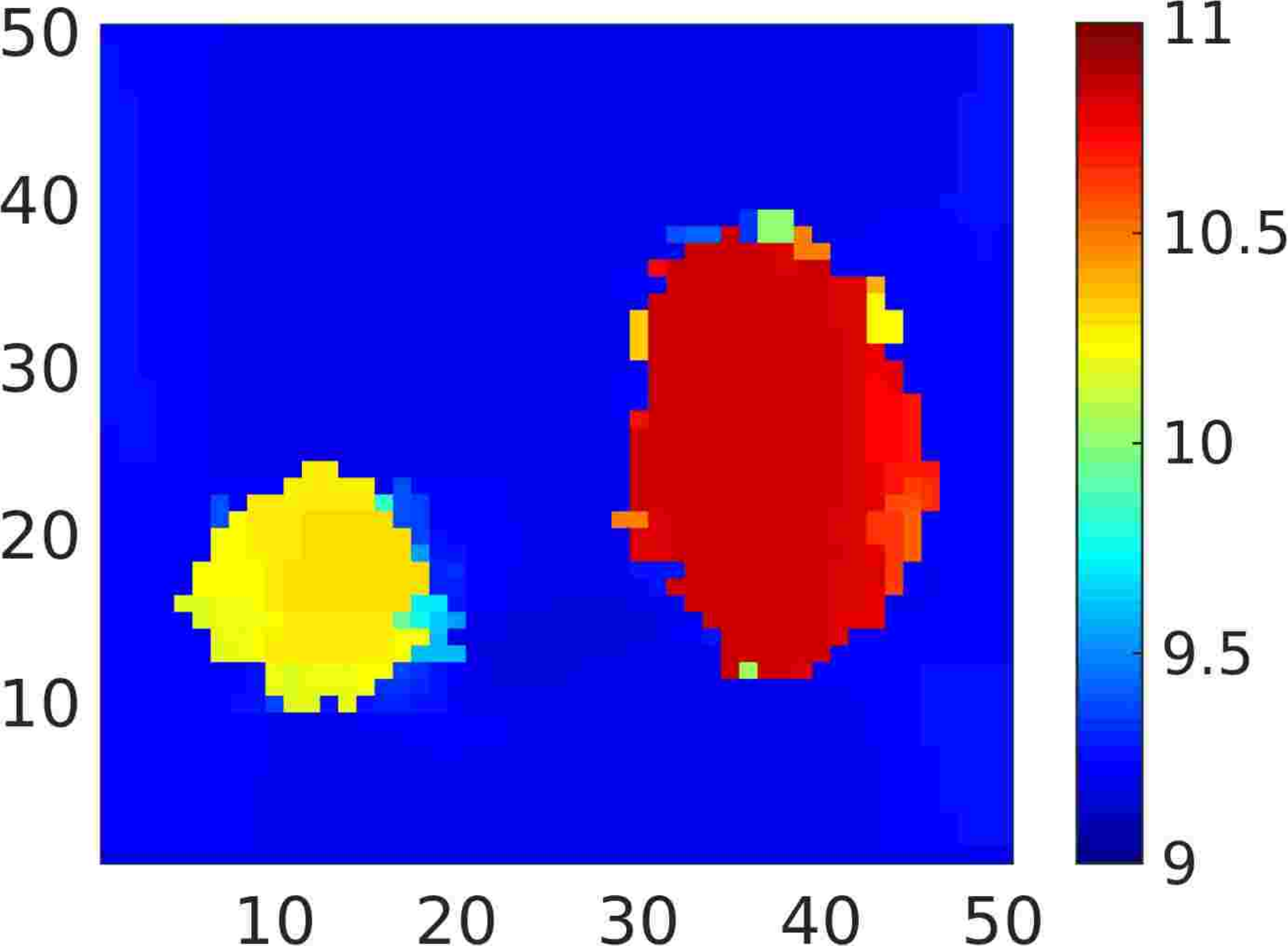}} 
		\hspace{0.3cm}
		\\
	  }
	  \caption{Posterior mean $\bmu_j$ for various  mixture components in log scale and their posterior  probabilities $q(s=j)$. The most active  components are $1$ and $3$. Mixture components $8,9$ are very similar to mixture components $4,2$ respectively and are therefore deleted/deactivated (based on ``Component Death'' criterion in Algorithm \ref{alg:adapt}, see also Table \ref{tab:KLDivMoG}).
	  }
	 \label{fig:PosteriorMeanMixtures}
\end{figure}

\begin{figure}[H]
	\centering{
		{\includegraphics[width=0.850\textwidth]{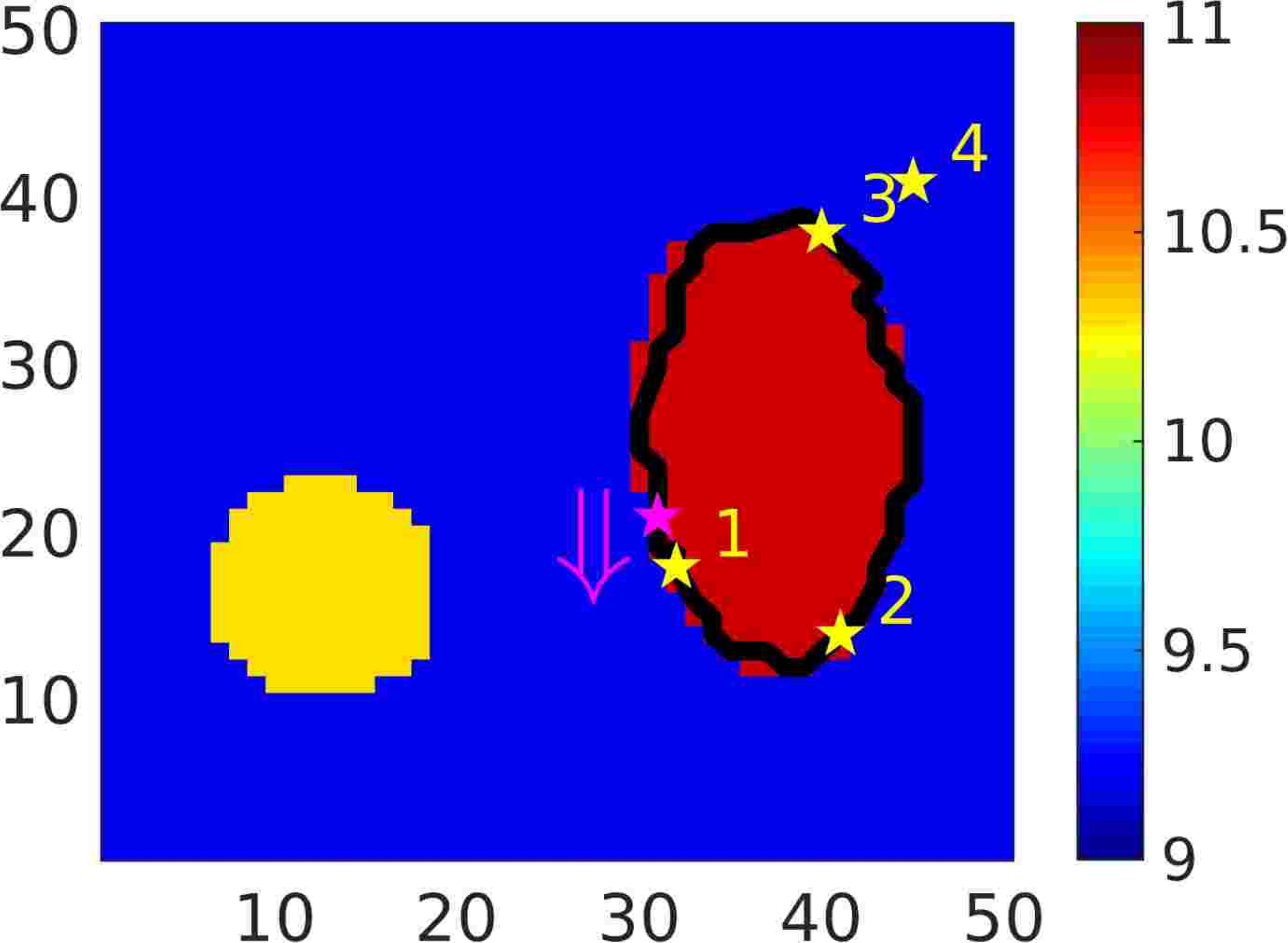}} 
		\hspace{0.1cm}
	  }
	  \caption{Posterior statistics for the elements along the black line are provided in \reffig{fig:CirclelargerInclusion}, starting at the magenta star and proceeding anti-clockwise around the inclusion. Posterior statistics for the elements marked by yellow stars ($1-4$) are supplied in \reffig{fig:MixturesSelectedParameters}. The background shows the ground truth in log scale.}
	 \label{fig:CirclelargerInclusionMarking}
\end{figure}

\begin{figure}[H]{
	\centering
	\captionsetup[subfigure]{labelformat=empty}
		\vspace{-1cm}
		\subfloat[][{Mixture component 1}] 
		{\includegraphics[width=0.70\textwidth]{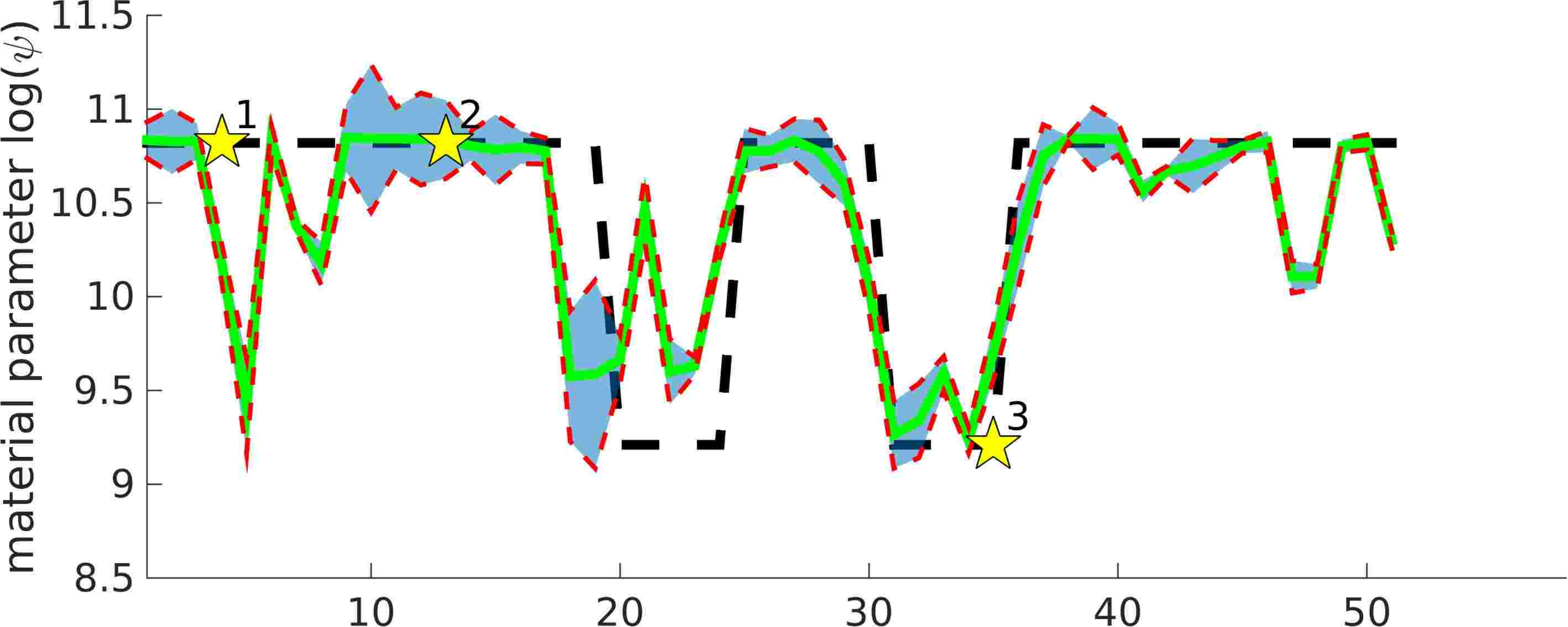}} \\
		\hspace{-0.cm}
		\subfloat[][{Mixture component 3}] 
		{\includegraphics[width=0.70\textwidth]{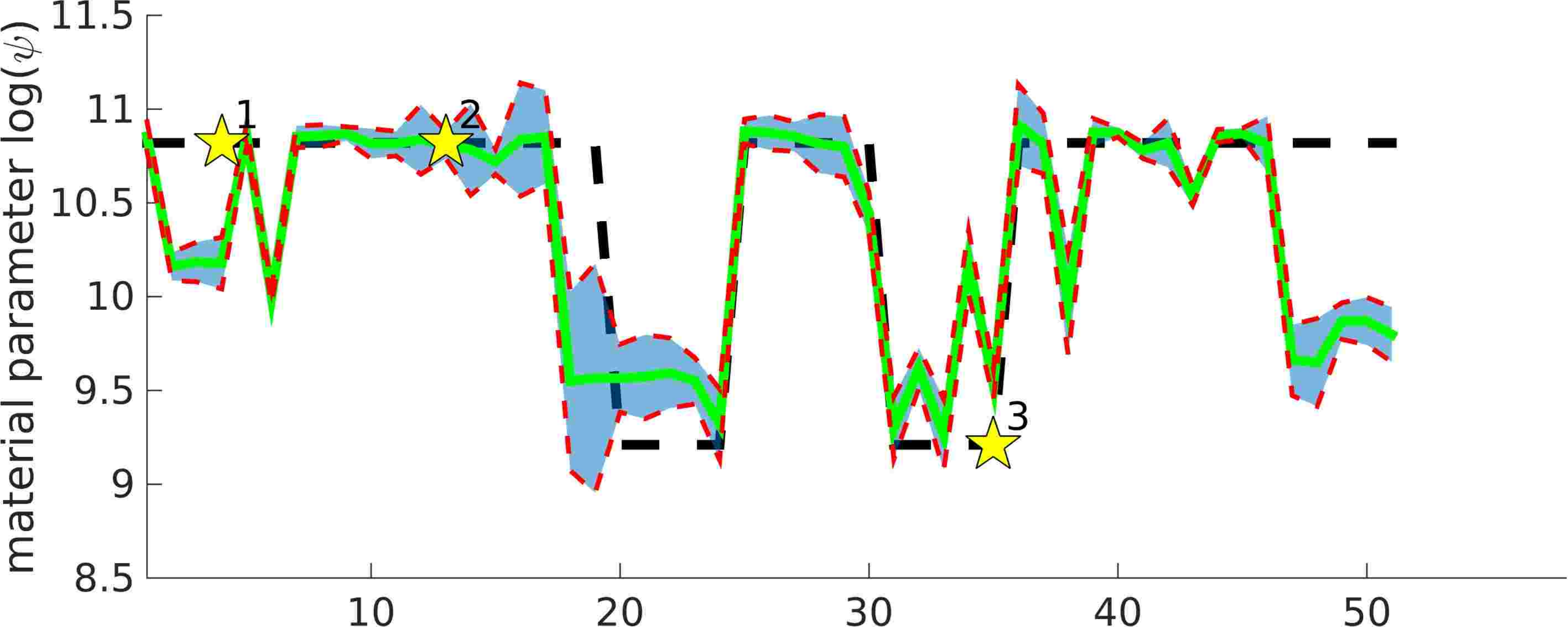}} \\
		\hspace{-0.2cm}
		\subfloat[][{Mixture component 6}] 
		{\includegraphics[width=0.70\textwidth]{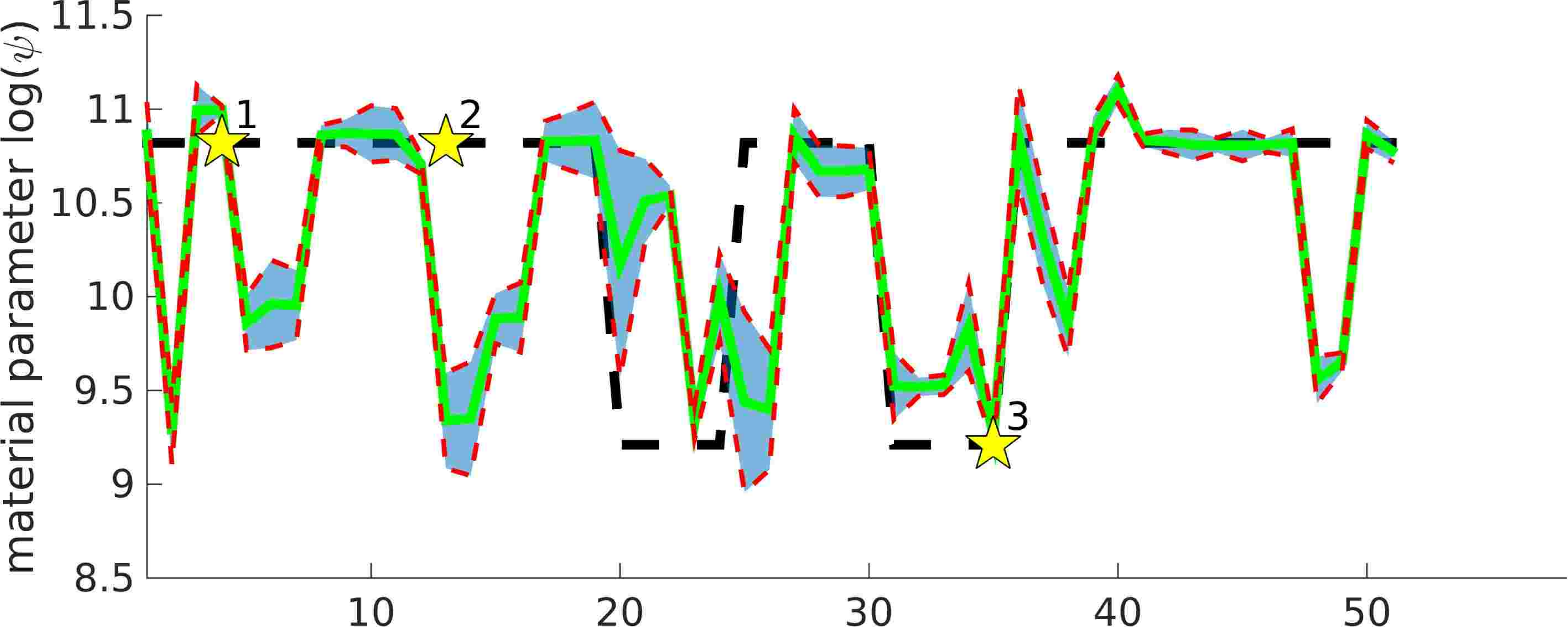}} \\
		  \hspace{-0.2cm}
		\subfloat[][{Mixture of Gaussians}] 
		{\includegraphics[width=0.70\textwidth]{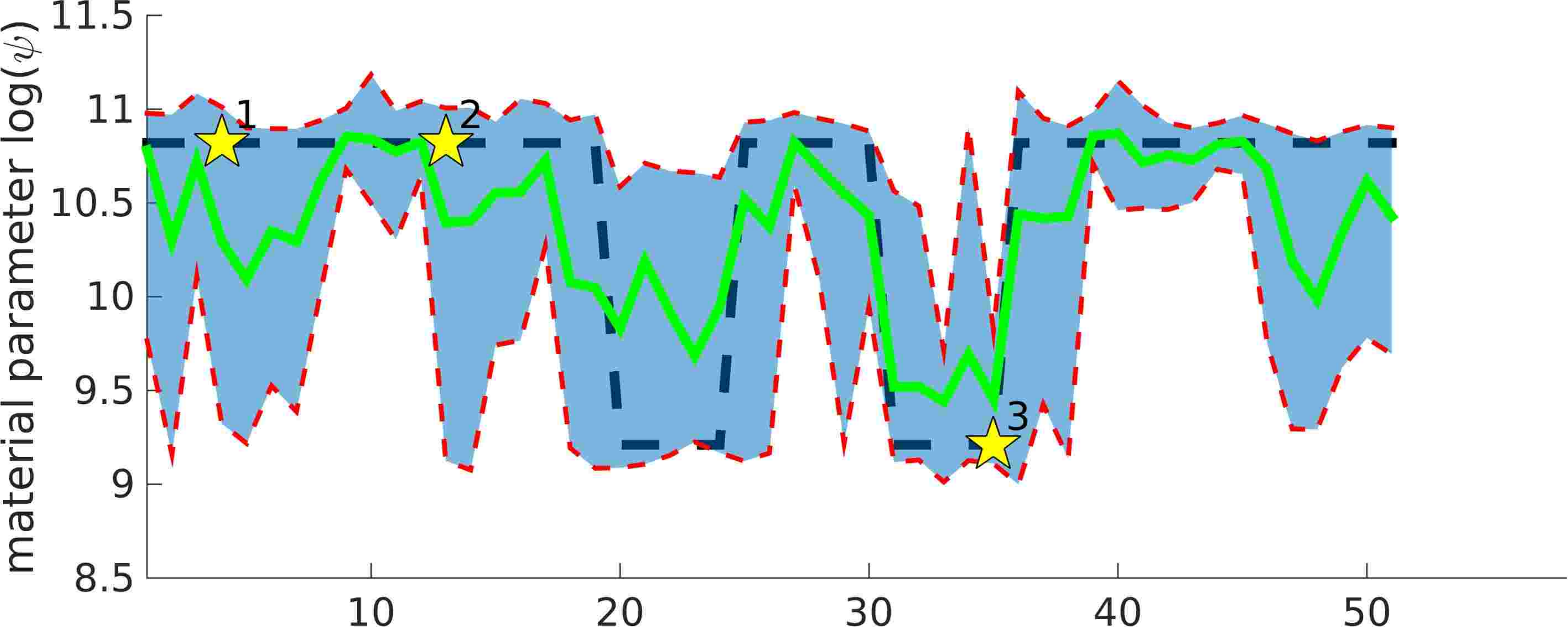}}  \\ 
	  }
	  \caption{Posterior statistics along the black line of \reffig{fig:CirclelargerInclusionMarking}. 
	 The ground truth is indicated by a black, dashed line. The posterior means are drawn with a green line {\color[rgb]{0,1,0}{---}} and credible intervals corresponding to $1\%$, $99\%$ percentiles in red {\color[rgb]{1,0,0}{- - -}}. 
}
	 \label{fig:CirclelargerInclusion}
\end{figure}

\begin{figure}[H]{
	\centering
	\captionsetup[subfigure]{labelformat=empty}
		\subfloat[][{Selected parameter 1}] 
		{\includegraphics[width=0.50\textwidth]{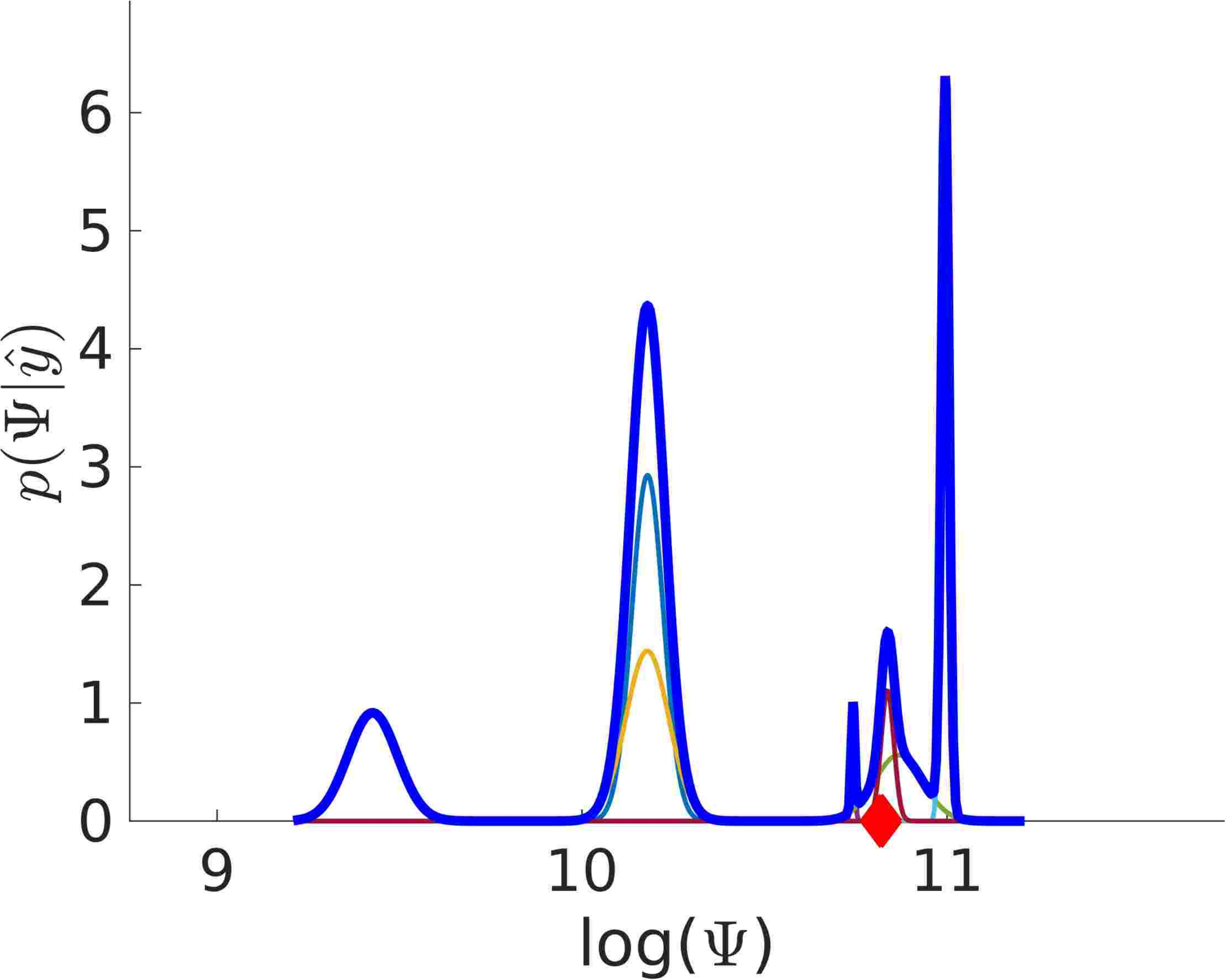}} 
		\hspace{0.1cm}
		\subfloat[][{Selected parameter 2}] 
		{\includegraphics[width=0.50\textwidth]{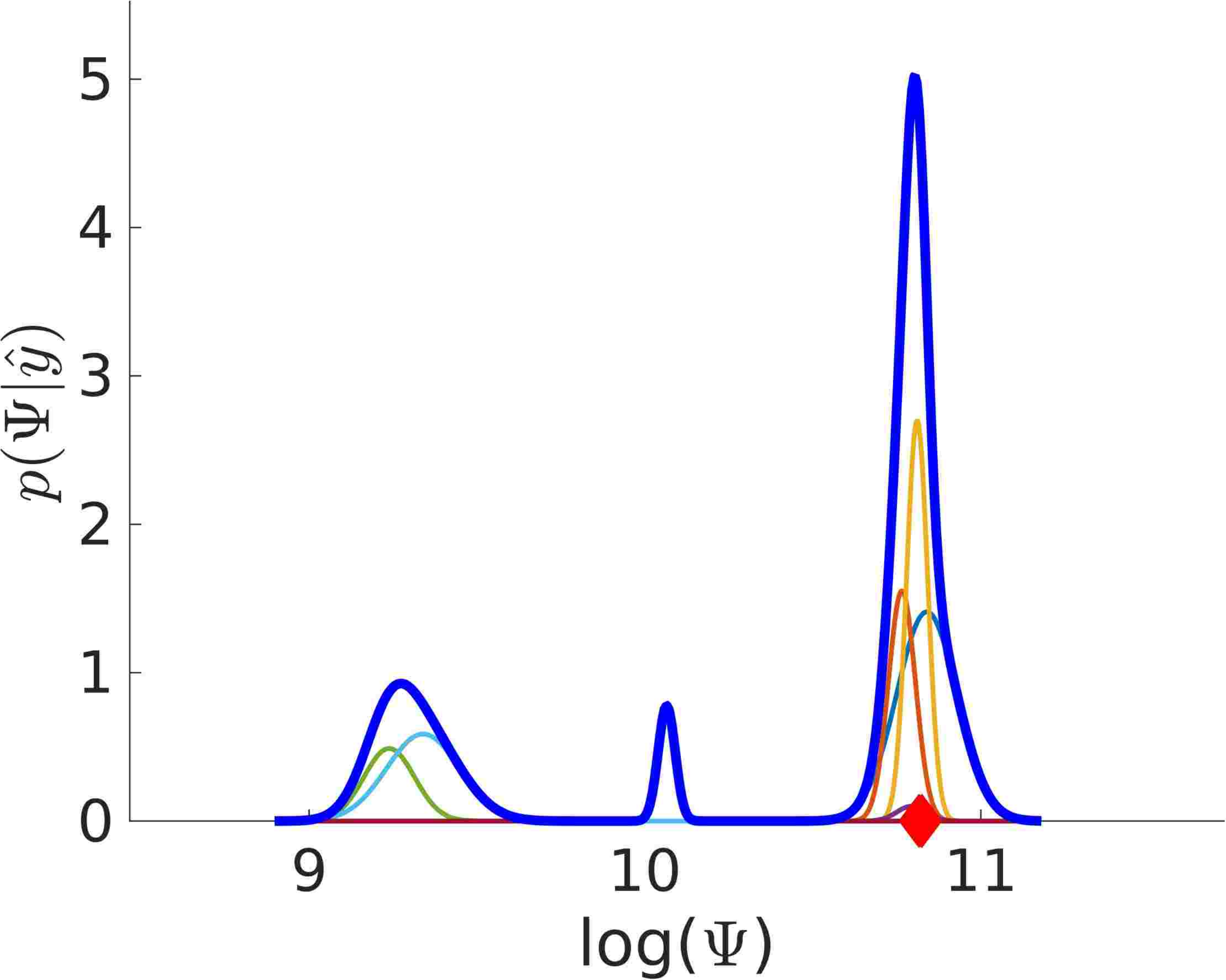}} 
		\hspace{0.1cm}
		\subfloat[][{Selected parameter 3}] 
		{\includegraphics[width=0.50\textwidth]{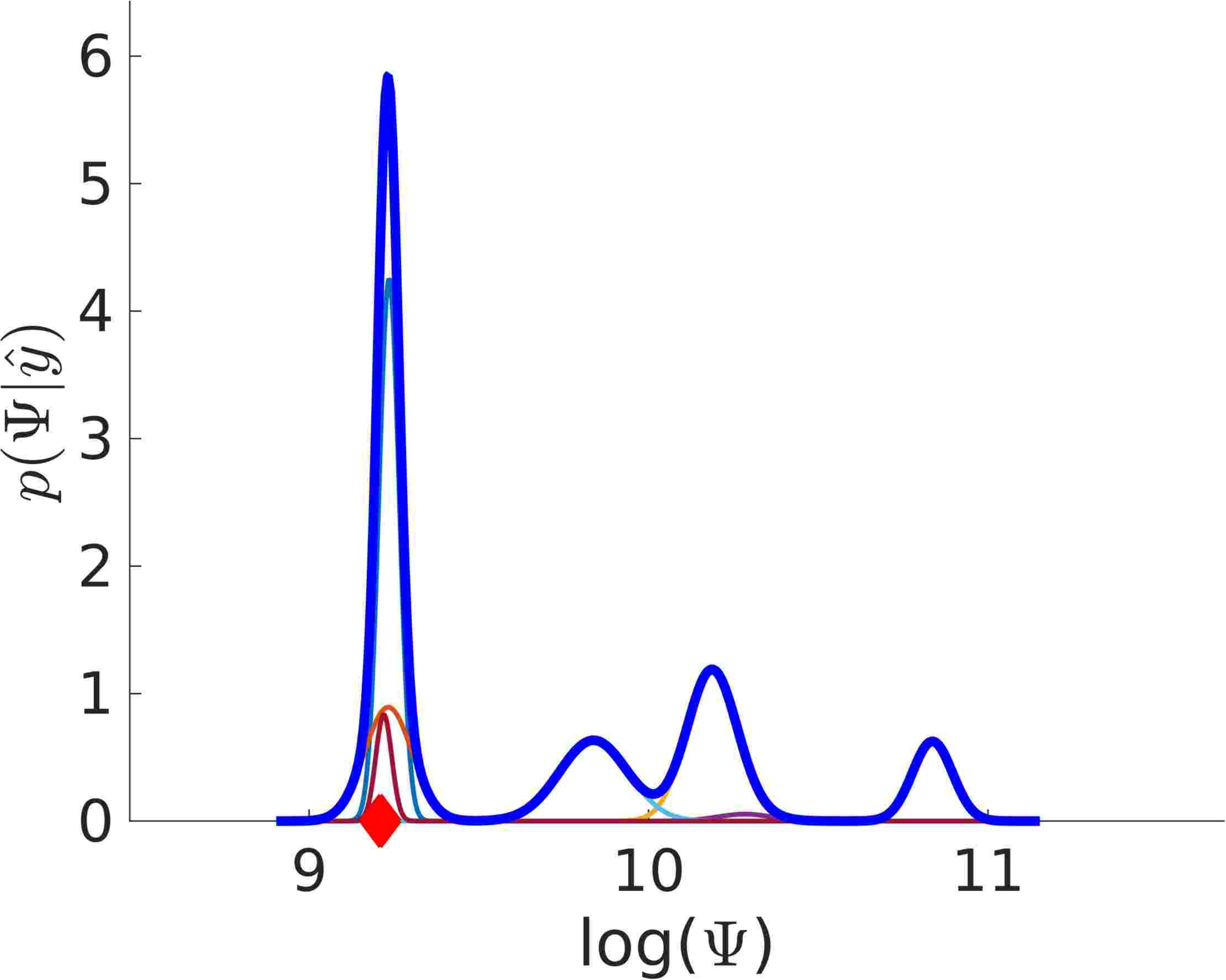}} 
		 \hspace{0.3cm}
		 \subfloat[][{Selected parameter 4}] 
		{\includegraphics[width=0.50\textwidth]{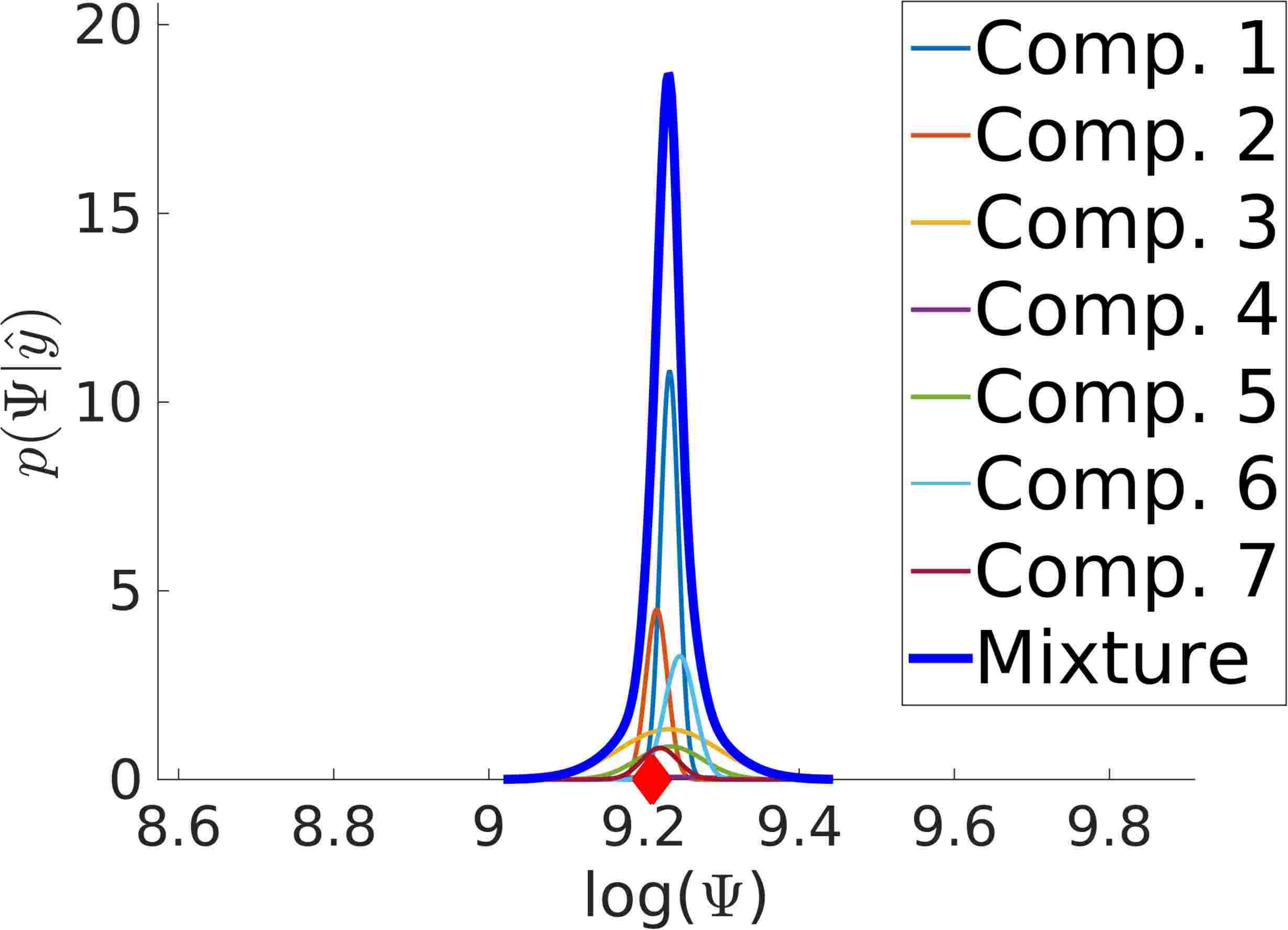}} 
		 \hspace{0.3cm}
	  }
	  \caption{Posterior probability densities of the log of the  elastic modulus, $\log(\Psi)$, of the elements $1$ through $4$ marked by yellow stars in \reffig{fig:CirclelargerInclusionMarking}. The ground truth values are indicated with red rhombuses. The probability densities (Gaussians) associated with each of the $7$  mixture components are multiplied by the corresponding  posterior probabilities  $q(s)$ and are shown by different colors. The combined, mixture of Gaussian is plotted in with a blue line. }
	 \label{fig:MixturesSelectedParameters}
\end{figure}

Upon convergence, seven ($S=7$) distinct mixture components were identified, which jointly approximate the posterior. The mean $\bs{\mu}_j$ of each component is shown in \reffig{fig:PosteriorMeanMixtures} where the posterior responsibilities $q(s)$ are also reported. The numbering of the components relates to the order in which they were found by the algorithm. We observe that all mixture components identify the bulk of the two inclusions and most differences pertain to their boundaries (see also Figures \ref{fig:CirclelargerInclusionMarking}, \ref{fig:CirclelargerInclusion}, \ref{fig:MixturesSelectedParameters}). The shape of the boundaries has been found to play a defining role in distinguishing between malignant and benign tumors and metrics have been developed that provide a good diagnostic signature using this information \cite{liu_noninvasive_2015}. Apart from the seven active components, the means of two additional  mixture components  ($s=8, s=9$) which were deactivated (based on the  ``Component Death'' criterion in 
Algorithm \ref{alg:adapt}),  are shown. 

In  Table \ref{tab:KLDivMoG}, we also report the  (normalized) KL-divergence between all pairs of these nine components. One notes that component $8$ was deleted because it was too similar to component $4$ (from \refeq{eq:kldist} $d_{4,8}=0.33 \times 10^{-2} < d_{min}=0.01$) and component $9$ was too similar to component $2$ ($d_{2,9}=0.56 \times 10^{-2} < d_{min}=0.01$).  

\begin{table}[h]
    \begin{center}
      \begin{tabular}{ | c | c | c | c| c| c| c| c | c| c|c | c|}
	\hline
	    & s=1 &  s=2 & s=3 & s=4 & s=5 & s=6 & s=7 &  s=8 & s=9 \\ \hline\hline
	s=1 & 0	& 12.05 &  9.87 & 14.33 & 17.10 & 16.96 & 15.02 & 14.82 & 12.50 \\ \hline
	s=2 & 	& 0     & 16.46 & 15.86 & 21.18 & 19.72 & 16.88 & 16.54 &   \bf{0.56}\\ \hline
	s=3 & 	&       & 0     & 11.06 & 16.43 & 17.23 & 17.06 & 11.45 & 18.16  \\ \hline
	s=4 & 	& 	&	&0	& 12.74 & 12.80 & 16.31 & \bf{0.33} &  16.68  \\ \hline
	s=5 & 	& 	&	&	&0	& 12.62 & 17.99 & 13.73 & 23.47  \\ \hline
	s=6 & 	& 	&	&	&	& 0 	& 11.13 & 13.25 & 20.52  \\ \hline
	s=7 & 	& 	&	&	&	&  	&   0	& 16.72 & 19.51 \\ \hline
	s=8 & 	& 	&	&	&	&  	&   	& 0 	& 18.44 \\ \hline
	\hline
    
      \end{tabular}   
      \caption{Normalized  KL divergences (\refeq{eq:kldist}) between all pairs of the  mixture components. All values shown should be multiplied with  $\times 10^{-2}$.
      }
      \label{tab:KLDivMoG}  
    \end{center}  
\end{table}

\begin{figure}[H]{
	\centering
	\captionsetup[subfigure]{labelformat=empty}
		\subfloat[][{ $q(s=1) =0.318$}] 
		{\includegraphics[width=0.31\textwidth]{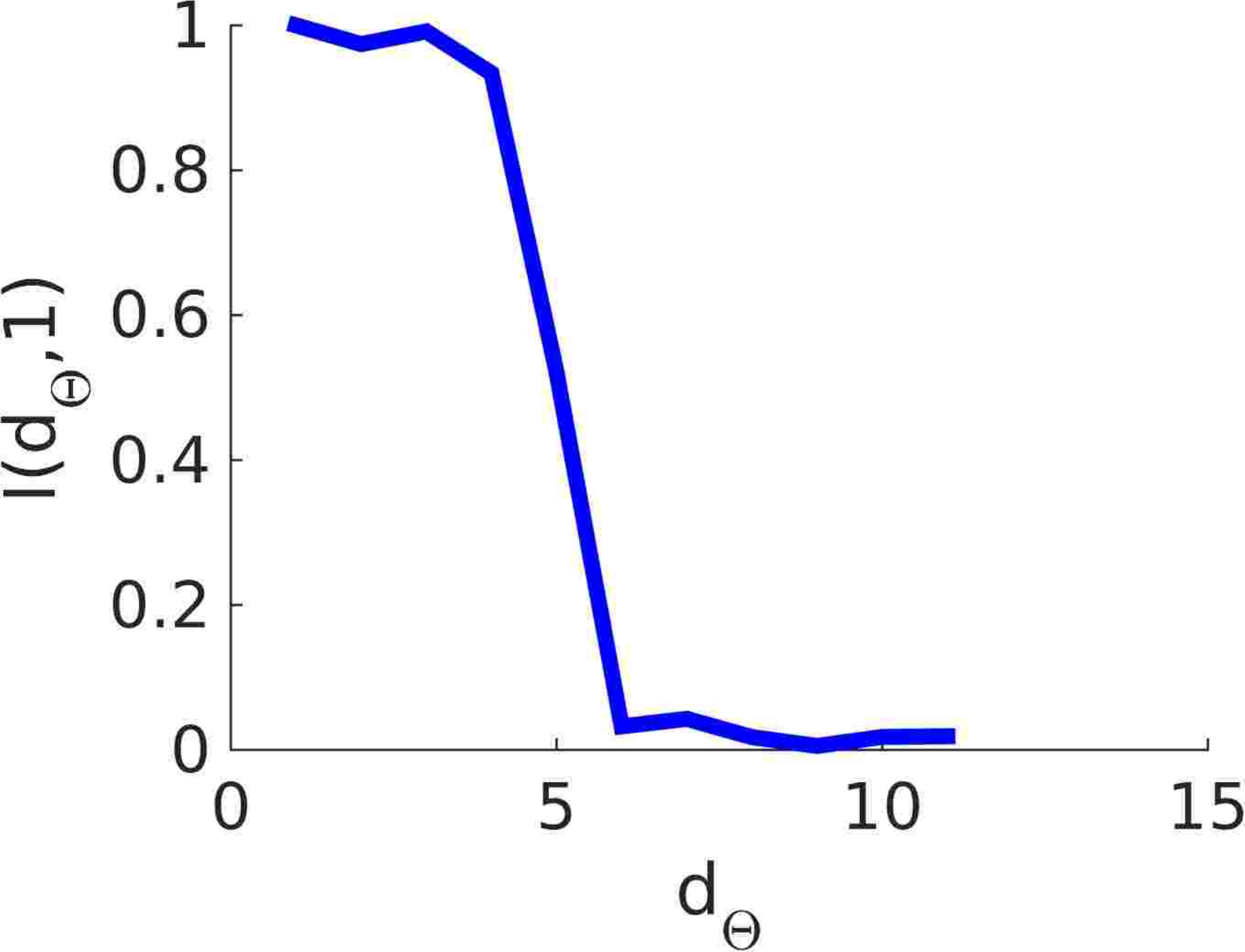}} 
		\hspace{0.1cm}
		\subfloat[][{ $q(s=3) =0.213$}] 
		{\includegraphics[width=0.31\textwidth]{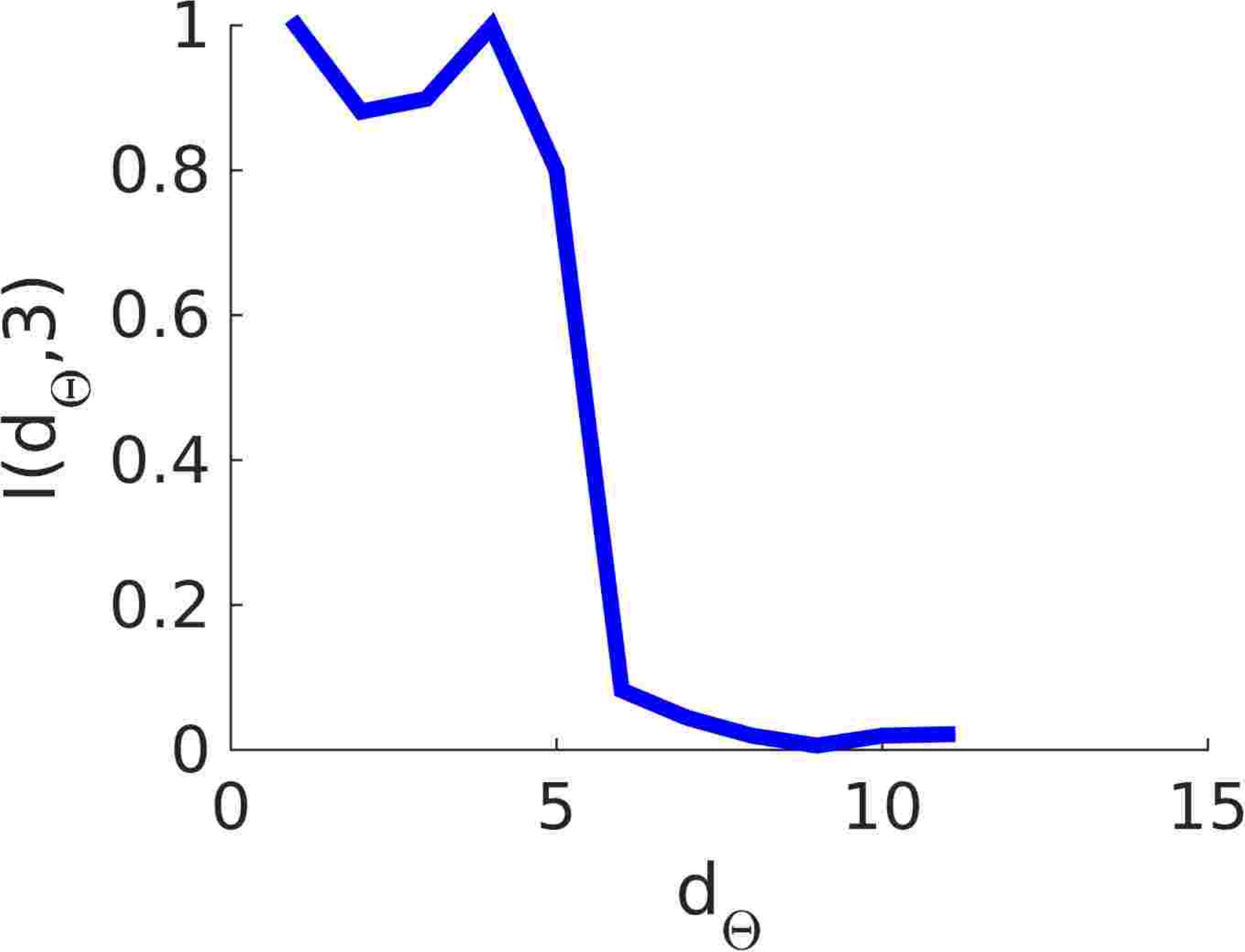}} 
		\hspace{0.1cm}
		\subfloat[][{ $q(s=6) =0.160$}] 
		{\includegraphics[width=0.31\textwidth]{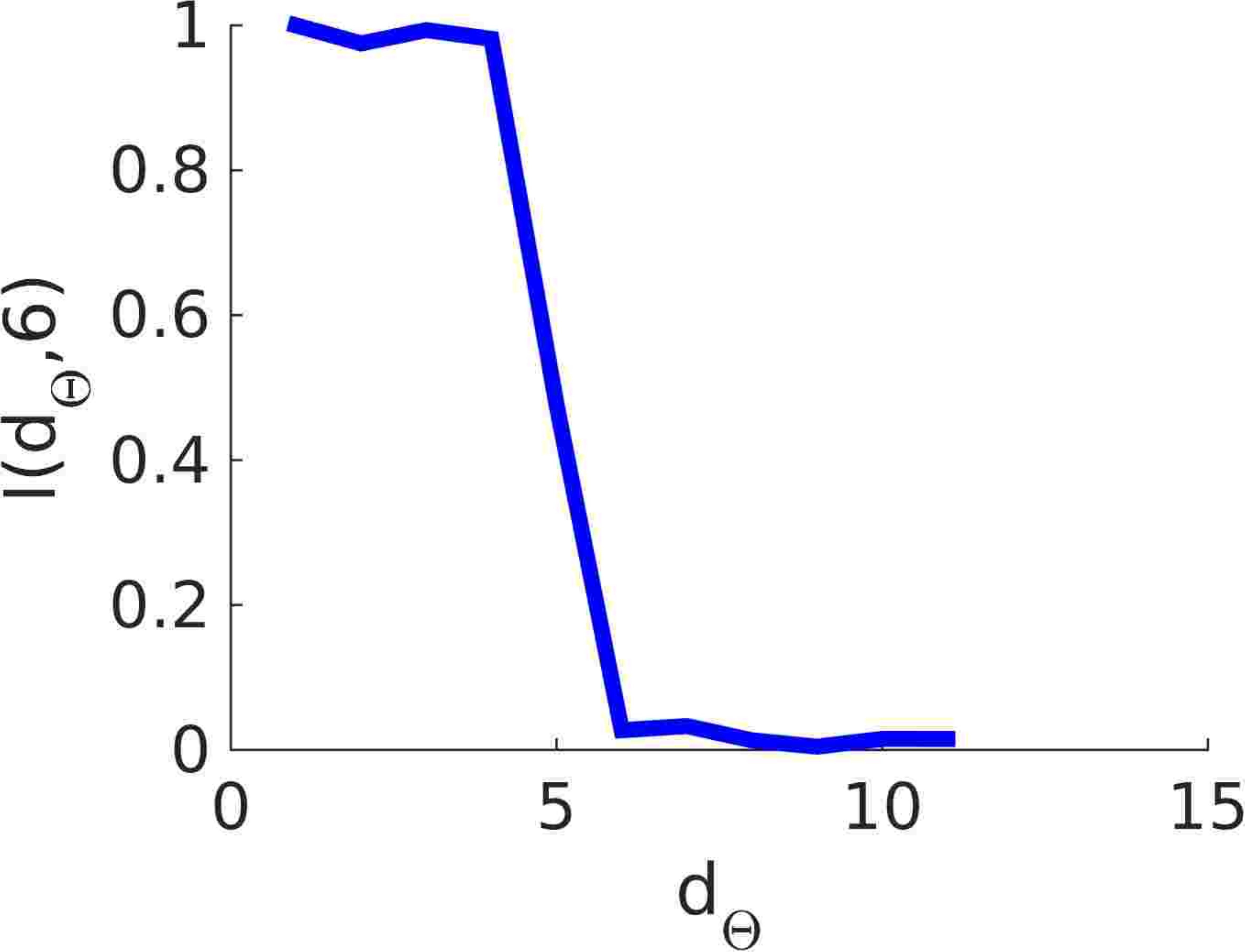}} 
		\hspace{0.3cm}
	  }
	  \caption{Information gain $I(\dth,j)$ for three mixture components (see also \ref{app:IGBases}). 
	  }
	 \label{fig:IG}
\end{figure}

With regards to the covariance of each mixture component and the identification of the lower-dimensional subspaces, we employ the information-theoretic criterion previously discussed in order to adaptively determine the  number of reduced-dimensions $\dth$. To that end we use the relative information gains $I(\dth,j)$ (\refeq{eq:infgain}, see also \ref{app:IGBases}) which are for the three most active mixture components depicted in \reffig{fig:IG}).
We note that $I(\dth,j)$ drops to relatively small values after a small number of reduced coordinates (with $\dth = 8$, it drops to $1\%$).  In the following results we used $\dth = 11$. We discuss in \refsec{sec:other}  the behavior of the proposed scheme in cases in which the problem is not amenable to such a dimensionality reduction.
%
%
%

\begin{figure}[H]{
	\centering
	\captionsetup[subfigure]{labelformat=empty}
		\subfloat[][{Mean of the mixture}] 
		{\includegraphics[width=0.450\textwidth]{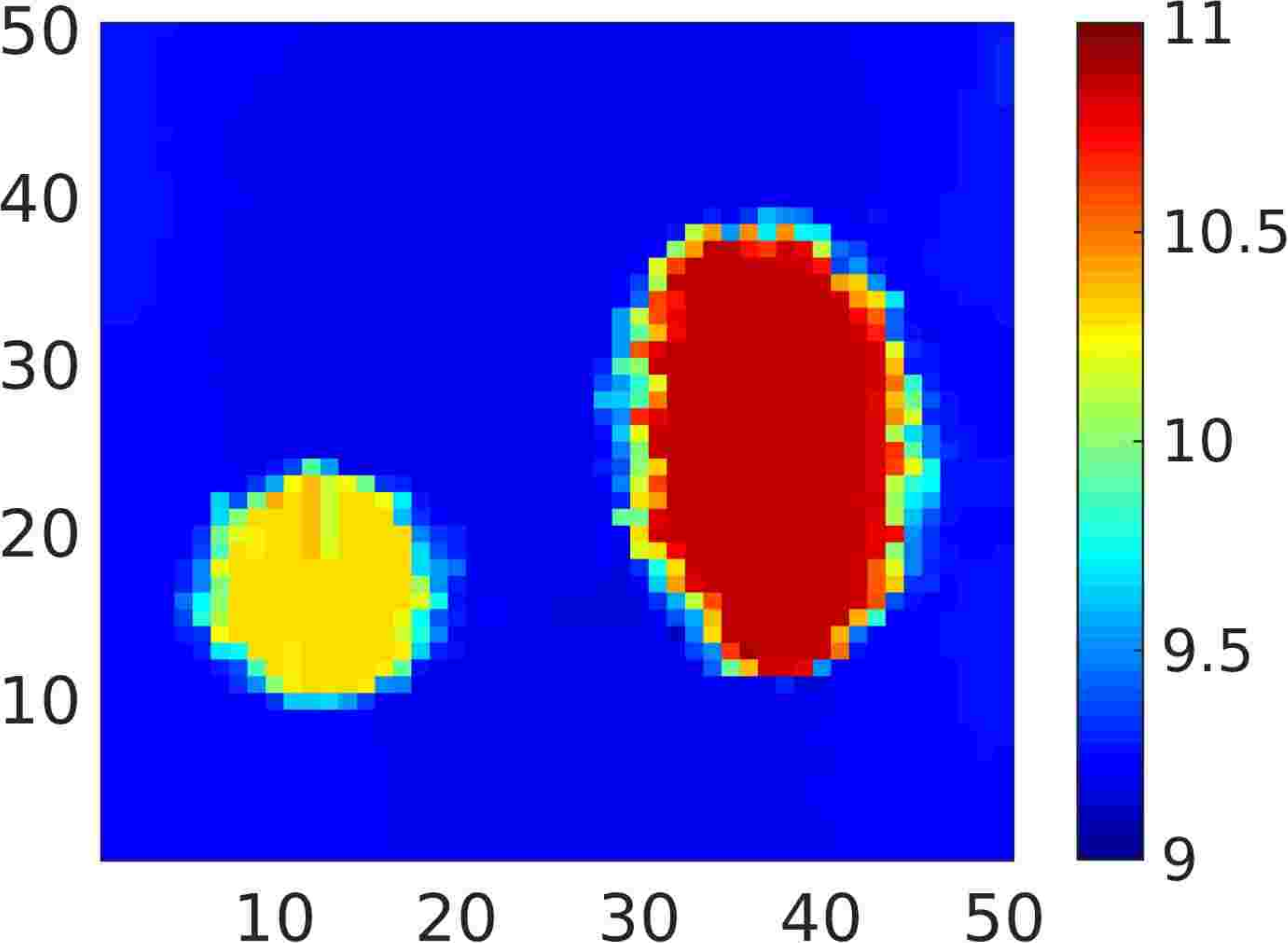}} 
		\hspace{0.1cm}
		\subfloat[][{Standard deviation of the mixture}] 
		{\includegraphics[width=0.450\textwidth]{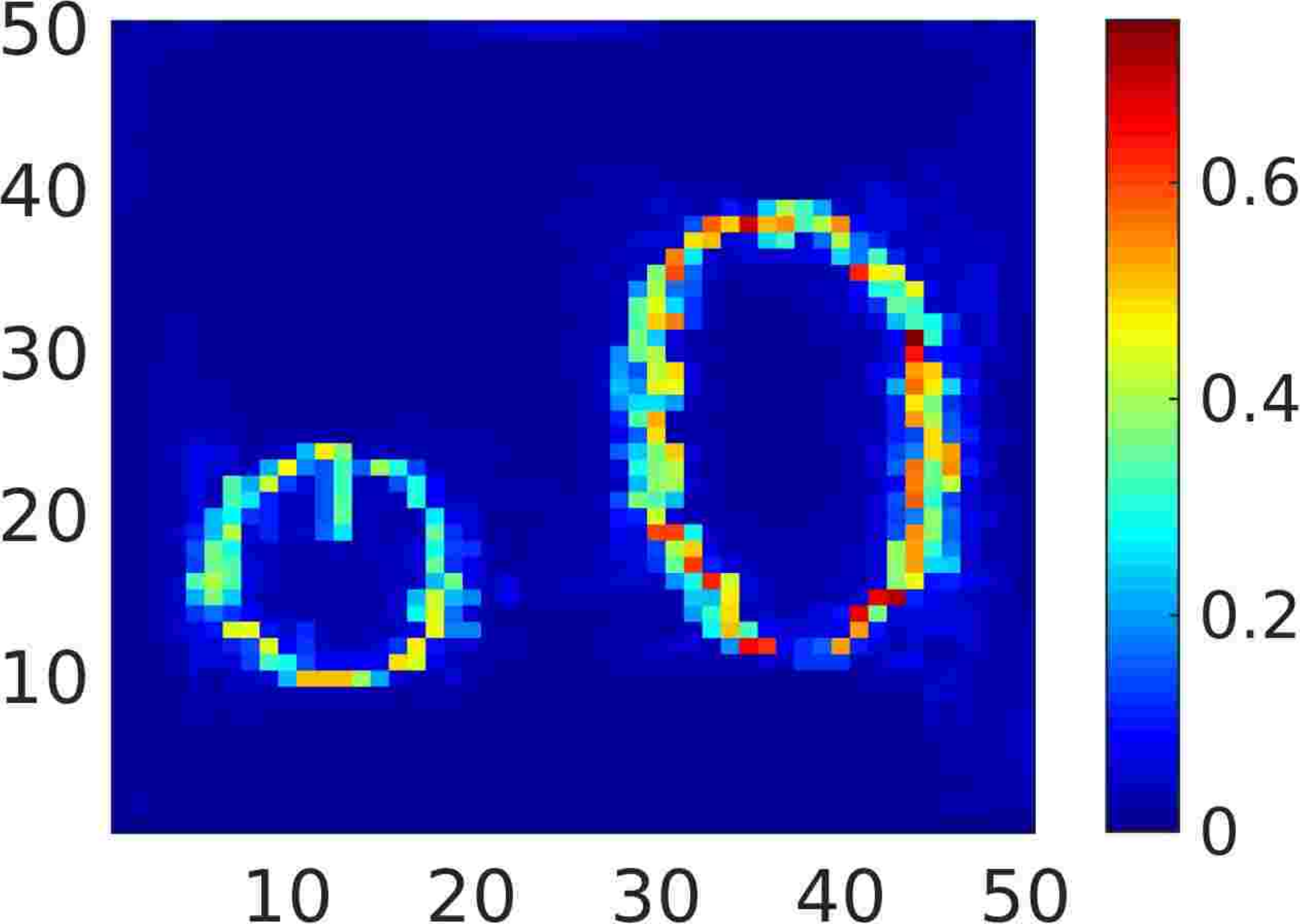}} 
		\hspace{0.1cm}
	  }
	  \caption{Approximate posterior mean and posterior standard deviation as computed from the mixture of Gaussians in \refeq{eq:qpost1} (in log scale).}
	 \label{fig:PosteriorMeanStdMixtures}
\end{figure}

We defer further discussions on the individual mixture components in order to discuss the overall approximate posterior.
The posterior mean and standard deviation of the mixture of Gaussians (\refeq{eq:qpost1}) are shown in \reffig{fig:PosteriorMeanStdMixtures}.
As expected, the posterior variance is largest at the boundaries of the inclusions.
%

\begin{figure}[H]
	\centering{
		{\includegraphics[width=0.850\textwidth]{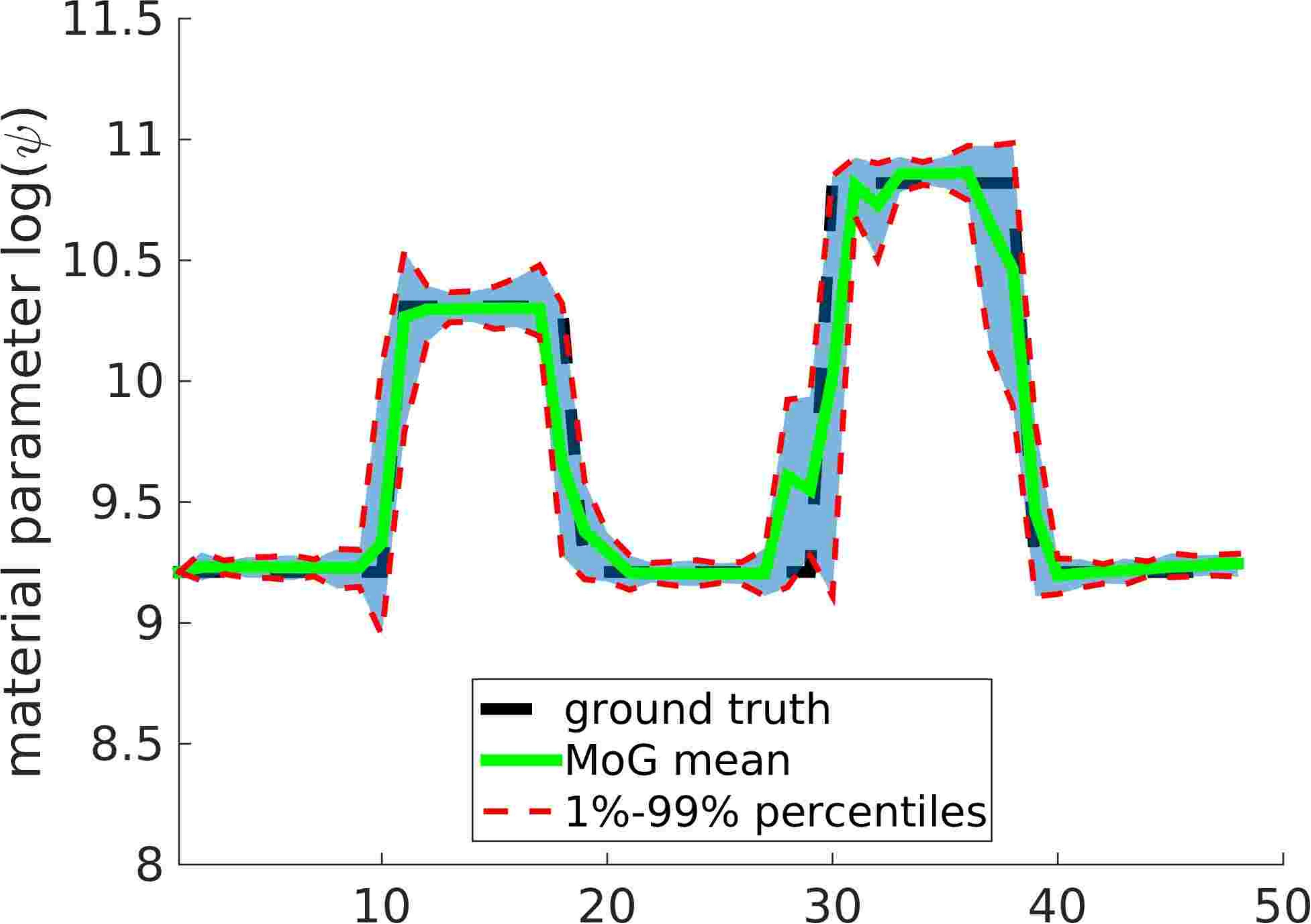}} 
		\hspace{0.1cm}
	  }
	  \caption{Posterior mean and credible intervals corresponding to  $1\%$ and $99\%$- (dashed lines), along the 
	  diagonal from $(0, 0)$ to $(50, 50)$.  
	  }
	 \label{fig:DiagonalCutMixture}
\end{figure}

\reffig{fig:DiagonalCutMixture} depicts the posterior mean and $1\%-99\%$ credible intervals  
along the diagonal of the problem domain i.e. from $(0, 0)$ to $(50, 50)$. We note that the posterior quantiles envelop the ground truth.

For validation purposes we performed Importance Sampling as described in \refsec{sec:IS} in order to assess the overall accuracy of the approximation (a total of $M=5000$ were generated). The Effective Sample Size (\refeq{eq:ess}) was $ESS=0.48$  which indicates that the identified mixture of low-dimensional Gaussians provides a very good approximation to the actual posterior.  In comparison, MCMC simulations performed using a Metropolis-adjusted Langevin scheme (MALA, \cite{robert_monte_2013})  exhibited very long correlation lengths resulting in $ESS_{MCMC} < 10^{-3}$ \footnote{Due to the computational expense, the MALA simulation results were actually obtained  on a coarser discretization of the forward problem resulting in only $100$ unknowns (in contrast to the $2500$ in the target problem). 
The step sizes in the proposals were adapted to ensure that, on average, $60\%$ of the moves were accepted \cite{roberts_optimal_1998}. The resulting  $ESS_{MCMC}$ was $10^{-3}$.  While additional fine-tuning could improve upon this, we doubt that, for the actual problem which has $25$ times more unknowns,  it will ever reach the ESS of the proposed approximation.}.

\begin{figure}[h]{
	\centering
	\captionsetup[subfigure]{labelformat=empty}
		\subfloat[][{$\bs{\mu}_1$}] 
		{\includegraphics[width=0.30\textwidth]{FiguresIF/PlotYoungModulusMeanVBMoG0.pdf}} 
		\hspace{0.1cm}
		\subfloat[][{$\bs{\mu}_{mixture}$}] 
		{\includegraphics[width=0.3\textwidth]{FiguresIF/PlotYoungModulusMeanMixture.pdf}} 
		\hspace{0.1cm}
		\subfloat[][{$\bs{\mu}_{IS}$}] 
		{\includegraphics[width=0.3\textwidth]{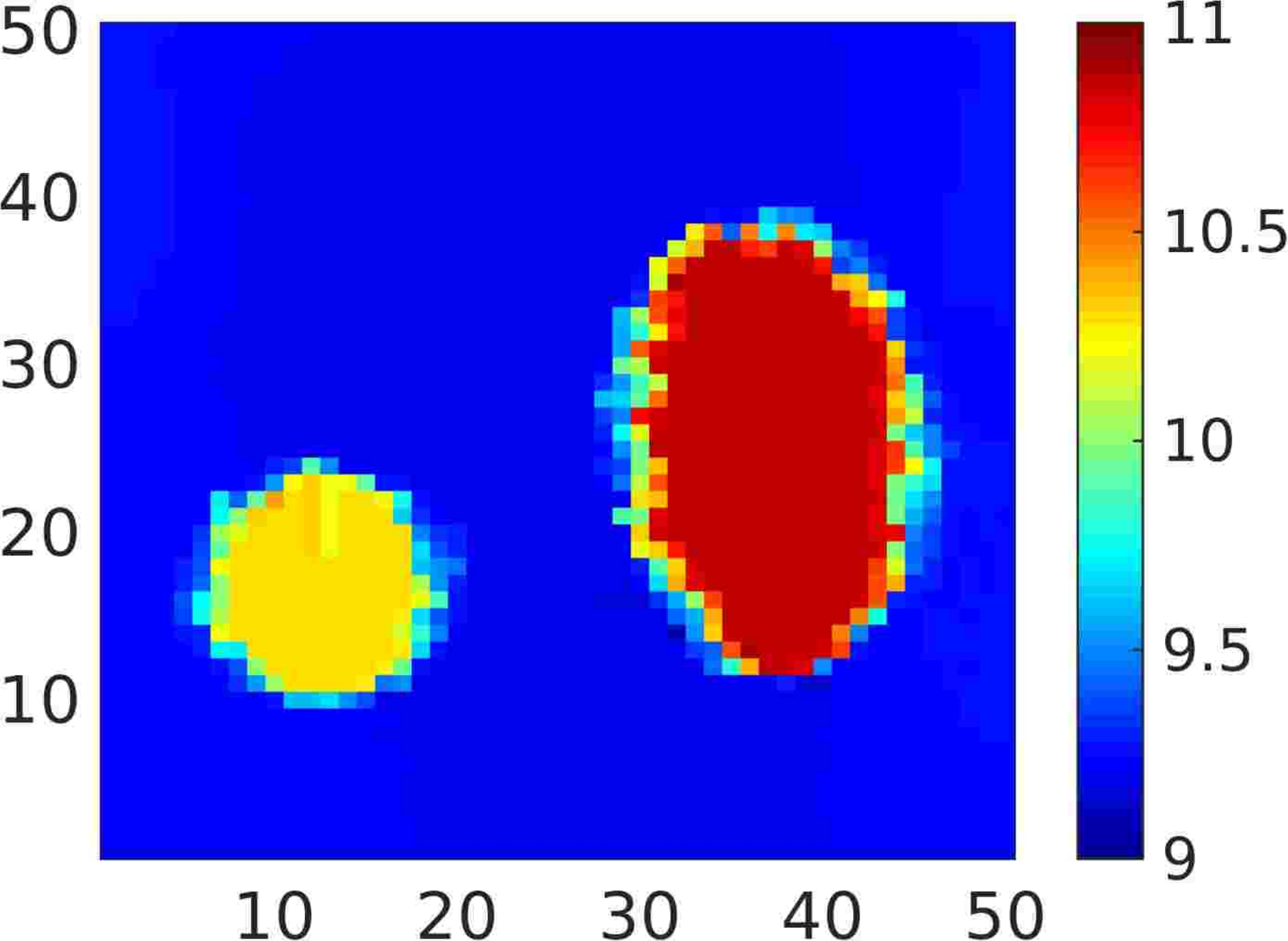}} 
		  \hspace{0.3cm}
	  }
	  \caption{Comparison of the posterior mean found with a single mixture component ($\bs{\mu}_1$, left), with that found with a mixture of Gaussians ($\bs{\mu}_{mixture}$, middle) and the exact mean estimated with IS ($\bs{\mu}_{IS}$, right). Depictions are in log scale. 
	  }
	 \label{fig:ISMean}
\end{figure}

\begin{figure}[H]{
	\centering
	\captionsetup[subfigure]{labelformat=empty}
		\subfloat[][{$\bs{\sigma}_1$}] 
		{\includegraphics[width=0.30\textwidth]{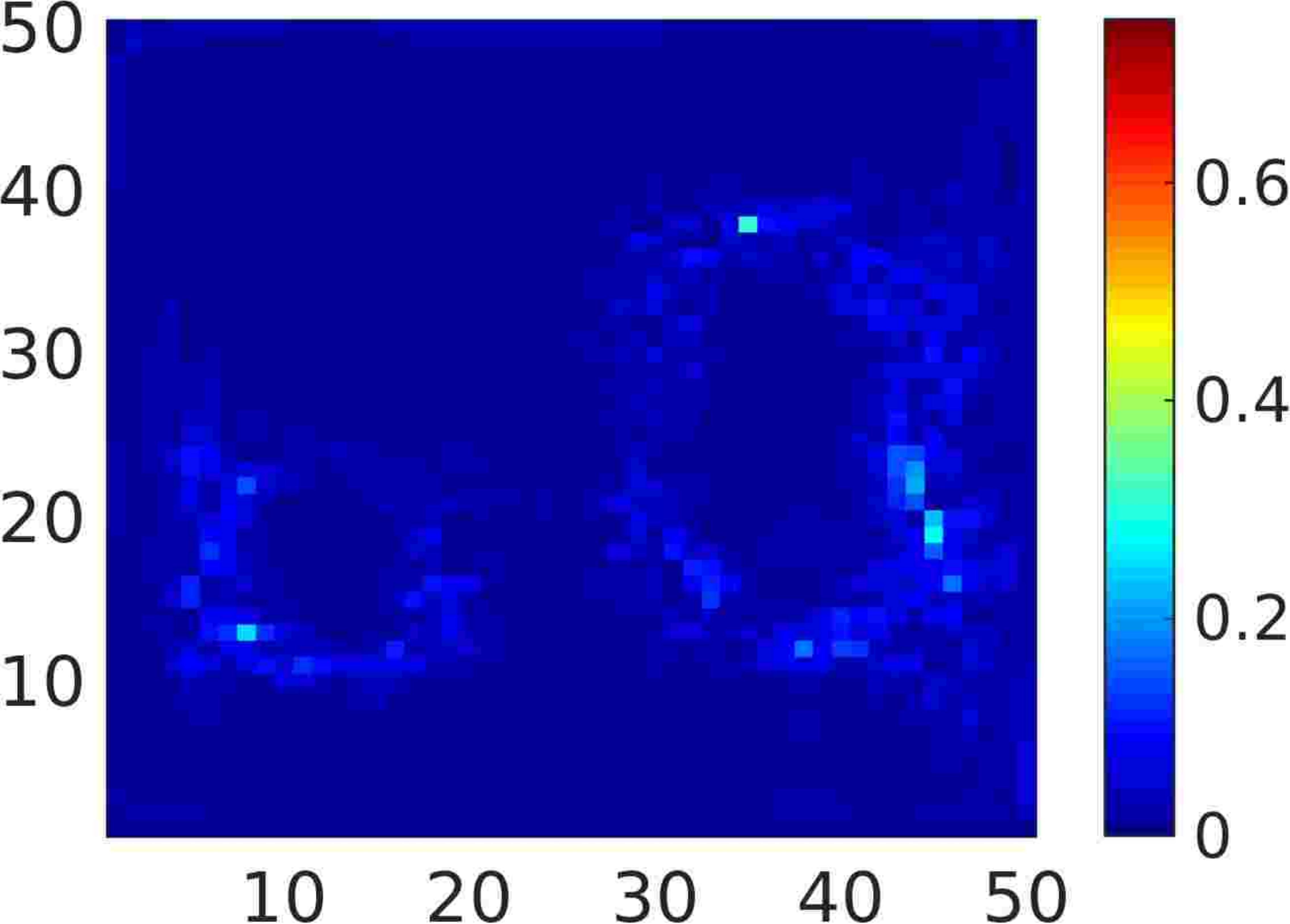}} 
		\hspace{0.1cm}
		\subfloat[][{$\bs{\sigma}_{mixture}$}] 
		{\includegraphics[width=0.3\textwidth]{FiguresIF/PlotYoungModulusSTDMixture.pdf}} 
		\hspace{0.1cm}
		\subfloat[][{$\bs{\sigma}_{IS}$}] 
		{\includegraphics[width=0.3\textwidth]{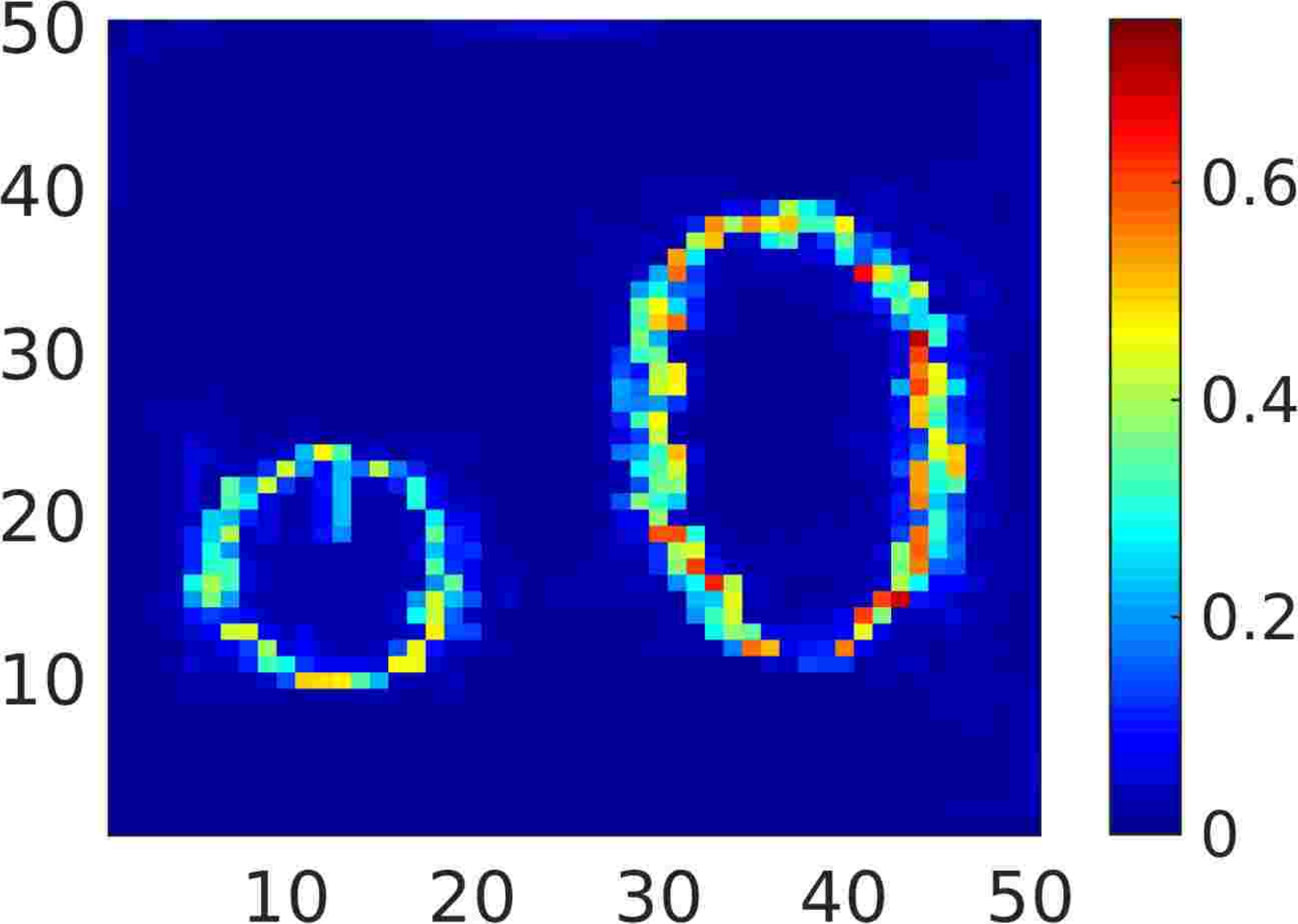}} 
		  \hspace{0.3cm}
	  }
	  \caption{Comparison of the standard deviation of $\bs{\Psi}$ found with a single mixture component ($\bs{\sigma}_1$, left), with that found with a mixture of Gaussians ($\bs{\sigma}_{mixture}$, middle) and the exact values estimated with IS ($\bs{\sigma}_{IS}$, right). Depictions are in log scale. 
}
	 \label{fig:ISStd}
\end{figure}

In Figures \ref{fig:ISMean} and \ref{fig:ISStd}, the approximate posterior mean and standard deviation are compared with the (asymptotically) exact values estimated by IS. Furthermore in these figures we plot the posterior mean and standard deviation found solely on the basis of the most prominent mixture component i.e. $s=1$. While, visually,  the differences in the mean are not that striking (they are primarily concentrated at the boundaries of the inclusions), we observe that the posterior variance is clearly underestimated by a single component. 
In terms of the  Euclidean norm (across the whole problem domain), we obtained that $\frac{|| \bs{\mu}_1 -\bs{\mu}_{IS} ||}{ || \bs{\mu}_{mixture} -\bs{\mu}_{IS} ||}=5$ where $\bs{\mu}_{IS}$ is the exact mean obtained with $IS$ and $\bs{\mu}_{mixture}$ is the approximate mean obtained from the mixture of Gaussians in \refeq{eq:qpost1}. Similarly for the standard deviation, we obtained that $\frac{|| \bs{\sigma}_1 -\bs{\sigma}_{IS} ||}{ || \bs{\sigma}_{mixture} -\bs{\sigma}_{IS} ||}=6$ where $\bs{\sigma}_1,\bs{\sigma},\bs{\sigma}_{IS}$ are the vectors of  standard deviation across the whole problem domain, obtained with a single component, the mixture and IS respectively.

\begin{figure}[H]{
	\centering
		{\includegraphics[width=0.850\textwidth]{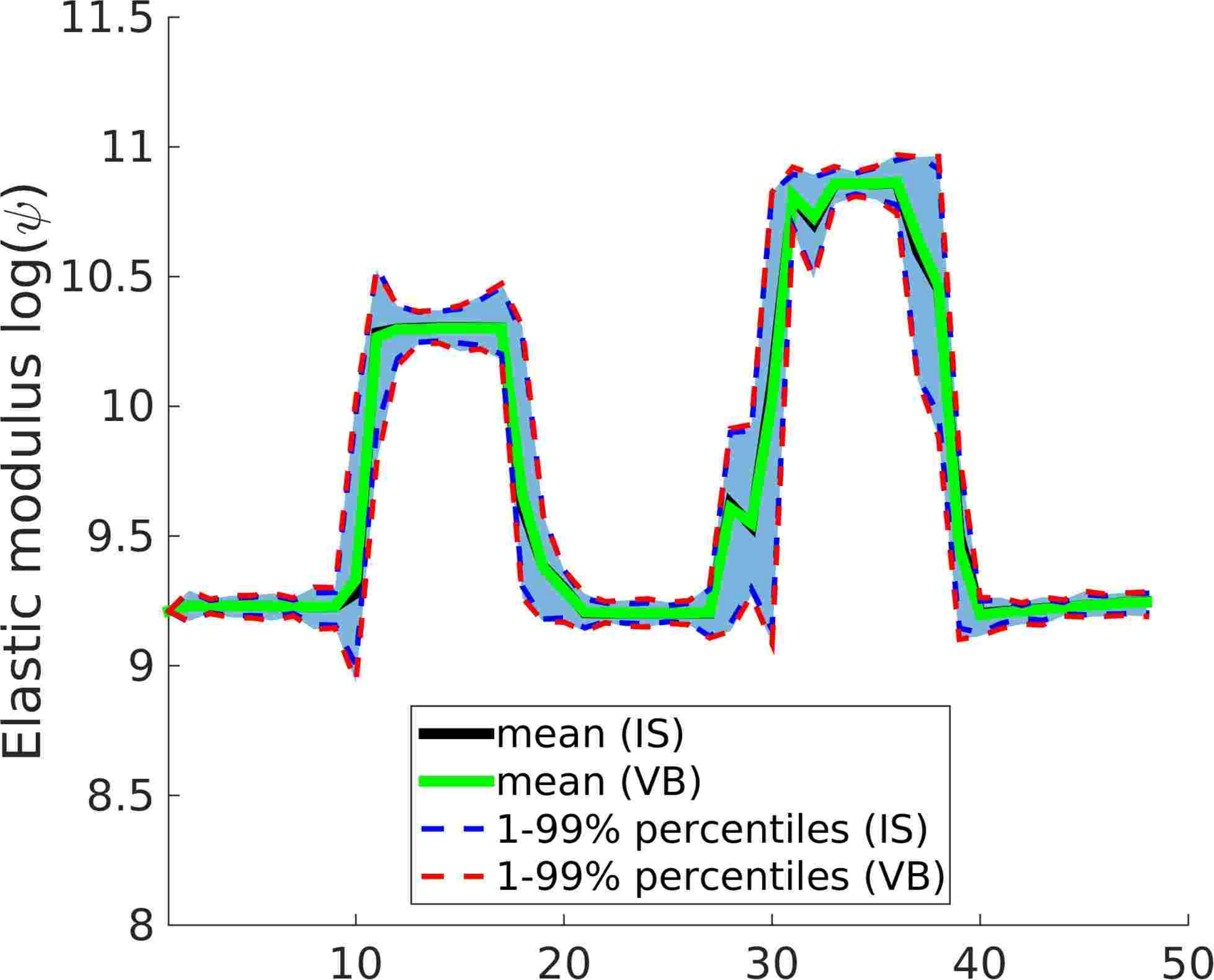}} 
		\hspace{0.1cm}
	  }
	  \caption{Posterior mean and credible intervals along the  
	  diagonal cut from$(0, 0)$ to $(50, 50)$ 
	  for mixture of Gaussians. Comparing the results with the results obtained by Importance Sampling (IS), we can see that they fit well to each other.}
	 \label{fig:DiagonalCutMixtureIS}
\end{figure}

\reffig{fig:DiagonalCutMixtureIS} offers another view of the results along the diagonal of the problem domain and compares also the $1\%$ and $99\%$ credible intervals where again very good agreement with the (asymptotically) exact values found with IS is observed.

\begin{figure}[H]{
	\centering
	\captionsetup[subfigure]{labelformat=empty}
		\subfloat[][{ $q(s=1) =0.318$ }] 
		{\includegraphics[width=0.30\textwidth]{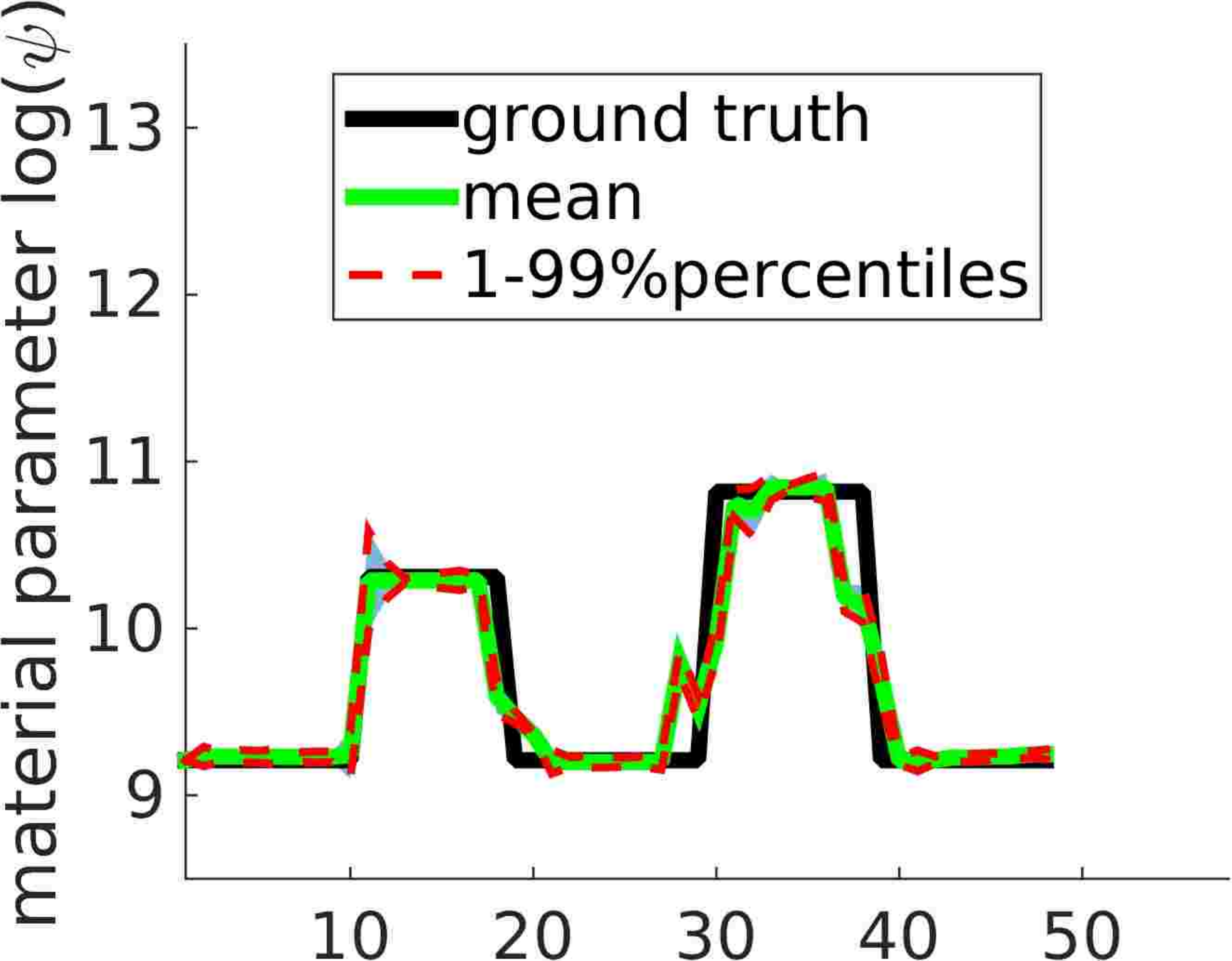}} 
		\hspace{0.1cm} 
		\subfloat[][{ $q(s=3) =0.213$}] 
		{\includegraphics[width=0.30\textwidth]{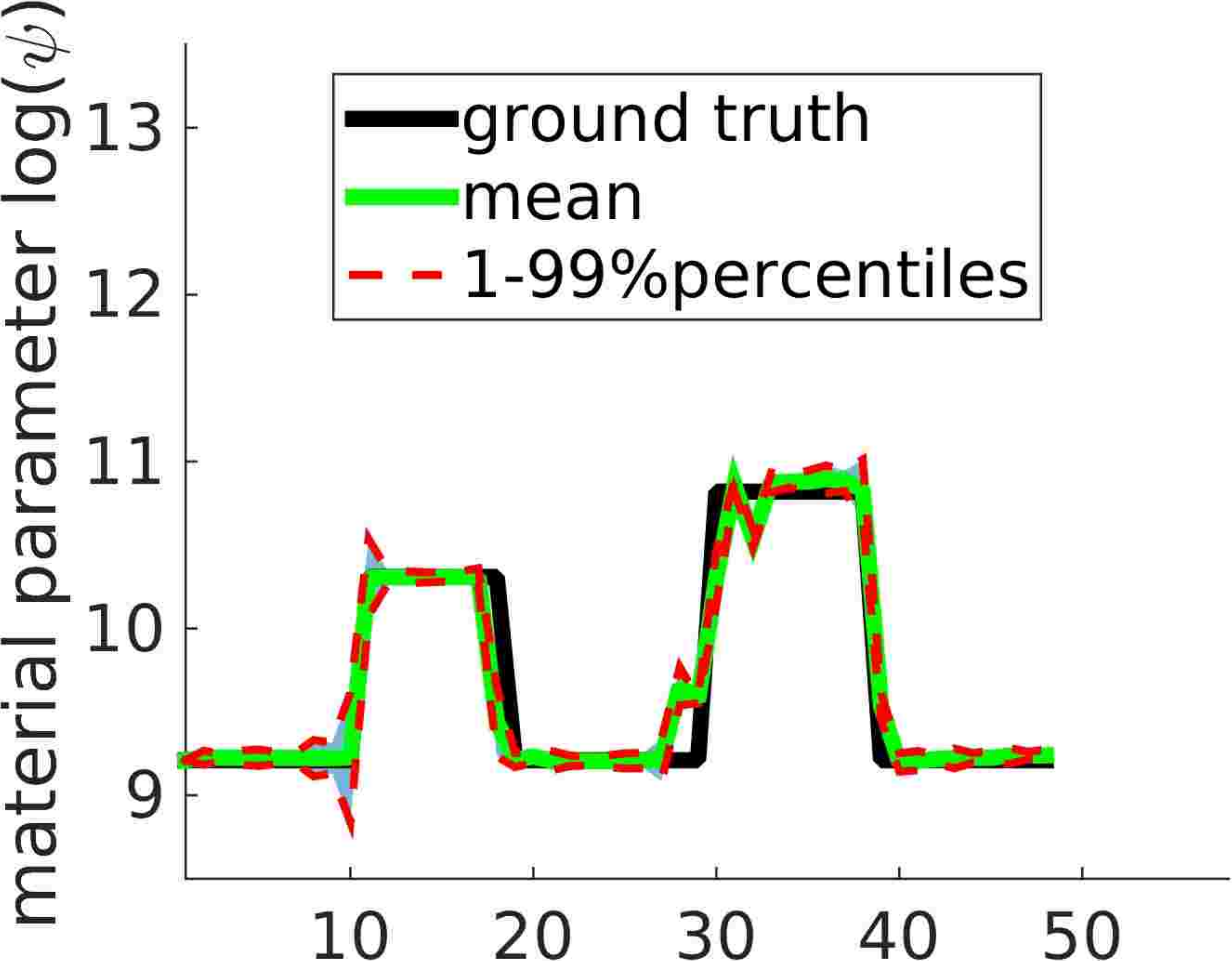}} 
		\hspace{0.1cm}
		\subfloat[][{ $q(s=6) =0.160$}] 
		{\includegraphics[width=0.30\textwidth]{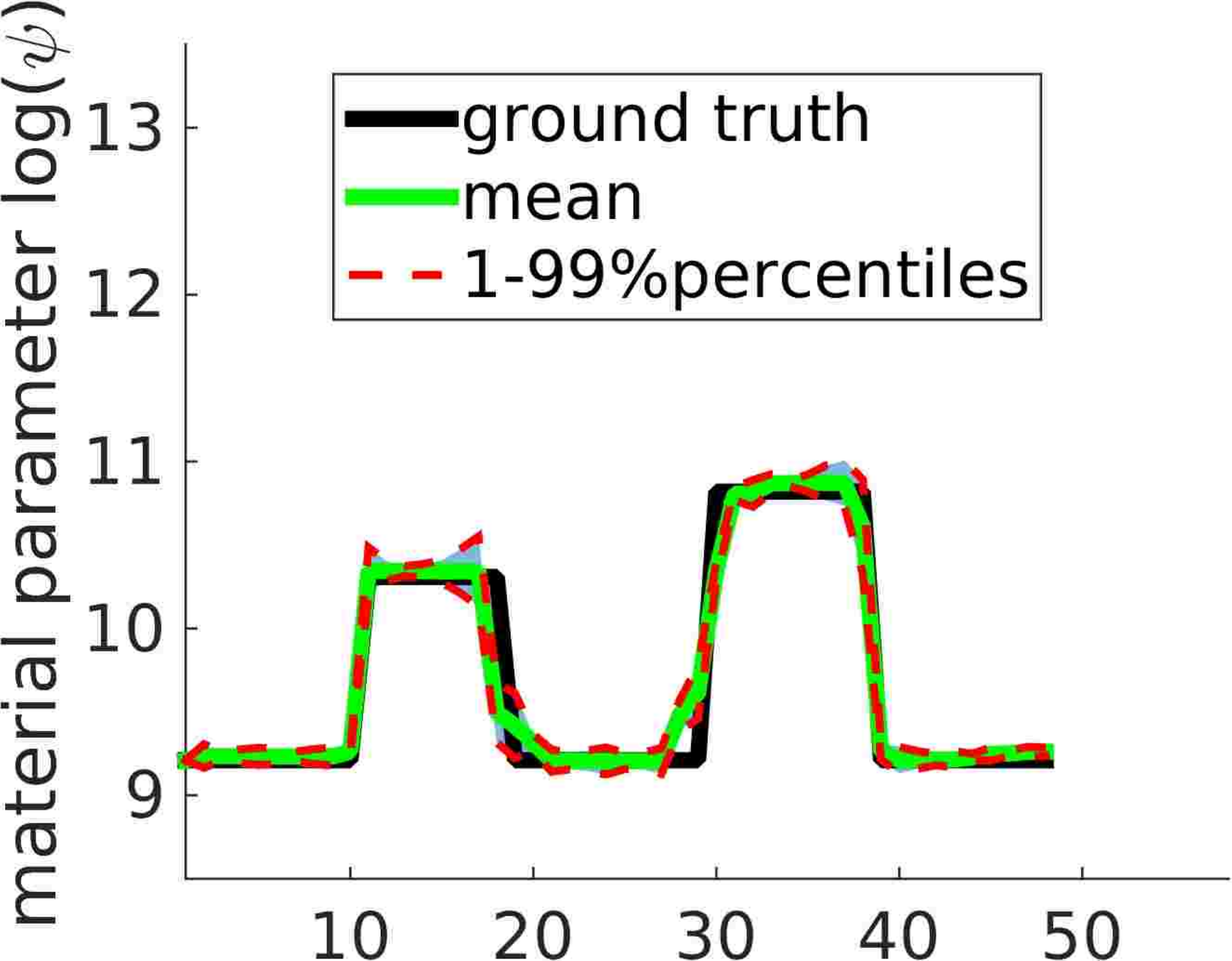}} 
		  \hspace{0.3cm}
	  }
	  \caption{Posterior mean and $1\%-99\%$ credibility intervals (dashed lines) along the diagonal cut from (0, 0) to (50, 50) for different mixture components.}
	 \label{fig:CutMixtures}
\end{figure}

It is interesting to contrast this with \reffig{fig:CutMixtures} which depicts the posterior along  the same diagonal of the problem domain computed solely from each of the most prominent components i.e. from $q_s(\bpsi)$ for $s=1,3,6$. We note again that away from the boundaries, strong similarities are observed but none of the components by itself can fully capture or envelop the ground truth (compare also with \reffig{fig:DiagonalCutMixture}). 

\begin{figure}[H]{
	\centering
	\captionsetup[subfigure]{labelformat=empty}
		\subfloat[][{ $q(s=1) =0.318$ }] 
		{\includegraphics[width=0.30\textwidth]{FiguresIF/StdMoG0.pdf}} 
		\hspace{0.1cm}
		\subfloat[][{ $q(s=3) =0.213$}] 
		{\includegraphics[width=0.30\textwidth]{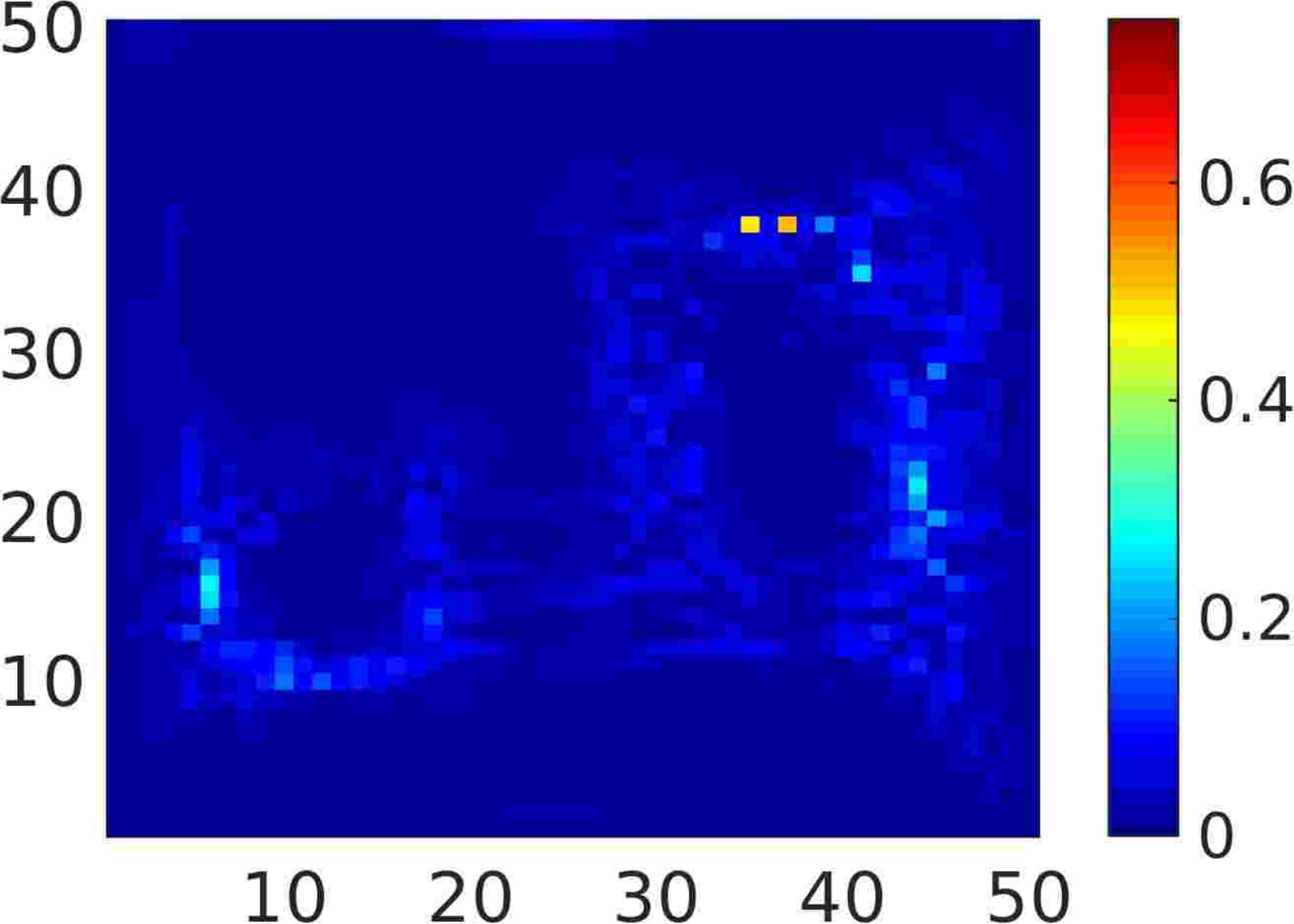}} 
		\hspace{0.1cm}
		\subfloat[][{ $q(s=6) =0.160$}] 
		{\includegraphics[width=0.30\textwidth]{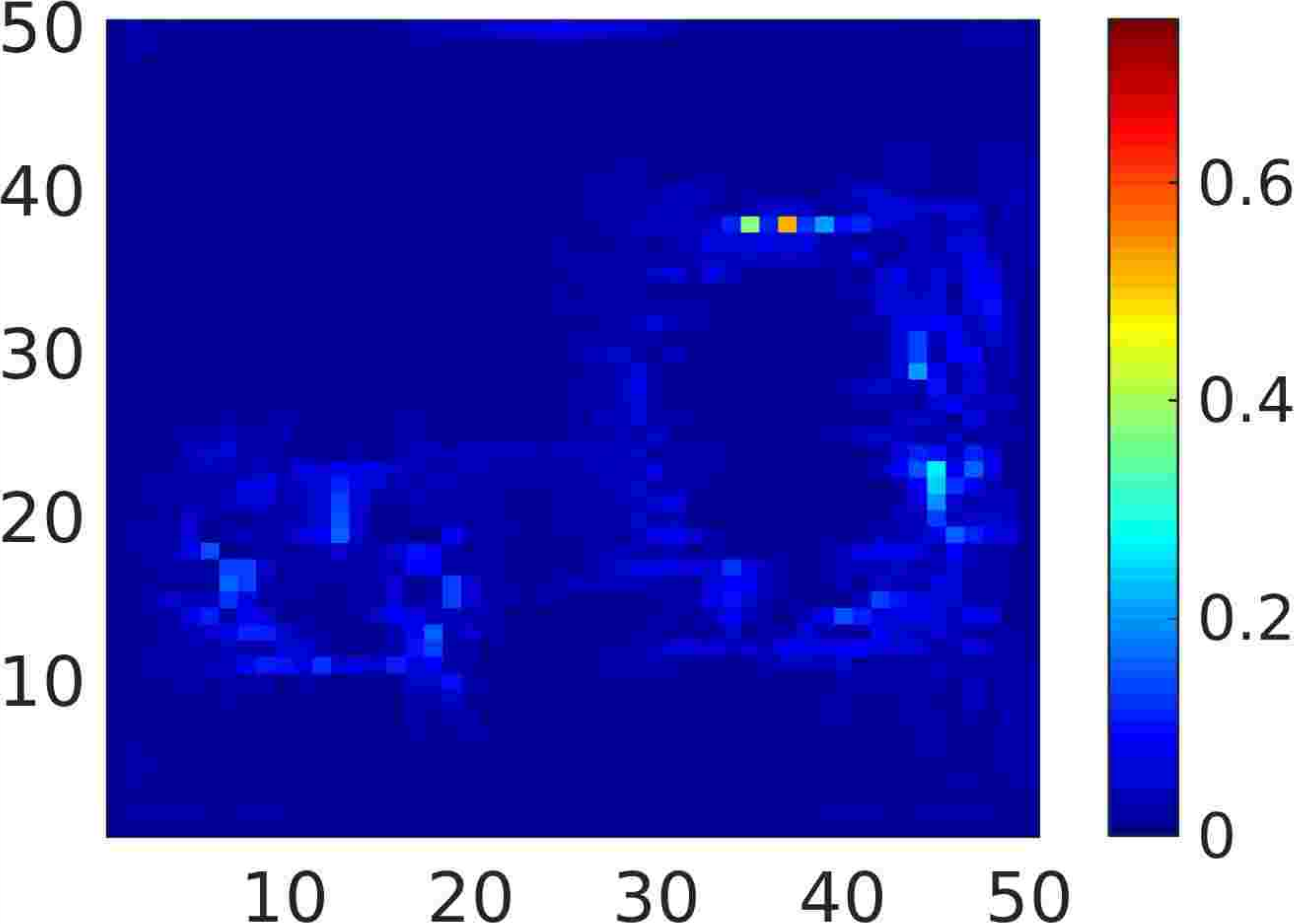}} 
		  \hspace{0.3cm}
	  }
	  \caption{Standard deviation for selected mixture components in log scale.}
	 \label{fig:StdMixtures}
\end{figure}

We provide further details on the most prominent mixture components i.e. $1$, $3$ and $6$. \reffig{fig:StdMixtures} depicts the posterior standard deviation of $\bpsi$ as computed by using each of these components individually i.e. from $q_s(\bpsi)$ in \refeq{eq:postmix2} for $s=1,3,6$.  All components yield small variance for the surrounding tissue and the interior of the inclusions while the posterior uncertainty is concentrated on the boundaries of the inclusions. 

\begin{figure}[H]{
	\vspace{-0.5cm}
	\caption*{\hspace{-0.6cm}\textbf{Mixture comp. 1} 	\hspace{1.5cm}	\textbf{Mixture comp. 3} 	\hspace{1.3cm}	\textbf{$\Delta$ (Mixture 1 - 3)}}
	\captionsetup[subfigure]{position=bottom}
	
	\begin{minipage}{\linewidth}
		\subfloat[][{$\lambda_1^{-1} = 9.885 \times  10^{-1}$ }]
		{\includegraphics[width=0.3\textwidth]{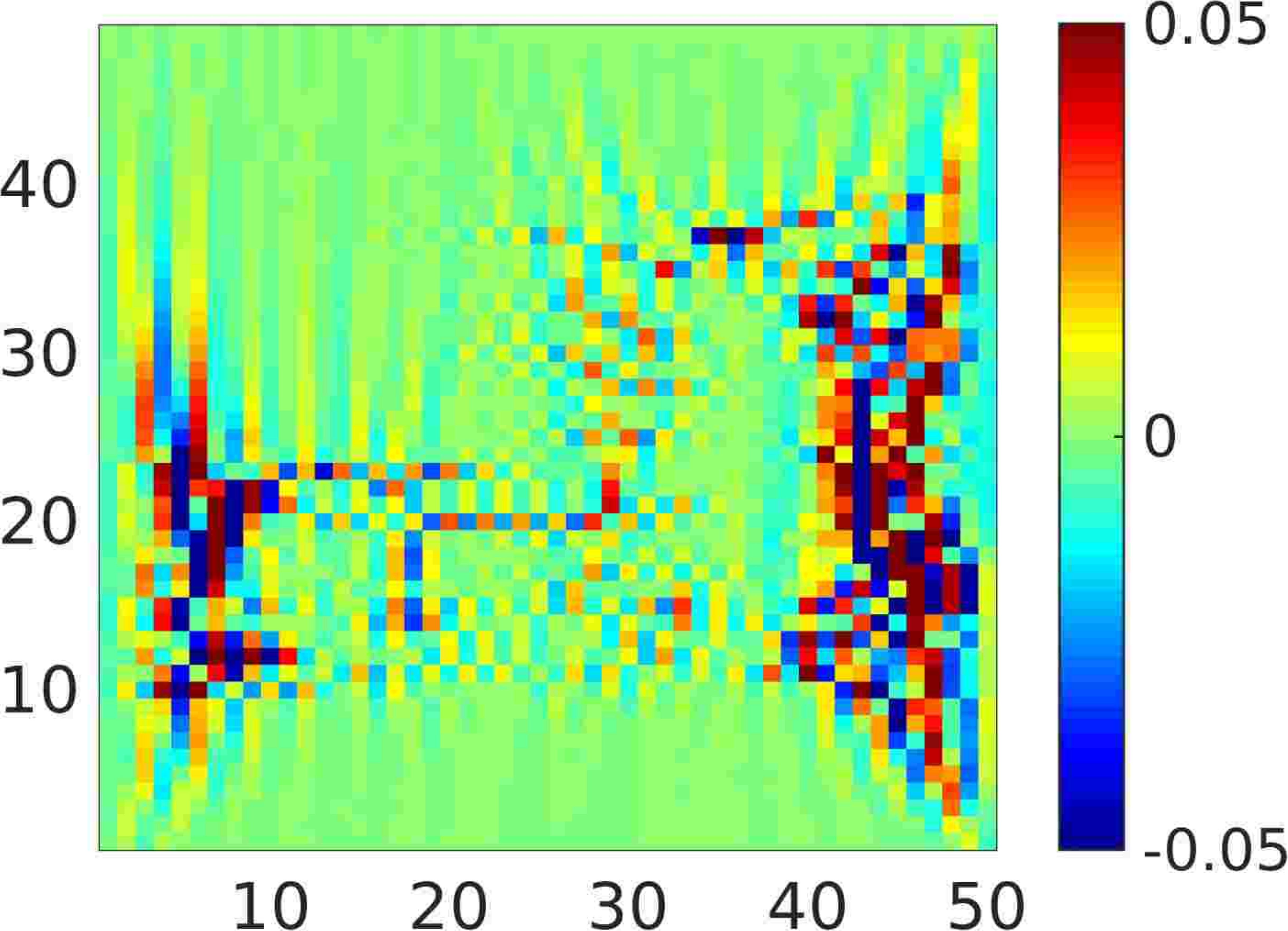} } 
		\hspace{0.1cm}
		\subfloat[][{$\lambda_1^{-1} = 9.920 \times  10^{-1}$ }]
		{\includegraphics[width=0.3\textwidth]{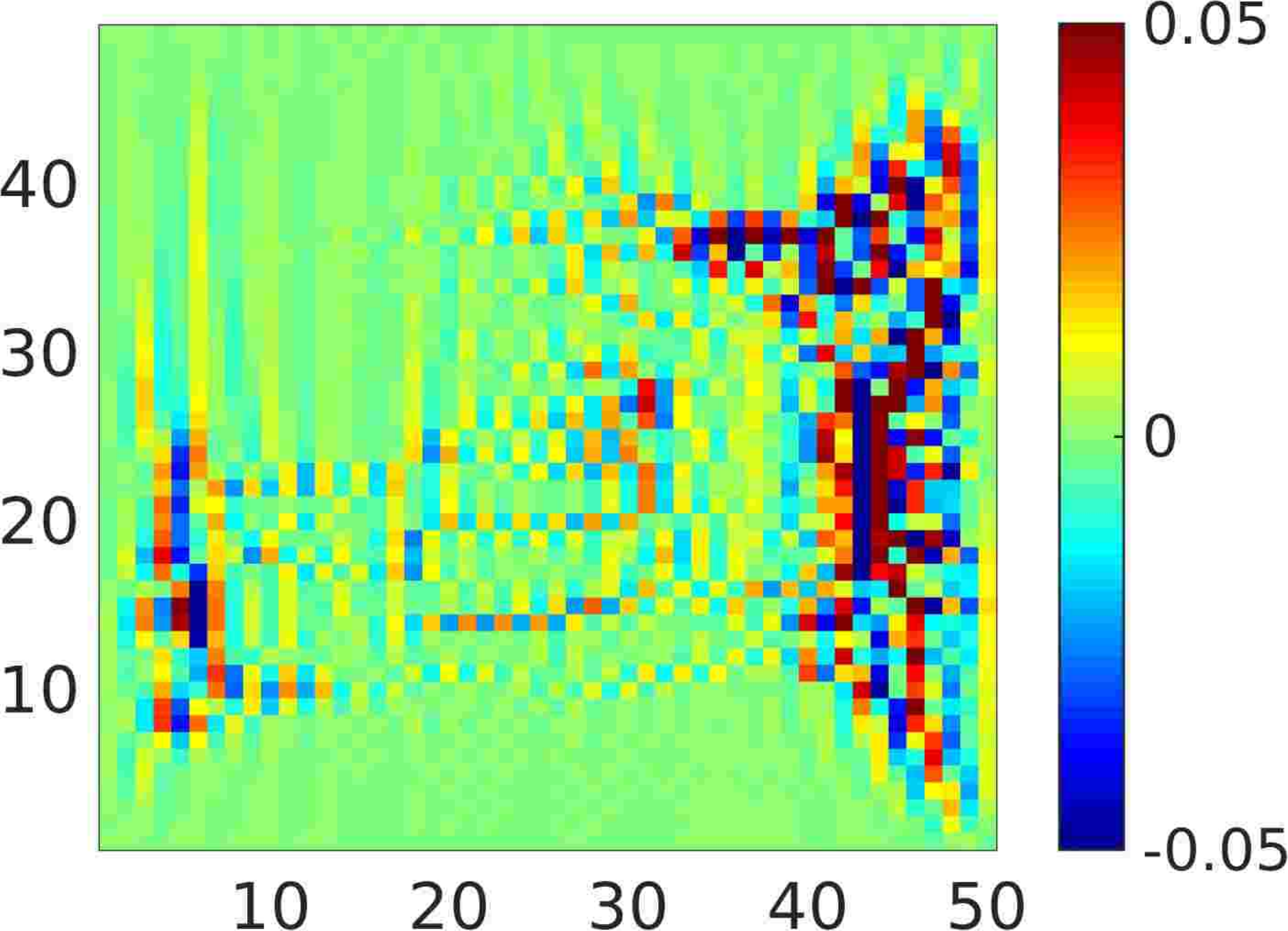} } 
		\hspace{0.1cm}
		{\includegraphics[width=0.3\textwidth]{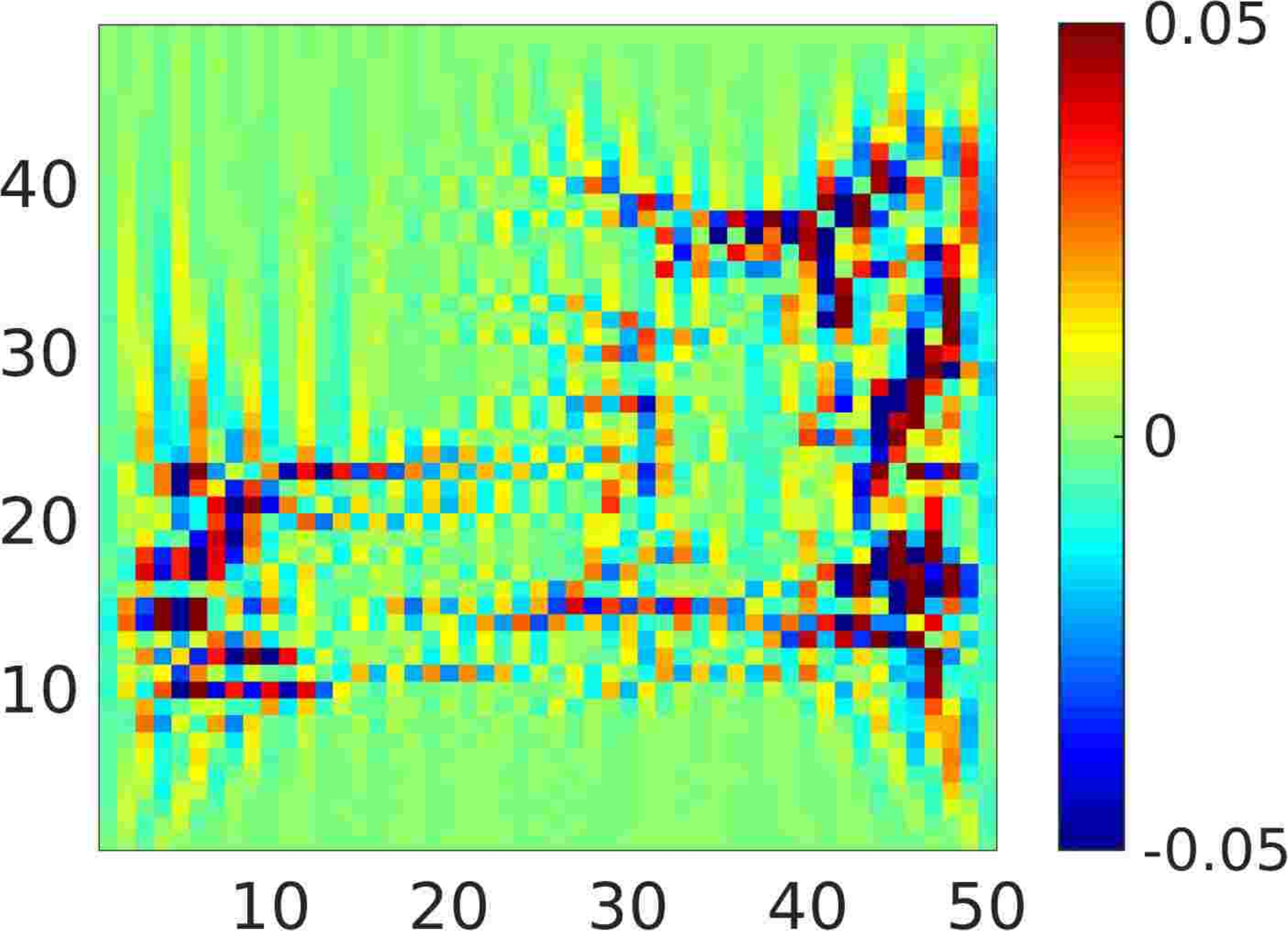} } 	
	\end{minipage}
	\par\bigskip
	
	\begin{minipage}{\linewidth}	
 		\vspace{-0.2cm}
		\subfloat[][{$\lambda_2^{-1}= 9.328 \times  10^{-1}$ }]
		{\includegraphics[width=0.3\textwidth]{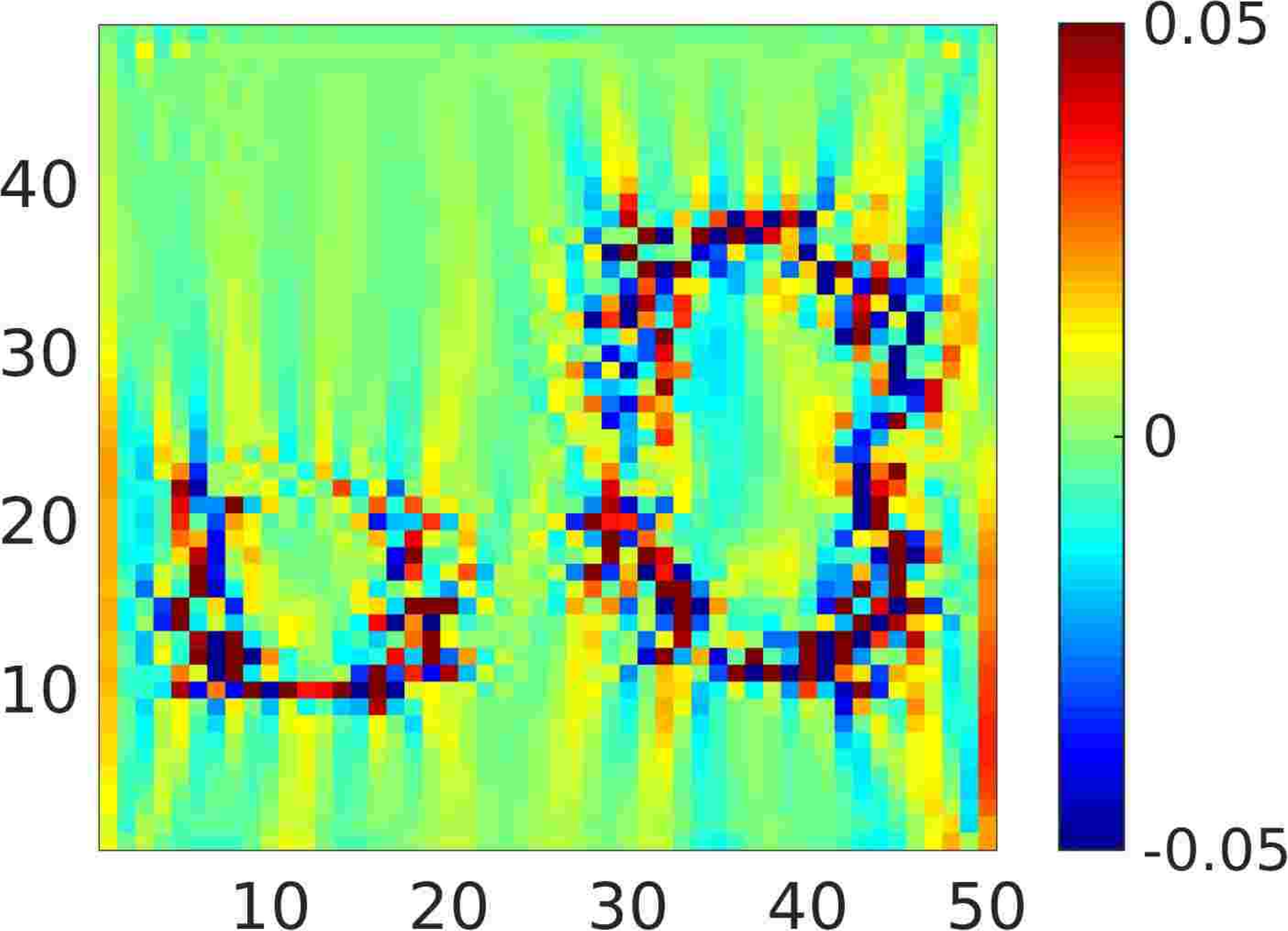} }  		
 		\hspace{0.1cm}
		\subfloat[][{$\lambda_2^{-1}= 9.785 \times  10^{-1}$ }]
		{\includegraphics[width=0.3\textwidth]{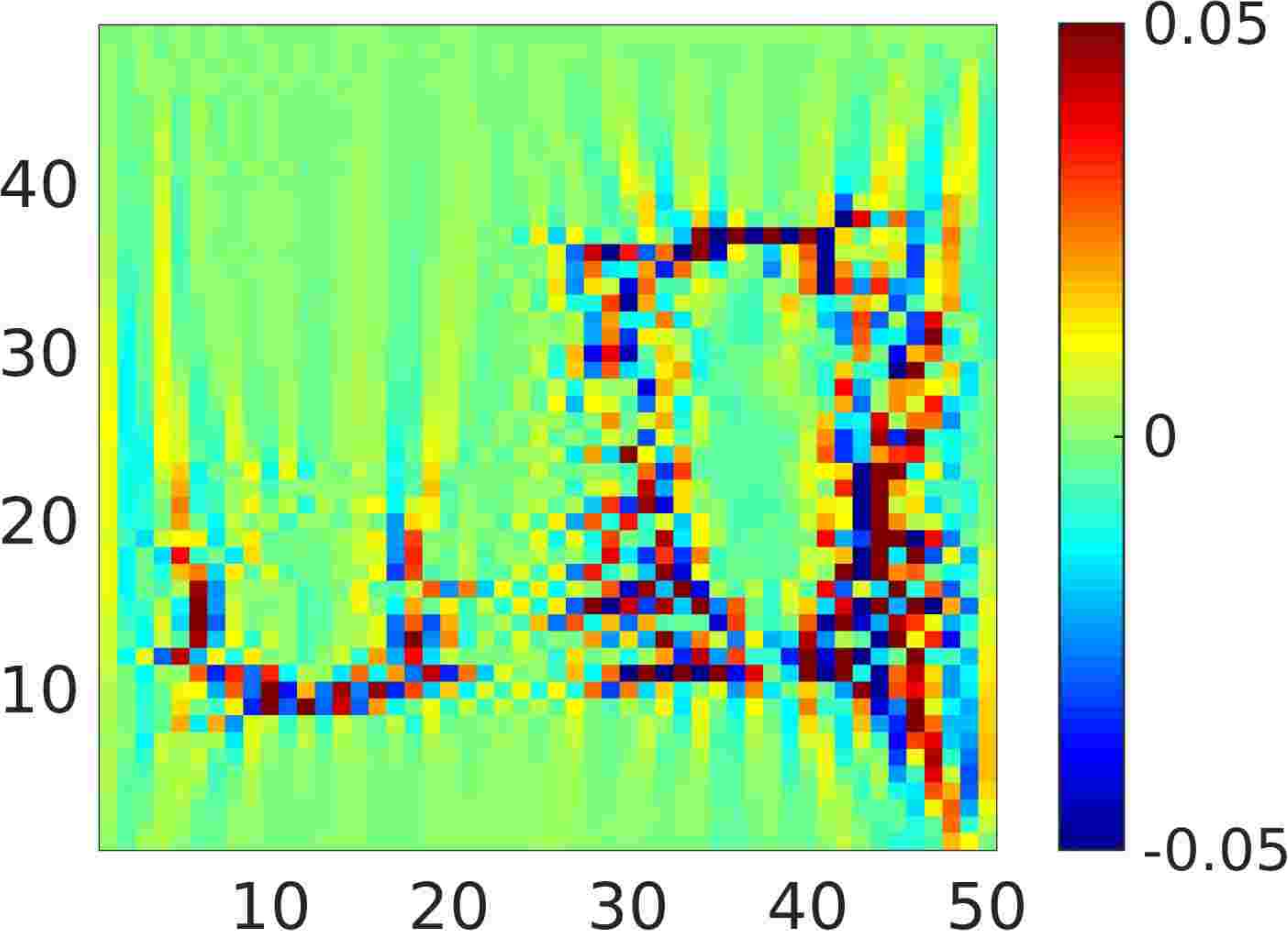} } 
		\hspace{0.1cm}		
		{\includegraphics[width=0.3\textwidth]{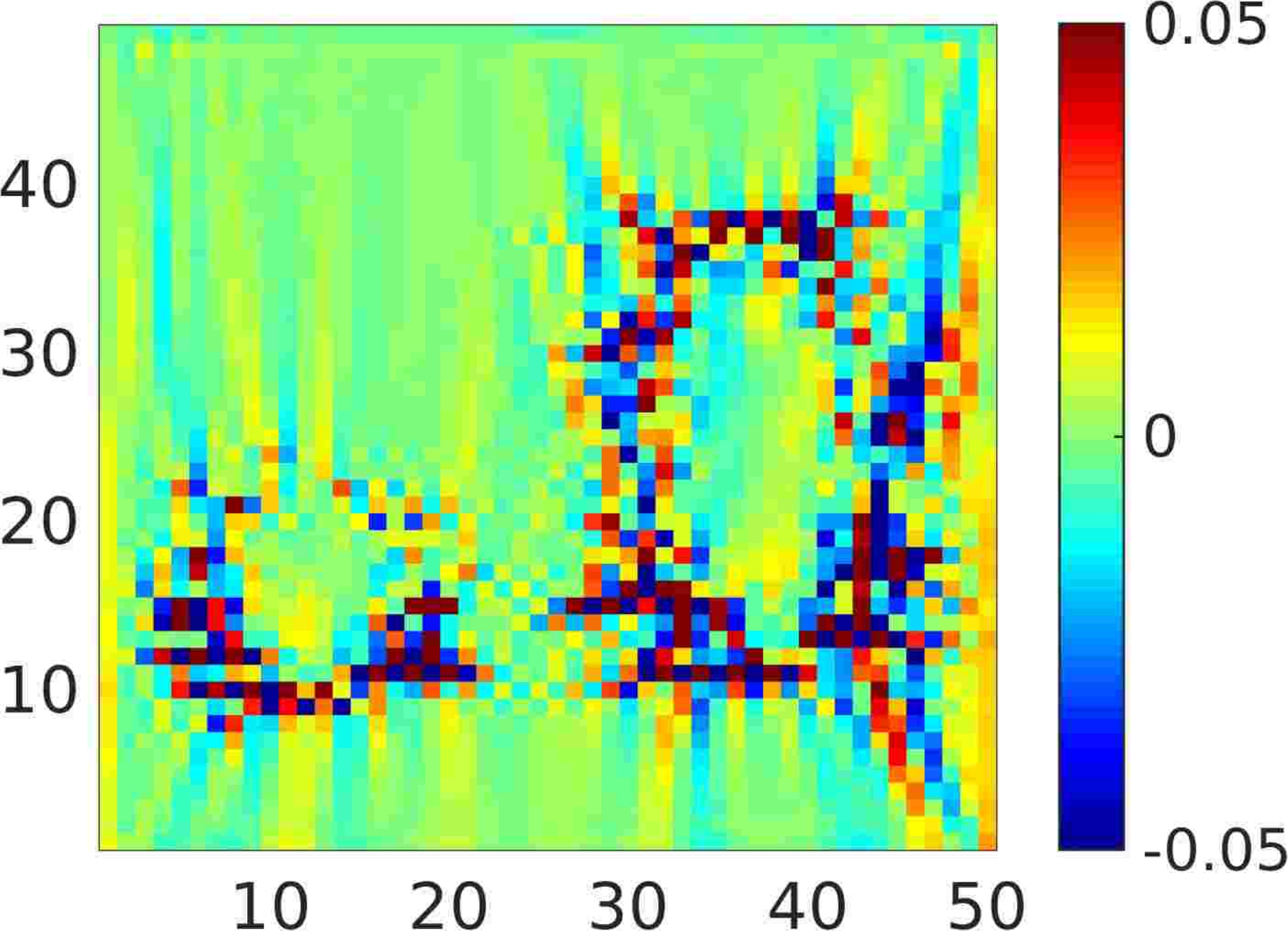} } 
	\end{minipage}
	\par\bigskip	
	
	\begin{minipage}{\linewidth}		
 		\vspace{-0.2cm}
		\subfloat[][{$\lambda_3^{-1}= 5.269 \times  10^{-1}$ }]
		{\includegraphics[width=0.3\textwidth]{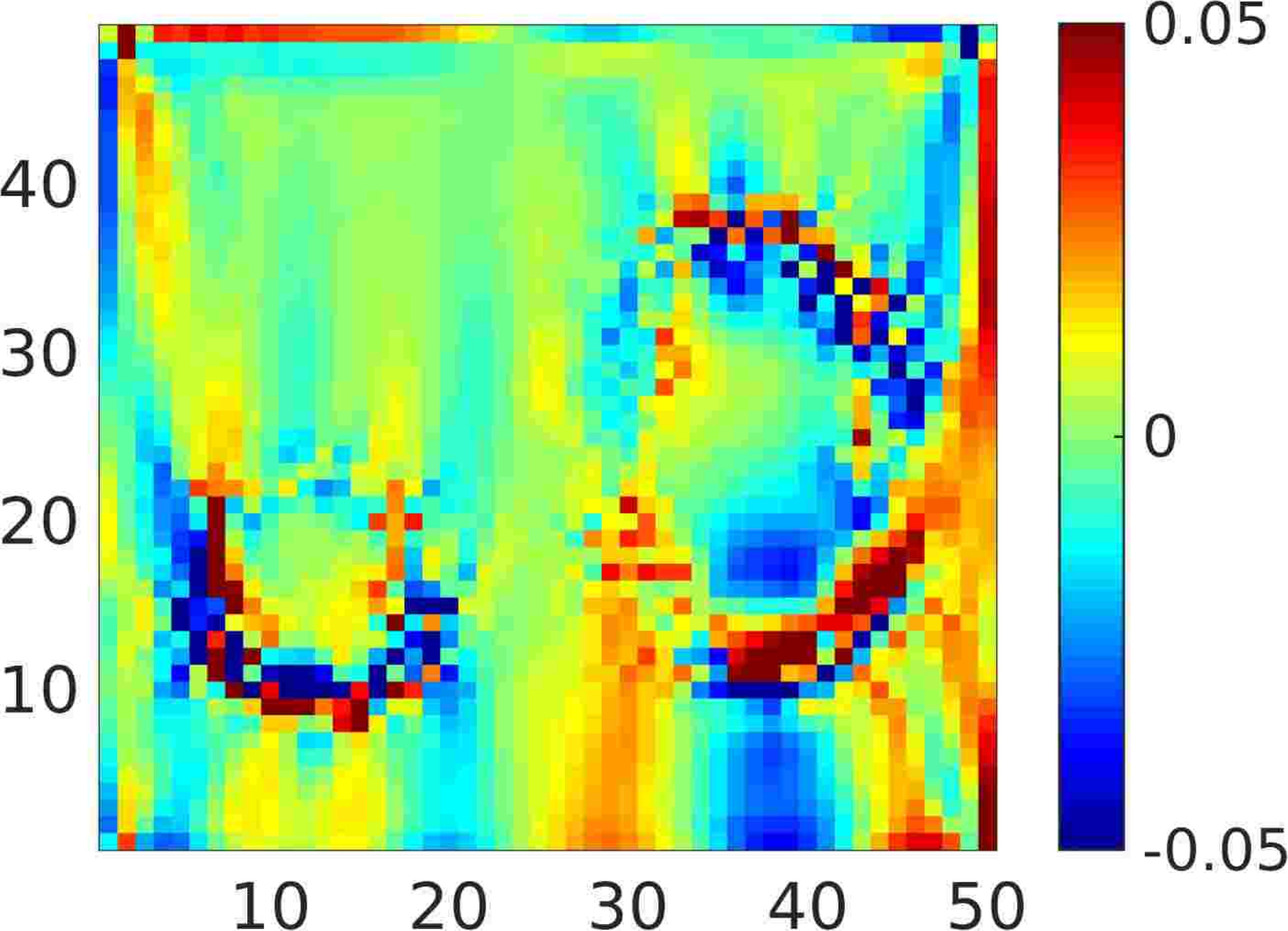} }  		
 		\hspace{0.1cm}
		\subfloat[][{$\lambda_3^{-1}= 9.344 \times  10^{-1}$ }]
		{\includegraphics[width=0.3\textwidth]{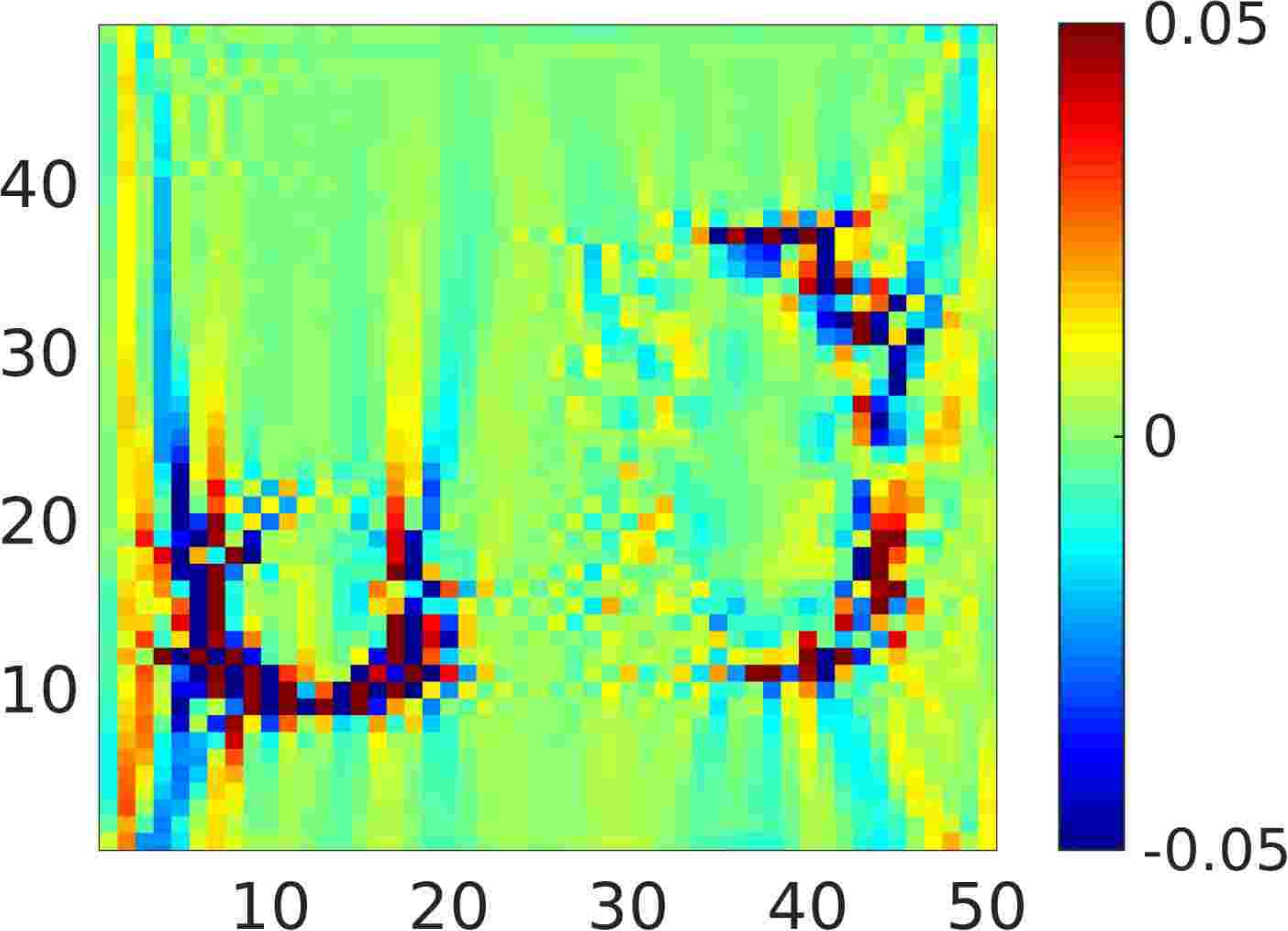} } 
		\hspace{0.1cm}		
		{\includegraphics[width=0.3\textwidth]{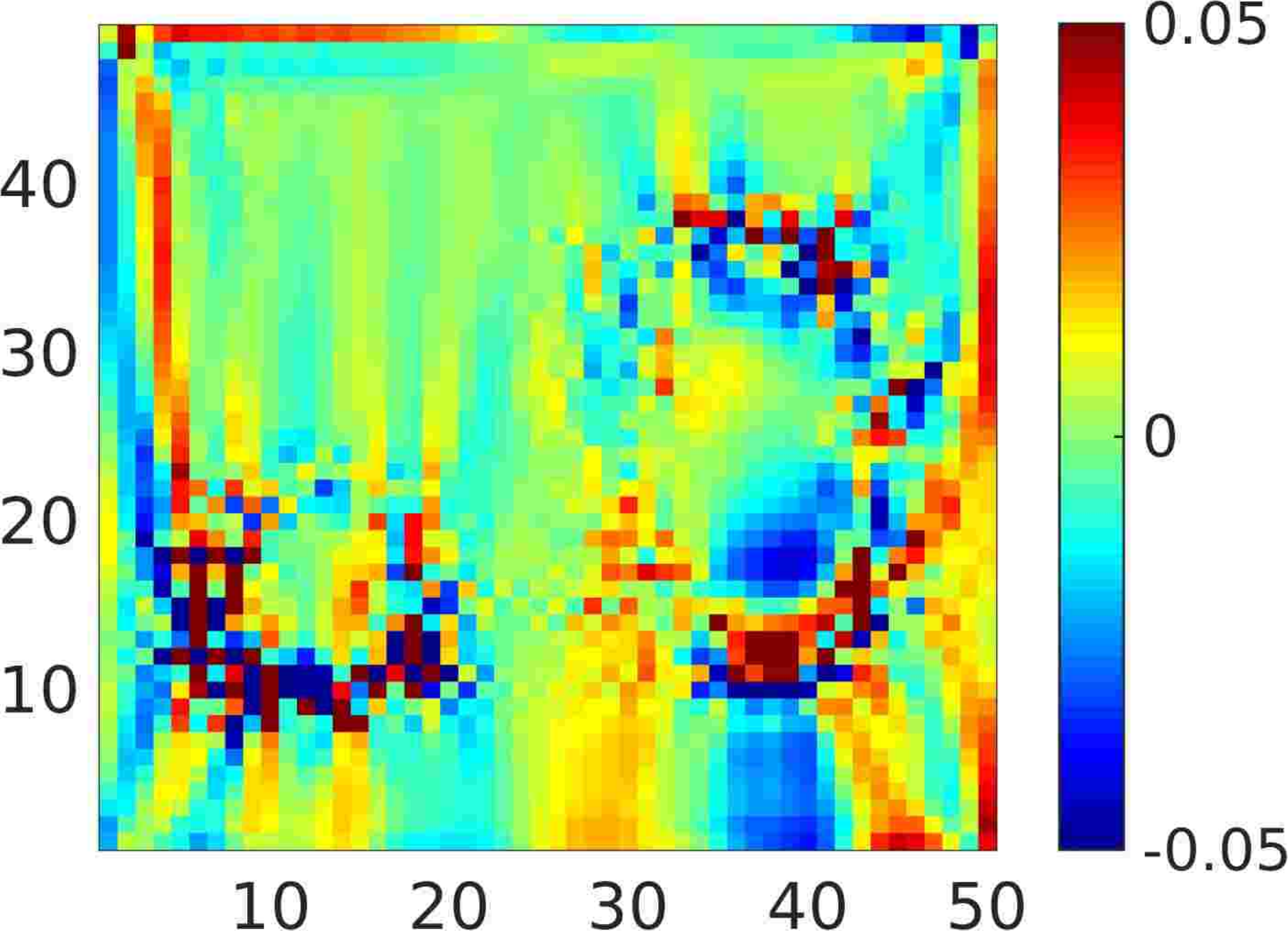} } 
	\end{minipage}
	\par\bigskip	
	
	\begin{minipage}{\linewidth}		
 		\vspace{-0.2cm}
		\subfloat[][{$\lambda_4^{-1}= 7.823 \times  10^{-2}$ }]
		{\includegraphics[width=0.3\textwidth]{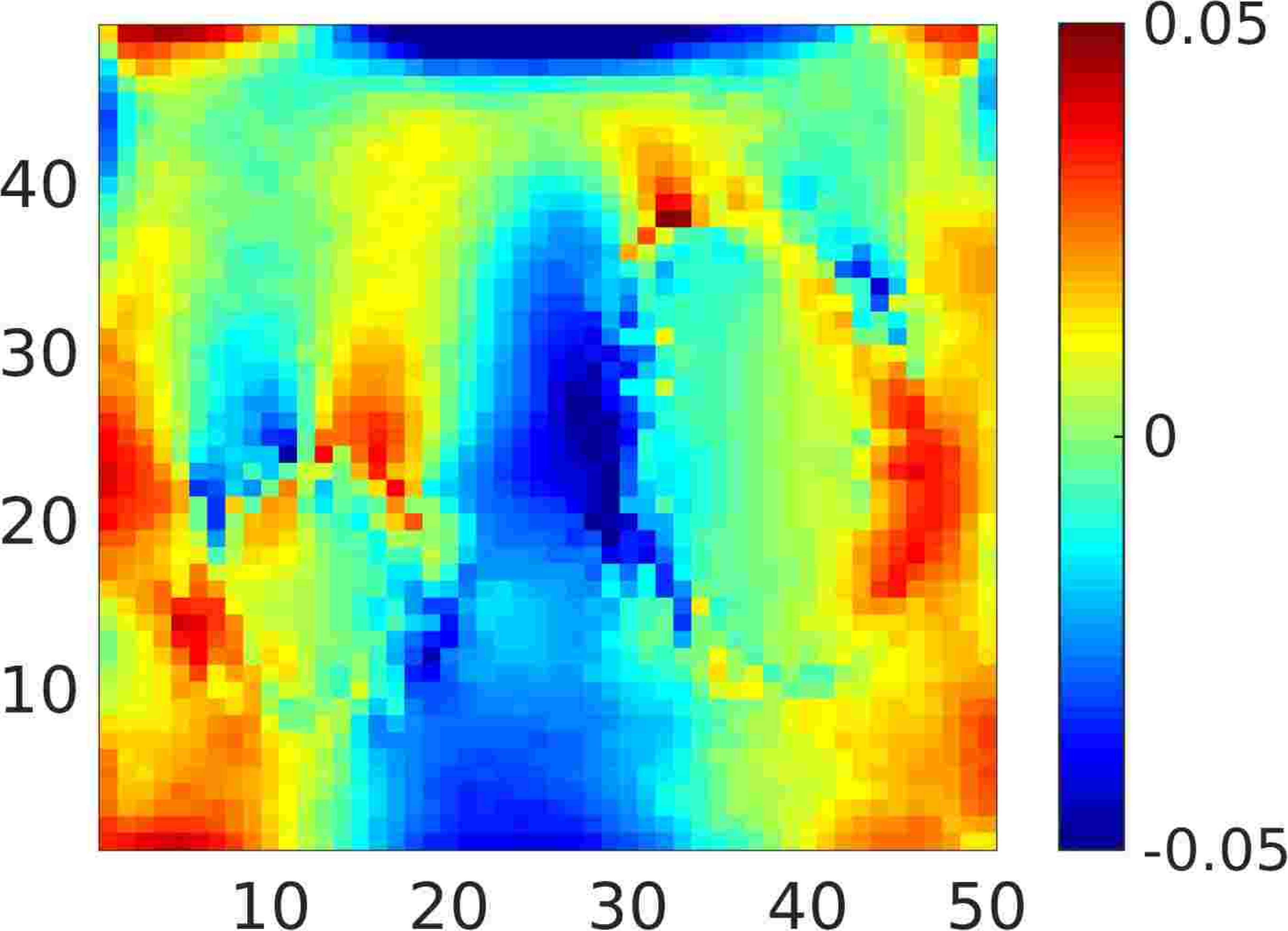} }  		
 		\hspace{0.1cm}
		\subfloat[][{$\lambda_4^{-1}= 2.822 \times  10^{-1}$ }]
		{\includegraphics[width=0.3\textwidth]{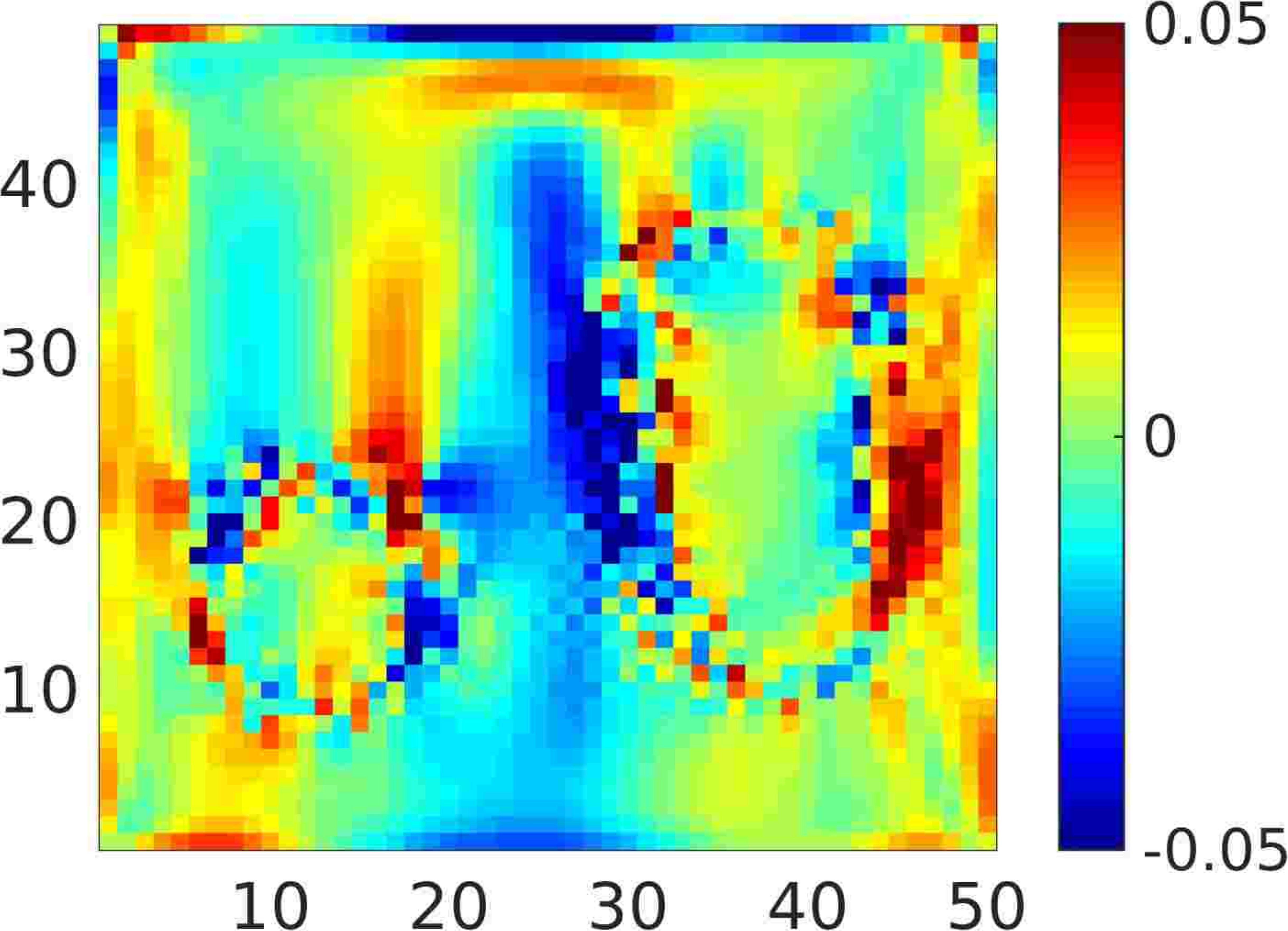} } 
		\hspace{0.1cm}		
		{\includegraphics[width=0.3\textwidth]{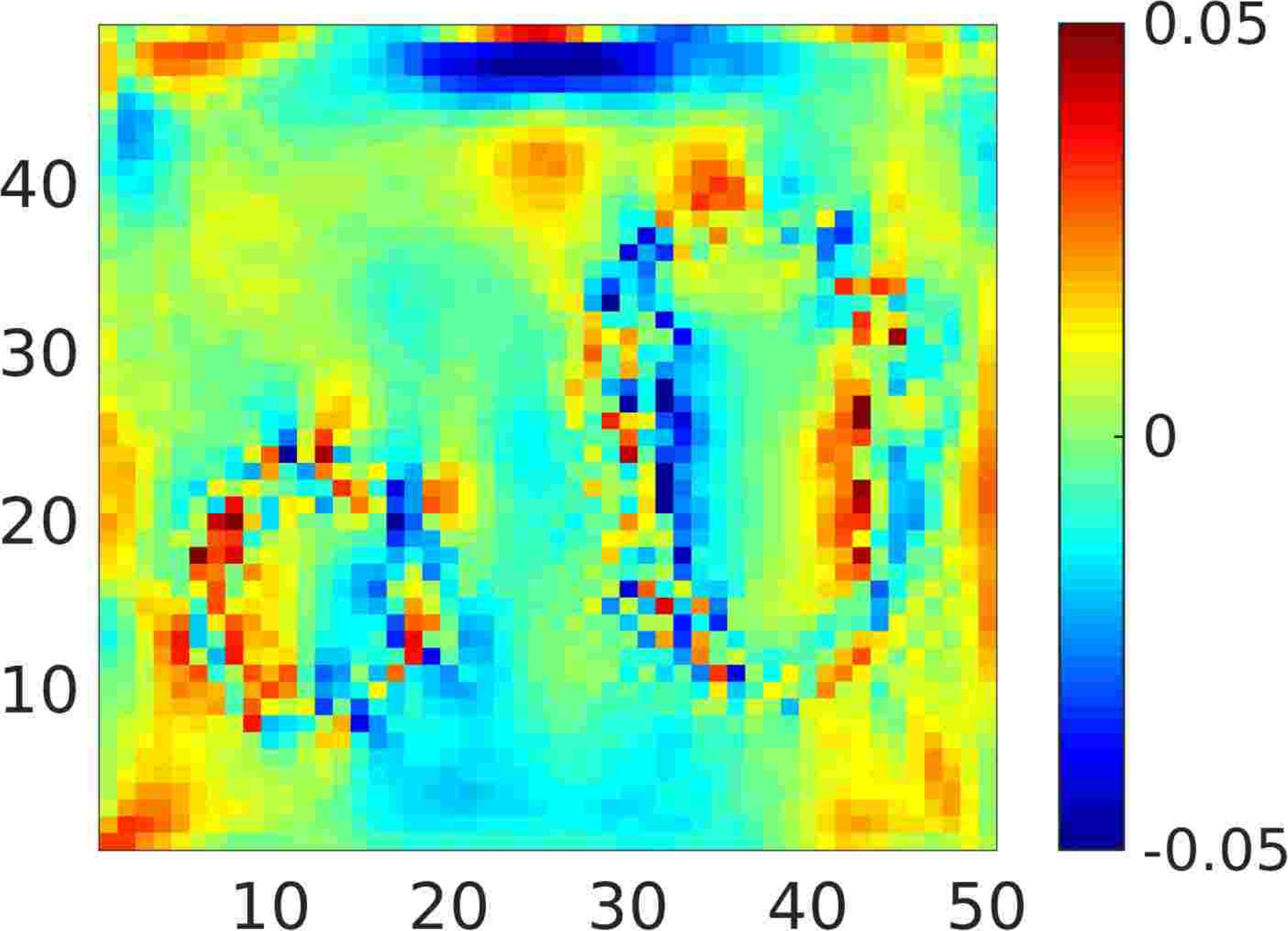} } 
	\end{minipage}
	\par\bigskip	

	  }
	  \vspace{-0.2cm}
	  \caption{The first few basis vectors of $\bs{W}_s$ for mixture components $s=1$ and $s=3$ are shown in the first and second column. In the third column,  the difference between the basis vectors in the first two columns, is plotted. The differences are more pronounced in the vicinity of the boundary of the inclusions.}
	  \label{fig:BASESLarge}
\end{figure}

\reffig{fig:BASESLarge} depicts the first four columns (basis vectors) of $\bs{W}_s$ for $s=1,3$ and the corresponding posterior variances $\lambda_{s,i}^{-1}$. The third column is perhaps the most informative,  showing the differences between these vectors. These differences are most pronounced around the  inclusions but, most importantly,  reveal that the posterior variance is concentrated along different subspaces for different mixture components (see also \reffig{fig:MoGReducedBases}). 
We note also with regards to $q(\bet|s)$ i.e.  the posterior of the residual noise in the representation of the unknowns, that, for all mixture components i.e. $s=1, \ldots 7$,  $\lambda_{\eta,s}^{-1}$ was found approximately the same and equal to $4 \times 10^{-3}$, 
 which is one or two orders of magnitude smaller than the variance associated with $\bt$ and has as a result a minimal overall influence.

%

In order to gain further insight we provide inference results along the boundary of the elliptical inclusion. In particular we consider the elements along the black line in \reffig{fig:CirclelargerInclusionMarking} and pay special attention to the elements marked with yellow stars from $1$ to $4$. We have purposely selected the black line to lie partly in the interior and partly in the exterior of the inclusion.
In \reffig{fig:CirclelargerInclusion}, we plot the posterior mean  along this black line (including  credible intervals corresponding to $1\%$ and $99\%$) as obtained exclusively from  one  of the three most prominent components (from $q_s(\bpsi)$ in \refeq{eq:postmix2} for $s=1,3,6$) as well as by the mixture of Gaussians (\refeq{eq:qpost}). As it can now be seen more clearly, the individual components are only partially capable of capturing the ground truth. At times, points are misclassified in terms of whether they belong to the inclusion or not. On the other hand, the approximation provided by the mixture of Gaussians, averages over the individual components and  leads to a posterior mean that is closer to the ground truth. More importantly, and especially in the regions where transitions from the inclusions to the surrounding tissue are present, the posterior uncertainty is larger to account for this ambiguity.
%


In \reffig{fig:MixturesSelectedParameters} we plot the  posterior statistics for the elastic modulus of the  elements $1$ through $4$ marked by yellow stars in \reffig{fig:CirclelargerInclusionMarking}. The ground truth values are indicated with red rhombuses. We note that each of the mixture components gives rise to a Gaussian with, in general, a different mean/variance. These Gaussians reflect the uncertainty of the material properties at these points and are synthesized in the mixture. Interestingly, at element $4$ which is further away from the boundary, all mixture components give rise to Gaussians with very similar  means and variances, yielding a unimodal posterior when combined in the mixture.

\begin{figure}[H]{
	\centering
		{\includegraphics[width=0.850\textwidth]{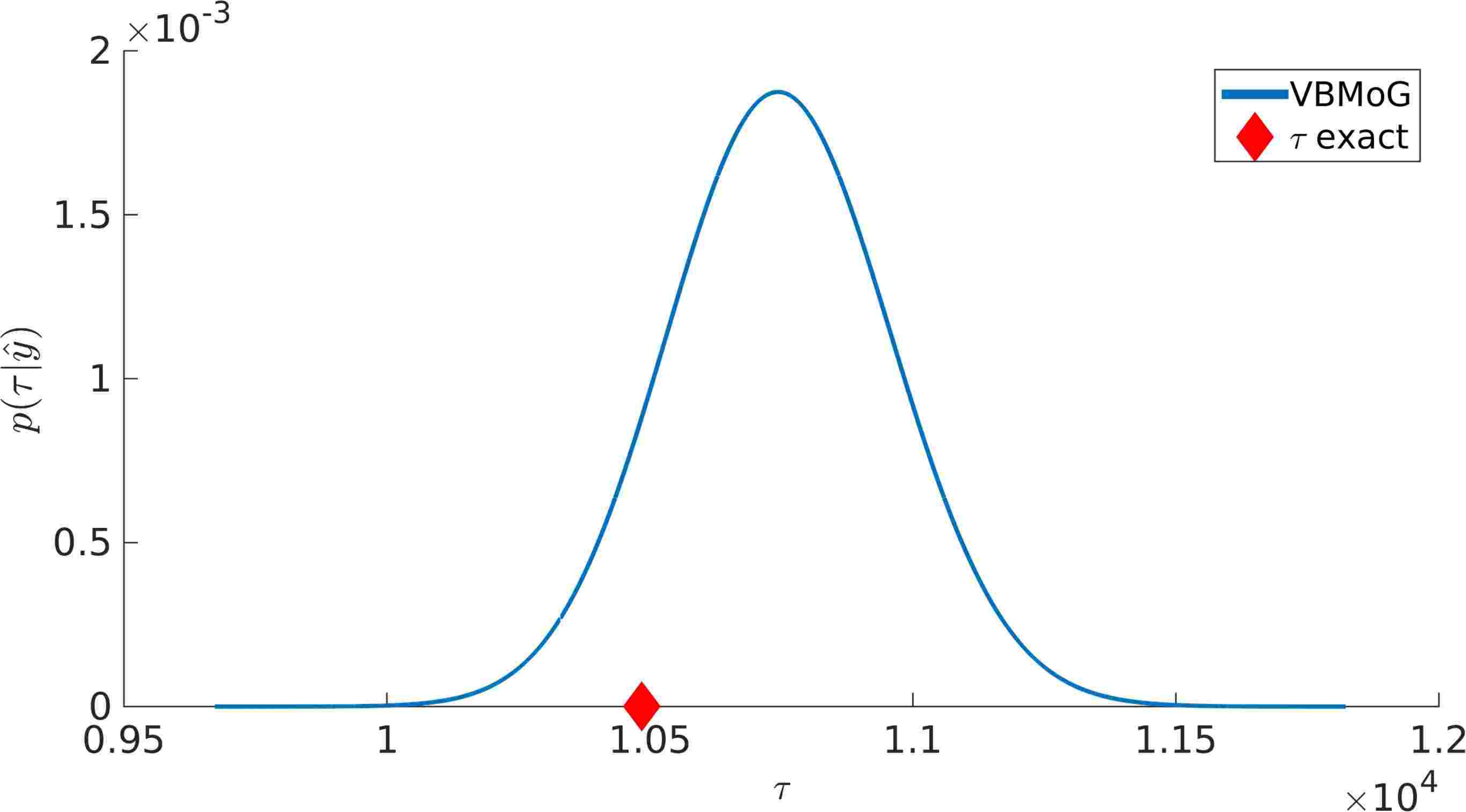}} 
		\hspace{0.1cm}
	  }
	  \caption{Approximate posterior $q(\tau)$ of the noise precision $\tau$. The ground truth is indicated with the red rhombus
	  }
	 \label{fig:Taudistribution}
\end{figure}

Apart from  the posterior probability distributions of the material parameters, we note also that noise precision was treated as an unknown and its (approximate) posterior was computed via Variational inference (\refeq{eq:qopt1}). This is plotted in \reffig{fig:Taudistribution}.
%
%

\subsection{Other examples/configurations}
\label{sec:other}

The previous results have demonstrated the capability of the proposed method not only to identify multiple modes but also to learn, for each mode, a different lower-dimensional subspace where posterior uncertainty is concentrated. 
 There are of course problems where the posterior is either unimodal (or at least one mode is much more prominent than the rest) or such that the posterior variance is distributed equally over a large number of dimensions (i.e. the posterior is not amenable to dimensionality reduction). In the context of the elastography problems examined the former scenario can take place when the noise in the data is relatively low. Then the data provide very strong evidence that preclude or make  the presence of multiple modes of comparable significance unlikely. The second scenario can appear when the available data is limited and/or very noisy. In this case, even if the posterior consists of a single mode, it is very likely that a large number of directions (if not all) will be characterized by large posterior variance as the signature of the data is very weak.  In the following two subsections we examine such settings in order to demonstrate the ability of the proposed framework to adapt and 
provide good approximations even though these might consist of a single mode or they do not encompass any significant dimensionality reduction.  
%

\subsubsection*{Example 2a: Only dimensionality reduction}

\begin{figure}[H]{
	\centering
	\captionsetup[subfigure]{labelformat=empty}
		\subfloat[][{Posterior mean ($S=1$)}] 
		{\includegraphics[width=0.30\textwidth]{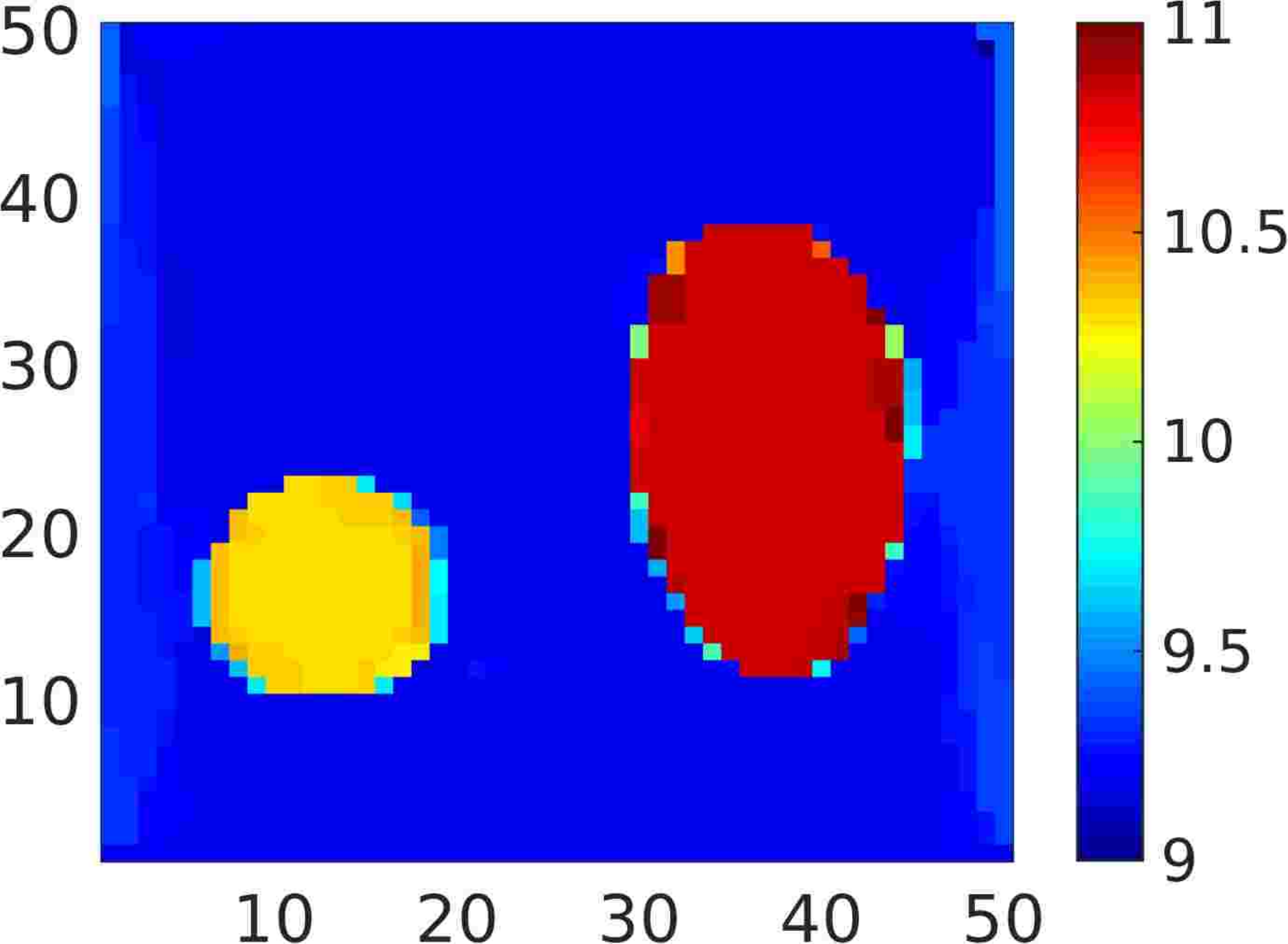}} 
		\hspace{0.1cm}
		\subfloat[][{Posterior standard deviation ($S=1$)}] 
		{\includegraphics[width=0.32\textwidth]{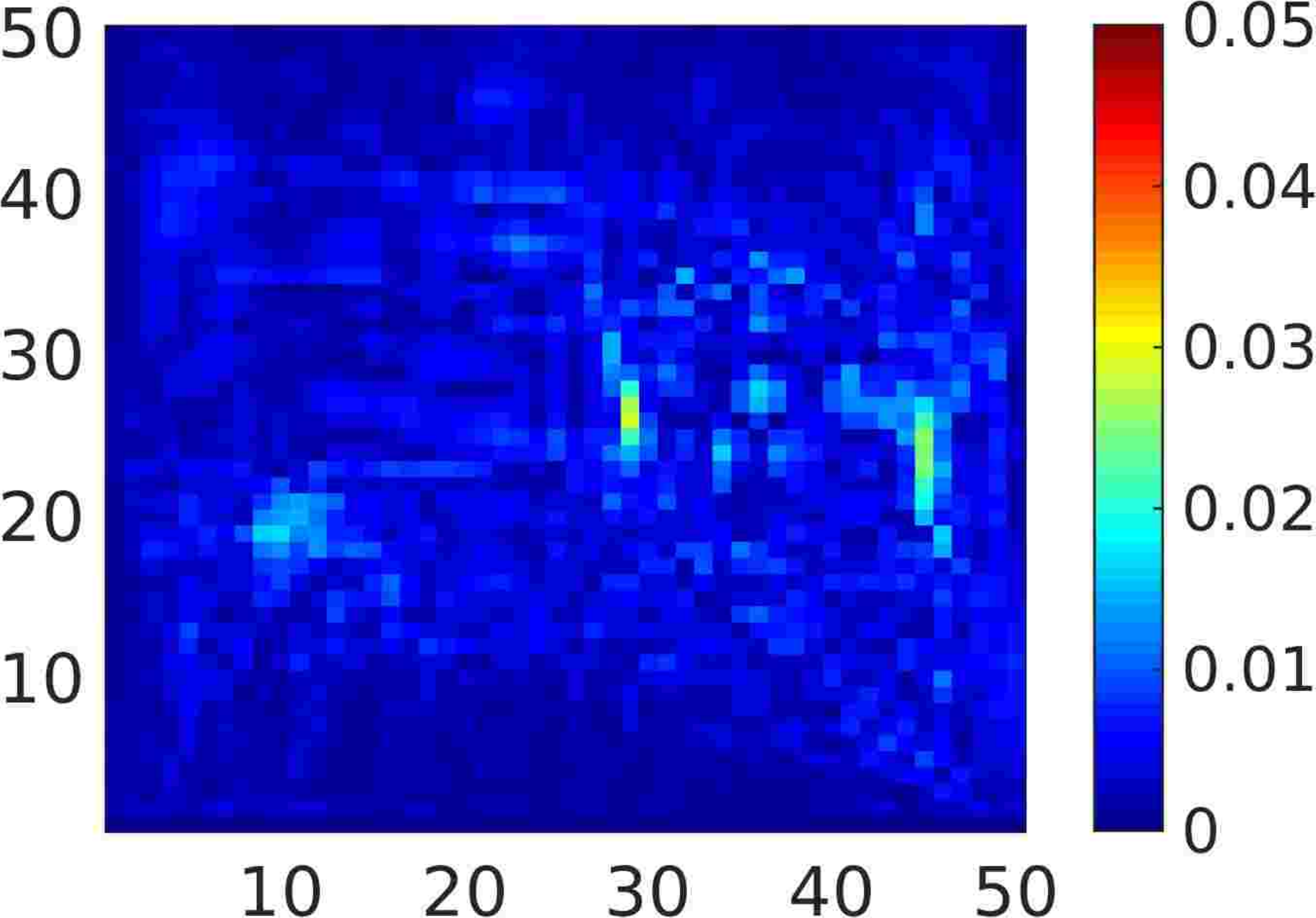}} 
		\hspace{0.1cm}
		\subfloat[][{Information gain}] 
		{\includegraphics[width=0.30\textwidth]{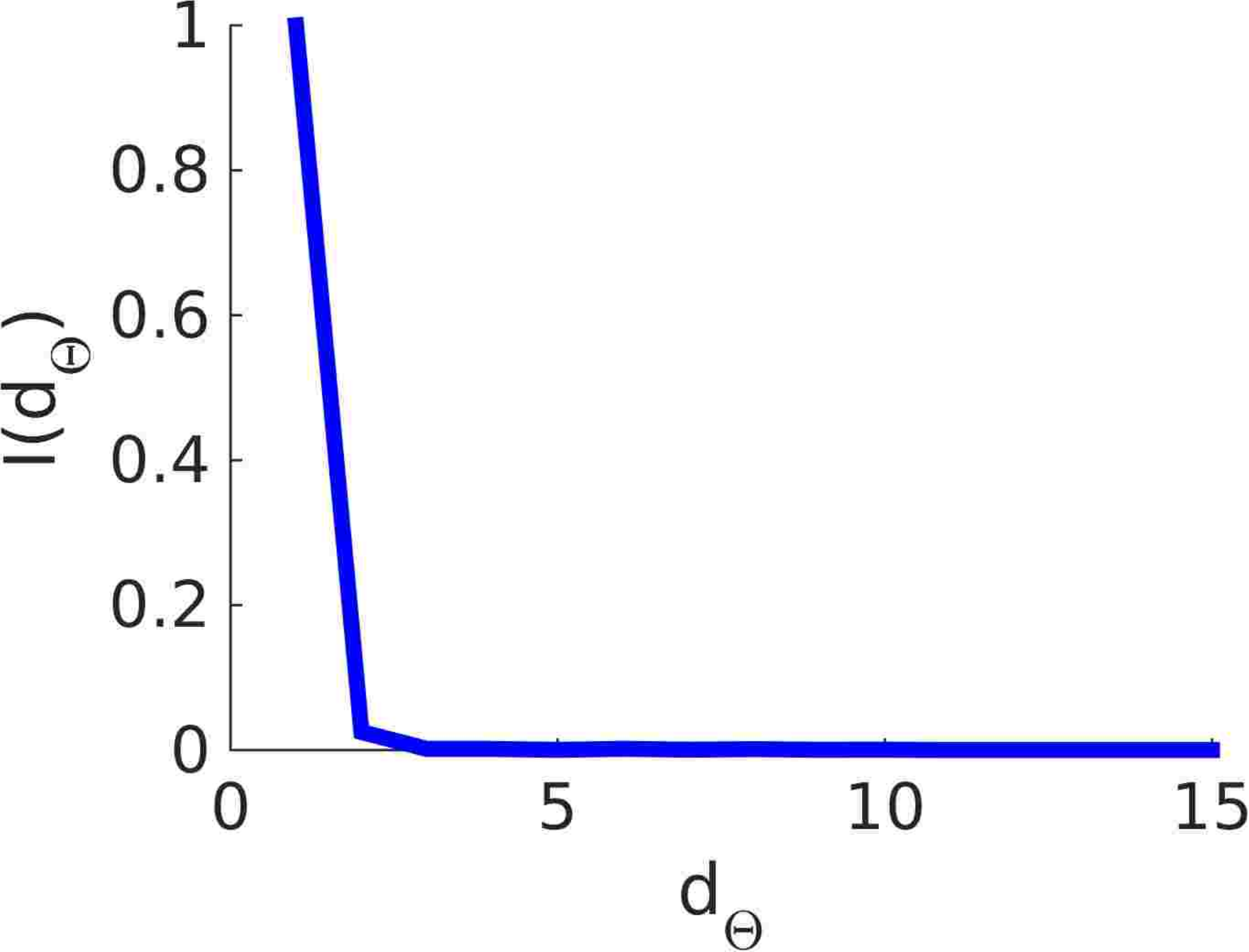}} 
		\hspace{0.1cm}
	  }
	  \caption{ Example 2a: On the left panel, the posterior mean of the material parameters is plotted and in the middle panel the posterior standard deviation (in log scale). The right panel depicts the information gain as a function of $\dth$. 
}
	 \label{fig:PosteriorMeanStdUnimodal}
\end{figure}

We consider the same problem (i.e. the same material properties and forward model) but instead contaminate the data with much less noise resulting in  a SNR of $1 \times 10^{4}$ (in contrast to $1 \times 10^{3}$  previously). 
In such a case a single mixture component is found. Despite multiple proposals (a total of $100$ were attempted) the identified components are either deleted because they violate the KL-based similarity criterion (\refeq{eq:compdeath1}) or they have negligible posterior probability $q(s) \ll q_{min}=10^{-3}$. The Gaussian identified has a mean that is extremely close to the ground truth as it can be seen in \reffig{fig:PosteriorMeanStdUnimodal}. The posterior variance across the problem domain is much smaller than in the previous setting (\reffig{fig:PosteriorMeanStdUnimodal}) and is, as expected, due to the low levels of noise, concentrated along very few dimensions. Hence, the  information gain metric $I(\dth)$ decays extremely rapidly and we can adequately approximate the posterior with less than $5$  
 reduced coordinates $\bt$ (\reffig{fig:PosteriorMeanStdUnimodal}). 

\subsubsection*{Example 2b: Only multimodality}

\begin{figure}[H]{
	\vspace{-2.1cm}
	\centering
	\captionsetup[subfigure]{labelformat=empty}
		\hspace{-1.5cm}
		\subfloat[][{ $q(s=1) =0.00409$ }] 
		{\includegraphics[width=0.25\textwidth]{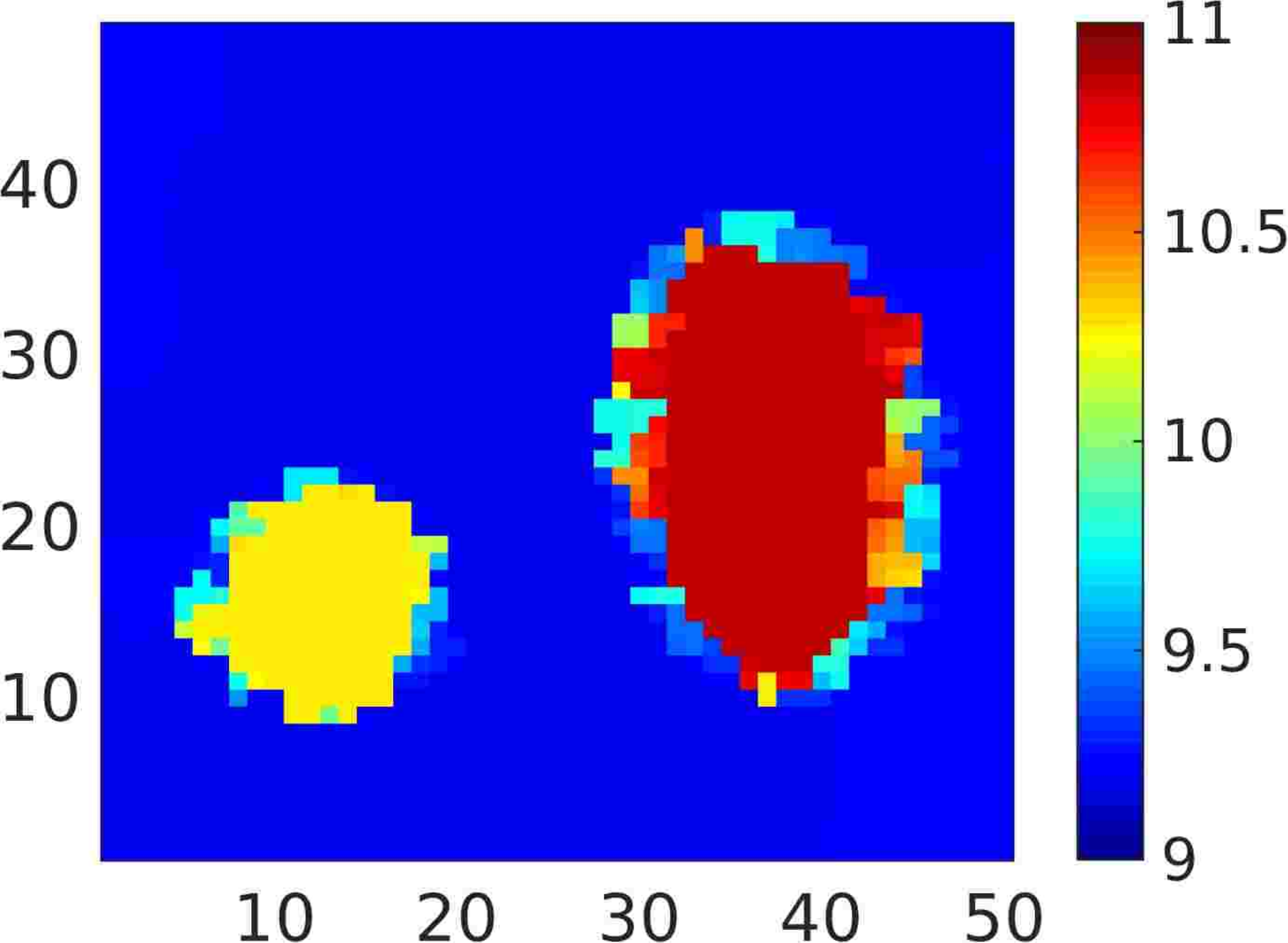}} 
		\hspace{0.1cm}
		\subfloat[][{ $q(s=2) =0.0276$}] 
		{\includegraphics[width=0.25\textwidth]{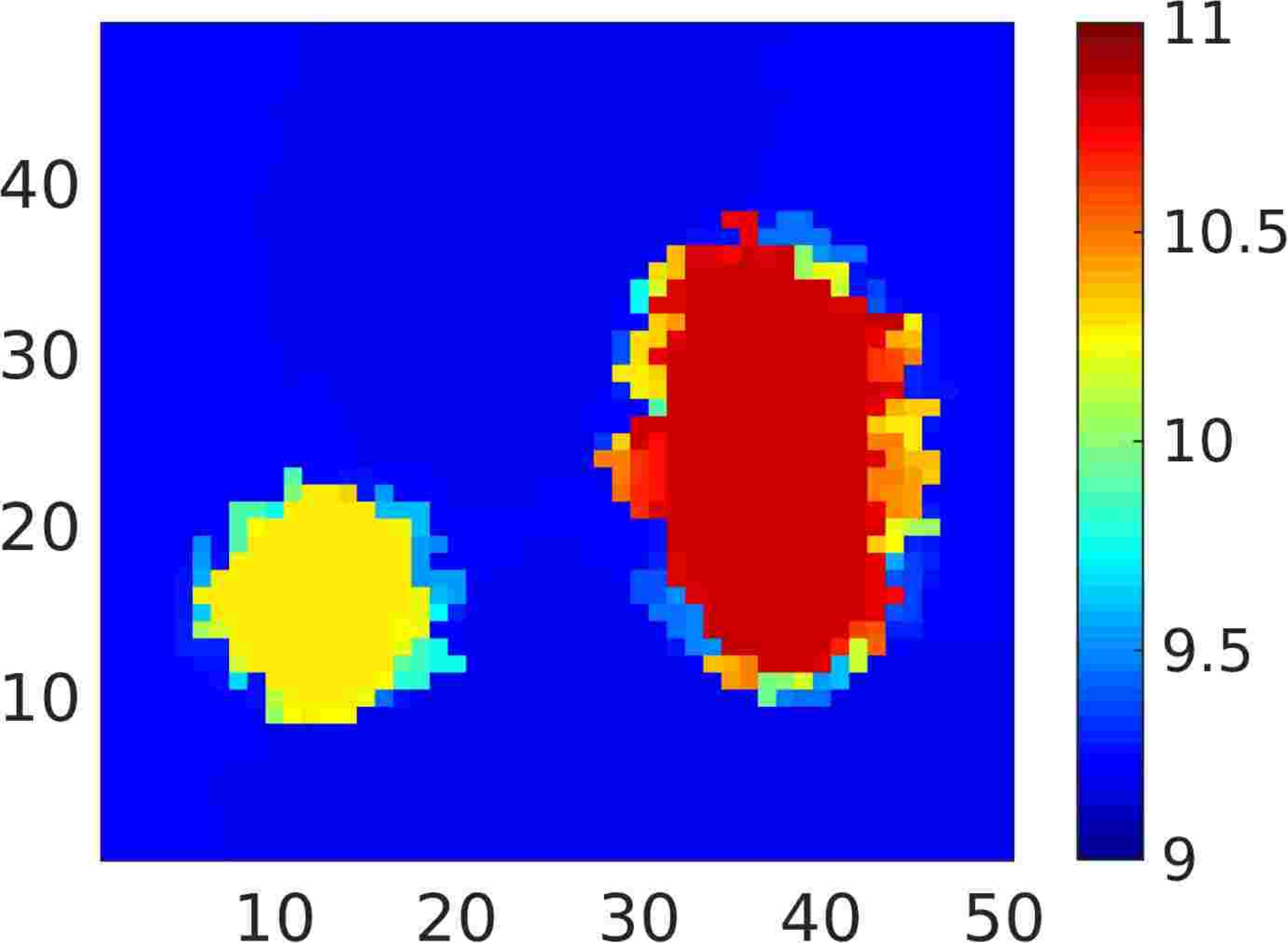}} 
		\hspace{0.1cm}
		\subfloat[][{ $q(s=3) =0.0698$}] 
		{\includegraphics[width=0.25\textwidth]{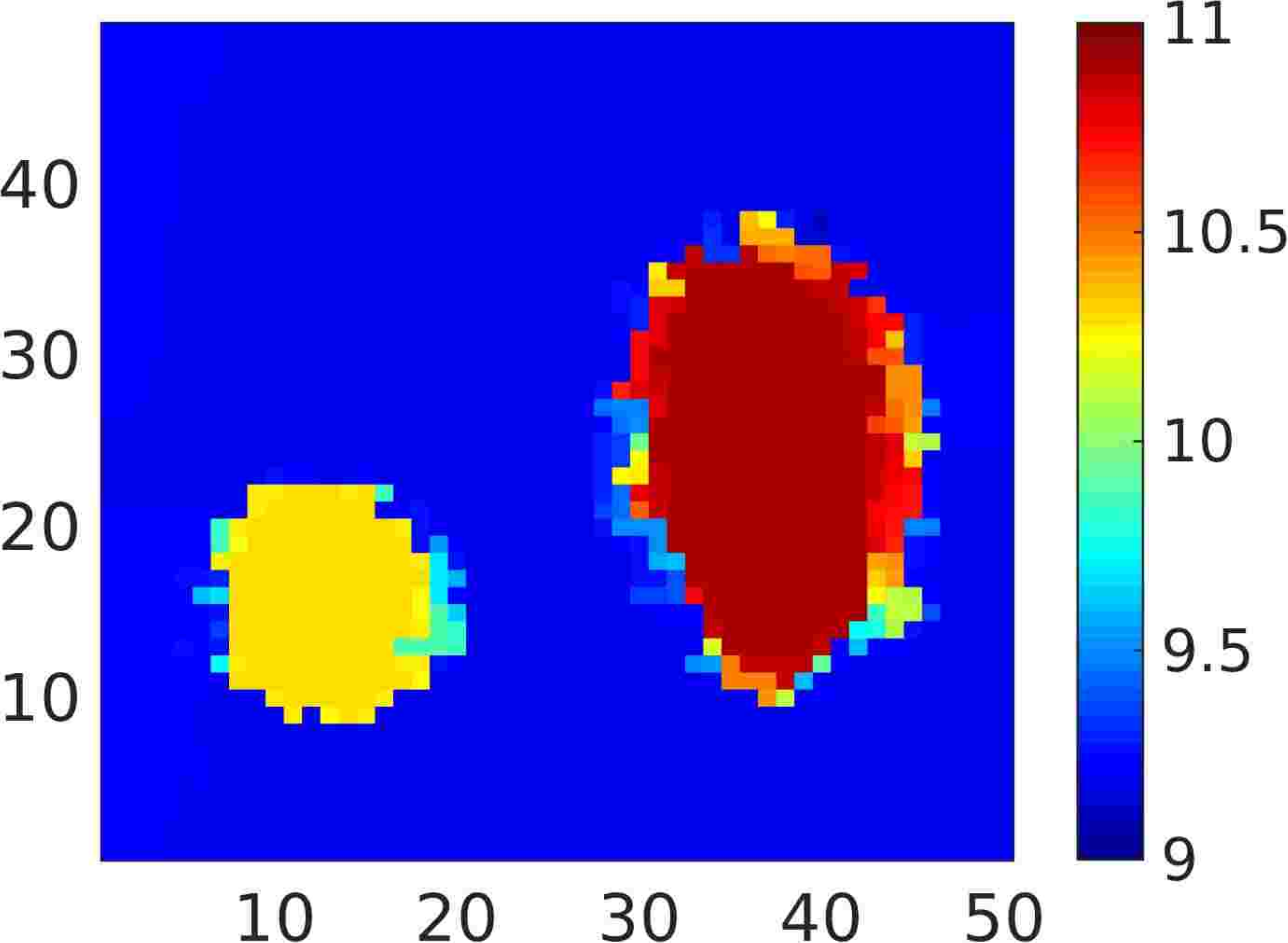}} 
		\hspace{0.1cm}
		\subfloat[][{ $q(s=4) =0.0893$ }] 
		{\includegraphics[width=0.25\textwidth]{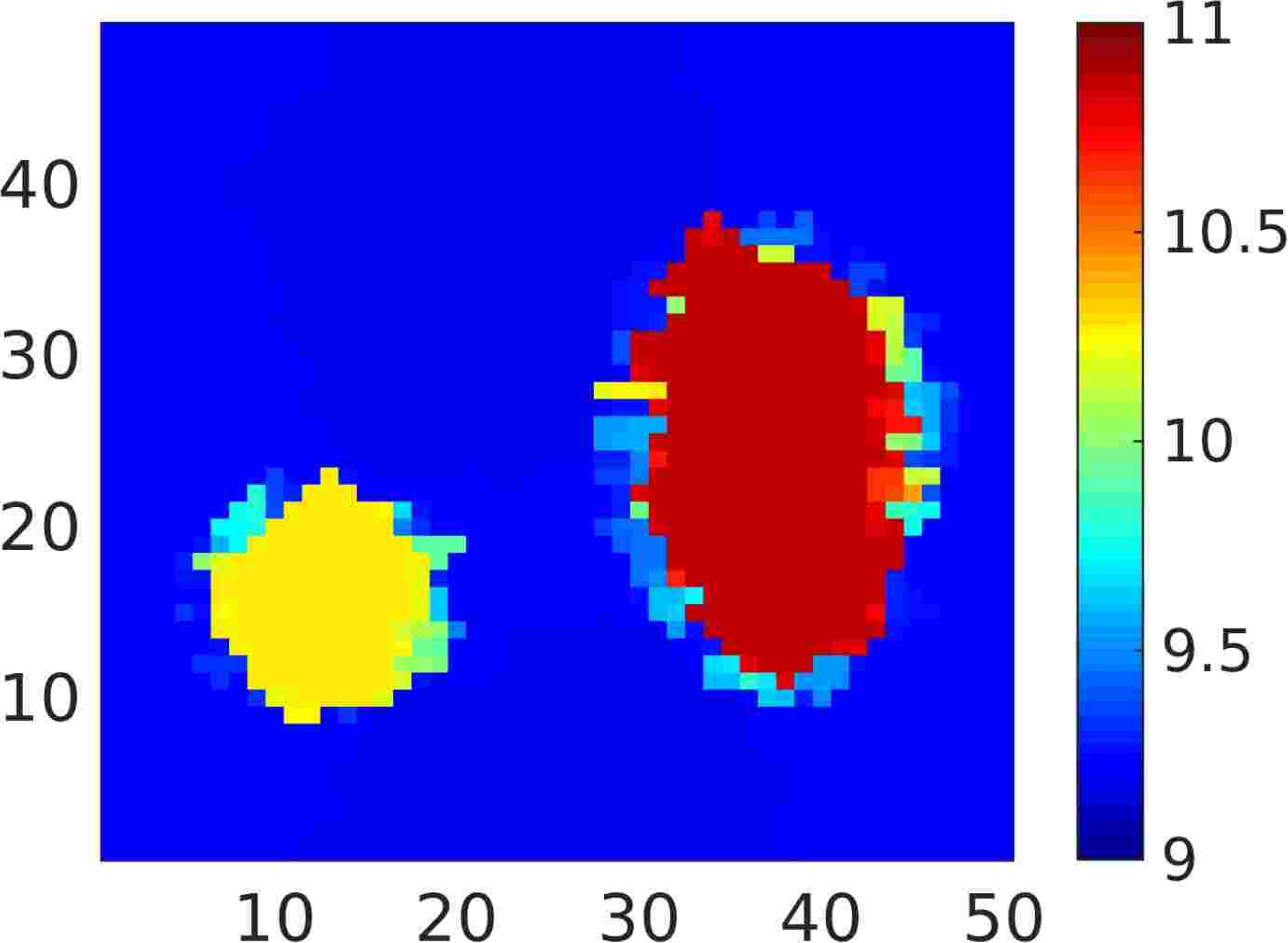}} 
		\hspace{-0.5cm}
		\\
		\hspace{-1.5cm}
		\subfloat[][{ $q(s=5) =0.0095$}] 
		{\includegraphics[width=0.25\textwidth]{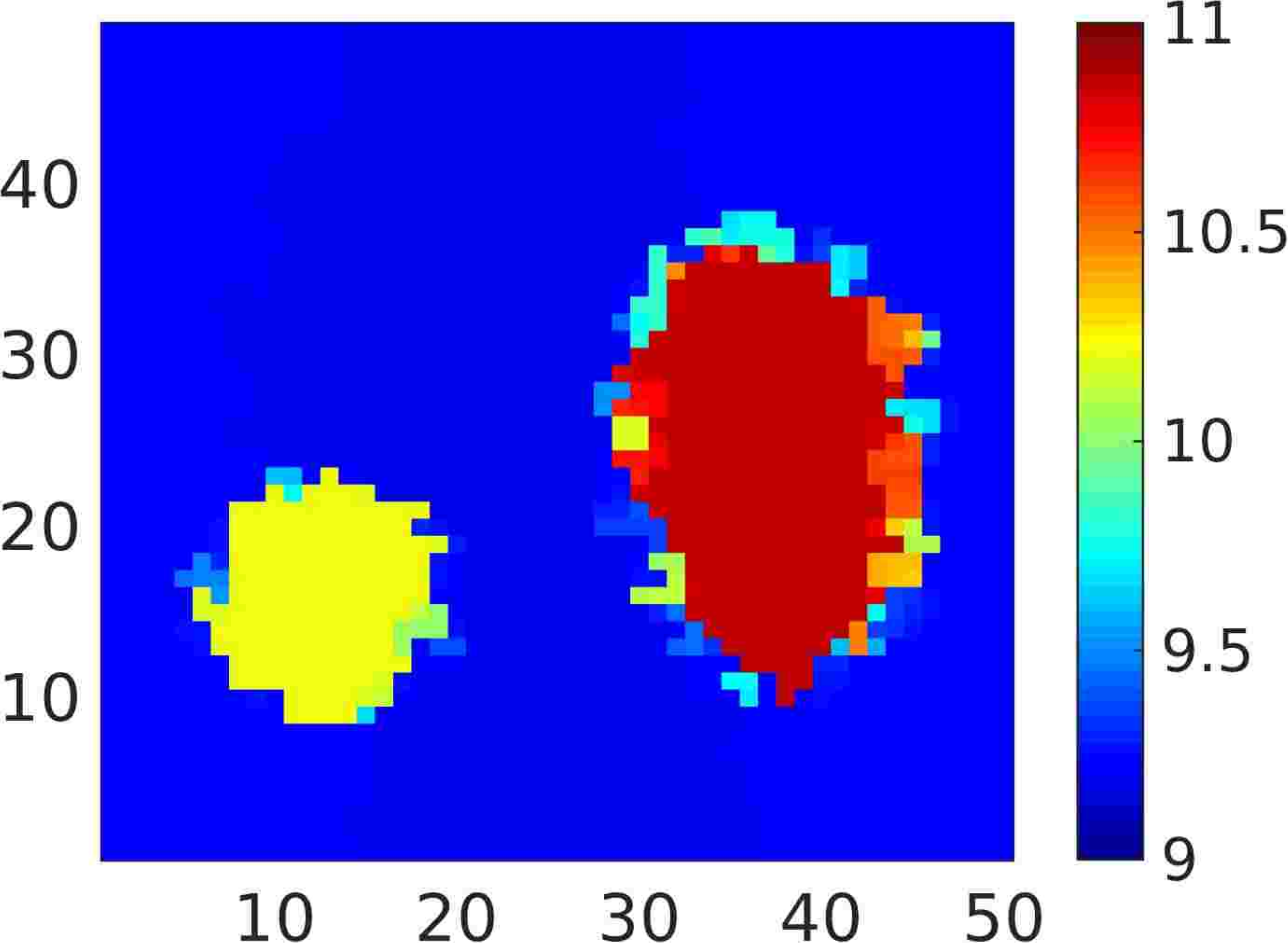}} 
		\hspace{0.1cm}
		\subfloat[][{ $q(s=6) =0.104$}] 
		{\includegraphics[width=0.25\textwidth]{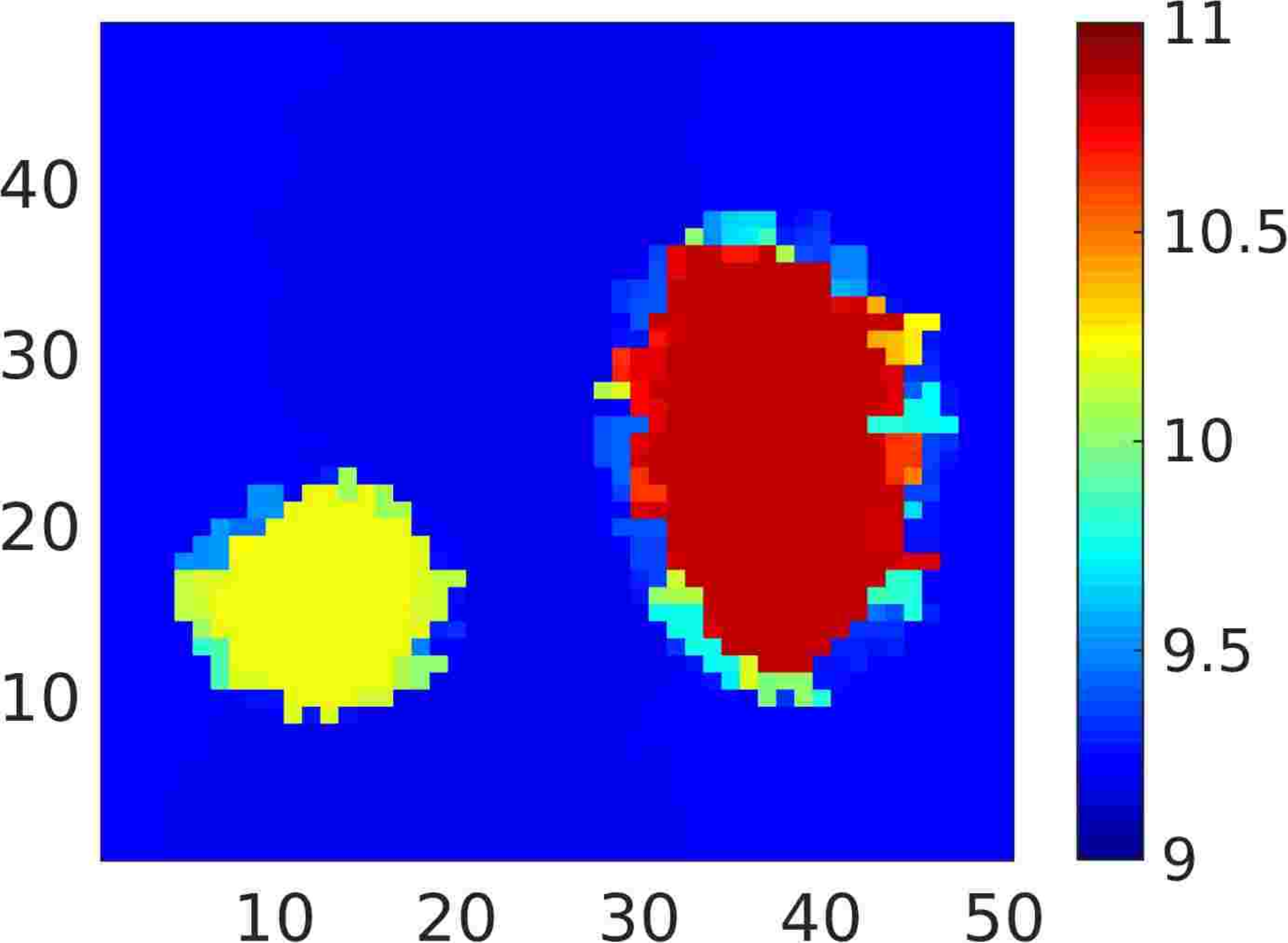}} 
		\hspace{0.1cm}
		\subfloat[][{ $q(s=7) =0.00777$ }] 
		{\includegraphics[width=0.25\textwidth]{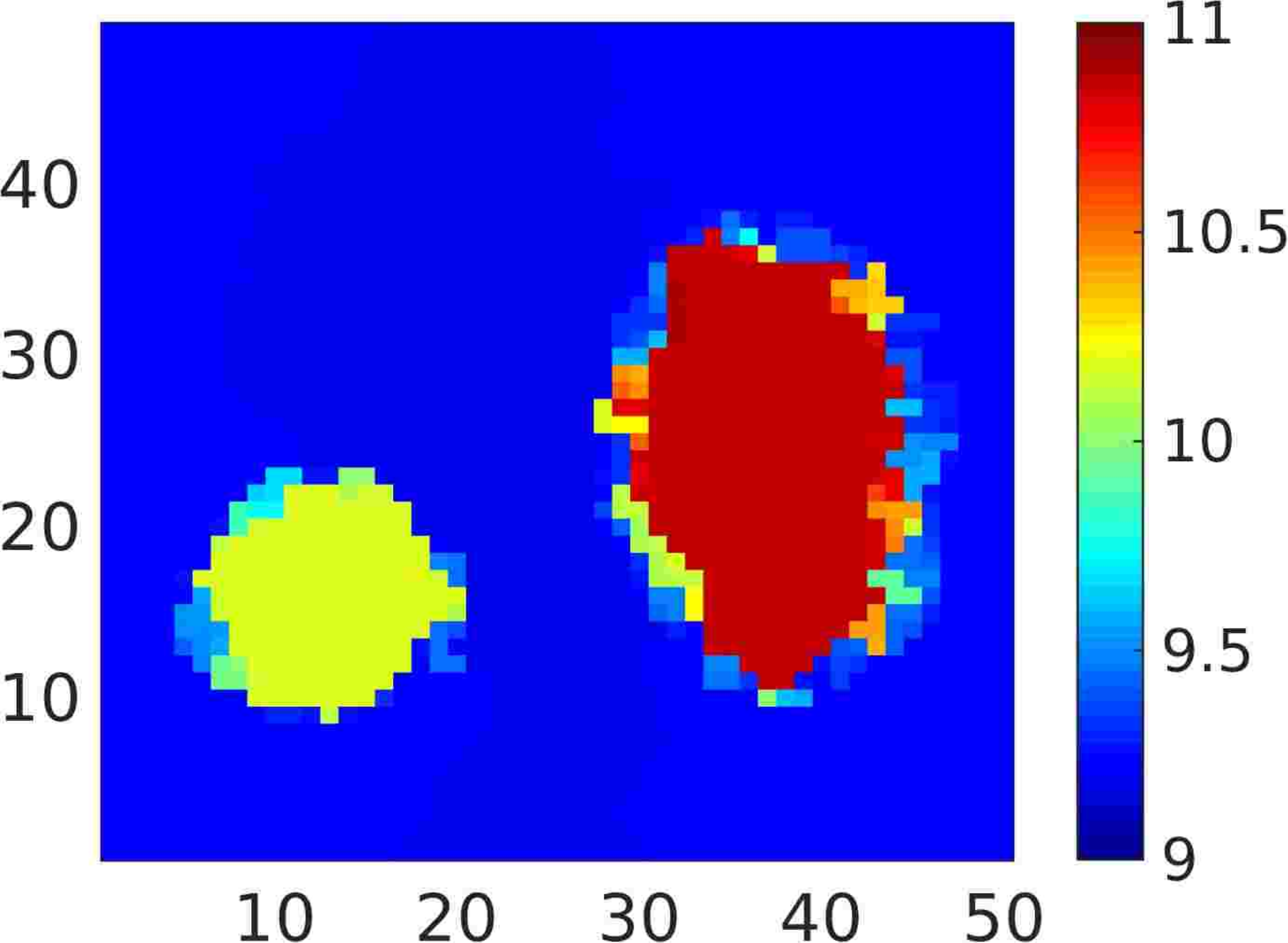}} 
		\hspace{0.1cm}
		\subfloat[][{$q(s=8) = 0.00606$}] 
		{\includegraphics[width=0.25\textwidth]{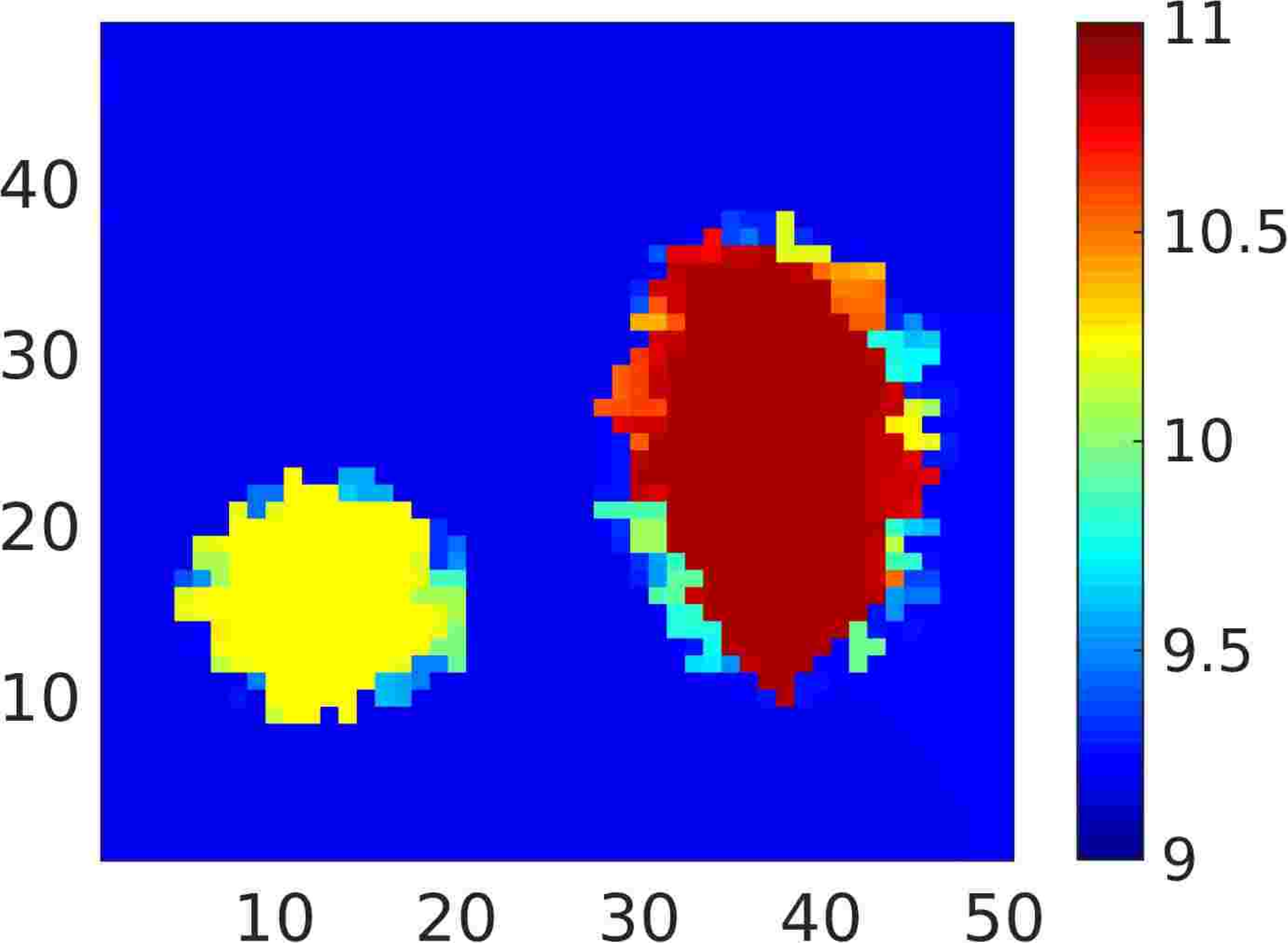}} 
		\hspace{-0.5cm}
		\\					
		\hspace{-1.5cm}		
		\subfloat[][{ $q(s=9)= 0.0247$}] 
		{\includegraphics[width=0.25\textwidth]{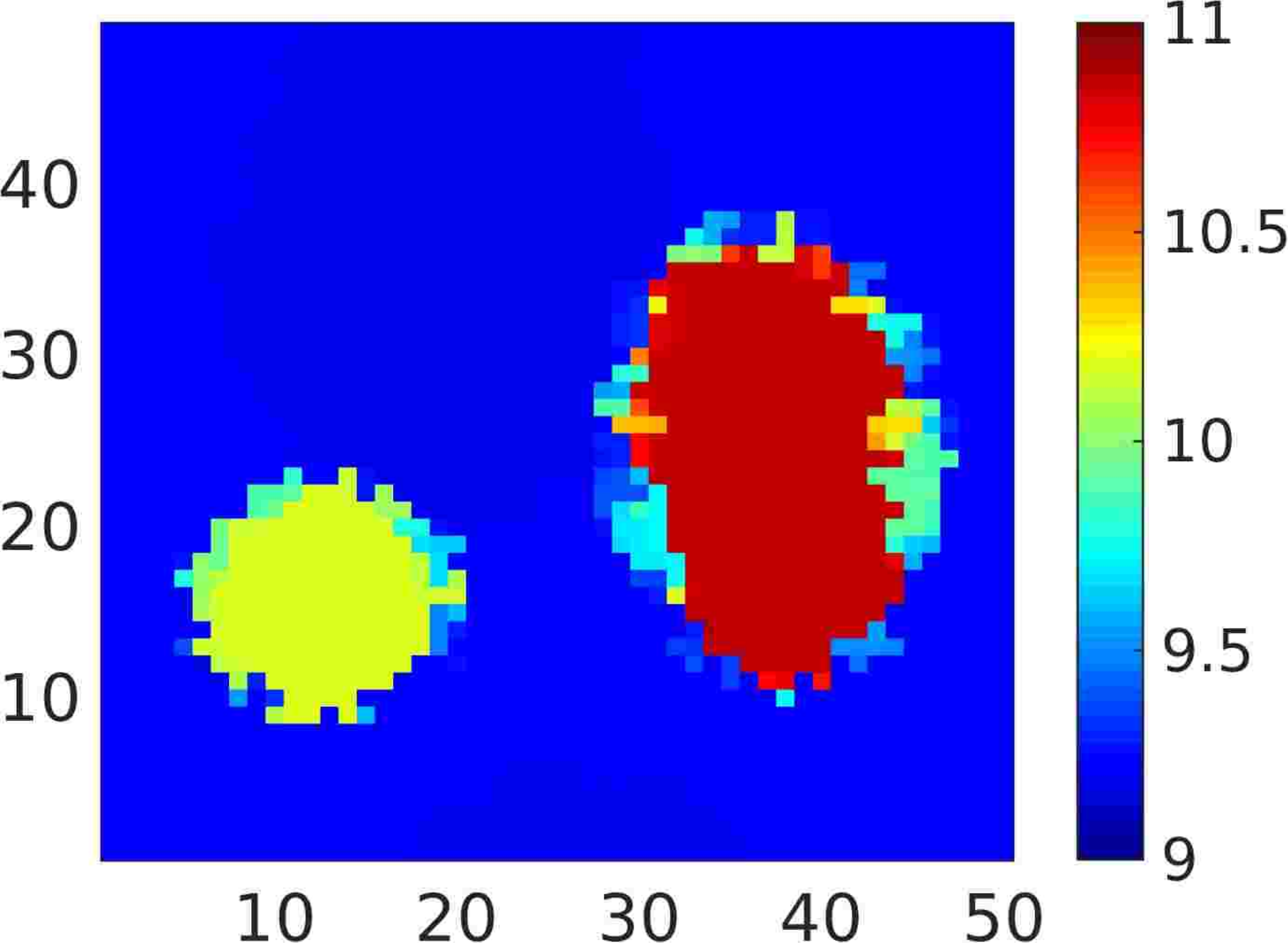}} 
		\hspace{0.1cm}									
		\subfloat[][{ $q(s=10) =0.00276$ }] 
		{\includegraphics[width=0.25\textwidth]{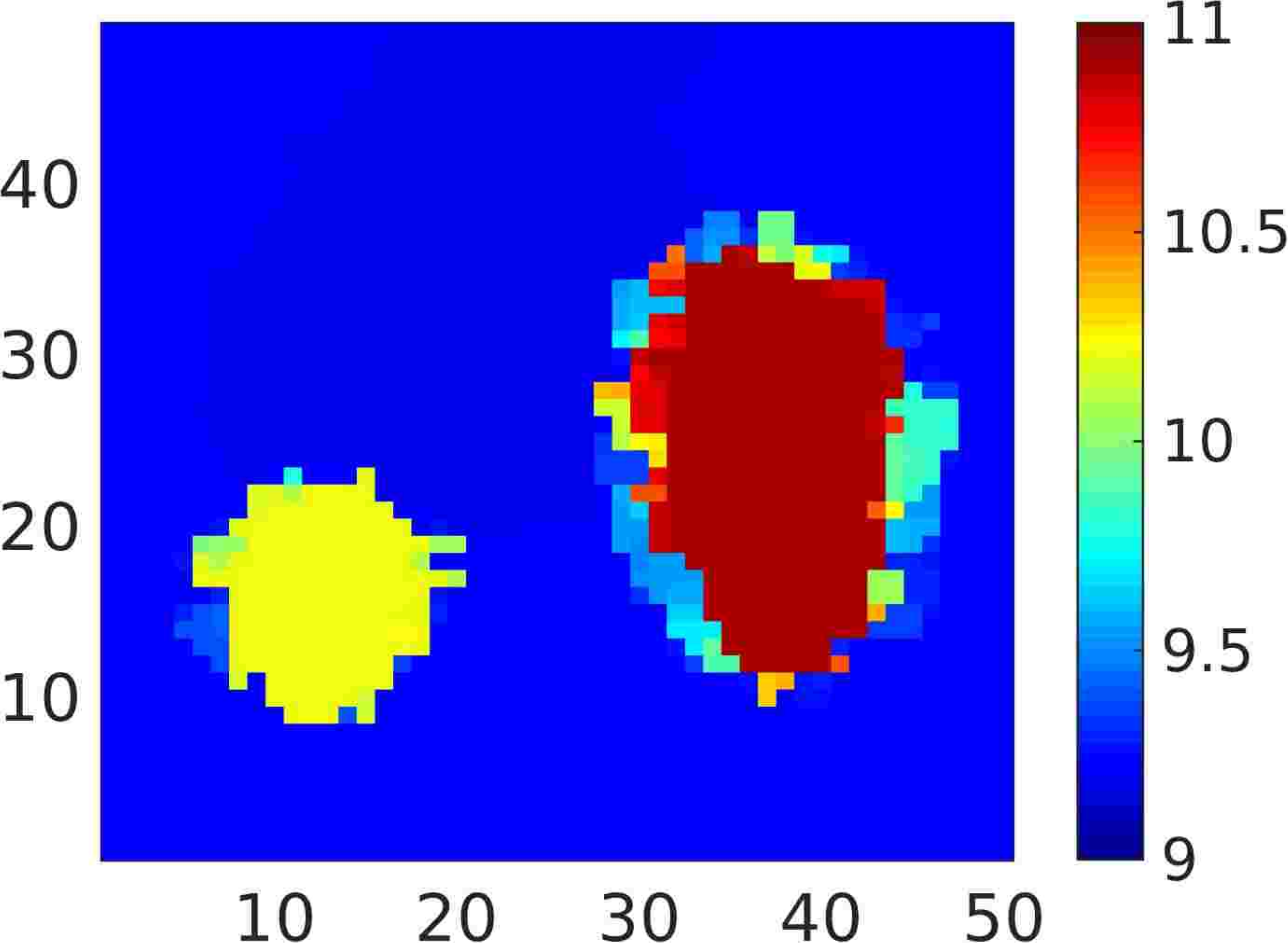}} 
		\hspace{0.1cm}
		\subfloat[][{ $q(s=11) =0.107$}] 
		{\includegraphics[width=0.25\textwidth]{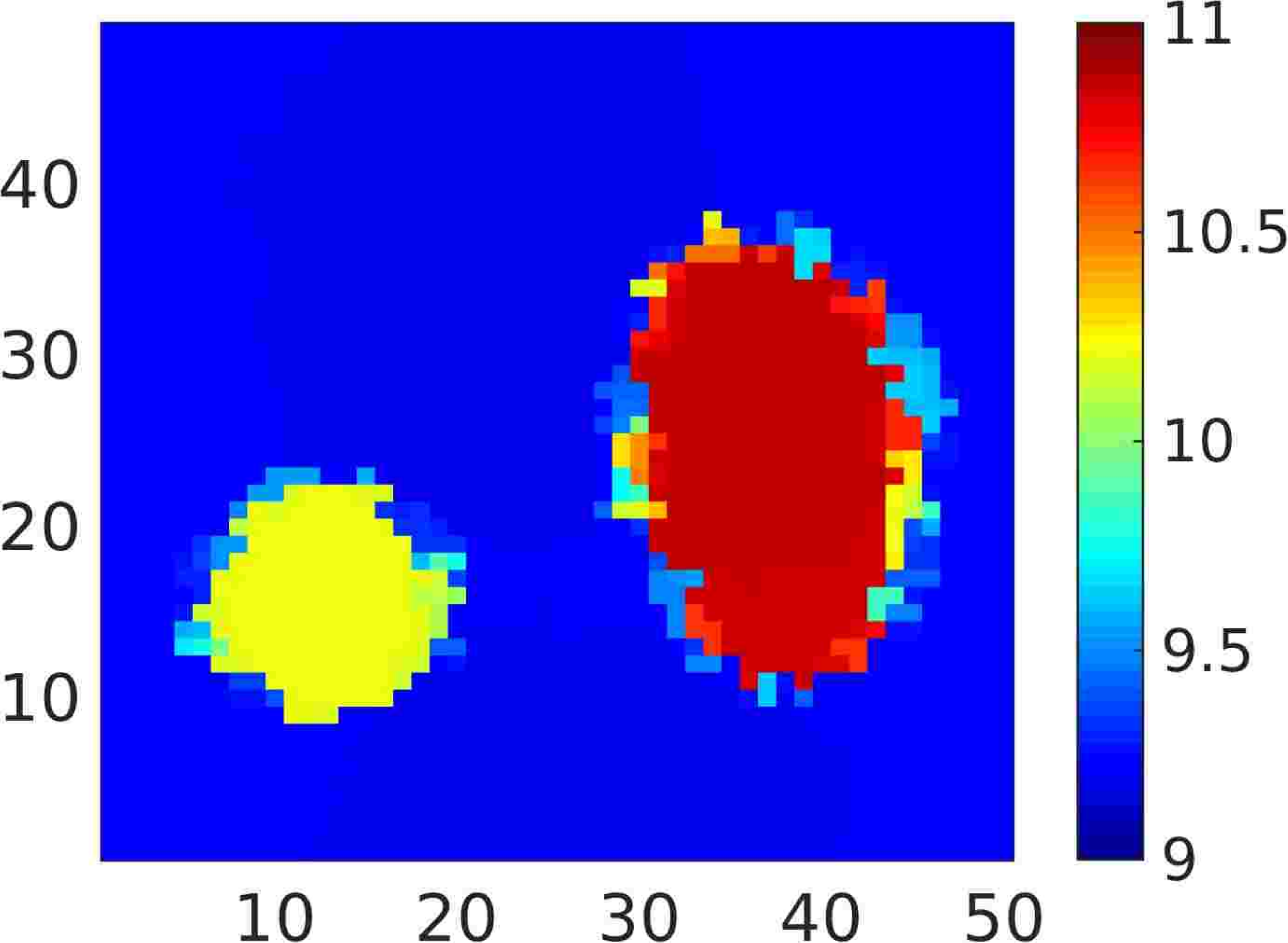}} 
		\hspace{0.1cm}
		\subfloat[][{ $q(s=12) =0.0523$}] 
		{\includegraphics[width=0.25\textwidth]{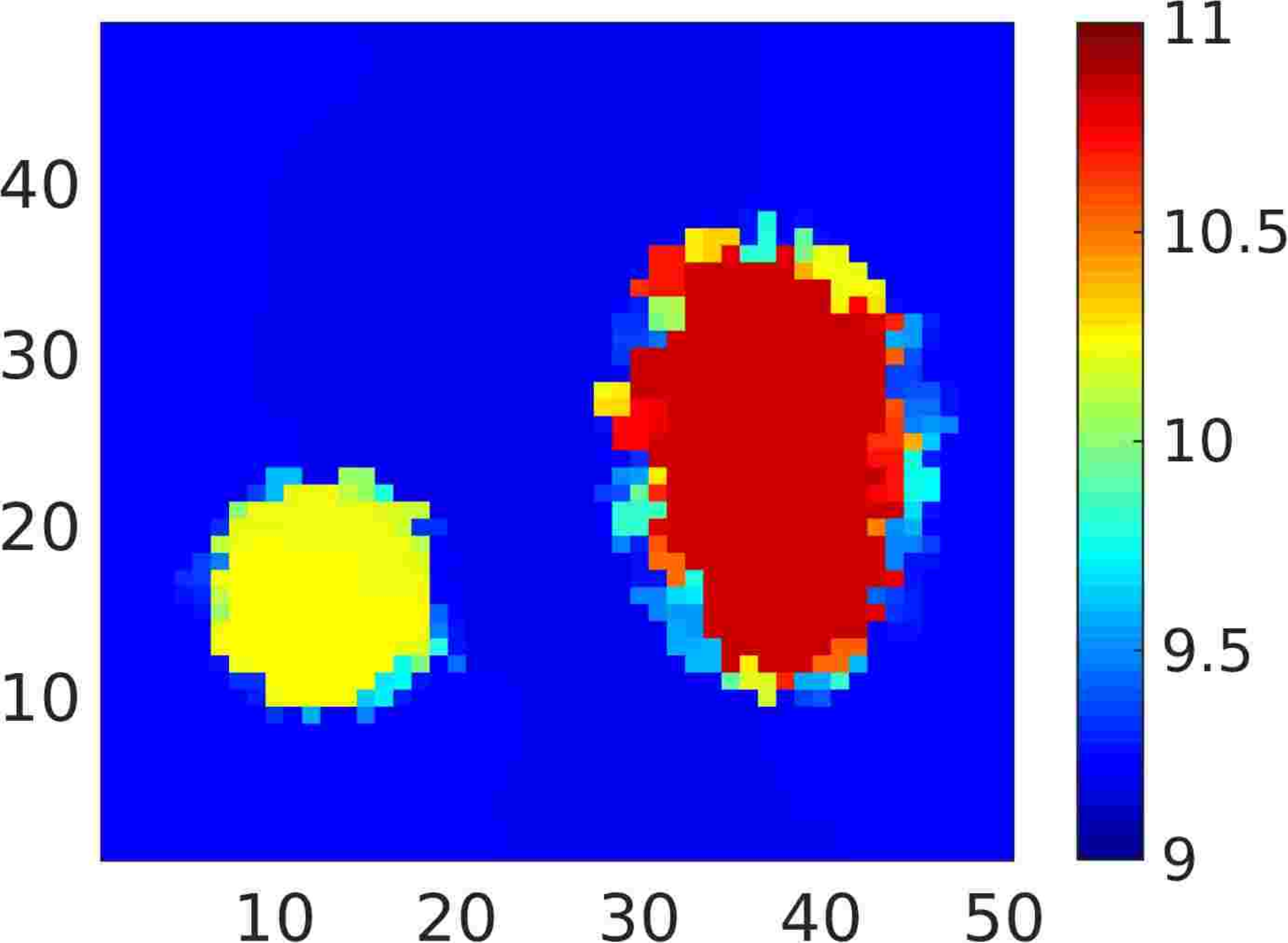}} 
		\hspace{-0.5cm}
		\\
		\hspace{-1.5cm}			
		\subfloat[][{ $q(s=13) =0.0267$ }] 
		{\includegraphics[width=0.25\textwidth]{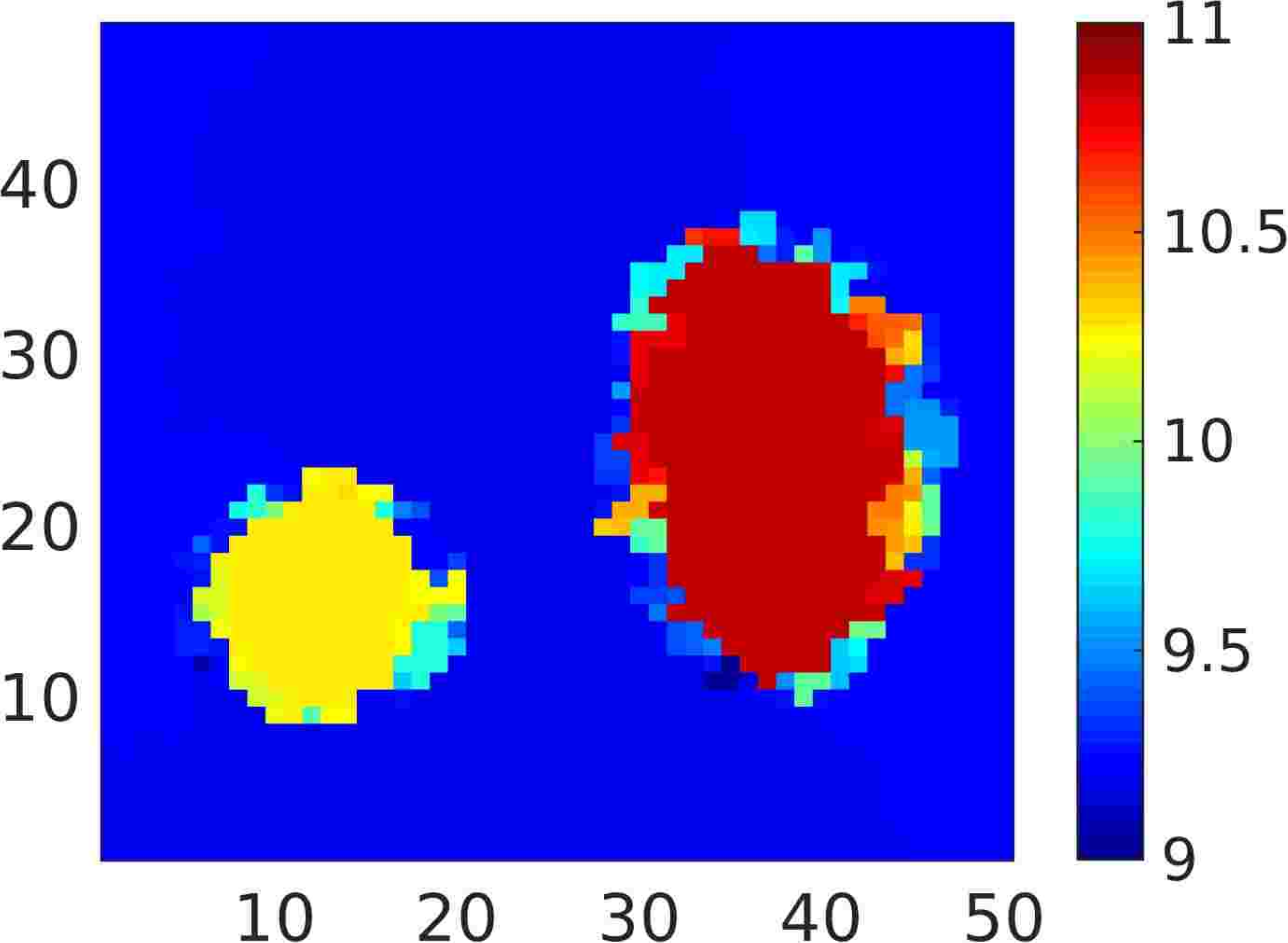}} 
		\hspace{0.1cm}
		\subfloat[][{ $q(s=14) =0.0462$}] 
		{\includegraphics[width=0.25\textwidth]{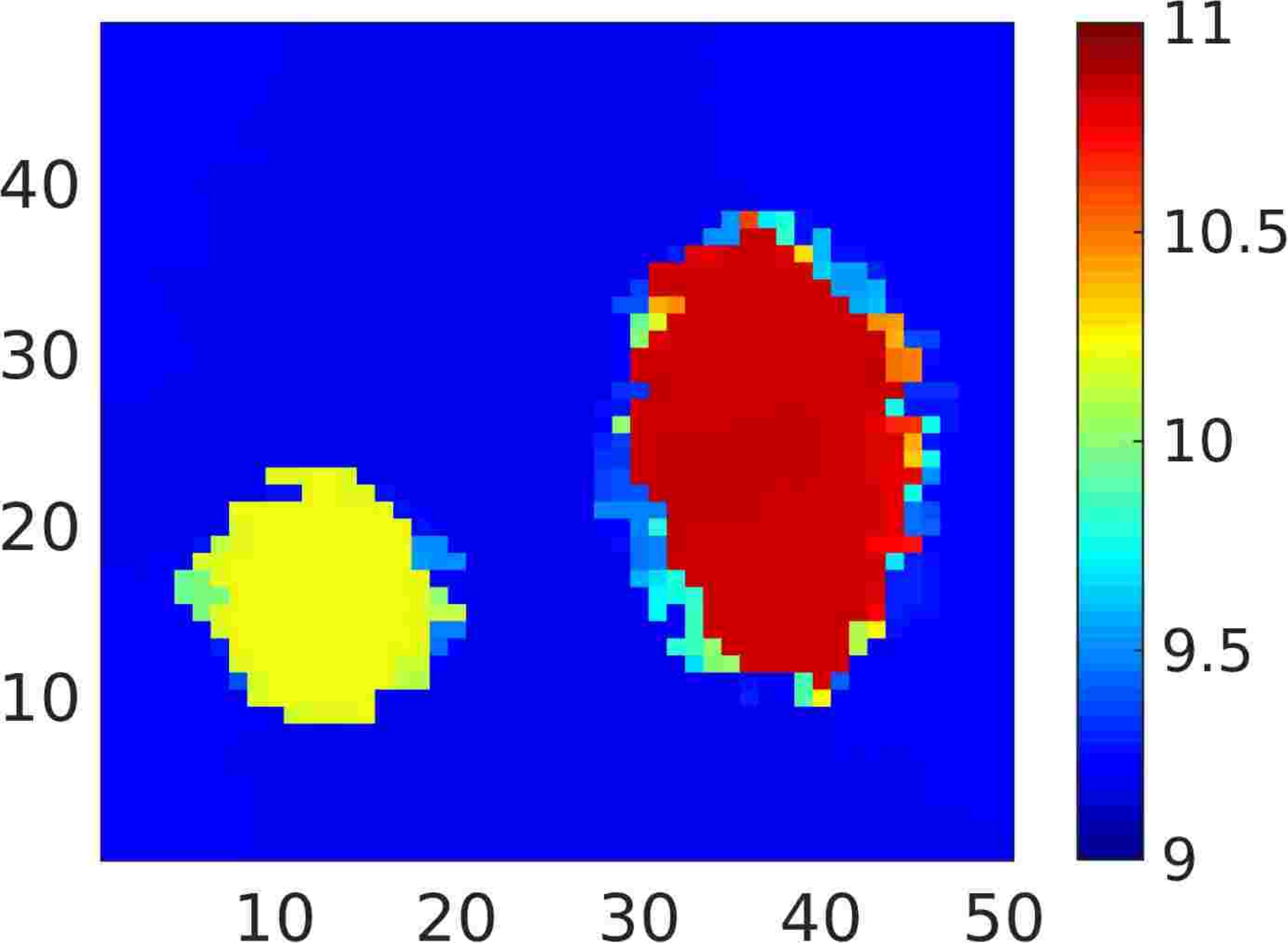}} 
		\hspace{0.1cm}
		\subfloat[][{$q(s=15) = 0.0868$}] 
		{\includegraphics[width=0.25\textwidth]{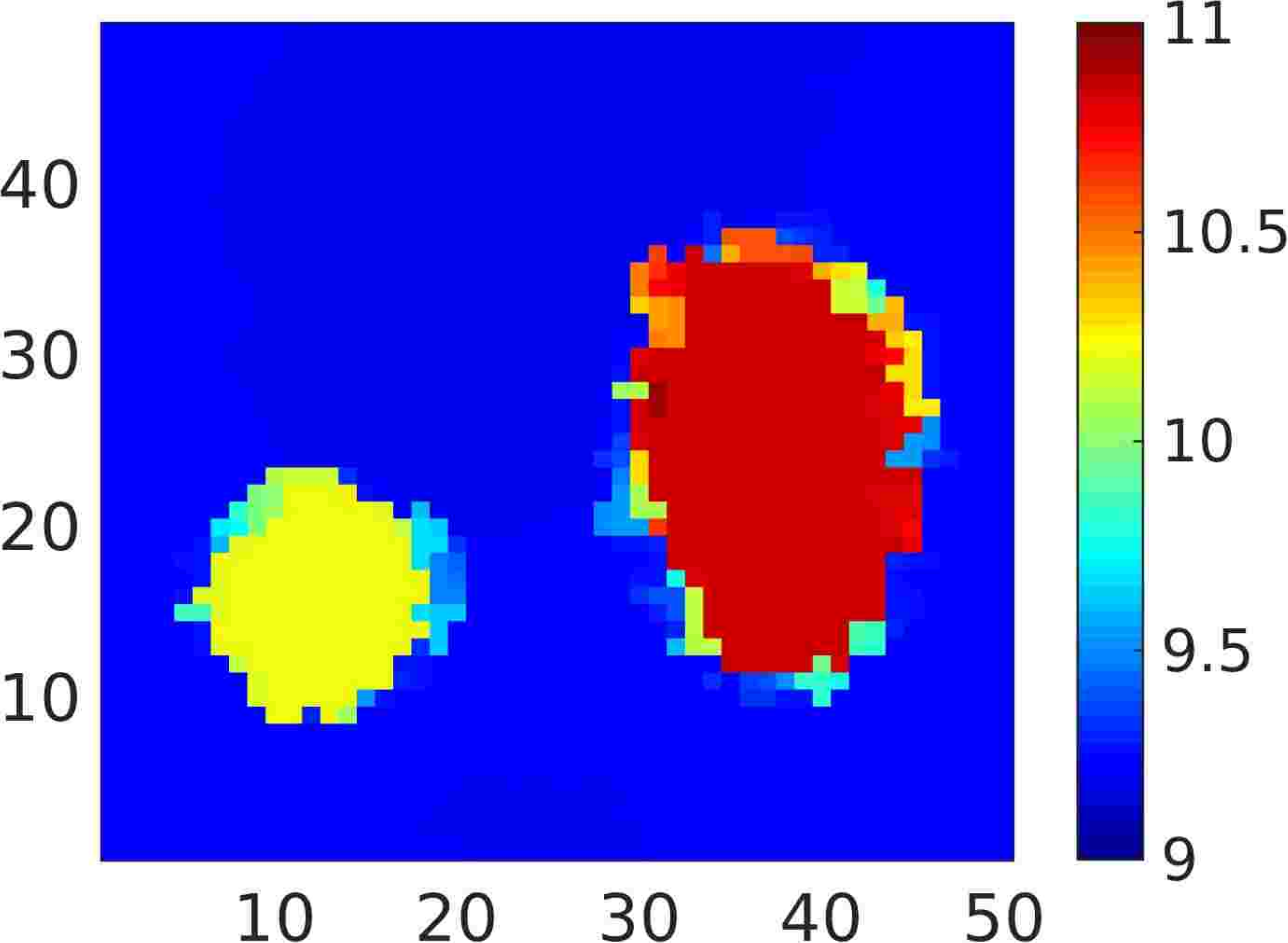}} 
		\hspace{0.1cm}
		\subfloat[][{ $q(s=16) =0.101$}] 
		{\includegraphics[width=0.25\textwidth]{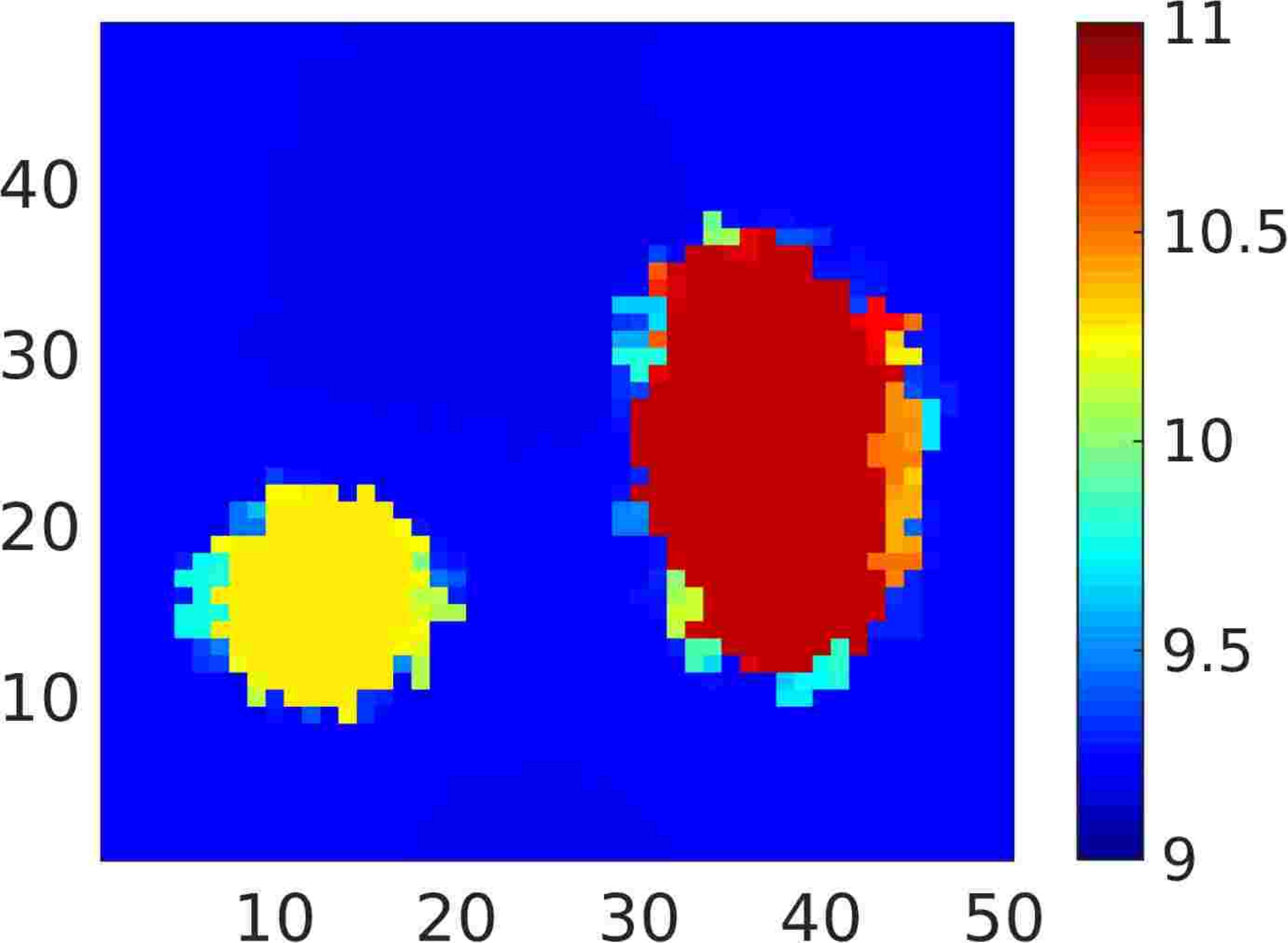}} 
		\hspace{-0.5cm}
		\\								
		\hspace{-3.5cm}	
		\subfloat[][{\footnotesize$q(s=17) = 0.222$}] 
		{\includegraphics[width=0.21\textwidth]{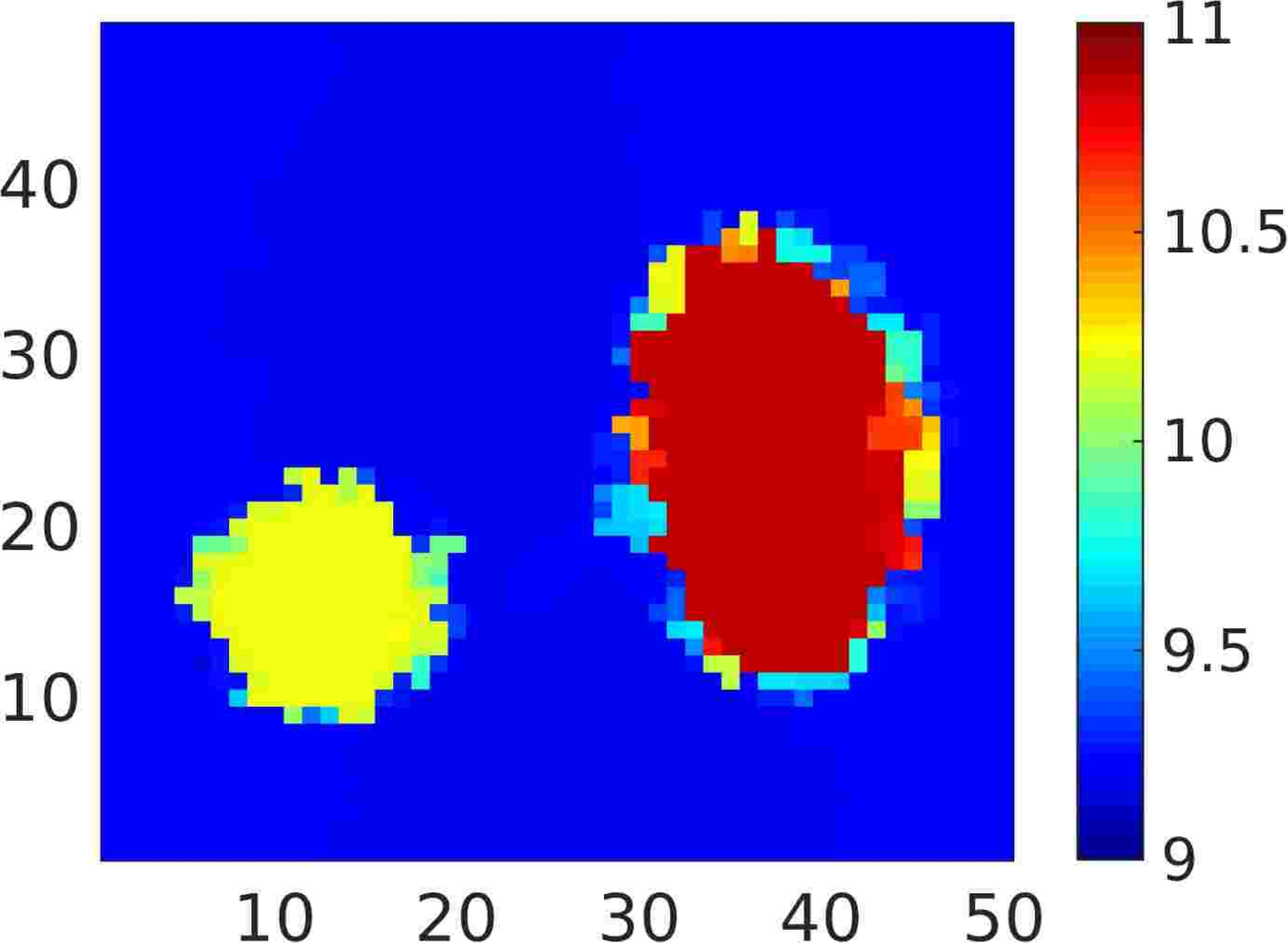}} 
		\hspace{0.1cm}
		\subfloat[][{\footnotesize$q(s=18) =0.0017$}] 
		{\includegraphics[width=0.21\textwidth]{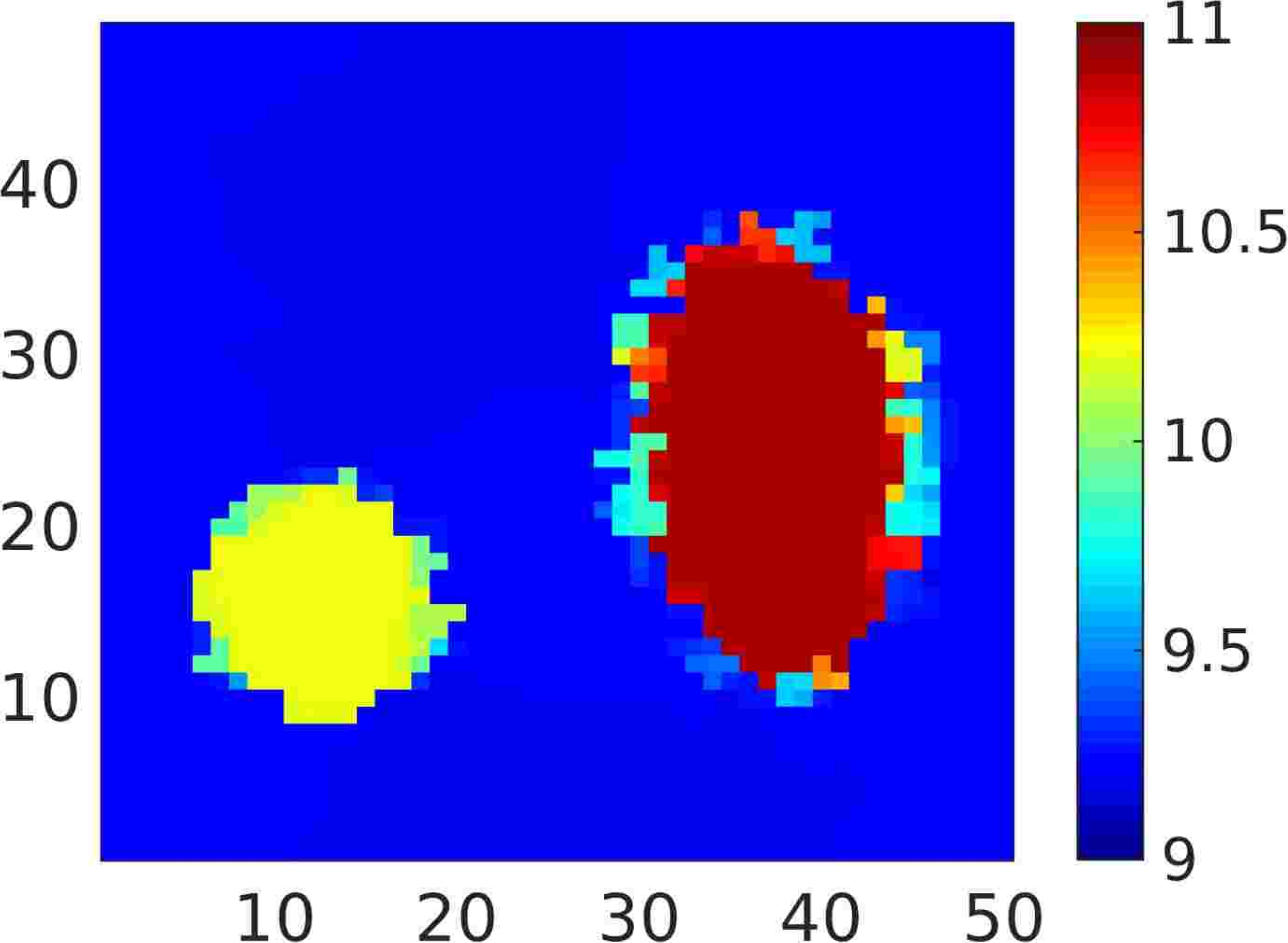}} 
		\hspace{0.1cm}
		\subfloat[][{\footnotesize$q(s=19) =0.0032$}] 
		{\includegraphics[width=0.21\textwidth]{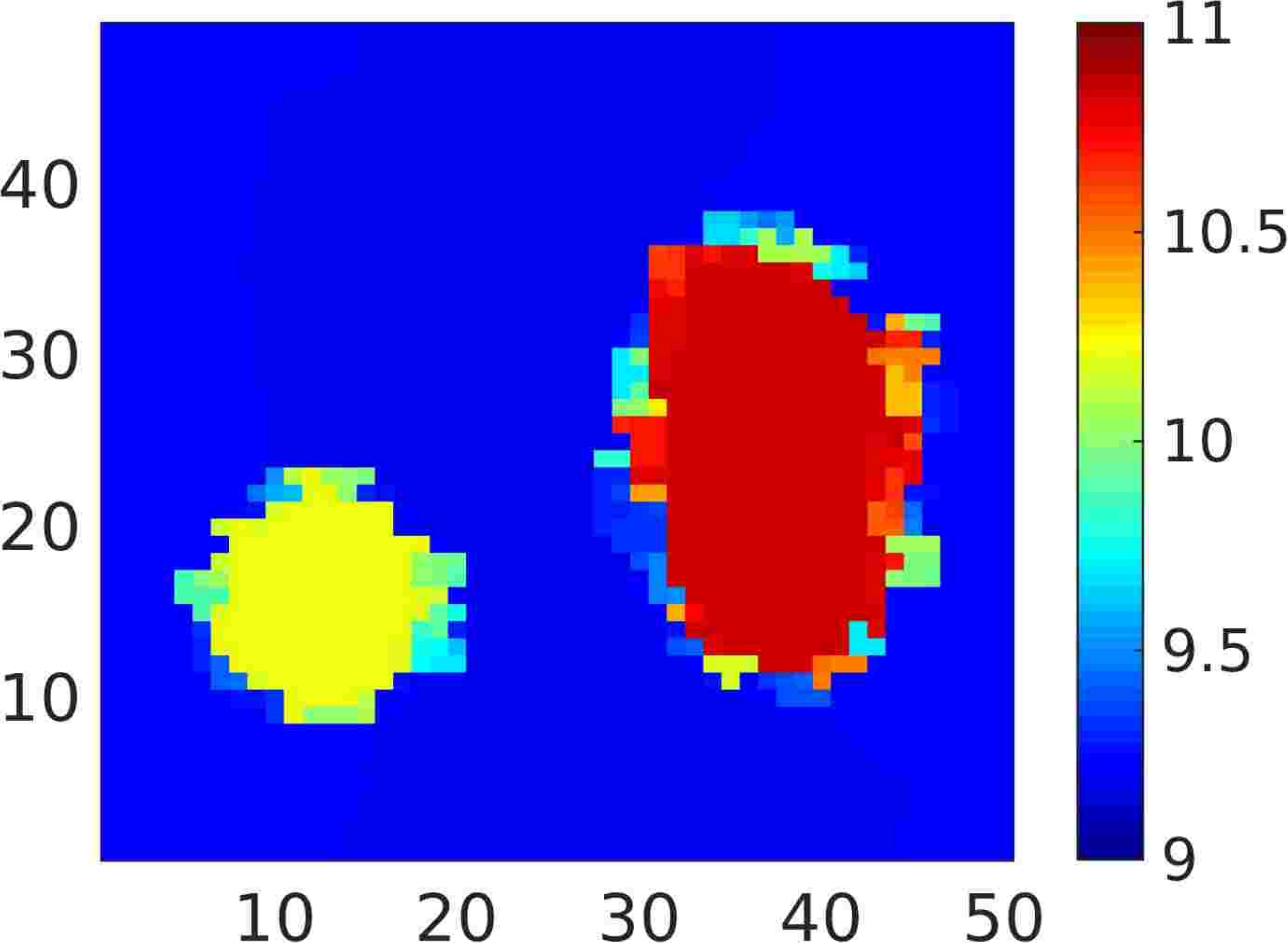}} 
		\hspace{0.1cm}
		\subfloat[][{\footnotesize$q(s=20) =0.0053$}] 
		{\includegraphics[width=0.21\textwidth]{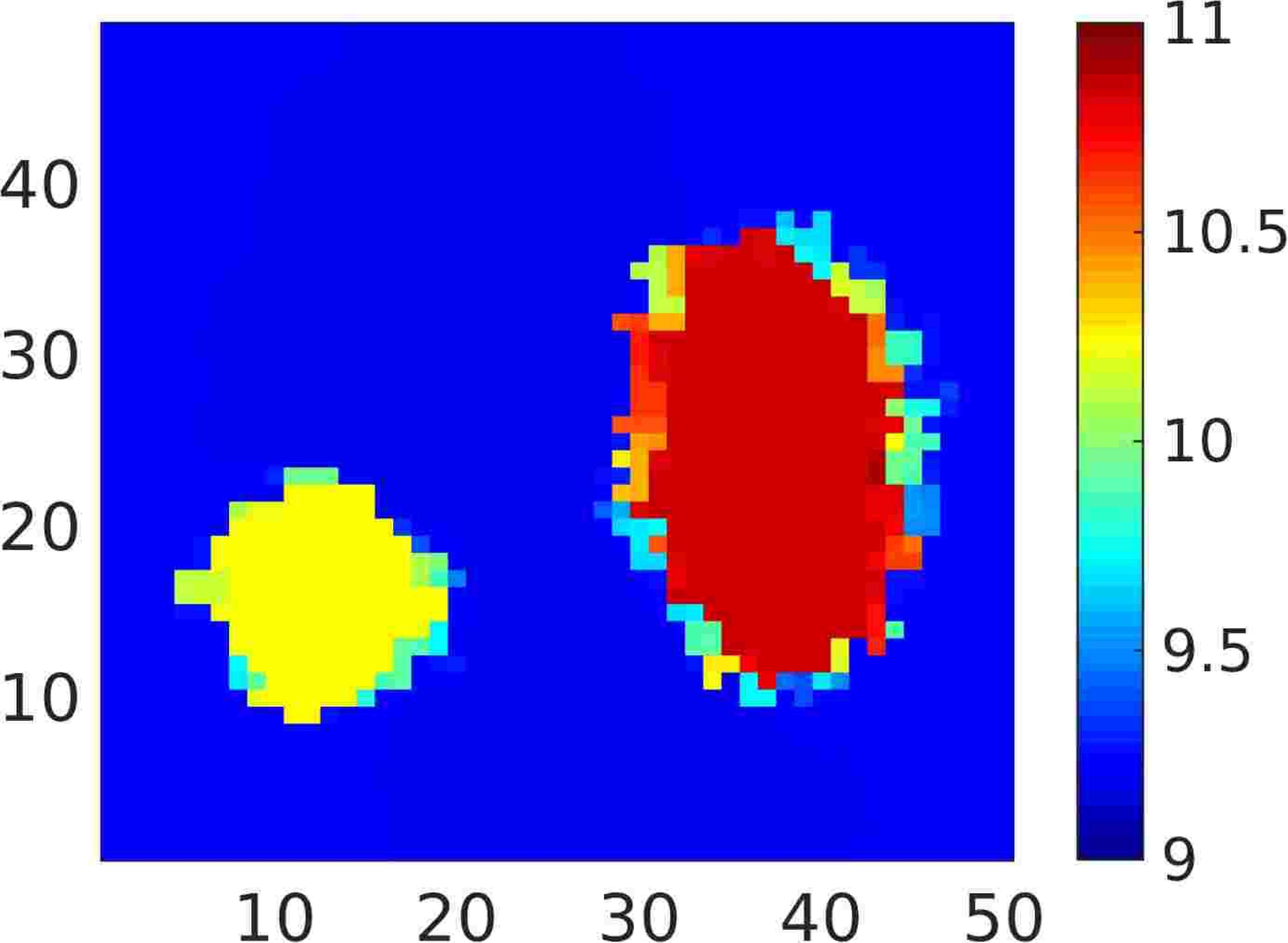}} 
		\hspace{0.1cm}
		\subfloat[][{\footnotesize$q(s=21) = 0.0032$}] 
		{\includegraphics[width=0.21\textwidth]{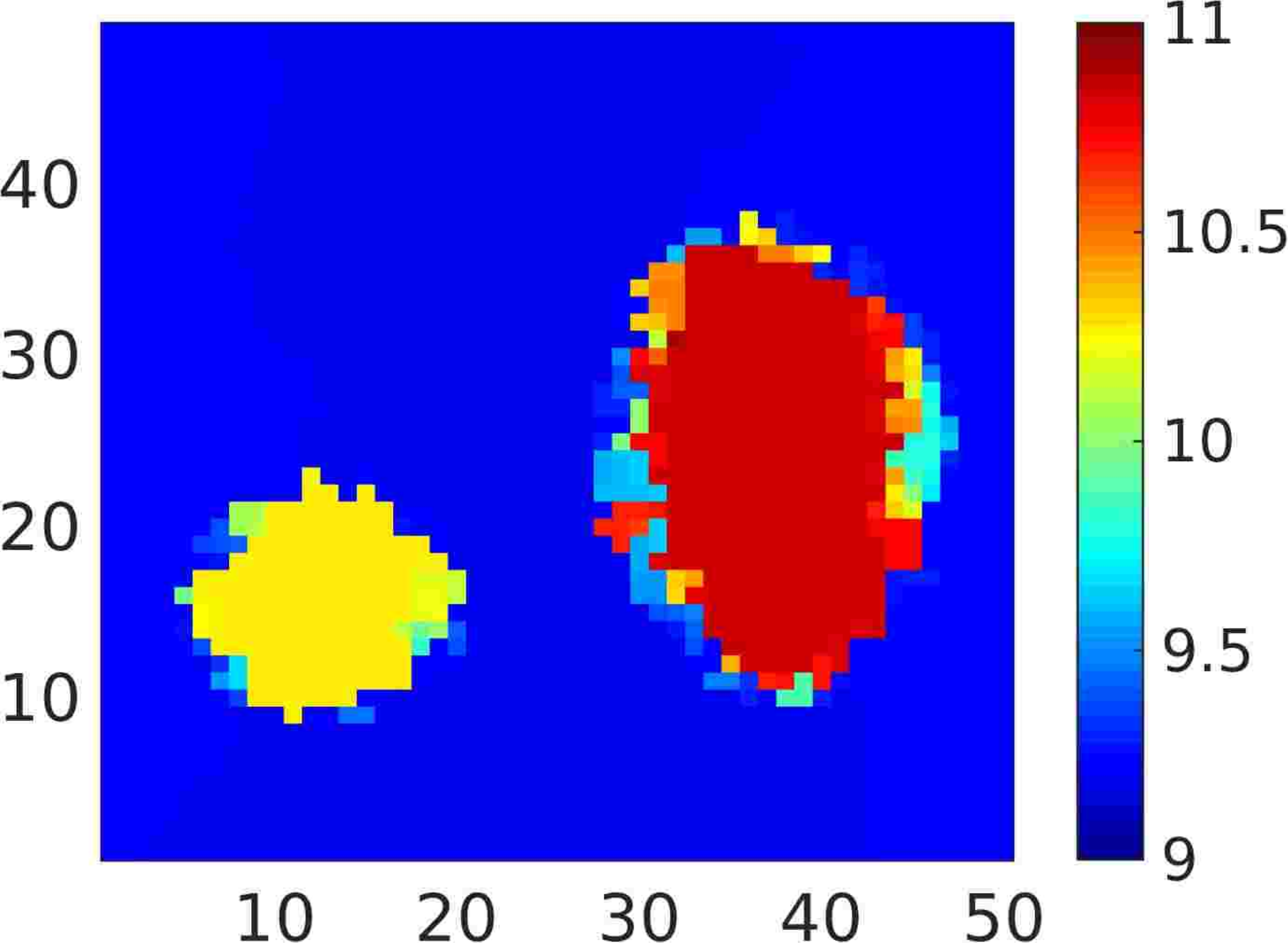}} 
		\hspace{-2.5cm}		
		
	  }
	  \caption{ Example 2b: Posterior mean $\bmu_j$ and posterior probabilities $q(s=j)$ of each of the $S=21$ mixture components identified (in log scale). }
	 \label{fig:PosteriorMeanMixturesNoDimRed}
\end{figure}

We consider again  the same problem (i.e. the same material properties and forward model) but instead contaminate the data with much more noise resulting in  a SNR of $5 \times 10^{2}$ (in contrast to $1 \times 10^{3}$  previously) and assume that only half of the displacements are available i.e.  $d_y = 2550$ (in contrast to $d_y=5100$ before). 
The proposed algorithm was employed and identified  $21$ active mixture components (in contrast to the $7$ before). The means $\bs{\mu}_j$ of all these components are depicted in   \reffig{fig:PosteriorMeanMixturesNoDimRed}) where  the posterior probabilities $q(s)$ are also  reported. As expected, the presence of more noise and the reduction in the available data have lead to more modes in the posterior. 
%

\begin{figure}[H]{
	\centering
	\captionsetup[subfigure]{labelformat=empty}
		\subfloat[][{$q(s=6) =0.104$}] 
		{\includegraphics[width=0.31\textwidth]{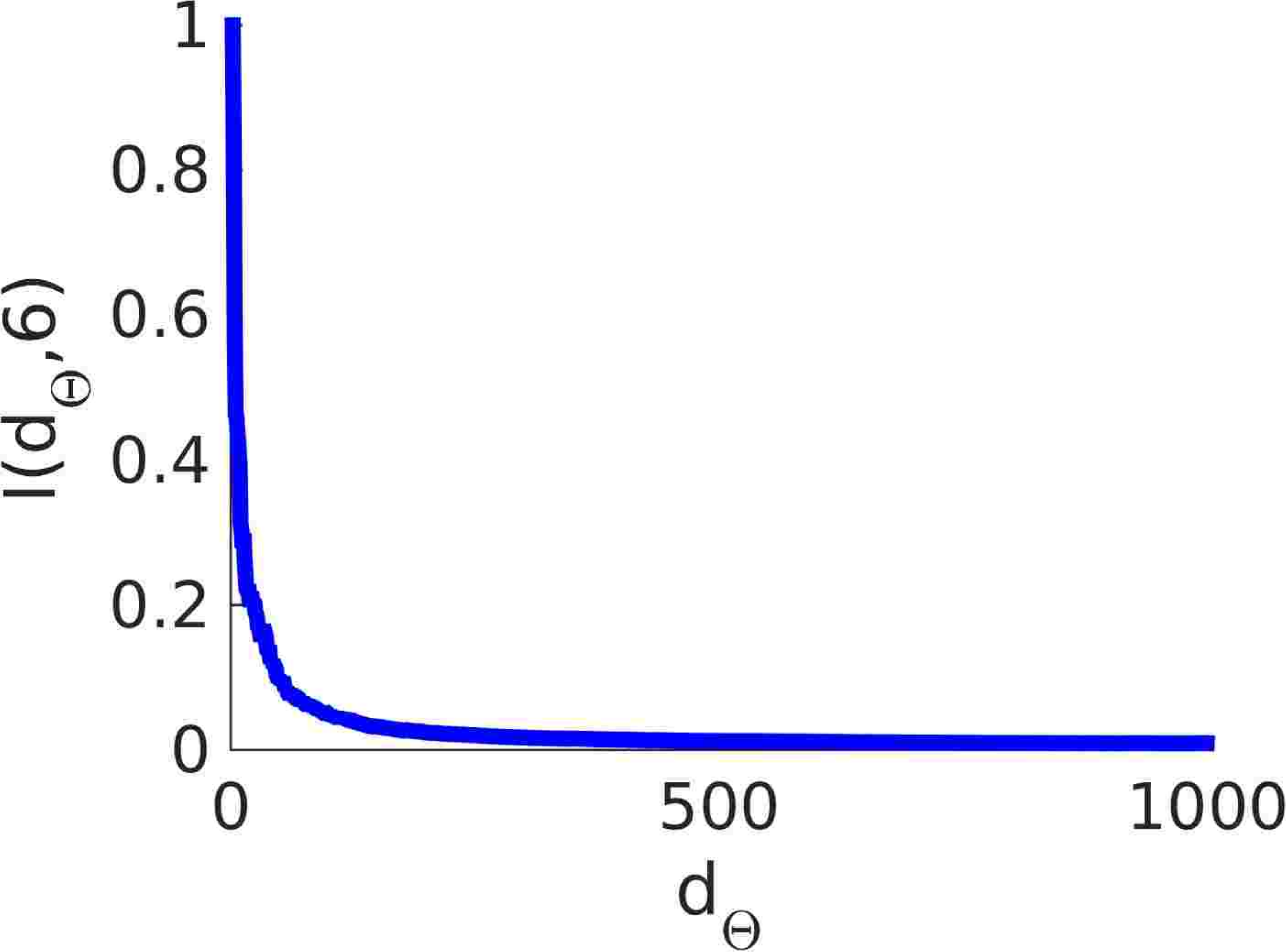}} 
		\hspace{0.1cm}
		\subfloat[][{$q(s=11) =0.107$}] 
		{\includegraphics[width=0.31\textwidth]{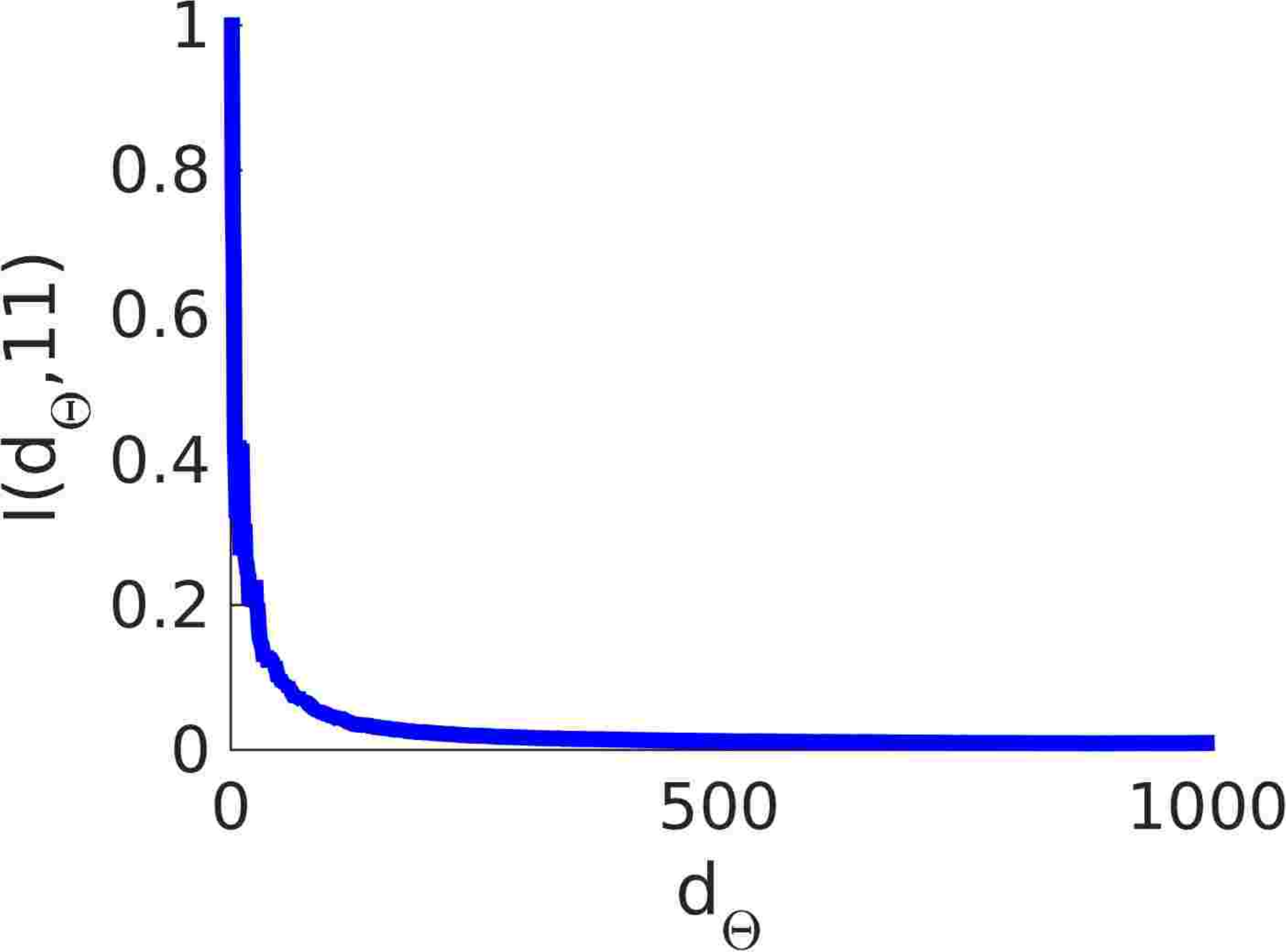}} 
		\hspace{0.1cm}
		\subfloat[][{$q(s=17) = 0.222$}] 
		{\includegraphics[width=0.31\textwidth]{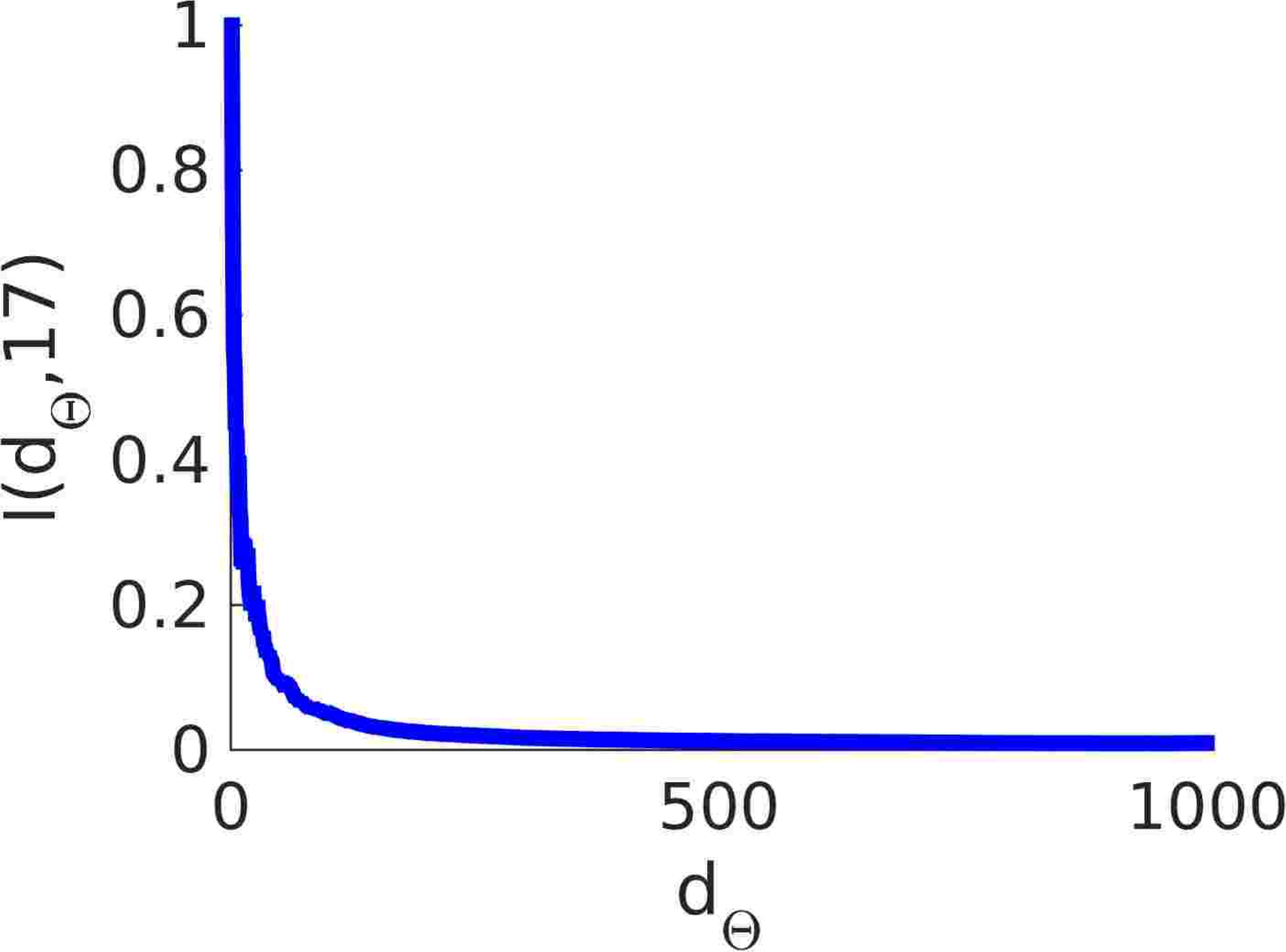}} 
		\hspace{0.3cm}
	  }
	  \caption{Information gain $I(\dth,j)$ for the $3$ (out of the $21$) mixture components with the largest posterior probability $q(s)$. 
	  }
	 \label{fig:IGNoDimRed}
\end{figure}

Moreover, as one would expect, none of these modes is particularly amenable to dimensionality reduction as the posterior variance is large and distributed along multiple dimensions. In fact by employing the information gain metric (\reffig{fig:IGNoDimRed})  we found that for most modes at least $\dth\approx 750$ reduced coordinates were  necessary to represent the variance accurately.

\begin{figure}[H]{
	\centering
	\captionsetup[subfigure]{labelformat=empty}
		\subfloat[][{Posterior mean ($S=21$)}] 
		{\includegraphics[width=0.48\textwidth]{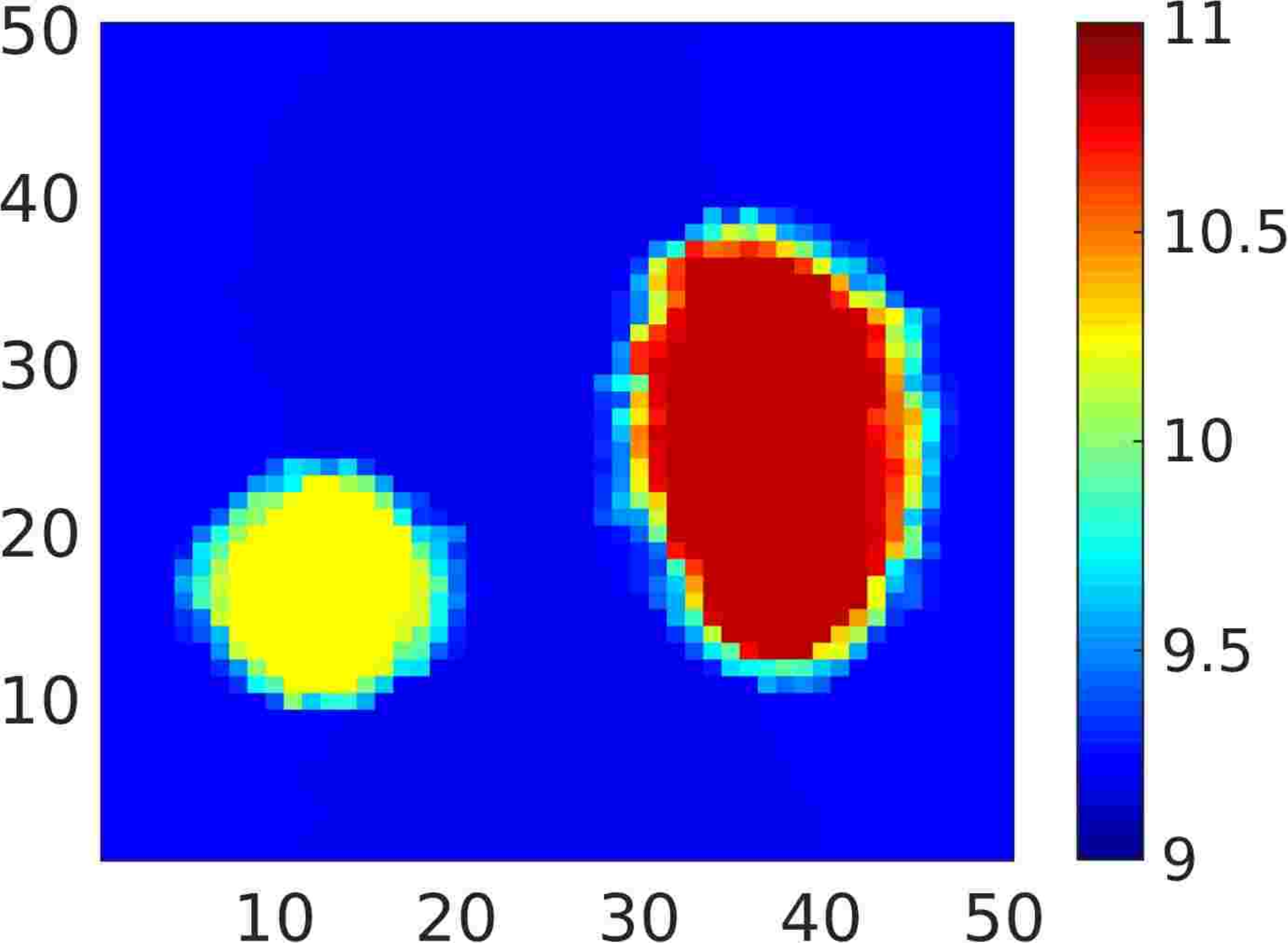}} 
		\hspace{0.1cm}
		\subfloat[][{Posterior standard deviation ($S=21$)}] 
		{\includegraphics[width=0.48\textwidth]{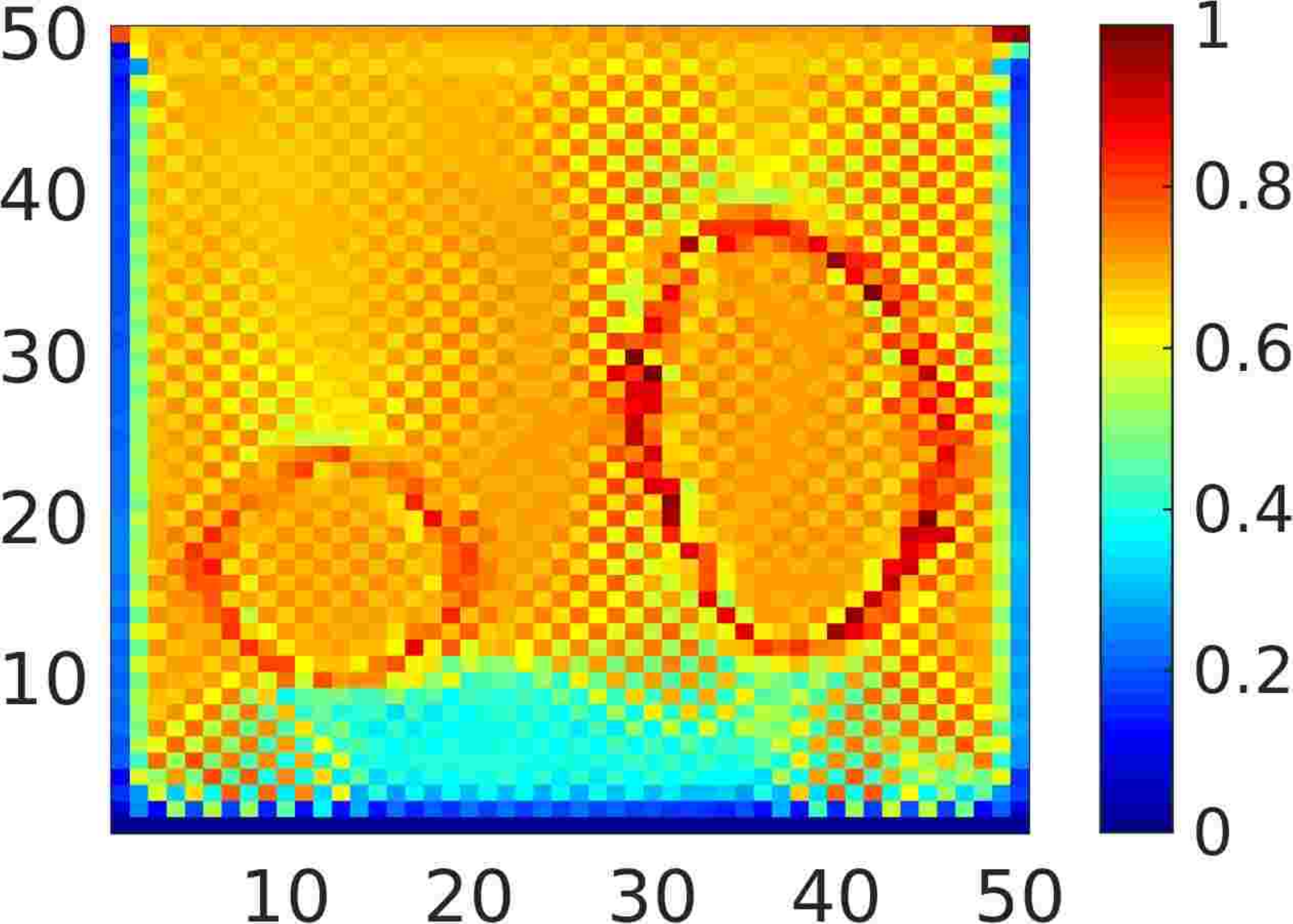}} 
	  }
	  \caption{
	  Example 2b: On the left panel, the posterior mean of the material parameters is plotted and in the right panel the posterior standard deviation (in log scale).}
	 \label{fig:PosteriorMeanStdSNR5e2}
\end{figure}

Nevertheless, the posterior mean estimated from the mixture of these $21$ Gaussians (\refeq{eq:qpost1}) is very close to the ground truth, see \reffig{fig:PosteriorMeanStdSNR5e2}. Understandably however, the posterior variance across the problem domain (\reffig{fig:PosteriorMeanStdSNR5e2}) is much larger.
%

\section{Conclusions}
We presented a novel Variational Bayesian framework for the solution  of high-dimensional inverse problems with computationally-demanding forward models and high-dimensional vector of unknowns. The strategy advocated addresses two fundamental challenges in the context of such problems. Firstly, the poor performance of existing inference tools in high-dimensions by identifying lower dimensional subspaces where most of the posterior variance  is concentrated. Secondly, it is capable of capturing multimodal posteriors by making use of a mixture of multivariate Gaussians. Each of the Gaussians  is associated with a different mean and covariance and provides an accurate  local approximation. We validated the proposed strategy with Importance Sampling and as demonstrated in the numerical examples, the bias introduced by the approximations is small and can be very efficiently corrected (i.e. with very few forward model calls).

In the context of the motivating application, i.e. static, nonlinear elastography, it was shown that the multimodal approximation computed can provide a more accurate picture to the analyst or medical practitioner   from which better diagnostic decisions can be drawn. In particular,
 the model proposed can better capture the spatial heterogeneity of material parameters which is a strong indicator of malignancy in tumors \cite{liu_noninvasive_2015}. This is especially  manifested in the boundaries of the inclusions (tumors) which can be better classified as well as in quantifying their effect in the results.  
 
%
%

The method advocated is applicable to other problems characterized by high-dimensional vectors of unknowns such as those involving spatially varying model parameters. It does not make use of any 
 particular features of the forward model solver and requires the computation of first-order derivatives which can be handled with an adjoint formulation. While the number  of forward model solutions, which is the primary metric of computational efficiency in our setting, increases with the number of identified posterior modes and depends on the stopping criteria employed,  we have demonstrated that a few hundred forward calls are usually enough in the applications of interest. 
 Furthermore, our algorithm is readily able to handle unimodal posteriors as well as densities which are not amenable to dimensionality reductions (e.g. due to large noise or sparse data).
 
 We finally note that a restriction of the uncertainty quantification strategy proposed pertains to the forward model itself. Another source of uncertainty, which is largely  unaccounted for, is {\em model  uncertainty}. Namely, the parameters which are calibrated, are associated with a particular forward model (in our case a system of (discretized) PDEs) but one cannot be certain about the validity of the model employed. In general, there will be deviations between the physical reality where measurements are made, and the idealized mathematical/computational description. A critical extension therefore, particularly in the context of biomedical applications, would be in the direction of identifying sources of model error and being able to quantify them in the final results.

 \appendix

\section{Variational lower bound}
\label{app:Lowerbound}
The lower bound from \refeq{eq:fvarhatElaborated} combined with the optimal probability distributions $q^{opt}$, \refeq{eq:qopt1}, is:
\be
\begin{array}{llr}
\mathcal{\hat{F}}(q^{opt}(\bt, \bet, \tau, s), \bs{T}) 
	  & = \tcg { \frac{d_y}{2}< \log \tau>_{\tau}  }   &  (\textrm{\em \footnotesize  $<\log p(\bs{\hat{y}} | s, \bt, \bet, \tau, \bs{T})>$})\\
	  & \tcbl{ - \frac{<\tau >_{\tau}}{2} \sum_s q^{opt}(s) || \hat{\bs{y}}- \bs{y}(\bs{\mu}_s)||^2  			}	& \\
	  & \tcb{ - \frac{<\tau >_{\tau}}{2} \sum_s q^{opt}(s) \bs{W}_s^T \bs{G}_s^T \bs{G}_s \bs{W}_s: <\bt \bt^T >_{\bt|s} 	}	&  \\
	  & \tcbr{ - \frac{<\tau >_{\tau}}{2} \sum_s q^{opt}(s)\bs{G}_s^T \bs{G}_s : <\bet\bet^T>_{\bet|s} }	 & \\
	  & \tcbl{ + \sum_s q^{opt}(s) \log \frac{1}{S}  }						&  (\textrm{\em \footnotesize  $<\log p_s(s)>$}) \\
 	  & \tcg {   + (a_0-1)<\log \tau>_{\tau} }   \tcr{-b_0 <\tau>_{\tau}}  				& (\textrm{\em \footnotesize  $<\log p_{\tau}(\tau)>$}) \\
 	  & + \sum_s q^{opt}(s) ( \tcbl{ \frac{1}{2} \log | \bs{\Lambda}_{0,s}|}    \tcb{-\frac{1}{2}  \bs{\Lambda}_0: <\bt \bt^T>_{\bt|s}) }& (\textrm{\em \footnotesize  $<\log p_{\Theta}(\bt|s)>$}) \\
 	  &  + \sum_s q^{opt}(s) ( \tcbl{ \frac{\dpsi}{2} \log \lambda_{0,\eta,s}}  \tcbr{ -\frac{\lambda_{0,\eta,s}}{2} \bs{I}:<\bet \bet^T>_{\bet|s}) }  & (\textrm{\em \footnotesize  $<\log p_{\eta}(\bet)>$}) \\
 	  & \tcbl{- \sum_s q^{opt}(s) \frac{1}{2}  | \bs{\Lambda}_s| } 					& (\textrm{\em \footnotesize  $-<\log q^{opt}(\bt| s)>$}) \\
 	  & \tcbl{- \sum_s q^{opt}(s) \frac{\dpsi}{2} \log \lambda_{\eta,s} } 				& (\textrm{\em \footnotesize  $-<\log q^{opt}(\bet| s)>$}) \\
 	  & \tcbl{- \sum_s q^{opt}(s) \log q^{opt}(s) 	}						& (\textrm{\em \footnotesize  $-<\log q^{opt}(s)>$})  \\
 	  & \tcg{ - (a-1)<\log \tau >_{\tau}  }  \tcr{+ b<\tau >_{\tau} + \log Z(a,b)}			& (\textrm{\em \footnotesize  $-<\log q^{opt}(\tau)>$}) 	 
\end{array}
\label{eq:fvarhatWithQ}
\ee
where $Z(a,b)=\frac{\Gamma(a)}{b^{a}}$ is the normalization constant of a $Gamma$ distribution with parameters $a,b$. 

Certain terms become constants and can be neglected. By reformulating, we can derive (see also \refeq{eq:taua}, \refeq{eq:thetaupd}, \refeq{eq:etaupd}): 
\be
  \tcg { ( (a_0-1) +  \frac{d_y}{2} - (a-1) )<\log \tau >_{\tau} }  = (a_0 + \frac{d_y}{2} - a)<\log \tau >_{\tau} = 0
\ee
\be
  \begin{split}
    &\tcb{- \frac{1}{2} \sum_s q^{opt}(s)  (   (  <\tau >_{\tau}  \bs{W}_s^T \bs{G}_s^T \bs{G}_s \bs{W}_s   +  \bs{\Lambda}_0)   : <\bt \bt^T >_{\bt|s}) } \\
    &= -\frac{1}{2} \sum_s q^{opt}(s)  \bs{\Lambda}_s : \bs{\Lambda}_s^{-1} \\
    &= -\frac{\dpsi}{2}
 \end{split} 
\ee
\be
  \begin{split}
    &\tcbr{ - \frac{1}{2} \sum_s q^{opt}(s)  ( <\tau >_{\tau}\bs{G}_s^T \bs{G}_s :<\bet\bet^T>_{\bet|s} + \lambda_{0,\eta,s}  \bs{I}:<\bet \bet^T>_{\bet|s} )} \\
    &=    - \frac{1}{2} \sum_s q^{opt}(s) \lambda_{\eta,s} \lambda_{\eta,s}^{-1} \dpsi \\
    &= -\frac{\dpsi}{2}.
   \end{split}
\ee
\be
   \begin{split} 
     &\tcr{-b_0 <\tau>_{\tau}+ b<\tau >_{\tau} + \log Z(a,b)}  	\\
     &= -b_0 <\tau>_{\tau} + b  \frac{a}{b}  + \log(\frac{\Gamma(a)}{b^{a}}) \\
     &= -b_0 <\tau>_{\tau} + a  + \log(\Gamma(a)) - a \log(b)     \\
     &= -b_0 <\tau>_{\tau} + a  + \log(\Gamma(a)) - a \log(\frac{a}{<\tau>_{\tau}}) \\
     &= -b_0 <\tau>_{\tau} + a  + \log(\Gamma(a)) - a \log(a)  + a \log(<\tau>_{\tau})\\
     &\propto a \log(<\tau>_{\tau})
   \end{split}
\ee
as $a$ from \refeq{eq:taua} is constant, $b_0 = 0$ and $<\tau >_{\tau}=\frac{a}{b}$.

Therefore \refeq{eq:fvarhatWithQ} becomes (neglecting constant terms and including \refeq{eq:taub}):
\be
      \begin{split}
      \mathcal{\hat{F}}(q^{opt}(\bt, \bet, \tau, s), \bs{T}) =&  \sum_s q^{opt}(s) [ - \frac{<\tau >_{\tau}}{2}|| \hat{\bs{y}}- \bs{y}(\bs{\mu}_s)||^2  \\
      &+   \frac{1}{2} \log\frac{| \bs{\Lambda}_{0,s}|}{| \bs{\Lambda}_s|}  +  \frac{\dpsi}{2} \log \frac{\lambda_{0,\eta,s}}{\lambda_{\eta,s}}  -  \log q^{opt}(s) ]\\	
      &+ a \log(<\tau>_{\tau}).
      \end{split}
\ee  

\section{Hierarchical prior specification and Maximization for  $\bmu$}
\label{app:muPrior}

      If $d_L$ is the total number of neighboring pairs and $k_m$ and
     $l_m$ the entries of $\bs{\mu}_j$ forming the neighboring pair $m$, then we define a prior:
     \be
	p( \mu_{j,k_m}-\mu_{j,l_m}| \phi_{j,m})= \sqrt{ \frac{\phi_{j,m}}{2\pi}} e^{ -\frac{\phi_{j,m}}{2} (\mu_{j,k_m}-\mu_{j,l_m})^2}.
	\label{eq:priormuj}
     \ee
     The hyperparameter $\phi_{j,m}>0$ controls the strength of the penalty i.e. small values of 
     $\phi_{j,m}$ induce a weaker penalty and vice versa \cite{bardsley_gaussian_2013}. In summary, by aggregating all neighboring pairs we obtain an improper prior of the form:
     \be
	p(\bs{\mu}_j | \bs{\Phi}_j) \propto |\bs{\Phi}_j|^{1/2} e^{-\frac{1}{2} \bs{\mu}_j^T \bs{L}^T \bs{\Phi}_j \bs{L} \bs{\mu}_j} 
	\label{eq:priormuj2}
     \ee
     where $\bs{L}$ the $d_L \times d_{\Psi}$ denotes the Boolean matrix that gives rise to  the vector of all $d_L$ jumps (such as the one in \refeq{eq:priormuj}) when multiplied with $\bs{\mu}_j$, and $\bs{\Phi}_j=diag(\phi_{j,m})$ the {\em diagonal $d_L \times d_L$ matrix} containing all the hyperparameters $\phi_{j,m}$ associated with each of these jumps. 
     
     We use a conjugate hyperprior for $\bphi_j$, a product of Gamma distributions:
     \be
	p(\bphi_j) = \prod_{m=1}^{d_L} Gamma(a_{\phi},b_{\phi}).
	\label{eq:priorphi}
     \ee
     The independence of the $\bphi_j$ in the prior is motivated by the absence of correlation (a priori) with respect to the locations of the jumps. We use $a_{\phi} = b_{\phi} =0$ which results in an Automatic Relevance Determination (ARD, refs) hyperprior.

%
%
%

Due to the analytical unavailability of $\log p(\bs{\mu}_j)$ and its derivatives $\frac{ \pa \log p(\bs{\mu}_j)}{\pa \bs{\mu}_j}$, we employ an Expectation-Maximization scheme which we describe briefly here for completeness \cite{dempster_maximum_1977,neal_view_1998}. Proceeding as in \refeq{eq:loglike1} i.e. by making use of Jensen's inequality and an arbitrary distribution $q(\bs{\Phi}_j)$ we can bound $\log p(\bs{\mu}_j)$ as follows:
 \be
 \begin{array}{ll}
   \log p(\bs{\mu}_j) &= \log \int p(\bs{\mu}_j|\bs{\Phi}_j) p(\bs{\Phi}_j)~d\bs{\Phi}_j \\
		      & \log \int \frac{ p(\bs{\mu}_j|\bs{\Phi}_j) p(\bs{\Phi}_j)}{q(\bs{\Phi}_j)}  q(\bs{\Phi}_j)~d\bs{\Phi}_j \\
		      & \ge \int q(\bs{\Phi}_j) \log \frac{  p(\bs{\mu}_j|\bs{\Phi}_j) p(\bs{\Phi}_j)}{q(\bs{\Phi}_j)}~d\bs{\Phi}_j  \\
		      & = E_{q(\bs{\Phi}_j)}[\log p(\bs{\mu}_j|\bs{\Phi}_j)]+E_{q(\bs{\Phi}_j)}[ \log \frac{ p(\bs{\Phi}_j)}{q(\bs{\Phi}_j)}  ].
 \end{array}
 \label{eq:emphi}
\ee
This inequality becomes an equality only when $q(\bs{\Phi}_j) \equiv p(\bs{\Phi}_j | \bs{\mu}_j )$ i.e. it is the actual posterior on $\bs{\Phi}_j$ given $\bs{\mu}_j$.
The latter can be readily established from Equations (\ref{eq:priormuj}) and (\ref{eq:priorphi}), from which it follows that $p(\bs{\Phi}_j | \bs{\mu}_j )=\prod_{m=1}^{d_L} Gamma(a_{\phi_{j,m}},b_{\phi_{j,m}})$ with:
\be
      a_{\phi_{j,m}}=a_{\phi}+\frac{1}{2}, \quad b_{\phi_{j,m}}=b_{\phi}+\frac{1}{2}  (\mu_{j,k_m}-\mu_{j,l_m})^2.
      \label{eq:phiupd}
\ee
This suggests a two-step procedure for computing $\log p(\bs{\mu}_j )$ and  $\frac{ \pa \log p(\bs{\mu}_j )}{\pa \bs{\mu}_j }$ for each $\bs{\mu}_j $:
\bi
\item[(E-step)] Find  $p(\bs{\Phi}_j  | \bs{\mu}_j  )=\prod_{m=1}^{d_L} Gamma(a_{\phi_{j,m}},b_{\phi_{j,m}})$ from \refeq{eq:phiupd}
\item[(M-step)] Find  $\log p(\bs{\mu}_j )$ and  $\frac{ \pa \log p(\bs{\mu}_j )}{\pa \bs{\mu}_j }$  from \refeq{eq:emphi} for $q(\bs{\Phi}_j) \equiv p(\bs{\Phi}_j | \bs{\mu}_j )$ as follows:
\be
    \begin{array}{ll}
    \log p(\bs{\mu}_j) 					& = E_{q(\bs{\Phi}_j)}[\log p(\bs{\mu}_j|\bs{\Phi}_j)]= -\frac{1}{2} \bs{\mu}_j^T \bs{L}^T <\bs{\Phi}_j> \bs{L} \bs{\mu}_j  \\
    \frac{ \pa \log p(\bs{\mu}_j)}{\pa \bs{\mu}_j} 	& = \frac{ \pa }{\pa \bs{\mu}_j}E_{q(\bs{\Phi}_j)}[\log p(\bs{\mu}_j|\bs{\Phi}_j)] \\
							& = E_{q(\bs{\Phi}_j)}[\frac{\pa }{\pa \bs{\mu}_j} \log p(\bs{\mu}_j|\bs{\Phi}_j)] \\
							& = - \bs{L}^T <\bs{\Phi}_j> \bs{L} \bs{\mu}_j
    \end{array}
    \label{eq:pmu}
\ee
where $<\bs{\Phi}_j>=E_{q(\bs{\Phi}_j)}[diag(\phi_{j,m})]=diag( \frac{a_{\phi_{j,m}}}{b_{\phi_{j,m}}})$.
\ei

The determination of the derivatives of $\frac{\pa \mathcal{F}_{\mu_j}}{\pa \bmu_j}$ requires also $\bs{G}_j=\frac{\pa \bs{y} }{\pa \bs{\Psi}_j }|_{\bpsi_j=\bs{\mu}_j}$, which depends on $\bs{\mu}_j$. To avoid second-order derivatives and their high computational costs we linearize \refeq{eq:muupd} and assume that $\bs{G}_j$ remains constant in the vicinity of the current guess.\footnote{We only linearize \refeq{eq:muupd} \cite{franck_sparse_2015} for the purpose of updating the $\bmu_j$.} In particular, we denote $\bs{\mu}_j^{(n)}$  the value of $\bs{\mu}_j$ at iteration $n$. Then to find the increment $\Delta \bs{\mu}_j^{(n)}$, we specify the new objective $\mathcal{F}_{\mu_j}^{(n)}(\Delta \bs{\mu}_j^{(n)})$ as follows:
\be
    \begin{split}
    {F}_{\mu_j}^{(n)}(\Delta \bs{\mu}_j^{(n)}) = & {F}_{\mu_j}(\bs{\mu}_j^{(n)}+\Delta \bs{\mu}_j^{(n)})+\log p(\bs{\mu}_j^{(n)}+\Delta \bs{\mu}_j^{(n)}) \\
					       = &-\frac{<\tau>}{2} |\bs{\hat{y}} - \bs{y}(\bs{\mu}_j^{(n)}+\Delta \bs{\mu}_j^{(n)})|^2   \\
					      & -\frac{1}{2} (\bs{\mu}_j^{(n)}+\Delta \bs{\mu}_j^{(n)})^T \bs{L}^T <\bs{\Phi}_j> \bs{L} (\bs{\mu}_j^{(n)}+\Delta \bs{\mu}_j^{(n)})\\
					      \approx & -\frac{<\tau>}{2} |\bs{\hat{y}} - \bs{y}(\bs{\mu}_j^{(n)})-\bs{G}_j^{(n)} \Delta \bs{\mu}_j^{(n)}|^2 \\
					      & -\frac{1}{2} (\bs{\mu}_j^{(n)}+\Delta \bs{\mu}_j^{(n)})^T \bs{L}^T <\bs{\Phi}_j> \bs{L} (\bs{\mu}_j^{(n)}+\Delta \bs{\mu}_j^{(n)}).\\
    \end{split}
\ee
We note that there is no approximation of the $p(\bs{\mu}_j)$ prior term.
By keeping only the terms depending on $\Delta \bs{\mu}_j^{(n)}$ in the Equation above we obtain:
\be
    \begin{array}{ll}
    {F}_{\mu_j}^{(n)}(\Delta \bs{\mu}_j^{(n)}) = & -\frac{<\tau>}{2} (\Delta \bs{\mu}_j^{(n)} )^T (\bs{G}_j^{(n)} )^T \bs{G}_j^{(n)}~\Delta \bs{\mu}_j^{(n)}  +<\tau> (\bs{\hat{y}} - \bs{y}(\bs{\mu}_j^{(n)}))^T \bs{G}_j^{(n)}~\Delta \bs{\mu}_j^{(n)} \\
    & - \frac{1}{2}(\Delta \bs{\mu}_j^{(n)})^T \bs{L}^T <\bs{\Phi}_j> \bs{L} ~\Delta \bs{\mu}_j^{(n)} \\
    & - (\bs{\mu_j}^{(n)})^T\bs{L}^T <\bs{\Phi}_j> \bs{L} ~\Delta \bs{\mu}_j^{(n)}.
    \end{array}
\ee
This is concave and quadratic with respect to the unknown $\Delta \bs{\mu}_j^{(n)}$. The maximum can be found by setting $ \frac{\pa {F}_{\mu_j}^{(n)}(\Delta \bs{\mu}_j^{(n)}) }{\pa \Delta \bs{\mu}_j^{(n)}}=\bs{0}$ which yields:
\be
    (<\tau>  (\bs{G}_j^{(n)} )^T \bs{G}_j^{(n)}+\bs{L}^T <\bs{\Phi}_j> \bs{L}) \Delta \bs{\mu}_j^{(n)} = <\tau>  (\bs{\hat{y}} - \bs{y}(\bs{\mu}_j^{(n)}))^T \bs{G}_j^{(n)}  -\bs{L}^T <\bs{\Phi}_j> \bs{L} \bs{\mu}_j^{(n)}.
    \label{eq:muupd2}
\ee

\section{Determination of required number of basis vectors - Adaptive learning}
\label{app:IGBases}
An important  question is how many basis vectors in $\bs{W}_{j}\in \RR^{\dpsi\times\dth}$ should be considered for a mixture component $j$. We use an information-theoretic  criterion \cite{franck_sparse_2015} that measures the information gain of the approximated posterior to the prior beliefs. Specifically, if $p_{\dth}(\bt|s)$ (\refeq{eq:priortheta}) and $q_{\dth}(\bt|s)$ (\refeq{eq:qopt1}) denote the $\dth-$dimensional prior and posterior for a given $s=j$, we define the quantity $I(\dth,s)$ as follows: 
\be
 I(\dth,s) = \frac{KL(p_{\dth}(\bt|s)|| q_{\dth}(\bs{\Theta}|s)) - KL(p_{{\dth}-1}(\bt |s)|| q_{{\dth}-1}(\bs{\Theta}|s))}   {KL(p_{\dth}(\bt|s)|| q_{\dth}(\bs{\Theta}|s))}
 \label{eq:IGain}
\ee
which measures the (relative) information gain from $\dth-1$ to $\dth$ reduced coordinates. When the information gain falls below a threshold $I_{max}$, we assume that the information gain is marginal and the addition of reduced coordinates can be terminated. For all mixture components we consider the same $\dth$, chosen from the mixture component that requires the largest $\dth$. Therefore $\dth$ is determined when the information gain with respect to all mixture components falls below the threshold $I_{max}$, (in our examples we use $I_{max} = 1\%$):
\be
    max(I(\dth,s=1),I(\dth,s=2), ..., I(\dth,s=S))  ~{\leq} ~I_{max}.
\ee

The KL divergence between the two Gaussians,  $p_{\dth}(\bt|s)=\mathcal{N}(\bs{0},\bs{\Lambda}_{0,s}^{-1})$ and $q_{\dth}(\bt|s)=\mathcal{N}(\bs{0},\bs{\Lambda}_{s}^{-1})$, where $\bs{\Lambda}_{0,s}^{-1}$ and $\bs{\Lambda}_{s}^{-1}$ are diagonal, as described in \refsec{sec:Variational}, \refeq{eq:thetaupd2}, is given by:
\be
  KL(p_{\dth}(\bs{\Theta}|s)|| q_{\dth}(\bs{\Theta}|s)) = \frac{1}{2} \sum_{i=1}^{\dth} (-\log(\frac{\lambda_{s,i}}{\lambda_{0,s,i}})+\frac{\lambda_{s,i}}{\lambda_{0,s,i}} -1)
\ee
and (\refeq{eq:IGain}) becomes:
\be
I(\dth,s) = \frac{\sum_{i=1}^{\dth} (-\log(\frac{\lambda_{s,i}}{\lambda_{0,s,i}})+\frac{\lambda_{s,i}}{\lambda_{0,s,i}}-1)   -  \sum_{i=1}^{\dth-1} (-\log(\frac{\lambda_{s,i}}{\lambda_{0,s,i}})+\frac{\lambda_{s,i}}{\lambda_{0,s,i}} -1) }   {\sum_{i=1}^{\dth} (-\log(\frac{\lambda_{s,i}}{\lambda_{0,s,i}})+\frac{\lambda_{s,i}}{\lambda_{0,s,i}}-1)}.
\ee
Naturally, one could consider different values of $\dth$ for each mixture component   which could lead to additional savings.


\newpage
  \bibliographystyle{elsarticle-num}
  \bibliography{short}


%
%
%
%
\end{document}